\documentclass[11pt,a4paper]{article}
\usepackage{maple2e}
\usepackage[english]{babel}
 \def\emptyline{\vspace{12pt}}
\DefineParaStyle{Heading 1}
\DefineParaStyle{Heading 2}
\DefineParaStyle{Heading 3}
\DefineParaStyle{Heading 4}
\DefineParaStyle{Maple Output}
\DefineParaStyle{Maple Output}
\DefineParaStyle{Maple Plot}
\DefineParaStyle{Normal}
\DefineParaStyle{Warning}
\DefineCharStyle{2D Comment}
\DefineCharStyle{2D Math}
\DefineCharStyle{2D Output}
\DefineCharStyle{Hyperlink}

\begin{document}
\pagestyle{plain}

\title{Introduction to relativistic astrophysics and cosmology
through Maple}

\author{Vladimir L. Kalashnikov}

\maketitle

\begin{center}
\textit{Belarussian Polytechnical Academy}
\end{center}

\begin{center}
\docLink{mailto:kalashnikov\verb+_+vl@mailru.com}{kalashnikov\_vl@mailru.com}
\end{center}

\begin{center}
\docLink{http://www.geocities.com/optomaplev}{www.geocities.com/optomaplev}
\end{center}

\begin{abstract} 
The basics of the relativistic astrophysics
including the celestial mechanics in weak field, black holes and
cosmological models are illustrated and analyzed by means of Maple 6
\end{abstract}

\noindent
\textit{Application Areas/Subjects:} Science, Astrophysics, General
Relativity, Tensor Analysis, Differential geometry, Differential
equations

\clearpage

\tableofcontents

\newpage

\section{Introduction}

\emptyline
A rapid progress of the observational astrophysics, which resulted
from the active use of orbital telescopes, essentially intensifies the
astrophysical researches at the last decade and allows to choose the
more definite directions of further investigations. At the same time,
the development of high-performance computers advances in the
numerical astrophysics and cosmology. Against a background of these
achievements, there is the renascence of analytical and
semi-analytical approaches, which is induced by new generation of
high-efficient computer algebra systems.

\emptyline
\noindent
Here we present the pedagogical introduction to relativistic
astrophysics and cosmology, which is based on computational and
graphical resources of Maple 6. The pedagogical aims define the use
only standard functions despite the fact that there are the powerful
General Relativity (GR) oriented extensions like 
\textit{GRTensor} \cite{GRT}. The knowledge of basics of GR and differential geometry is
supposed. It should be noted, that our choice of metric signature (+2)
governs the definitions of Lagrangians and energy-momentum tensors. 

\emptyline
\noindent
The computations in this worksheet take about of 6 min of CPU time (PIII-500) and 9 Mb of memory.

\section{Relativistic celestial mechanics in weak gravitational
field}

\emptyline

\subsection{Introduction}

\emptyline
The first results in the GR-theory were obtained without exact
knowledge of the field equations (Einstein's equations for
space-time geometry). The leading idea was the equivalence principle
based on the equality of inertial and gravitational masses. We will
demonstrate here that the natural consequences of this principle are
the Schwarzschild metric and the basic experimental effects of
GR-theory in weak gravitational field, i. e. planet's orbit precession
and light ray deflection  (see \cite{Sommerfeld}).

\emptyline

\subsection{Schwarzschild metric}

\emptyline
Let us consider the centrally symmetric gravitational field, which is
produced by mass \textit{M}. The small cell 
${K_{\infty }}$
 falls along \textit{x}-axis on the central mass. In the agreement
with the equivalence principle, the uniformly accelerated motion
locally compensates the gravitational force hence there is no a
gravitational field in the free falling system 
${K_{\infty }}$%
. This results in the locally Lorenzian metric with linear element
(\textit{c} is the velocity of light):

\emptyline
\begin{center}
\textit{d
$s^{2}$
 = d
${x_{\infty }}^{2}$
 + d
${y_{\infty }}^{2}$
 + d
${z_{\infty }}^{2}$
 - 
$c^{2}$
d
${t_{\infty }}^{2}$
}
\end{center}

\emptyline
\noindent
The velocity \textit{v} and radial coordinate \textit{r} are measured
in the spherical system \textit{K}, which are connected with central
mass. It is natural, the observer in this motionless system "feels" the
gravitational field. Since the first system moves relatively second
one there are the following relations between coordinates:

\emptyline
\begin{center}
\textit{d
${x_{\infty }}$
= 
$\frac {\mathit{dr}}{\sqrt{1 - \beta ^{2}}}$
} (
$\beta $
 = 
$\frac {v}{c}$
)
\end{center}

\emptyline
\begin{center}
\textit{d
${t_{\infty }}$
=
$\sqrt{1 - \beta ^{2}}$
dt}
\end{center}

\emptyline
\begin{center}
\textit{d
${y_{\infty }}$
=rd
$\theta $
}
\end{center}

\emptyline
\begin{center}
\textit{d
${z_{\infty }}$
=r
$\mathrm{sin}(\theta )$
d
$\phi $
}
\end{center}

\emptyline
\noindent
The first and second relations are the Lorentzian length shortening
and time slowing down in the moving system. As result, 
${K_{\infty }}$
 from \textit{K} looks as:

\begin{center}
\textit{d
$s^{2}$
 = 
$(1 - \beta ^{2})^{(-1)}$
d
$r^{2}$
 + 
$r^{2}$
}(\textit{d
$\theta ^{2}$
 + 
$\mathrm{sin}(\theta )^{2}$
d
$\phi ^{2}$
}) \textit{- 
$c^{2}$
}(1 \textit{- 
$\beta ^{2}$
})\textit{d
$t^{2}$
}
\end{center}

\emptyline
\noindent
The sense of the additional terms in metric has to connect with the
characteristics of gravitational field. What is the energy of 
${K_{\infty }}$
 in \textit{K}? If the mass of 
${K_{\infty }}$
 is \textit{m}, and 
${m_{0}}$
 is the rest mass, the sum of kinetic and potential energies is:

\emptyline

\begin{mapleinput}
\mapleinline{active}{1d}{restart:
 \indent with(plots):
 \indent \indent (m - m0)*c^2 - G*M*m/r=0;#energy conservation law\\
  (we suppose that the Newtonian law of gravitation\\ 
  is correct in the first approximation), G is\\ 
  the gravitational constant
   \indent \indent \indent \%/(m*c^2):
    \indent \indent subs( m=m0/sqrt(1-beta^2),\% ):# relativistic mass
     \indent expand(\%);
      solve( \%,sqrt(1-beta^2) ):
       \indent sqrt(1-beta^2) = expand(\%);
        \indent \indent 1-beta^2 =
taylor((1-subs(op(2,\%)=alpha/r,rhs(\%)))^2,alpha=0,2);\\
#alpha=G*M/c^2, we use the first-order approximation \\
on alpha}{%
}
\end{mapleinput}

\begin{maplelatex}
\[
(m - \mathit{m0})\,c^{2} - {\displaystyle \frac {G\,M\,m}{r}} =0
\]
\end{maplelatex}

\begin{maplelatex}
\[
1 - \sqrt{1 - \beta ^{2}} - {\displaystyle \frac {G\,M}{c^{2}\,r}
} =0
\]
\end{maplelatex}

\begin{maplelatex}
\[
\sqrt{1 - \beta ^{2}}=1 - {\displaystyle \frac {G\,M}{c^{2}\,r}} 
\]
\end{maplelatex}

\begin{maplelatex}
\[
1 - \beta ^{2}=1 - 2\,{\displaystyle \frac {1}{r}} \,\alpha  + 
\mathrm{O}(\alpha ^{2})
\]
\end{maplelatex}

\emptyline
\noindent
The last results in:

\begin{center}
\textit{d
$s^{2}$
 = 
$\frac {\mathit{dr}^{2}}{1 - \frac {2\,\alpha }{r}}$
 + 
$r^{2}$
}(\textit{d
$\theta ^{2}$
} + 
$\mathrm{sin}(\theta )^{2}$
\textit{d
$\phi ^{2}$
})\textit{ -} 
$c^{2}$
(1 \textit{-} 
$\frac {2\,\alpha }{r}$
)\textit{d
$t^{2}$
}
\end{center}

\emptyline
\noindent
This linear element describes the so-called \underline{Schwarzschild
metric}. In the first-order approximation:

\emptyline
\begin{mapleinput}
\mapleinline{active}{1d}{taylor(
d(r)^2/(1-2*alpha/r)+r^2*(d(theta)^2+sin(theta)^2*\\
d(phi)^2)-c^2*(1-2*alpha/r)*d(t)^2, alpha=0, 2 ):
 \indent convert( \% , polynom ):
  \indent \indent metric := d(s)^2 = collect( \% , \{d(r)^2, d(t)^2\} );}{%
}
\end{mapleinput}

\mapleresult
\maplemultiline{
\mathit{metric} :=
 \mathrm{d}(s)^{2}=\\
 ( - c^{2} + {\displaystyle 
\frac {2\,c^{2}\,\alpha }{r}} )\,\mathrm{d}(t)^{2} + (1 + 
{\displaystyle \frac {2\,\alpha }{r}} )\,\mathrm{d}(r)^{2} + r^{2
}\,(\mathrm{d}(\theta )^{2} + \mathrm{sin}(\theta )^{2}\,\mathrm{
d}(\phi )^{2})
}

\subsection{Equations of motion}

\emptyline
Let us consider a motion of the small unit mass in Newtonian
gravitational potential 
$\Phi $
 = \textit{- 
$\frac {G\,M}{r}$
}. \underline{Lagrangian} describing the motion in centrally
symmetric field is:

\emptyline
\begin{mapleinput}
\mapleinline{active}{1d}{L :=\\
(diff(r(t),t)^2 + r(t)^2*diff(theta(t),t)^2)/2\\
+ G*M/r(t);}{%
}
\end{mapleinput}

\mapleresult
\begin{maplelatex}
\[
L := {\displaystyle \frac {1}{2}} \,({\frac {\partial }{\partial 
t}}\,\mathrm{r}(t))^{2} + {\displaystyle \frac {1}{2}} \,\mathrm{
r}(t)^{2}\,({\frac {\partial }{\partial t}}\,\theta (t))^{2} + 
{\displaystyle \frac {G\,M}{\mathrm{r}(t)}} 
\]
\end{maplelatex}

\emptyline
\noindent
The transition to Schwarzschild metric transforms the time and radial
differentials of coordinates (see, for example, \cite{Kai-Chia Cheng,N.T.
Roseveare}):
 \textit{dr --\TEXTsymbol{>} }(1 \textit{+ 
$\frac {\alpha }{r}$
})\textit{dr} and \textit{dt --\TEXTsymbol{>}} 
$(1 + \frac {\alpha }{r})^{(-1)}$
 \textit{dt }(it should be noted, that the replacement will be
performed for differentials, not coordinates, and we use the weak
field approximation for square-rooting):

\emptyline
\begin{mapleinput}
\mapleinline{active}{1d}{L_n := collect(\\
 expand(\\
  subs(\\
   \{diff(r(t),t)=gamma(r(t))^2*diff(r(t),t),\\          
diff(theta(t),t)=gamma(r(t))*diff(theta(t),t)\}, L ) ),\\
\{diff(r(t),t)^2, diff(theta(t),t)^2\});# modified\\
Lagrangian, gamma(r) = 1+alpha/r(t)}{%
}
\end{mapleinput}

\mapleresult
\begin{maplelatex}
\[
\mathit{L\_n} := {\displaystyle \frac {1}{2}} \,\gamma (\mathrm{r
}(t))^{4}\,({\frac {\partial }{\partial t}}\,\mathrm{r}(t))^{2}
 + {\displaystyle \frac {1}{2}} \,\mathrm{r}(t)^{2}\,\gamma (
\mathrm{r}(t))^{2}\,({\frac {\partial }{\partial t}}\,\theta (t))
^{2} + {\displaystyle \frac {G\,M}{\mathrm{r}(t)}} 
\]
\end{maplelatex}

\emptyline
\noindent
Next step for the obtaining of the equations of motion from Lagrangian
is the calculation of the force 
${\frac {\partial }{\partial x}}\,\mathrm{L}(x, \,y)$
 and the momentum 
${\frac {\partial }{\partial y}}\,\mathrm{L}(x, \,y)$
, where \textit{y}=
${\frac {\partial }{\partial t}}\,x$
:

\emptyline
\begin{mapleinput}
\mapleinline{active}{1d}{e1 := Diff(Lagrangian(r, Diff(r,t)), r) = \\
diff(subs(r(t) = r,L_n),r);#first component of force
 \indent e2 := Diff(Lagrangian(r, Diff(r,t)), Diff(r,t)) = \\
 subs(x=diff(r(t),t), diff(subs(diff(r(t),t)=x, L_n),\\
 x));#first component of momentum
  \indent \indent e3 := Diff(Lagrangian(theta, Diff(theta,t)), theta) =
diff(subs(theta(t) = theta, L_n), theta);#second\\
component of force
   \indent \indent \indent e4 := Diff(Lagrangian(theta, Diff(theta,t)),\\
    Diff(theta,t)) = subs(y=diff(theta(t),t),\\
    diff(subs(diff(theta(t),t)=y, L_n), y));#second\\
      component of momentum}{%
}
\end{mapleinput}

\mapleresult
\begin{maplelatex}
\maplemultiline{
\mathit{e1} := {\frac {\partial }{\partial r}}\,\mathrm{
Lagrangian}(r, \,{\frac {\partial }{\partial t}}\,r)= \\
2\,\gamma (r)^{3}\,({\frac {\partial }{\partial t}}\,r)^{2}\,(
{\frac {\partial }{\partial r}}\,\gamma (r)) + r\,\gamma (r)^{2}
\,({\frac {\partial }{\partial t}}\,\theta (t))^{2} + r^{2}\,
\gamma (r)\,({\frac {\partial }{\partial t}}\,\theta (t))^{2}\,(
{\frac {\partial }{\partial r}}\,\gamma (r)) \\
- {\displaystyle 
\frac {G\,M}{r^{2}}}  }
\end{maplelatex}

\begin{maplelatex}
\[
\mathit{e2} := \mathrm{Diff}(\mathrm{Lagrangian}(r, \,{\frac {
\partial }{\partial t}}\,r), \,{\frac {\partial }{\partial t}}\,r
)=\gamma (\mathrm{r}(t))^{4}\,({\frac {\partial }{\partial t}}\,
\mathrm{r}(t))
\]
\end{maplelatex}

\begin{maplelatex}
\[
\mathit{e3} := {\frac {\partial }{\partial \theta }}\,\mathrm{
Lagrangian}(\theta , \,{\frac {\partial }{\partial t}}\,\theta )=
\mathrm{r}(t)^{2}\,\gamma (\mathrm{r}(t))^{2}\,({\frac {\partial 
}{\partial t}}\,\theta )\,({\frac {\partial ^{2}}{\partial \theta
 \,\partial t}}\,\theta )
\]
\end{maplelatex}

\begin{maplelatex}
\[
\mathit{e4} := \mathrm{Diff}(\mathrm{Lagrangian}(\theta , \,
{\frac {\partial }{\partial t}}\,\theta ), \,{\frac {\partial }{
\partial t}}\,\theta )=\mathrm{r}(t)^{2}\,\gamma (\mathrm{r}(t))
^{2}\,({\frac {\partial }{\partial t}}\,\theta (t))
\]
\end{maplelatex}

\emptyline
\noindent
The equations of motion are the so-called \underline{Euler-Lagrange
equations}\\
\noindent 
${\frac {\partial ^{2}}{\partial t\,\partial y}}\,\mathrm{L}(x, 
\,y)  - 
{\frac {\partial }{\partial x}}\,\mathrm{L}(x, \,y)$
 = 0 (in fact, these equations are the second Newton's law and result from
the law of least action). Now let us write the equations of motion in
angular coordinates. Since \textit{e3 = 0} due to an equality to zero
of the mixed derivative, we have from \textit{e4 }the equation of
motion 
${\frac {\partial ^{2}}{\partial t\,\partial {y_{2}}}}\,\mathrm{L
}(r, \,\theta , \,{y_{1}}, \,{y_{2}})$
\textit{  -  
${\frac {\partial }{\partial \theta }}\,\mathrm{L}(r, \,\theta , 
\,{y_{1}}, \,{y_{2}})$
} = 0 (
${y_{1}}$
= 
${\frac {\partial }{\partial t}}\,r$
, 
${y_{2}}$
=
${\frac {\partial }{\partial t}}\,\theta $
 ) in the form:

\emptyline

\begin{mapleinput}
\mapleinline{active}{1d}{Eu_Lagr_1 := Diff(rhs(e4),t) = 0;}{%
}
\end{mapleinput}

\mapleresult
\begin{maplelatex}
\[
\mathit{Eu\_Lagr\_1} := {\frac {\partial }{\partial t}}\,\mathrm{
r}(t)^{2}\,\gamma (\mathrm{r}(t))^{2}\,({\frac {\partial }{
\partial t}}\,\theta (t))=0
\]
\end{maplelatex}

\emptyline
\noindent
Hence 
\emptyline
\begin{center}
$\mathrm{r}(t)^{2}\,\gamma (\mathrm{r}(t))^{2}\,({\frac {
\partial }{\partial t}}\,\theta (t)) = \textit{H} = \textit{const}$
\end{center}

\emptyline
\begin{mapleinput}
\mapleinline{active}{1d}{sol_1 := Diff(theta(t),t) = solve(\\
gamma^2*diff(theta(t),t)/u(theta)^2 = H,\\
 diff(theta(t),t) );#u=1/r is the new variable}{%
}
\end{mapleinput}

\mapleresult
\begin{maplelatex}
\[
\mathit{sol\_1} := {\frac {\partial }{\partial t}}\,\theta (t)=
{\displaystyle \frac {H\,\mathrm{u}(\theta )^{2}}{\gamma ^{2}}} 
\]
\end{maplelatex}

\emptyline
\noindent
The introduced replacement \textit{u=
$\frac {1}{r}$
}  leads to the next relations:

\emptyline
\begin{mapleinput}
\mapleinline{active}{1d}{Diff(r(t), t) = diff(1/u(t),t);
 \indent Diff(r(t), t) = diff(1/u(theta),theta)*\\
  diff(theta(t),t);# change of variables
  \indent \indent sol_2 := Diff(r(t), t) =\\
    subs(diff(theta(t),t) = rhs(sol_1),\\
rhs(\%));# the result of substitution of\\
  above obtained Euler-Lagrange equation}{%
}
\end{mapleinput}

\mapleresult
\begin{maplelatex}
\[
{\frac {\partial }{\partial t}}\,\mathrm{r}(t)= - {\displaystyle 
\frac {{\frac {\partial }{\partial t}}\,\mathrm{u}(t)}{\mathrm{u}
(t)^{2}}} 
\]
\end{maplelatex}

\begin{maplelatex}
\[
{\frac {\partial }{\partial t}}\,\mathrm{r}(t)= - {\displaystyle 
\frac {({\frac {\partial }{\partial \theta }}\,\mathrm{u}(\theta 
))\,({\frac {\partial }{\partial t}}\,\theta (t))}{\mathrm{u}(
\theta )^{2}}} 
\]
\end{maplelatex}

\begin{maplelatex}
\[
\mathit{sol\_2} := {\frac {\partial }{\partial t}}\,\mathrm{r}(t)
= - {\displaystyle \frac {({\frac {\partial }{\partial \theta }}
\,\mathrm{u}(\theta ))\,H}{\gamma ^{2}}} 
\]
\end{maplelatex}

\emptyline
\noindent
The last result will be used for the manipulations with second
Euler-Lagrange equation 
${\frac {\partial ^{2}}{\partial t\,\partial {y_{1}}}}\,\mathrm{L
}(r, \,\theta , \,{y_{1}}, \,{y_{2}})$
\textit{  -  
${\frac {\partial }{\partial r}}\,\mathrm{L}(r, \,\theta , \,{y_{
1}}, \,{y_{2}})$
} = 0. We have for the right-hand side of \textit{e2}:

\emptyline
\begin{mapleinput}
\mapleinline{active}{1d}{Diff( subs( \{diff(r(t),t)=rhs(\%),gamma(r(t))=gamma\},\\
 rhs(e2)), t);}{%
}
\end{mapleinput}

\mapleresult
\begin{maplelatex}
\[
{\frac {\partial }{\partial t}}\,( - \gamma ^{2}\,({\frac {
\partial }{\partial \theta }}\,\mathrm{u}(\theta ))\,H)
\]
\end{maplelatex}

\emptyline
\noindent
and can rewrite this expression:

\emptyline
\begin{mapleinput}
\mapleinline{active}{1d}{-2*gamma*H*diff(1+alpha/r(t),t)*diff(u(theta),theta) -\\
H*gamma^2*diff(u(theta),theta\$2)*diff(theta(t),t);#from\\
 the previous expression, definition of gamma and\\
  definition of derivative of product
  \indent first_term := subs(\{r(t)=1/u(theta),\\
diff(theta(t),t)=rhs(sol_1)\},\\
subs(diff(r(t),t) = rhs(sol_2),\%));# this is \\
a first term in second Euler-Lagrange equation}{%
}
\end{mapleinput}

\mapleresult
\begin{maplelatex}
\[
2\,{\displaystyle \frac {\gamma \,H\,\alpha \,({\frac {\partial 
}{\partial t}}\,\mathrm{r}(t))\,({\frac {\partial }{\partial 
\theta }}\,\mathrm{u}(\theta ))}{\mathrm{r}(t)^{2}}}  - H\,\gamma
 ^{2}\,({\frac {\partial ^{2}}{\partial \theta ^{2}}}\,\mathrm{u}
(\theta ))\,({\frac {\partial }{\partial t}}\,\theta (t))
\]
\end{maplelatex}

\begin{maplelatex}
\[
\mathit{first\_term} :=  - 2\,{\displaystyle \frac {H^{2}\,\alpha
 \,\mathrm{u}(\theta )^{2}\,({\frac {\partial }{\partial \theta 
}}\,\mathrm{u}(\theta ))^{2}}{\gamma }}  - H^{2}\,({\frac {
\partial ^{2}}{\partial \theta ^{2}}}\,\mathrm{u}(\theta ))\,
\mathrm{u}(\theta )^{2}
\]
\end{maplelatex}

\emptyline
\noindent
\textit{e1} results in:

\emptyline

\begin{mapleinput}
\mapleinline{active}{1d}{2*gamma(r)^3*diff(r(t),t)^2*diff(gamma(r),r) + \\
r*gamma(r)^2*diff(theta(t),t)^2 + \\
r^2*gamma(r)*diff(theta(t),t)^2*diff(gamma(r),r) - \\
G*M/(r^2);# This is e1
\indent subs(\\
\{diff(r(t),t) = rhs(sol_2), diff(gamma(r),r) =\\
 diff(1+alpha/r,r),\\
diff(theta(t),t) = rhs(sol_1)\},\\
\%):# we used the expressions for diff(r(t),t), gamma(r)\\
 and the first equation of motion
\indent \indent subs(gamma(r) = gamma, \%):
\indent \indent \indent second_term := subs(r = 1/u(theta), \%);}{%
}
\end{mapleinput}

\mapleresult
\begin{maplelatex}

\maplemultiline{
2\,\gamma (r)^{3}\,({\frac {\partial }{\partial t}}\,\mathrm{r}(t
))^{2}\,({\frac {\partial }{\partial r}}\,\gamma (r)) + r\,\gamma
 (r)^{2}\,({\frac {\partial }{\partial t}}\,\theta (t))^{2} + r^{
2}\,\gamma (r)\,({\frac {\partial }{\partial t}}\,\theta (t))^{2}
\,({\frac {\partial }{\partial r}}\,\gamma (r))\\
 - {\displaystyle 
\frac {G\,M}{r^{2}}} } 
\end{maplelatex}

\begin{maplelatex}

\maplemultiline{
\mathit{second\_term} :=  \\
- 2\,{\displaystyle \frac {H^{2}\,
\alpha \,\mathrm{u}(\theta )^{2}\,({\frac {\partial }{\partial 
\theta }}\,\mathrm{u}(\theta ))^{2}}{\gamma }}  + {\displaystyle 
\frac {\mathrm{u}(\theta )^{3}\,H^{2}}{\gamma ^{2}}}  - 
{\displaystyle \frac {H^{2}\,\mathrm{u}(\theta )^{4}\,\alpha }{
\gamma ^{3}}}  - G\,M\,\mathrm{u}(\theta )^{2} }
\end{maplelatex}

\emptyline
\noindent
And finally:

\emptyline
\begin{mapleinput}
\mapleinline{active}{1d}{Eu_Lagr_2 := expand( simplify(first_term -\\
second_term)/u(theta)^2/H^2);}{%
}
\end{mapleinput}

\mapleresult
\begin{maplelatex}
\[
\mathit{Eu\_Lagr\_2} :=  - ({\frac {\partial ^{2}}{\partial 
\theta ^{2}}}\,\mathrm{u}(\theta )) - {\displaystyle \frac {
\mathrm{u}(\theta )}{\gamma ^{2}}}  + {\displaystyle \frac {
\mathrm{u}(\theta )^{2}\,\alpha }{\gamma ^{3}}}  + 
{\displaystyle \frac {G\,M}{H^{2}}} 
\]
\end{maplelatex}

\emptyline
\noindent
In the first-order approximation:

\emptyline

\begin{mapleinput}
\mapleinline{active}{1d}{gamma^n = taylor((1+alpha*u)^n, alpha=0,2);}{%
}
\end{mapleinput}

\mapleresult
\begin{maplelatex}
\[
\gamma ^{n}=1 + n\,u\,\alpha  + \mathrm{O}(\alpha ^{2})
\]
\end{maplelatex}

\emptyline
\noindent
So

\emptyline

\begin{mapleinput}
\mapleinline{active}{1d}{taylor( subs(gamma = 1+alpha*u(theta),Eu_Lagr_2),\\
  alpha=0,2 ):
 \indent basic_equation := convert(\%, polynom) = 0;}{%
}
\end{mapleinput}

\mapleresult
\begin{maplelatex}
\[
\mathit{basic\_equation} :=  - ({\frac {\partial ^{2}}{\partial 
\theta ^{2}}}\,\mathrm{u}(\theta )) - \mathrm{u}(\theta ) + 
{\displaystyle \frac {G\,M}{H^{2}}}  + 3\,\mathrm{u}(\theta )^{2}
\,\alpha =0
\]
\end{maplelatex}

\subsection{Light ray deflection}

\emptyline
In the beginning the obtained equation will be used for the search of
the light ray deflection in the vicinity of a star. The fundamental
consideration has to be based on the condition of null geodesic line
\textit{d
$s^{2}$
}=0 for light, but we simplify a problem and consider the trajectory
of the particle moving from the infinity. In this case \textit{H=
$\infty $
}.

\emptyline
\begin{mapleinput}
\mapleinline{active}{1d}{eq_def := subs(G*M/(H^2)=0,basic_equation);}{%
}
\end{mapleinput}

\mapleresult
\begin{maplelatex}
\[
\mathit{eq\_def} :=  - ({\frac {\partial ^{2}}{\partial \theta ^{
2}}}\,\mathrm{u}(\theta )) - \mathrm{u}(\theta ) + 3\,\mathrm{u}(
\theta )^{2}\,\alpha =0
\]
\end{maplelatex}

\emptyline
The free propagation (
$\alpha $
=0) results in:

\emptyline
\begin{mapleinput}
\mapleinline{active}{1d}{subs(alpha=0, lhs(eq_def)) = 0;
 \indent sol := dsolve(\{\%, u(0) = 1/R, D(u)(0) = 0\}, u(theta));
 # theta is measured from perihelion, where r = R}{%
}
\end{mapleinput}

\mapleresult
\begin{maplelatex}
\[
 - ({\frac {\partial ^{2}}{\partial \theta ^{2}}}\,\mathrm{u}(
\theta )) - \mathrm{u}(\theta )=0
\]
\end{maplelatex}

\begin{maplelatex}
\[
\mathit{sol} := \mathrm{u}(\theta )={\displaystyle \frac {
\mathrm{cos}(\theta )}{R}} 
\]
\end{maplelatex}

\emptyline
That is \textit{r=
$\frac {R}{\mathrm{cos}(\theta )}$
} . The last expression corresponds to the straight ray passing
through point 
$\theta $
=0, \textit{r}=\textit{R}. To find the corrected solution in the
gravitational field let substitute the obtained solution into
eliminated term in \textit{eq\_def}:

\emptyline
\begin{mapleinput}
\mapleinline{active}{1d}{-op(1,lhs(eq_def)) - op(2,lhs(eq_def)) =\\
subs(u(theta)=rhs(sol),op(3,lhs(eq_def)));
 \indent sol := dsolve(\\
 \{\%, u(0) = 1/R, D(u)(0) = 0\}, u(theta));#corrected\\ 
 solution in the presence of field}{%
}
\end{mapleinput}

\begin{maplelatex}
\mapleresult
\[
({\frac {\partial ^{2}}{\partial \theta ^{2}}}\,\mathrm{u}(\theta
 )) + \mathrm{u}(\theta )=3\,{\displaystyle \frac {\mathrm{cos}(
\theta )^{2}\,\alpha }{R^{2}}} 
\]
\end{maplelatex}

\begin{maplelatex}
\[
\mathit{sol} := \mathrm{u}(\theta )={\displaystyle \frac {( - 
\alpha  + R)\,\mathrm{cos}(\theta )}{R^{2}}}  - {\displaystyle 
\frac {1}{2}} \,{\displaystyle \frac {\alpha \,\mathrm{cos}(2\,
\theta )}{R^{2}}}  + {\displaystyle \frac {{\displaystyle \frac {
3}{2}} \,\alpha }{R^{2}}} 
\]
\end{maplelatex}

\emptyline
The equation for asymptote
$\lim _{\theta \rightarrow (\frac {\pi }{2})}\,\mathrm{u}(\theta 
)$
  describes the observed ray. Then angle of ray deflection is 
$\frac {2\,R}{r}$
 (symmetrical deflection before and after perihelion). Hence

\emptyline
\begin{mapleinput}
\mapleinline{active}{1d}{simplify( 2*subs(\{theta=Pi/2, alpha=G*M/c^2\},\\
rhs(sol))*R );}{%
}
\end{mapleinput}

\mapleresult
\begin{maplelatex}
\[
4\,{\displaystyle \frac {G\,M}{R\,c^{2}}} 
\]
\end{maplelatex}

\emptyline
This is a correct expression for the light ray deflection in the
gravitational field. For sun we have:

\emptyline
\begin{mapleinput}
\mapleinline{active}{1d}{subs(\{kappa=0.74e-28, M=2e33, R=6.96e10\},\\
4*kappa*M/R/4.848e-6);
# in ["],where kappa = G/c^2 [cm/g], \\
4.848e-6 rad corresponds to 1"}{%
}
\end{mapleinput}

\mapleresult
\begin{maplelatex}
\[
1.754485794
\]
\end{maplelatex}

\emptyline
The ray trajectory within Pluto's orbit distance is presented below:

\emptyline
\begin{mapleinput}
\mapleinline{active}{1d}{K := (theta, alpha, R) -> 1 /\\
(-(alpha-R)*cos(theta)/(R^2)-1/2*alpha*cos(2*theta)/\\
(R^2)+3/2*alpha/(R^2)): 
  \indent S := theta -> K(theta, 2.125e-6, 1):#deflected ray
   \indent \indent SS := theta -> 1/cos(theta):#ray without deflection
  \indent p1 := polarplot(\\
  [S,theta->theta,Pi/2..-Pi/2],axes=boxed):
 p2 :=\\
polarplot([SS,theta->theta,Pi/2..-Pi/2],\\
axes=boxed,color=black):
  \indent display(p1,p2,view=-10700..10700,\\
  title=`deflection of light ray`);\\
#distance of propagation corresponds to\\
 100 AU=1.5e10 km, distance is normalyzed to Sun radius}{%
}
\end{mapleinput}

\mapleresult
\begin{center}
\mapleplot{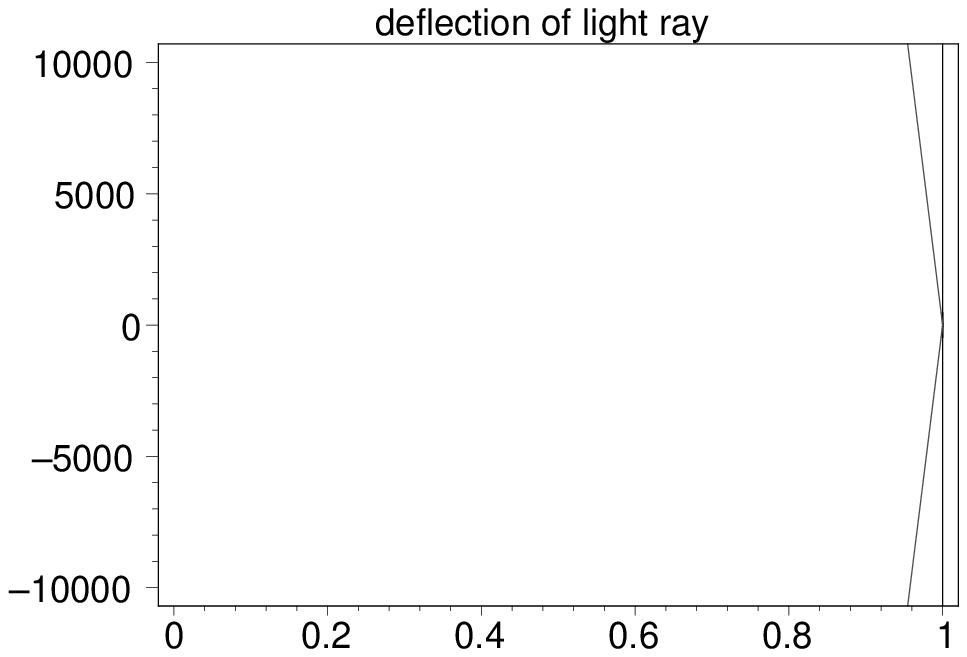}
\end{center}

\subsection{Planet's perihelion motion}

\emptyline
Now we return to \textit{basic\_equation}. Without relativistic
correction to the metric (
$\alpha $
=0) the solution is

\emptyline
\begin{mapleinput}
\mapleinline{active}{1d}{sol :=\\
dsolve(\{subs(\{G*M/(H^2)=k,alpha=0\},\\
basic_equation),u(0)=1/R,D(u)(0)=0\},u(theta));}{%
}
\end{mapleinput}

\mapleresult
\begin{maplelatex}
\[
\mathit{sol} := \mathrm{u}(\theta )=k - {\displaystyle \frac {(k
\,R - 1)\,\mathrm{cos}(\theta )}{R}} 
\]
\end{maplelatex}

\emptyline
This equation describes an elliptical orbit:

\emptyline
\begin{center}
    \textit{u=k}(\textit{1+e*
$\mathrm{cos}(\theta )$
}), 
\end{center}

\emptyline
where \textit{e = }(
$\frac {1}{k\,R}$
\textit{ - }1) is the eccentricity. For Mercury \textit{k }= 0.01,
\textit{e} = 0.2056, \textit{R} = 83.3

\emptyline
\begin{mapleinput}
\mapleinline{active}{1d}{K := (theta, k, R) -> 1/(k-(k*R-1)*cos(theta)/R):
 \indent S := theta -> K(theta, .01, 83.3):
  \indent \indent polarplot([S,theta->theta,0..2*Pi],\\
  title=`Orbit of Mercury`);}{%
}
\end{mapleinput}

\mapleresult
\begin{center}
\mapleplot{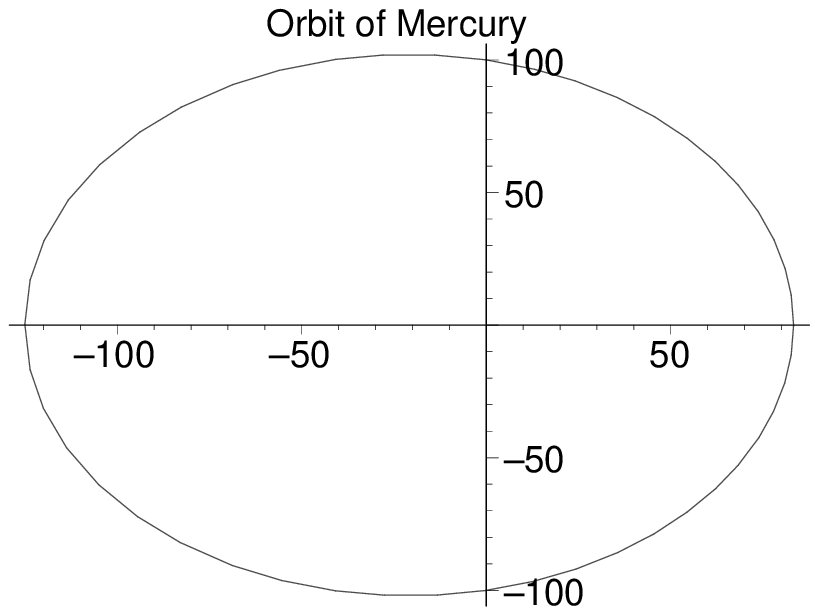}
\end{center}

\emptyline
The correction to this expression results from the substitution of
obtained solution in \textit{basic\_equation}.

\emptyline
\begin{mapleinput}
\mapleinline{active}{1d}{subs(\{G*M/(H^2)=k, 3*u(theta)^2*alpha=\\
3*alpha*(k*(1+e*cos(theta)))^2\}, basic_equation);
 \indent dsolve(\{\%,u(0)=1/R,D(u)(0)=0\},u(theta));}{%
}
\end{mapleinput}

\mapleresult
\begin{maplelatex}
\[
 - ({\frac {\partial ^{2}}{\partial \theta ^{2}}}\,\mathrm{u}(
\theta )) - \mathrm{u}(\theta ) + k + 3\,\alpha \,k^{2}\,(1 + e\,
\mathrm{cos}(\theta ))^{2}=0
\]
\end{maplelatex}

\begin{maplelatex}
\maplemultiline{
\mathrm{u}(\theta )=\\
3\,\alpha \,k^{2}\,e\,\mathrm{cos}(\theta )
 + 3\,\mathrm{sin}(\theta )\,\alpha \,k^{2}\,e\,\theta  + 
{\displaystyle \frac {3}{2}} \,\alpha \,k^{2}\,e^{2} - 
{\displaystyle \frac {1}{2}} \,\alpha \,k^{2}\,e^{2}\,\mathrm{cos
}(2\,\theta ) + k + 3\,\alpha \,k^{2} \\
\mbox{} - {\displaystyle \frac {(3\,\alpha \,k^{2}\,e\,R + \alpha
 \,k^{2}\,e^{2}\,R + k\,R + 3\,\alpha \,k^{2}\,R - 1)\,\mathrm{
cos}(\theta )}{R}}  }
\end{maplelatex}

\emptyline
Now it is possible to plot the corrected orbit (we choose the
exaggerated parameters for demonstration of orbit rotation):

\emptyline
\begin{mapleinput}
\mapleinline{active}{1d}{ K := (theta, k, R, alpha, e) -> 1 /\\
(3*alpha*k^2*e*cos(theta) + \\
3*sin(theta)*alpha*k^2*e*theta + \\
3/2*alpha*k^2*e^2 - 1/2*alpha*k^2*e^2*cos(2*theta) +\\
k + 3*alpha*k^2 - (3*alpha*k^2*e*R + alpha*k^2*e^2*R +\\
 k*R + 3*alpha*k^2*R - 1)*cos(theta)/R):
   \indent S := theta -> K(theta, .42, 1.5, 0.01, 0.6):
    \indent \indent polarplot([S,theta->theta,0..4.8*Pi],\\
    title=`rotation of orbit`);}{%
}
\end{mapleinput}

\mapleresult
\begin{center}
\mapleplot{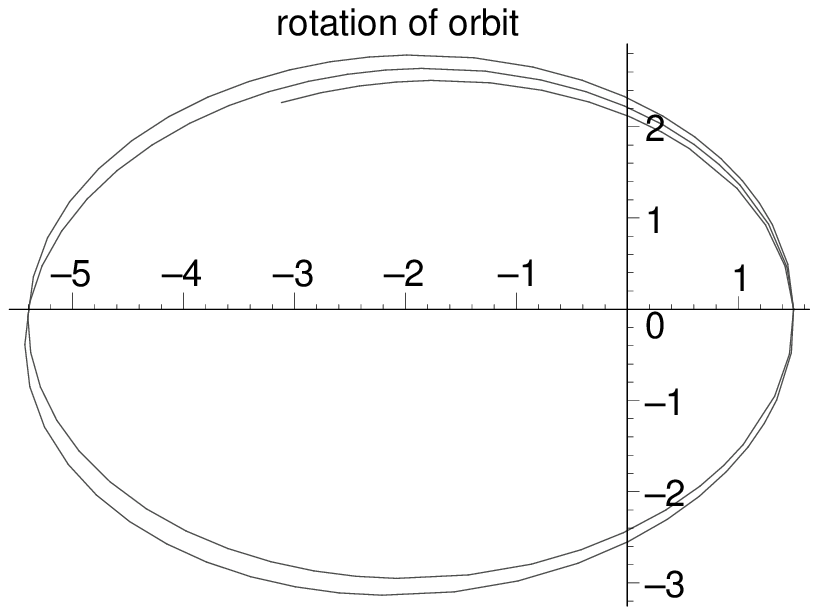}
\end{center}

\emptyline
Now we try to estimate the perihelion shift due to orbit rotation.
Let's suppose that the searched solution of \textit{basic\_equation}
differs from one in the plane space-time only due to ellipse rotation.
The parameter describing this rotation is 
$\omega $
: \textit{u}(
$\theta $
) = \textit{k}(1 + \textit{e
$\mathrm{cos}(\theta  - \omega \,\theta )$
}).

\emptyline
\begin{mapleinput}
\mapleinline{active}{1d}{subs(\{G*M/(H^2)=k, u(theta)=\\
k*(1+e*cos(theta-omega*theta))\},\\
basic_equation):#substitution of approximate solution
  \indent simplify(\%):
   \indent \indent lhs(\%):
    \indent \indent \indent collect(\%, cos(-theta+omega*theta)):
     \indent \indent coeff(\%, cos(-theta+omega*theta)):#the coefficient\\
      before this term gives algebraic equation for omega
    \indent subs(omega^2=0,\%):#omega is the small value and \\
    we don't consider the quadratic term
   solve(\% = 0, omega):
  \indent subs(\{alpha = G*M/c^2, k = 1/R/(1+e)\},2*Pi*\%);#result\\
   is expressed through the minimal distance R between\\
    planet and sun; 2*Pi corresponds to the transition to \\
    circle frequency of rotation
  \indent \indent subs(R=a*(1-e),\%);#result expressed through larger\\
   semiaxis a of an ellipse}{%
}
\end{mapleinput}

\mapleresult
\begin{maplelatex}
\[
6\,{\displaystyle \frac {\pi \,G\,M}{c^{2}\,R\,(1 + e)}} 
\]
\end{maplelatex}

\begin{maplelatex}
\[
6\,{\displaystyle \frac {\pi \,G\,M}{c^{2}\,a\,(1 - e)\,(1 + e)}
} 
\]
\end{maplelatex}

\emptyline
Hence for Mercury we have the perihelion shift during 100 years:

\emptyline
\begin{mapleinput}
\mapleinline{active}{1d}{subs(\{kappa=0.74e-28,M=2e33,a=57.9e11,e=0.2056\},\\
6*Pi*kappa*M/a/(1-e^2)*(100*365.26/87.97)/4.848e-6):\\
#here we took into account the periods of Earth's\\
 and Mercury's rotations   
  \indent evalf(\%);#["]}{%
}
\end{mapleinput}

\mapleresult
\begin{maplelatex}
\[
43.08704513
\]
\end{maplelatex}

\emptyline
Now we will consider the \textit{basic\_equation} in detail
\cite{J. L. Synge}. The implicit solutions of this equation are:

\emptyline
\begin{mapleinput}
\mapleinline{active}{1d}{dsolve( subs(G*M/H^2=k, basic_equation), u(theta));}{%
}
\end{mapleinput}

\mapleresult
\begin{maplelatex}
\maplemultiline{
{\displaystyle \int _{\ }^{\mathrm{u}(\theta )}} {\displaystyle 
\frac {1}{\sqrt{\mathit{\_C1} - \mathit{\_a}^{2} + 2\,k\,\mathit{
\_a} + 2\,\mathit{\_a}^{3}\,\alpha }}} \,d\mathit{\_a} - \theta 
 - \mathit{\_C2}=0,  \\
{\displaystyle \int _{\ }^{\mathrm{u}(\theta )}}  - 
{\displaystyle \frac {1}{\sqrt{\mathit{\_C1} - \mathit{\_a}^{2}
 + 2\,k\,\mathit{\_a} + 2\,\mathit{\_a}^{3}\,\alpha }}} \,d
\mathit{\_a} - \theta  - \mathit{\_C2}=0 }
\end{maplelatex}

\emptyline
\noindent
Hence, these implicit solutions result from the following equation (
$\beta $
 is the constant depending on the initial conditions):

\emptyline
\begin{mapleinput}
\mapleinline{active}{1d}{diff( u(theta), theta)^2 =\\
 2*M*u(theta)^3 - u(theta)^2 + 2*u(theta)*M/H^2 + beta;\\
 \# we use the units, where c=1, G=1 (see\\
 definition of the geometrical units in the next part)
 \indent f := rhs(\%):
  \indent \indent f = 2*M*\\
  ( u(theta) - u1 )*( u(theta) - u2 )*( u(theta) - u3 );}{%
}
\end{mapleinput}

\mapleresult
\begin{maplelatex}
\[
({\frac {\partial }{\partial \theta }}\,\mathrm{u}(\theta ))^{2}=
2\,M\,\mathrm{u}(\theta )^{3} - \mathrm{u}(\theta )^{2} + 
{\displaystyle \frac {2\,\mathrm{u}(\theta )\,M}{H^{2}}} + \beta
\]
\end{maplelatex}

\begin{maplelatex}
\[
2\,M\,\mathrm{u}(\theta )^{3} - \mathrm{u}(\theta )^{2} + 
{\displaystyle \frac {2\,\mathrm{u}(\theta )\,M}{H^{2}}}  + \beta
 =2\,M\,(\mathrm{u}(\theta ) - \mathit{u1})\,(\mathrm{u}(\theta )
 - \mathit{u2})\,(\mathrm{u}(\theta ) - \mathit{u3})
\]
\end{maplelatex}

\emptyline
\noindent
Here \textit{u1}, \textit{u2}, \textit{u3} are the roots of cubic
equation, which describes the "potential" defining the orbital motion:

\emptyline
\begin{mapleinput}
\mapleinline{active}{1d}{fun := rhs(\%);}{%
}
\end{mapleinput}

\mapleresult
\begin{maplelatex}
\[
\mathit{fun} := 2\,M\,(\mathrm{u}(\theta ) - \mathit{u1})\,(
\mathrm{u}(\theta ) - \mathit{u2})\,(\mathrm{u}(\theta ) - 
\mathit{u3})
\]
\end{maplelatex}

\emptyline
\noindent
The dependence of this function on \textit{u} leads to the different
types of motion. The confined motion corresponds to an elliptical
orbit

\emptyline
\begin{mapleinput}
\mapleinline{active}{1d}{plot(subs(\{u3=2, u2=1, u1=0.5, M=1/2\},fun),u=0.4..1.1,\\
title=`elliptical motion`, axes=boxed, view=0..0.08);}{%
}
\end{mapleinput}

\mapleresult
\begin{center}
\mapleplot{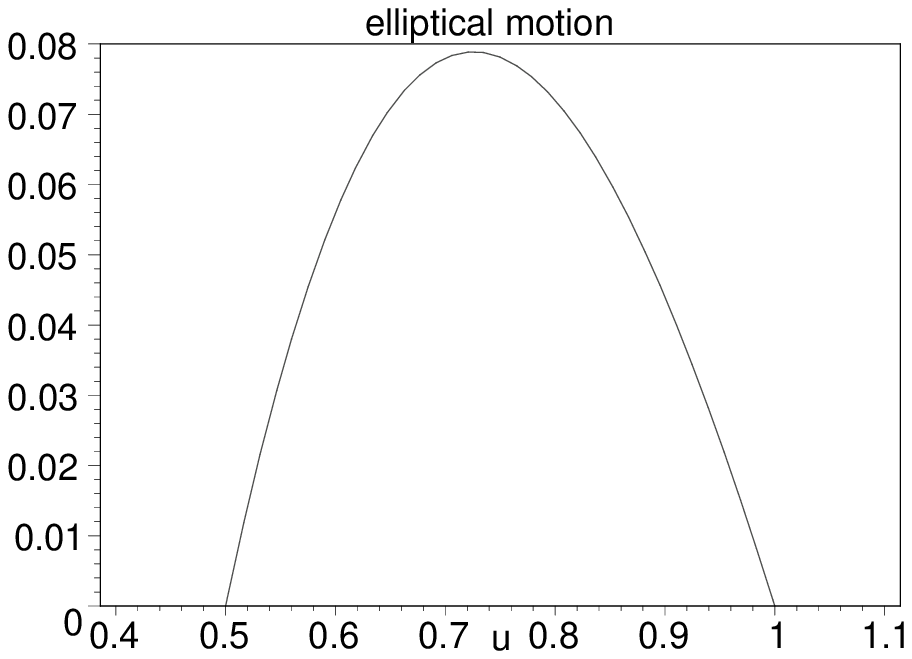}
\end{center}

\emptyline
\noindent
The next situation with \textit{u--}\TEXTsymbol{>}0
(\textit{r--}\TEXTsymbol{>}
$\infty $
) corresponds to an infinite motion:

\emptyline
\begin{mapleinput}
\mapleinline{active}{1d}{plot(subs(\{u3=2, u2=1, u1=-0.1, M=1/2\},fun),u=0..1.1,\\
title=`hyperbolical motion`, axes=boxed, view=0..0.6);}{%
}
\end{mapleinput}

\mapleresult
\begin{center}
\mapleplot{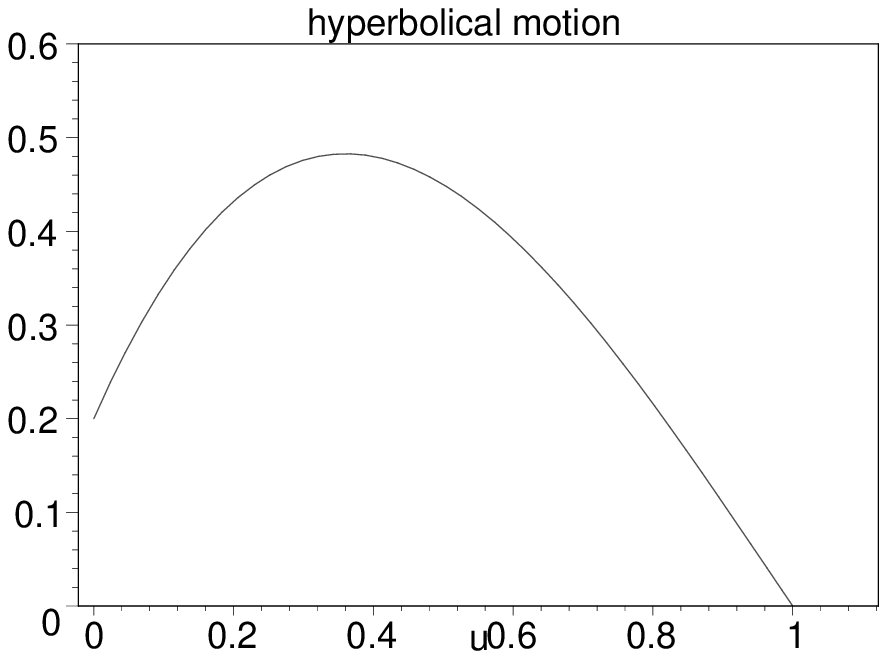}
\end{center}

\emptyline
\noindent
Now we return to the right-hand side of the modified basic equation.

\emptyline
\begin{mapleinput}
\mapleinline{active}{1d}{f;}{%
}
\end{mapleinput}

\mapleresult
\begin{maplelatex}
\[
2\,M\,\mathrm{u}(\theta )^{3} - \mathrm{u}(\theta )^{2} + 
{\displaystyle \frac {2\,\mathrm{u}(\theta )\,M}{H^{2}}} + \beta
\]
\end{maplelatex}

\emptyline
\noindent
In this expression, one can eliminate the second term by the
substitution \textit{u}(
$\theta $
)= \textit{y}(
$\theta $
)
$(\frac {2}{M})^{(\frac {1}{3})}$
 + 
$\frac {1}{6\,M}$
:

\emptyline
\begin{mapleinput}
\mapleinline{active}{1d}{collect( subs(u(theta) =\\
(2/M)^(1/3)*y(theta) + 1/(6*M), f),y(theta));}{%
}
\end{mapleinput}

\mapleresult
\begin{maplelatex}
\maplemultiline{
4\,\mathrm{y}(\theta )^{3} + \\
 \left(  \!  - {\displaystyle 
\frac {1}{6}} \,{\displaystyle \frac {2^{(1/3)}\,({\displaystyle 
\frac {1}{M}} )^{(1/3)}}{M}}  + {\displaystyle \frac {2\,2^{(1/3)
}\,({\displaystyle \frac {1}{M}} )^{(1/3)}\,M}{H^{2}}}  \! 
 \right) \,\mathrm{y}(\theta ) - {\displaystyle \frac {1}{54}} \,
{\displaystyle \frac {1}{M^{2}}}  + \beta  + {\displaystyle 
\frac {{\displaystyle \frac {1}{3}} }{H^{2}}} 
}
\end{maplelatex}

\emptyline
\noindent
This substitution reduced our equation to canonical form for the \QTR{Hyperlink}{Weierstrass P function} \cite{E. T. Whittaker}:

\emptyline
\begin{mapleinput}
\mapleinline{active}{1d}{g2 = simplify(coeff(\%, y(theta)));
 \indent g3 = -simplify(coeff(\%\%, y(theta), 0));
  \indent \indent diff( y(theta), theta)^2 =\\
   4*y(theta)^3 - g2*y(theta) - g3;}{%
}
\end{mapleinput}

\mapleresult
\begin{maplelatex}
\[
\mathit{g2}={\displaystyle \frac {1}{6}} \,{\displaystyle \frac {
2^{(1/3)}\,({\displaystyle \frac {1}{M}} )^{(1/3)}\,( - H^{2} + 
12\,M^{2})}{M\,H^{2}}} 
\]
\end{maplelatex}

\begin{maplelatex}
\[
\mathit{g3}= - {\displaystyle \frac {1}{54}} \,{\displaystyle 
\frac { - H^{2} + 54\,\beta \,M^{2}\,H^{2} + 18\,M^{2}}{M^{2}\,H
^{2}}} 
\]
\end{maplelatex}

\begin{maplelatex}
\[
({\frac {\partial }{\partial \theta }}\,\mathrm{y}(\theta ))^{2}=
4\,\mathrm{y}(\theta )^{3} - \mathit{g2}\,\mathrm{y}(\theta ) - 
\mathit{g3}
\]
\end{maplelatex}

\emptyline
\noindent
that results in:

\emptyline
\begin{mapleinput}
\mapleinline{active}{1d}{y(theta) = WeierstrassP(theta, g2, g3);}{%
}
\end{mapleinput}

\mapleresult
\begin{maplelatex}
\[
\mathrm{y}(\theta )=\mathrm{WeierstrassP}(\theta , \,\mathit{g2}
, \,\mathit{g3})
\]
\end{maplelatex}

\emptyline
\noindent
In the general case, the potential in the form of three-order
polynomial produces the solution in the form of \QTR{Hyperlink}{Jacobi sn-function}:

\emptyline
\begin{mapleinput}
\mapleinline{active}{1d}{Orbit := proc(f, x)
\indent print(`Equation in the form: u'(theta)^2 =\\
a[0]*u^3+a[1]*u^2+a[2]*u+a[3]`): 
\indent \indent degree(f,x):
  \indent \indent \indent if(\% = 3) then
   \indent \indent a[0] := coeff(f, x^3):# coefficients of polynomial
    \indent a[1] := coeff(f, x^2):
     a[2] := coeff(f, x):
      \indent a[3] := coeff(f, x, 0):
       \indent \indent sol := solve(f = 0, x):
  \indent \indent \indent print(`Roots of polynomial u[1] < u[2] < u[3]:`):
   \indent \indent print(sol[1], sol[2], sol[3]):
    \indent solution := u[1] + (u[2]-u[1])*JacobiSN(theta*sqrt(\\
    2*M*(u[3]-u[1]) )/2 + delta,\\
    sqrt((u[2]-u[1])/(u[3]-u[1])))^2:
 print(`Result through Jacobi sn - function`):   
 \indent print(u(theta) = solution):
   \indent \indent else
    \indent \indent \indent print(`The polynomial degree is not 3`)
     \indent \indent fi
  \indent end:}{%
}
\end{mapleinput}

\mapleresult
\begin{maplelatex}
\maplemultiline{
\mathit{Equation\ in\ the\ form:}\\
\mathit{u^{\prime }(theta)\symbol{94}2
\ =\ a[0]*u\symbol{94}3+a[1]*u\symbol{94}2+a[2]*u+a[3]}
}
\end{maplelatex}

\begin{maplelatex}
\[
\mathit{Roots\ of\ polynomial\ u[1]\ <\ u[2]\ <\ u[3]:}
\]
\end{maplelatex}

\begin{maplelatex}
\maplemultiline{
{\displaystyle \frac {1}{6}} \,{\displaystyle \frac {\mathrm{\%1}
^{(1/3)}}{M\,H}}  - {\displaystyle \frac {1}{6}} \,
{\displaystyle \frac { - H^{2} + 12\,M^{2}}{M\,H\,\mathrm{\%1}^{(
1/3)}}}  + {\displaystyle \frac {{\displaystyle \frac {1}{6}} }{M
}} ,  \\
 - {\displaystyle \frac {1}{12}} \,{\displaystyle \frac {\mathrm{
\%1}^{(1/3)}}{M\,H}}  + {\displaystyle \frac {{\displaystyle 
\frac {1}{12}} \,( - H^{2} + 12\,M^{2})}{M\,H\,\mathrm{\%1}^{(1/3
)}}}  + {\displaystyle \frac {{\displaystyle \frac {1}{6}} }{M}} 
 +\\
  {\displaystyle \frac {1}{2}} \,I\,\sqrt{3}\, \left(  \! 
{\displaystyle \frac {1}{6}} \,{\displaystyle \frac {\mathrm{\%1}
^{(1/3)}}{M\,H}}  + {\displaystyle \frac {{\displaystyle \frac {1
}{6}} \,( - H^{2} + 12\,M^{2})}{M\,H\,\mathrm{\%1}^{(1/3)}}}  \! 
 \right) ,  \\
 - {\displaystyle \frac {1}{12}} \,{\displaystyle \frac {\mathrm{
\%1}^{(1/3)}}{M\,H}}  + {\displaystyle \frac {{\displaystyle 
\frac {1}{12}} \,( - H^{2} + 12\,M^{2})}{M\,H\,\mathrm{\%1}^{(1/3
)}}}  + {\displaystyle \frac {{\displaystyle \frac {1}{6}} }{M}} 
 -\\
  {\displaystyle \frac {1}{2}} \,I\,\sqrt{3}\, \left(  \! 
{\displaystyle \frac {1}{6}} \,{\displaystyle \frac {\mathrm{\%1}
^{(1/3)}}{M\,H}}  + {\displaystyle \frac {{\displaystyle \frac {1
}{6}} \,( - H^{2} + 12\,M^{2})}{M\,H\,\mathrm{\%1}^{(1/3)}}}  \! 
 \right)  \\
\mathrm{\%1} := H^{3} - 54\,\beta \,M^{2}\,H^{3} - 18\,H\,M^{2}
 \\
\mbox{} + 6\,\sqrt{3}\,\sqrt{ - M^{2}\,H^{2} + 16\,M^{4} - H^{6}
\,\beta  + 27\,H^{6}\,\beta ^{2}\,M^{2} + 18\,H^{4}\,\beta \,M^{2
}}\,M }
\end{maplelatex}

\begin{maplelatex}
\[
\mathit{Result\ through\ Jacobi\ sn\ -\ function}
\]
\end{maplelatex}

\begin{maplelatex}
\[
\mathrm{u}(\theta )={u_{1}} + ({u_{2}} - {u_{1}})\,\mathrm{
JacobiSN}({\displaystyle \frac {1}{2}} \,\theta \,\sqrt{2}\,
\sqrt{M\,({u_{3}} - {u_{1}})} + \delta , \,\sqrt{{\displaystyle 
\frac {{u_{2}} - {u_{1}}}{{u_{3}} - {u_{1}}}} })^{2}
\]
\end{maplelatex}

\emptyline
\noindent
As 
${u_{1}}$
 = 
$\frac {1}{{r_{1}}}$
 and 
${u_{2}}$
 = 
$\frac {1}{{r_{2}}}$
 are the small values for the planets (
${r_{1}}$
 and 
${r_{2}}$
 are the perihelion and aphelion points) and 
${u_{1}}$
 + 
${u_{2}}$
 + 
${u_{3}}$
 = 
$\frac {1}{2\,M}$
, we have 2\textit{M
${u_{3}}$
}=1 and

\emptyline
\begin{mapleinput}
\mapleinline{active}{1d}{u(theta) - u[1] =\\
 (u[2] - u[1])*JacobiSN(1/2*theta+delta,0)^2;}{%
}
\end{mapleinput}

\mapleresult
\begin{maplelatex}
\[
\mathrm{u}(\theta ) - {u_{1}}=({u_{2}} - {u_{1}})\,\mathrm{sin}(
{\displaystyle \frac {1}{2}} \,\theta  + \delta )^{2}
\]
\end{maplelatex}

\emptyline
\noindent
that is the equation of orbital motion (see above) with \textit{e }= 
$\frac {{u_{2}} - {u_{1}}}{{u_{2}} + {u_{1}}}$
. 
${u_{1}}$
 \TEXTsymbol{>} 0 corresponds to the elliptical motion, 
${u_{1}}$
 \TEXTsymbol{<} 0 corresponds to hyperbolical motion. The period of
the orbital motion in the general case:

\emptyline
\begin{mapleinput}
\mapleinline{active}{1d}{(u[2]-u[1])/(u[3]-u[1]):
  kernel := 2/sqrt((1-t^2)*(1-t^2*\%)):#2 in the numerator\\
   corresponds to sn^2-period
    \indent int(kernel, t=0..1);}{%
}
\end{mapleinput}

\mapleresult
\begin{maplelatex}
\[
2\,\mathrm{EllipticK}(\sqrt{ - {\displaystyle \frac {{u_{2}} - {u
_{1}}}{ - {u_{3}} + {u_{1}}}} })
\]
\end{maplelatex}

\emptyline
\noindent
and can be found approximately for small 
${u_{1}}$
 and 
${u_{2}}$
 as result of  expansion:

\emptyline
\begin{mapleinput}
\mapleinline{active}{1d}{series(2*EllipticK(x), x=0,4):
 \indent convert(\%,polynom):
  \indent \indent subs(x = sqrt( 2*M*(u[2]-u[1]) ), \%);}{%
}
\end{mapleinput}

\mapleresult
\begin{maplelatex}
\[
\pi  + {\displaystyle \frac {1}{2}} \,\pi \,M\,({u_{2}} - {u_{1}}
)
\]
\end{maplelatex}

\emptyline
\noindent
From the expression for \textit{u}(
$\theta $
) and obtained expression for the period of 
$\mathit{sn}^{2}$
 we have the change of angular coordinate over period:

\emptyline
\begin{mapleinput}
\mapleinline{active}{1d}{\%/(1/2*sqrt( 2*M*(u[3]-u[1]) )):
 \indent simplify(\%);}{%
}
\end{mapleinput}

\mapleresult
\begin{maplelatex}
\[
{\displaystyle \frac {1}{2}} \,{\displaystyle \frac {\pi \,(2 + M
\,{u_{2}} - M\,{u_{1}})\,\sqrt{2}}{\sqrt{ - M\,( - {u_{3}} + {u_{
1}})}}} 
\]
\end{maplelatex}

\emptyline
\noindent
But 
$\sqrt{2\,M\,({u_{3}} - {u_{1}})}$
= 
$\sqrt{1 - 2\,M\,({u_{1}} - {u_{2}} - {u_{1}})}$
  $\approx$ 1 + 
$M\,(2\,{u_{1}} + {u_{2}})$
 . And the result is  

\emptyline
\begin{mapleinput}
\mapleinline{active}{1d}{2*Pi*(1 + M*(u[2]-u[1])/2)*(1 + M*(2*u[1]+u[2])):
 \indent series(\%, u[1]=0,2):
  \indent \indent convert(\%,polynom):
   \indent \indent \indent series(\%, u[2]=0,2):
    \indent \indent \indent \indent convert(\%,polynom):
     \indent \indent \indent expand(\%):
      \indent \indent expand(\%-op(4,\%)):
       \indent factor(\%);}{%
}
\end{mapleinput}

\mapleresult
\begin{maplelatex}
\[
\pi \,(2 + 3\,M\,{u_{1}} + 3\,M\,{u_{2}})
\]
\end{maplelatex}

\emptyline
\noindent
The deviation of the period from 2
$\pi $
 causes the shift of the perihelion over one rotation of the planet
around massive star.

\emptyline
\begin{mapleinput}
\mapleinline{active}{1d}{\indent \indent \indent simplify(\%-2*Pi):
   \indent \indent subs(\{u[1]=1/r[1], u[2]=1/r[2]\}, \%):
  \indent subs(\{r[1]=a*(1+e), r[2]=a*(1-e)\},G*\%/c^2):# we\\
   returned the constants G and c
simplify(\%);
}{%
}
\end{mapleinput}

\mapleresult
\begin{maplelatex}
\[
 - 6\,{\displaystyle \frac {G\,\pi \,M}{a\,(1 + e)\,( - 1 + e)\,c
^{2}}} 
\]
\end{maplelatex}

\emptyline
\noindent
This result coincides with the expression, which was obtained on the
basis of approximate solution of \textit{basic\_equation.} From the
Kepler's low we can express this result through orbital period:

\emptyline
\begin{mapleinput}
\mapleinline{active}{1d}{subs(M=4*Pi^2*a^3/T^2, \%):# T is the orbital period
 \indent simplify(\%);}{%
}
\end{mapleinput}

\mapleresult
\begin{maplelatex}
\[
 - 24\,{\displaystyle \frac {G\,\pi ^{3}\,a^{2}}{T^{2}\,(1 + e)\,
( - 1 + e)\,c^{2}}} 
\]
\end{maplelatex}

\subsection{Conclusion}

\emptyline
So, we found the expression for the Schwarzschild metric from the
equivalence principle without introducing of the Einstein's equations.
On this basis and from Euler-Lagrange equations we obtained the main
experimental consequences of GR-theory: the light ray deflection and
planet's perihelion motion.
\emptyline

\section{Relativistic stars and black holes}

\emptyline

\subsection{Introduction}

\emptyline
\noindent
The most wonderful prediction of GR-theory is the existence of black
holes, which are the objects with extremely strong gravitational
field. The investigation of these objects is the test of our
understanding of space-time structure. We will base our consideration
on the analytical approach that demands to consider only symmetrical
space-times. But this restriction does not decrease the significance
of the obtained data because of the rich structure of analytical
results and possibilities of clear interpretation clarify the physical
basis of the phenomenon in the strongly curved space-time. The basic
principles can be found in \cite{C. W. Misner,V. Frolov}.

\emptyline

\subsection{Geometric units}

\emptyline
\noindent
The very useful normalization in GR utilizes the so-called geometric
units. Since the left-hand side of the Einstein equations describes
the curvature tensor (its dimension is 
$\mathit{cm}^{( - 2)}$
), the right-hand side is to have same dimension. Let's the
gravitational constant is 
\textit{G}
 = 
$6.673\,10^{( - 8)}$
$\frac {\mathit{cm}^{3}}{g\,s^{2}}$
 = 1 and the light velocity is\textit{ c }= 
$2.998\,10^{10}$
$\frac {\mathit{cm}}{s}$
 = 1,

\emptyline
\noindent
then

\emptyline
\textit{G/
$c^{2}$
= 
$.7425\,10^{( - 28)}$
} \textit{cm/g} = 1

$c^{5}$
\textit{/
G
 }= 
$3.63\,10^{59}$
 \textit{erg/s} = 1 (power unit)

\textit{G/c} = 
$2.23\,10^{( - 18)}$
 \textit{Hz*
$\mathit{cm}^{2}$
}/\textit{g }= 1 (characteristic of absorption)

$c^{2}$
/
$\sqrt{\textit{G} }$
 = 
$3.48\,10^{24}$
 CGSE units (field strength)

\textit{h}/2
$\pi $
 = 
$1.054\,10^{( - 27)}$
 \textit{g*
$\mathit{cm}^{2}$
/s = 
$2.612\,10^{( - 66)}$
} 
$\mathit{cm}^{2}$
    
elementary charge\textit{ }\textit{   e = 
$1.381\,10^{( - 34)}$
} \textit{cm}

1 \textit{ps} = 
$3.0856\,10^{18}$
 \textit{cm}

1\textit{ eV = 
$1.324\,10^{( - 61)}$
} \textit{cm}

\emptyline
\noindent
There are the following extremal values of length, time, mass and density, which are useful in the context of the consederation of GR validity:

\emptyline
$\sqrt{\frac {\textit{G} \,\textit{h}}{2\,\pi \,c^{3}}}$
  = 
$1.616\,10^{( - 33)}$
 \textit{cm} (Planck length)

$\sqrt{\frac {\textit{G} \,\textit{h}}{2\,\pi \,c^{5}}}$
 = 
$5.391\,10^{( - 44)}$
 \textit{s} (Planck time)

$\sqrt{\frac {h\,c}{2\,\pi \,\textit{G} }}$
 = 
$2.177\,10^{( - 5)}$
 \textit{g} (Planck mass)

$\frac {2\,\pi \,c^{5}}{h\,\textit{G} ^{2}}$
 = 
$5.157\,10^{93}$
 \textit{g}/
$\mathit{cm}^{3}$
 (Planck density)

\subsection{Relativistic star}

\emptyline
\noindent
As stated above the first realistic metric was introduced by
Schwarzschild for description of the spherically symmetric and static
curved space. Let us introduce the spherically symmetric metric in the
following form:

\emptyline
\begin{mapleinput}
\mapleinline{active}{1d}{restart:
 \indent with(tensor):
  \indent \indent with(plots):
   \indent \indent \indent with(linalg):
    \indent \indent \indent \indent with(difforms):

coord := [t, r, theta, phi]:# spherical coordinates,\\
 which will be designated in text as [0,1,2,3]
 \indent g_compts :=\\
  array(symmetric,sparse,1..4,1..4):# metric components
  \indent \indent g_compts[1,1] := -exp(2*Phi(r)):# component\\
   of interval attached to d(t)^2 
   \indent g_compts[2,2] := exp(2*Lambda(r)):# component\\
    of interval attached to d(r)^2
     g_compts[3,3] := r^2:# component of interval\\
      attached to d(theta)^2 
      \indent g_compts[4,4] := r^2*sin(theta)^2:# component \\
      of interval attached to d(phi)^2

g := create([-1,-1], eval(g_compts));# covariant\\
 metric tensor 

\indent ginv := invert( g, 'detg' );# contravariant\\
 metric tensor}{%
}
\end{mapleinput}

\begin{maplelatex}
\maplemultiline{
g :=\\
 \mathrm{table(}[\mathit{compts}= \left[ 
{\begin{array}{cccc}
 - e^{(2\,\Phi (r))} & 0 & 0 & 0 \\
0 & e^{(2\,\Lambda (r))} & 0 & 0 \\
0 & 0 & r^{2} & 0 \\
0 & 0 & 0 & r^{2}\,\mathrm{sin}(\theta )^{2}
\end{array}}
 \right] ,\\
  \,\mathit{index\_char}=[-1, \,-1]])
}
\end{maplelatex}

\begin{maplelatex}
\maplemultiline{
\mathit{ginv} :=\\
 \mathrm{table(}[\mathit{compts}= \left[ 
{\begin{array}{cccc}
 - {\displaystyle \frac {1}{e^{(2\,\Phi (r))}}}  & 0 & 0 & 0 \\
 [2ex]
0 & {\displaystyle \frac {1}{e^{(2\,\Lambda (r))}}}  & 0 & 0 \\
 [2ex]
0 & 0 & {\displaystyle \frac {1}{r^{2}}}  & 0 \\ [2ex]
0 & 0 & 0 & {\displaystyle \frac {1}{r^{2}\,\mathrm{sin}(\theta )
^{2}}} 
\end{array}}
 \right] ,\\
  \,\mathit{index\_char}=[1, \,1]])
}
\end{maplelatex}

\emptyline
\noindent
Now we can use the standard Maple procedure for Einstein tensor
definition

\emptyline
\begin{mapleinput}
\mapleinline{active}{1d}{# intermediate values
D1g := d1metric( g, coord ):
 \indent D2g := d2metric( D1g, coord ):
  \indent \indent Cf1 := Christoffel1 ( D1g ):
   \indent \indent \indent RMN := Riemann( ginv, D2g, Cf1 ):
    \indent \indent RICCI := Ricci( ginv, RMN ):
     \indent RS := Ricciscalar( ginv, RICCI ):}{%
}
\end{mapleinput}

\begin{mapleinput}
\mapleinline{active}{1d}{Estn := Einstein( g, RICCI, RS ):# Einstein tensor
 \indent displayGR(Einstein,Estn);# Its nonzero components}{%
}
\end{mapleinput}

\mapleresult
\begin{maplelatex}
\[
\mathit{The\ Einstein\ Tensor}
\]
\end{maplelatex}

\begin{maplelatex}
\[
\mathit{non-zero\ components\ :}
\]
\end{maplelatex}

\begin{maplelatex}
\[
\mathit{\ G11}= - {\displaystyle \frac {e^{(2\,\Phi (r))}\,(2\,(
{\frac {\partial }{\partial r}}\,\Lambda (r))\,r + e^{(2\,\Lambda
 (r))} - 1)}{r^{2}\,e^{(2\,\Lambda (r))}}} 
\]
\end{maplelatex}

\begin{maplelatex}
\[
\mathit{\ G22}= - {\displaystyle \frac {2\,({\frac {\partial }{
\partial r}}\,\Phi (r))\,r - e^{(2\,\Lambda (r))} + 1}{r^{2}}} 
\]
\end{maplelatex}

\begin{maplelatex}
\maplemultiline{
\mathit{\ G33}= - {\frac {r}{e^{(2\,\Lambda (r))}}}\\
\left[ {\frac {\partial }{
\partial r}}\,\Phi (r) - {\frac {\partial }{\partial r}}\,
\Lambda (r) + r\,{\frac {\partial ^{2}}{\partial r^{2}}}\,\Phi 
(r) + r\,({\frac {\partial }{\partial r}}\,\Phi (r))^{2} - r\,
{\frac {\partial }{\partial r}}\,\Phi (r)\,{\frac {\partial }{
\partial r}}\,\Lambda (r) \right] } 
\end{maplelatex}

\begin{maplelatex}
\maplemultiline{
\mathit{\ G44}= - {\frac {\mathrm{sin}(\theta )^{2}
\,r}{e^{(2\,\Lambda (r))}}}\\
\left[{\frac {\partial }{\partial r}}\,\Phi (r) - {\frac {
\partial }{\partial r}}\,\Lambda (r) + r\,{\frac {\partial ^{2}
}{\partial r^{2}}}\,\Phi (r) + r\,({\frac {\partial }{\partial r
}}\,\Phi (r))^{2} - r\,{\frac {\partial }{\partial r}}\,\Phi (r)
\,{\frac {\partial }{\partial r}}\,\Lambda (r) \right]
}
\end{maplelatex}

\begin{maplelatex}
\[
\mathit{character\ :\ [-1,\ -1]}
\]
\end{maplelatex}

\emptyline
\noindent
In the beginning we will consider the star in the form of drop of
liquid. In this case the energy-momentum tensor is 
${T_{\mu , \,\nu }}$
\textit{} = (
$p + \rho $
)\textit{ 
${u_{\mu }}$
${u_{\nu }}$}
 + \textit{p
${g_{\mu , \,\nu }}$
 }(all components of \textit{u} except for 
${u_{0}}$
 are equal to zero, and \textit{-}1\textit{ = 
$g^{(0, \,0)}$
${u_{0}}$
${u_{0}}$
}; the signature (-2) results in 
${u_{\alpha }}\,u^{\alpha }$
=1 and 
${T_{\mu , \,\nu }}$
\textit{} \textit{= }(
$p + \rho $
)\textit{ 
${u_{\mu }}$
${u_{\nu }}$
 - p
${g_{\mu , \,\nu }}$
}):

\emptyline
\begin{mapleinput}
\mapleinline{active}{1d}{T_compts :=\\
 array(symmetric,sparse,1..4,1..4):# energy-momentum\\
  tensor for drop of liquid
 \indent T_compts[1,1] := exp(2*Phi(r))*rho(r): 
  \indent \indent T_compts[2,2] := exp(2*Lambda(r))*p(r):
   \indent \indent \indent T_compts[3,3] := p(r)*r^2: 
    \indent \indent T_compts[4,4] := p(r)*r^2*sin(theta)^2:
     \indent T := create([-1,-1], eval(T_compts));}{%
}
\end{mapleinput}

\mapleresult
\begin{maplelatex}
\maplemultiline{
T :=\\
 \mathrm{table(}[\mathit{compts}= \left[ 
{\begin{array}{cccc}
e^{(2\,\Phi (r))}\,\rho (r) & 0 & 0 & 0 \\
0 & e^{(2\,\Lambda (r))}\,\mathrm{p}(r) & 0 & 0 \\
0 & 0 & \mathrm{p}(r)\,r^{2} & 0 \\
0 & 0 & 0 & \mathrm{p}(r)\,r^{2}\,\mathrm{sin}(\theta )^{2}
\end{array}}
 \right] ,  \\
\mathit{index\_char}=[-1, \,-1]]) }
\end{maplelatex}

\emptyline
\noindent
To write the \underline{Einstein equations} (sign corresponds to
\cite{W.Pauli})

\emptyline
\begin{center}
${G_{\mu , \,\nu }}$%
 = -8 $\pi $ ${T_{\mu , \,\nu }}$
\end{center}

\emptyline
\noindent
let's extract the tensor components:

\emptyline
\begin{mapleinput}
\mapleinline{active}{1d}{Energy_momentum := get_compts(T);
 \indent Einstein := get_compts(Estn);}{%
}
\end{mapleinput}

\mapleresult
\begin{maplelatex}
\maplemultiline{
\mathit{Energy\_momentum} :=\\
  \left[ 
{\begin{array}{cccc}
e^{(2\,\Phi (r))}\,\rho (r) & 0 & 0 & 0 \\
0 & e^{(2\,\Lambda (r))}\,\mathrm{p}(r) & 0 & 0 \\
0 & 0 & \mathrm{p}(r)\,r^{2} & 0 \\
0 & 0 & 0 & \mathrm{p}(r)\,r^{2}\,\mathrm{sin}(\theta )^{2}
\end{array}}
 \right] 
}
\end{maplelatex}

\begin{maplelatex}
\maplemultiline{
\mathit{Einstein} :=  \\
 \left[  \!  - {\displaystyle \frac {e^{(2\,\Phi (r))}\,(2\,(
{\frac {\partial }{\partial r}}\,\Lambda (r))\,r + \mathrm{\%1}
 - 1)}{r^{2}\,\mathrm{\%1}}} \,, \,0\,, \,0\,, \,0 \!  \right] 
 \\
 \left[  \! 0\,, \, - {\displaystyle \frac {2\,({\frac {\partial 
}{\partial r}}\,\Phi (r))\,r - \mathrm{\%1} + 1}{r^{2}}} \,, \,0
\,, \,0 \!  \right]  \\
 \left[ {\vrule height1.31em width0em depth1.31em} \right. \! 
 \! 0\,, \,0\,,  \\
 - {\displaystyle \frac {r\,(({\frac {\partial }{\partial r}}\,
\Phi (r)) - ({\frac {\partial }{\partial r}}\,\Lambda (r)) + r\,(
{\frac {\partial ^{2}}{\partial r^{2}}}\,\Phi (r)) + r\,({\frac {
\partial }{\partial r}}\,\Phi (r))^{2} - r\,({\frac {\partial }{
\partial r}}\,\Phi (r))\,({\frac {\partial }{\partial r}}\,
\Lambda (r)))}{\mathrm{\%1}}} \,, \,0 \! \! \left. {\vrule 
height1.31em width0em depth1.31em} \right]  \\
 \left[ {\vrule height1.31em width0em depth1.31em} \right. \! 
 \! 0\,, \,0\,, \,0\,, \, \\
 - {\displaystyle \frac {\mathrm{sin}(
\theta )^{2}\,r\,(({\frac {\partial }{\partial r}}\,\Phi (r)) - (
{\frac {\partial }{\partial r}}\,\Lambda (r)) + r\,({\frac {
\partial ^{2}}{\partial r^{2}}}\,\Phi (r)) + r\,({\frac {
\partial }{\partial r}}\,\Phi (r))^{2} - r\,({\frac {\partial }{
\partial r}}\,\Phi (r))\,({\frac {\partial }{\partial r}}\,
\Lambda (r)))}{\mathrm{\%1}}}  \\
 \! \! \left. {\vrule height1.31em width0em depth1.31em} \right] 
 \\
\mathrm{\%1} := e^{(2\,\Lambda (r))} }
\end{maplelatex}

\emptyline
\noindent
First Einstein equation for (0,0) - component is:

\emptyline
\begin{mapleinput}
\mapleinline{active}{1d}{8*Pi*Energy_momentum[1,1] + Einstein[1,1]:
 \indent expand(\%/exp(Phi(r))^2):
  \indent \indent eq1 := simplify(\%) = 0;#first Einstein equation}{%
}
\end{mapleinput}

\mapleresult
\begin{maplelatex}
\[
\mathit{eq1} :=  - {\displaystyle \frac { - 8\,\pi \,\rho (r)\,r
^{2} + 2\,e^{( - 2\,\Lambda (r))}\,({\frac {\partial }{\partial r
}}\,\Lambda (r))\,r + 1 - e^{( - 2\,\Lambda (r))}}{r^{2}}} =0
\]
\end{maplelatex}

\emptyline
\noindent
This equation can be rewritten as:

\emptyline
\begin{mapleinput}
\mapleinline{active}{1d}{eq1 :=\\
 -8*Pi*rho(r)*r^2+Diff(r*(1-exp(-2*Lambda(r))),r) = 0;}{%
}
\end{mapleinput}

\mapleresult
\begin{maplelatex}
\[
\mathit{eq1} :=  - 8\,\pi \,\rho (r)\,r^{2} + ({\frac {\partial 
}{\partial r}}\,r\,(1 - e^{( - 2\,\Lambda (r))}))=0
\]
\end{maplelatex}

\emptyline
\noindent
The formal substitution results in:

\emptyline
\begin{mapleinput}
\mapleinline{active}{1d}{eq1_n := subs(\\
 r*(1-exp(-2*Lambda(r))) = 2*m(r),\\
  lhs(eq1)) = 0;}{%
}
\end{mapleinput}

\mapleresult
\begin{maplelatex}
\[
\mathit{eq1\_n} :=  - 8\,\pi \,\rho (r)\,r^{2} + ({\frac {
\partial }{\partial r}}\,(2\,\mathrm{m}(r)))=0
\]
\end{maplelatex}

\begin{mapleinput}
\mapleinline{active}{1d}{dsolve(eq1_n,m(r));}{%
}
\end{mapleinput}

\mapleresult
\begin{maplelatex}
\[
\mathrm{m}(r)={\displaystyle \int } 4\,\pi \,\rho (r)\,r^{2}\,dr
 + \mathit{\_C1}
\]
\end{maplelatex}

\emptyline
\noindent
So, \textit{m(r)} is the mass inside sphere with radius \textit{r}. To
estimate \textit{\_C1} we have to express 
$\Lambda $
 from \textit{m}:

\emptyline
\begin{mapleinput}
\mapleinline{active}{1d}{r*(1-exp(-2*Lambda(r))) = 2*m(r):
 \indent expand(solve(\%, exp(-2*Lambda(r)) ));}{%
}
\end{mapleinput}

\mapleresult
\begin{maplelatex}
\[
1 - {\displaystyle \frac {2\,\mathrm{m}(r)}{r}} 
\]
\end{maplelatex}

\emptyline
\noindent
Hence (see expression for 
${g_{\mu , \,\nu }}$
 through 
$\Lambda $
) the Lorenzian metric in the absence of matter is possible only if
\textit{\_C1=0}.

\noindent
Second Einstein equation for  (1,1)-component is:

\emptyline
\begin{mapleinput}
\mapleinline{active}{1d}{eq2 := simplify( 8*Pi*Energy_momentum[2,2] +\\
 Einstein[2,2] ) = 0;#second Einstein equation}{%
}
\end{mapleinput}

\mapleresult
\begin{maplelatex}
\[
\mathit{eq2} := {\displaystyle \frac {8\,\pi \,e^{(2\,\Lambda (r)
)}\,\mathrm{p}(r)\,r^{2} - 2\,({\frac {\partial }{\partial r}}\,
\Phi (r))\,r + e^{(2\,\Lambda (r))} - 1}{r^{2}}} =0
\]
\end{maplelatex}

\begin{mapleinput}
\mapleinline{active}{1d}{eq2_2 := numer(\\
 lhs(\\
  simplify(\\
   subs(exp(2*Lambda(r)) = 1/(1-2*m(r)/r),eq2)))) = 0;}{%
}
\end{mapleinput}

\mapleresult
\begin{maplelatex}
\[
\mathit{eq2\_2} :=  - 8\,\pi \,r^{3}\,\mathrm{p}(r) + 2\,(
{\frac {\partial }{\partial r}}\,\Phi (r))\,r^{2} - 4\,({\frac {
\partial }{\partial r}}\,\Phi (r))\,r\,\mathrm{m}(r) - 2\,
\mathrm{m}(r)=0
\]
\end{maplelatex}

\emptyline
\noindent
As result we have:

\emptyline
\begin{mapleinput}
\mapleinline{active}{1d}{eq2_3 := Diff(Phi(r),r) = solve(eq2_2, diff(Phi(r),r) );}{%
}
\end{mapleinput}

\mapleresult
\begin{maplelatex}
\[
\mathit{eq2\_3} := {\frac {\partial }{\partial r}}\,\Phi (r)= - 
{\displaystyle \frac {4\,\pi \,r^{3}\,\mathrm{p}(r) + \mathrm{m}(
r)}{r\,( - r + 2\,\mathrm{m}(r))}} 
\]
\end{maplelatex}

\emptyline
\noindent
We can see, that the gradient of the gravitational potential 
$\Phi $
 is \underline{greater} than in the Newtonian case 
${\frac {\partial }{\partial r}}\,\Phi $
=
$\frac {m}{r^{2}}$
 , that is the pressure in GR is the source of gravitation. 

\noindent
For further analysis we have to define the relativist equation of
hydrodynamics (relativist Euler's equation):

\emptyline
\begin{mapleinput}
\mapleinline{active}{1d}{compts := array([u_t,u_r,u_th,u_ph]):
 \indent u := create([1], compts):# 4-velocity
  \indent \indent Cf2 := Christoffel2 ( ginv, Cf1 ):
   \indent (rho(r)+p(r))*get_compts(\\
   cov_diff( u, coord, Cf2 ))[1,2]/(u_t) =\\
   -diff(p(r),r);# radial component of Euler equation,\\
    u_r=u_th=u_ph=0
     eq3 := Diff(Phi(r),r) = solve(\%, diff(Phi(r),r) );}{%
}
\end{mapleinput}

\mapleresult
\begin{maplelatex}
\[
(\rho (r) + \mathrm{p}(r))\,({\frac {\partial }{\partial r}}\,
\Phi (r))= - ({\frac {\partial }{\partial r}}\,\mathrm{p}(r))
\]
\end{maplelatex}

\begin{maplelatex}
\[
\mathit{eq3} := {\frac {\partial }{\partial r}}\,\Phi (r)= - 
{\displaystyle \frac {{\frac {\partial }{\partial r}}\,\mathrm{p}
(r)}{\rho (r) + \mathrm{p}(r)}} 
\]
\end{maplelatex}

\emptyline
\noindent
As result we obtain the so-called \underline{Oppenheimer-Volkoff
equation} for hydrostatic equilibrium of star:

\emptyline
\begin{mapleinput}
\mapleinline{active}{1d}{Diff(p(r),r) = factor( solve(rhs(eq3) =\\
 rhs(eq2_3),diff(p(r),r)));}{%
}
\end{mapleinput}

\mapleresult
\begin{maplelatex}
\[
{\frac {\partial }{\partial r}}\,\mathrm{p}(r)={\displaystyle 
\frac {(4\,\pi \,r^{3}\,\mathrm{p}(r) + \mathrm{m}(r))\,(\rho (r)
 + \mathrm{p}(r))}{r\,( - r + 2\,\mathrm{m}(r))}} 
\]
\end{maplelatex}

\emptyline
\noindent
One can see that the pressure gradient is \underline{greater} than in
the classical limit (
${\frac {\partial }{\partial r}}\,p$
= \textit{-
$\frac {\rho \,m}{r^{2}}$
}) and this gradient is increased by pressure growth (numerator) and
\textit{r} decrease (denominator) due to approach to star center. So,
one can conclude that in our model the gravitation is stronger than in
Newtonian case.

\emptyline
\noindent
Out of star \textit{m}(\textit{r})\textit{=M}, \textit{p=}0
(\textit{M} is the full mass of star). Then

\emptyline
\begin{mapleinput}
\mapleinline{active}{1d}{diff(Phi(r),r) = subs(\{m(r)=M,p(r)=0\},rhs(eq2_3));
 \indent eq4 := dsolve(\%, Phi(r));}{%
}
\end{mapleinput}

\mapleresult
\begin{maplelatex}
\[
{\frac {\partial }{\partial r}}\,\Phi (r)= - {\displaystyle 
\frac {M}{r\,( - r + 2\,M)}} 
\]
\end{maplelatex}

\begin{maplelatex}
\[
\mathit{eq4} := \Phi (r)= - {\displaystyle \frac {1}{2}} \,
\mathrm{ln}(r) + {\displaystyle \frac {1}{2}} \,\mathrm{ln}(r - 2
\,M) + \mathit{\_C1}
\]
\end{maplelatex}

\emptyline
\noindent
The boundary condition

\emptyline
\begin{mapleinput}
\mapleinline{active}{1d}{0 = limit(rhs(eq4),r=infinity);}{%
}
\end{mapleinput}

\mapleresult
\begin{maplelatex}
\[
0=\mathit{\_C1}
\]
\end{maplelatex}

\emptyline
\noindent
results in

\emptyline
\begin{mapleinput}
\mapleinline{active}{1d}{subs(_C1=0,eq4);}{%
}
\end{mapleinput}

\mapleresult
\begin{maplelatex}
\[
\Phi (r)= - {\displaystyle \frac {1}{2}} \,\mathrm{ln}(r) + 
{\displaystyle \frac {1}{2}} \,\mathrm{ln}(r - 2\,M)
\]
\end{maplelatex}

\emptyline
\noindent
So, Schwarzschild metric out of star is:

\emptyline
\begin{mapleinput}
\mapleinline{active}{1d}{g_matrix := get_compts(g);}{%
}
\end{mapleinput}

\mapleresult
\begin{maplelatex}
\[
\mathit{g\_matrix} :=  \left[ 
{\begin{array}{cccc}
 - e^{(2\,\Phi (r))} & 0 & 0 & 0 \\
0 & e^{(2\,\Lambda (r))} & 0 & 0 \\
0 & 0 & r^{2} & 0 \\
0 & 0 & 0 & r^{2}\,\mathrm{sin}(\theta )^{2}
\end{array}}
 \right] 
\]
\end{maplelatex}

\begin{mapleinput}
\mapleinline{active}{1d}{coord := [t, r, theta, phi]:
 \indent sch_compts :=\\
  array(symmetric,sparse,1..4,1..4):# metric components
  \indent \indent sch_compts[1,1] := expand(\\
subs(Phi(r)=-1/2*ln(r)+1/2*ln(r-2*M),\\
g_matrix[1,1]) ):# coefficient of d(t)^2 in interval
   \indent \indent \indent sch_compts[2,2] := expand(\\
 subs(Lambda(r)=-ln(1-2*M/r)/2,\\
 g_matrix[2,2]) ):# coefficient of d(r)^2 in interval
    \indent \indent sch_compts[3,3] := g_matrix[3,3]:# coefficient\\
     of d(theta)^2 in interval  
     \indent sch_compts[4,4] := g_matrix[4,4]:# coefficient\\
      of d(phi)^2 in interval
      sch :=\\
       create([-1,-1], eval(sch_compts));# Schwarzschild metric}{%
}
\end{mapleinput}

\mapleresult
\begin{maplelatex}
\maplemultiline{
\mathit{sch} :=\\
 \mathrm{table(}[\mathit{compts}= \left[ 
{\begin{array}{cccc}
 - 1 + {\displaystyle \frac {2\,M}{r}}  & 0 & 0 & 0 \\ [2ex]
0 & {\displaystyle \frac {1}{1 - {\displaystyle \frac {2\,M}{r}} 
}}  & 0 & 0 \\ [2ex]
0 & 0 & r^{2} & 0 \\
0 & 0 & 0 & r^{2}\,\mathrm{sin}(\theta )^{2}
\end{array}}
 \right] ,\\
  \,\mathit{index\_char}=[-1, \,-1]])
}
\end{maplelatex}

\emptyline
\noindent
Now we consider the star, which is composed of an incompressible
substance 
$\rho $
 = 
$\rho $
0 = const (later we will use also the following approximation:
\textit{p}=(
$\gamma $
-1)
$\rho $
, where 
$\gamma $
=1 (dust), 4/3 (noncoherent radiation), 2 (hard matter) ).

\noindent
Then the mass is

\emptyline
\begin{mapleinput}
\mapleinline{active}{1d}{m(r) = int(4*Pi*rho0*r^2,r);
 \indent M = subs(r=R,rhs(\%));# full mass}{%
}
\end{mapleinput}

\mapleresult
\begin{maplelatex}
\[
\mathrm{m}(r)={\displaystyle \frac {4}{3}} \,\pi \,\rho 0\,r^{3}
\]
\end{maplelatex}

\begin{maplelatex}
\[
M={\displaystyle \frac {4}{3}} \,\pi \,\rho 0\,R^{3}
\]
\end{maplelatex}

\emptyline
\noindent
and the pressure is

\emptyline
\begin{mapleinput}
\mapleinline{active}{1d}{diff(p(r),r) = subs(\\
 \{rho(r)=rho0,m(r)=4/3*Pi*rho0*r^3\},\\
-(4*Pi*r^3*p(r)+m(r))*(rho(r)+p(r))/\\
(r*(r-2*m(r))) );# Oppenheimer-Volkoff equation
 \indent eq5 := dsolve(\%,p(r));}{%
}
\end{mapleinput}

\mapleresult
\begin{maplelatex}
\[
{\frac {\partial }{\partial r}}\,\mathrm{p}(r)= - {\displaystyle 
\frac {(4\,\pi \,r^{3}\,\mathrm{p}(r) + {\displaystyle \frac {4}{
3}} \,\pi \,\rho 0\,r^{3})\,(\rho 0 + \mathrm{p}(r))}{r\,(r - 
{\displaystyle \frac {8}{3}} \,\pi \,\rho 0\,r^{3})}} 
\]
\end{maplelatex}

\begin{maplelatex}
\maplemultiline{
\mathit{eq5} := \mathrm{p}(r)=\rho 0\,({\displaystyle \frac { - 6
\,\mathrm{\%1} + 2\,\sqrt{ - 3\,\mathrm{\%1} + 8\,\mathrm{\%1}\,
\pi \,\rho 0\,r^{2}}}{ - 3 + 8\,\pi \,\rho 0\,r^{2} - 9\,\mathrm{
\%1}}}  - 1),  \\
\mathrm{p}(r)=\rho 0\,({\displaystyle \frac { - 6\,\mathrm{\%1}
 - 2\,\sqrt{ - 3\,\mathrm{\%1} + 8\,\mathrm{\%1}\,\pi \,\rho 0\,r
^{2}}}{ - 3 + 8\,\pi \,\rho 0\,r^{2} - 9\,\mathrm{\%1}}}  - 1)
 \\
\mathrm{\%1} := e^{( - 16\,\mathit{\_C1}\,\pi \,\rho 0)} }
\end{maplelatex}

\begin{mapleinput}
\mapleinline{active}{1d}{\indent solve(\\
 simplify(subs(r=R,rhs(eq5[1])))=0,\\
  exp(-16*_C1*Pi*rho0));#boundary condition
   \indent solve(\\
    simplify(subs(r=R,rhs(eq5[2])))=0,\\
     exp(-16*_C1*Pi*rho0));#boundary condition}{%
}
\end{mapleinput}

\mapleresult
\begin{maplelatex}
\[
 - 3 + 8\,\pi \,\rho 0\,R^{2}
\]
\end{maplelatex}

\begin{maplelatex}
\[
 - 3 + 8\,\pi \,\rho 0\,R^{2}
\]
\end{maplelatex}

\begin{mapleinput}
\mapleinline{active}{1d}{\indent sol1 := simplify(subs(exp(-16*_C1*Pi*rho0) =
-3+8*Pi*rho0*R^2,rhs(eq5[1])));
 \indent sol2 := simplify(subs(exp(-16*_C1*Pi*rho0) =
-3+8*Pi*rho0*R^2,rhs(eq5[2])));}{%
}
\end{mapleinput}

\mapleresult
\begin{maplelatex}
\maplemultiline{
\mathit{sol1} :=  - {\displaystyle \frac {\rho 0\ }{4}}  \\
{\displaystyle \frac {- 3 + 12\,\pi \,\rho 0\,R^{2} + 
\sqrt{9 - 24\,\pi \,\rho 0\,R^{2} - 24\,\pi \,\rho 0\,r^{2} + 64
\,\pi ^{2}\,\rho 0^{2}\,r^{2}\,R^{2}} - 4\,\pi \,\rho 0\,r^{2}}{
 - 3 - \pi \,\rho 0\,r^{2} + 9\,\pi \,\rho 0\,R^{2}}}  }
\end{maplelatex}

\begin{maplelatex}
\maplemultiline{
\mathit{sol2} := {\displaystyle \frac {\rho 0\ }{4}}  \\
{\displaystyle \frac {3 - 12\,\pi \,\rho 0\,R^{2} + 
\sqrt{9 - 24\,\pi \,\rho 0\,R^{2} - 24\,\pi \,\rho 0\,r^{2} + 64
\,\pi ^{2}\,\rho 0^{2}\,r^{2}\,R^{2}} + 4\,\pi \,\rho 0\,r^{2}}{
 - 3 - \pi \,\rho 0\,r^{2} + 9\,\pi \,\rho 0\,R^{2}}}  }
\end{maplelatex}

\emptyline
\noindent
In the center of star we have

\emptyline
\begin{mapleinput}
\mapleinline{active}{1d}{simplify(subs(r=0,sol1));
 \indent simplify(subs(r=0,sol2));}{%
}
\end{mapleinput}

\mapleresult
\begin{maplelatex}
\[
 - {\displaystyle \frac {1}{12}} \,{\displaystyle \frac {\rho 0\,
( - 3 + 12\,\pi \,\rho 0\,R^{2} + \sqrt{9 - 24\,\pi \,\rho 0\,R^{
2}})}{ - 1 + 3\,\pi \,\rho 0\,R^{2}}} 
\]
\end{maplelatex}

\begin{maplelatex}
\[
 - {\displaystyle \frac {1}{12}} \,{\displaystyle \frac {\rho 0\,
( - 3 + 12\,\pi \,\rho 0\,R^{2} - \sqrt{9 - 24\,\pi \,\rho 0\,R^{
2}})}{ - 1 + 3\,\pi \,\rho 0\,R^{2}}} 
\]
\end{maplelatex}

\emptyline
\noindent
The pressure is infinity when the radius is equal to

\emptyline
\begin{mapleinput}
\mapleinline{active}{1d}{\indent R_crit1 =\\
simplify(solve(expand(denom(\%)/12)=0,R)[1],\\
radical,symbolic);
 \indent R_crit2 =\\
simplify(solve(expand(denom(\%\%)/12)=0,R)[2],\\
radical,symbolic);}{%
}
\end{mapleinput}

\mapleresult
\begin{maplelatex}
\[
\mathit{R\_crit1}={\displaystyle \frac {1}{3}} \,{\displaystyle 
\frac {\sqrt{3}\,\sqrt{\pi \,\rho 0}}{\pi \,\rho 0}} 
\]
\end{maplelatex}

\begin{maplelatex}
\[
\mathit{R\_crit2}= - {\displaystyle \frac {1}{3}} \,
{\displaystyle \frac {\sqrt{3}\,\sqrt{\pi \,\rho 0}}{\pi \,\rho 0
}} 
\]
\end{maplelatex}

\begin{mapleinput}
\mapleinline{active}{1d}{plot3d(\{subs(R=1,sol1),subs(R=1,sol2)\},\\
r=0..1,rho0=0..0.1,axes=boxed,title=`pressure`);#only\\
 positive solution has a physical sense}{%
}
\end{mapleinput}

\mapleresult
\begin{center}
\mapleplot{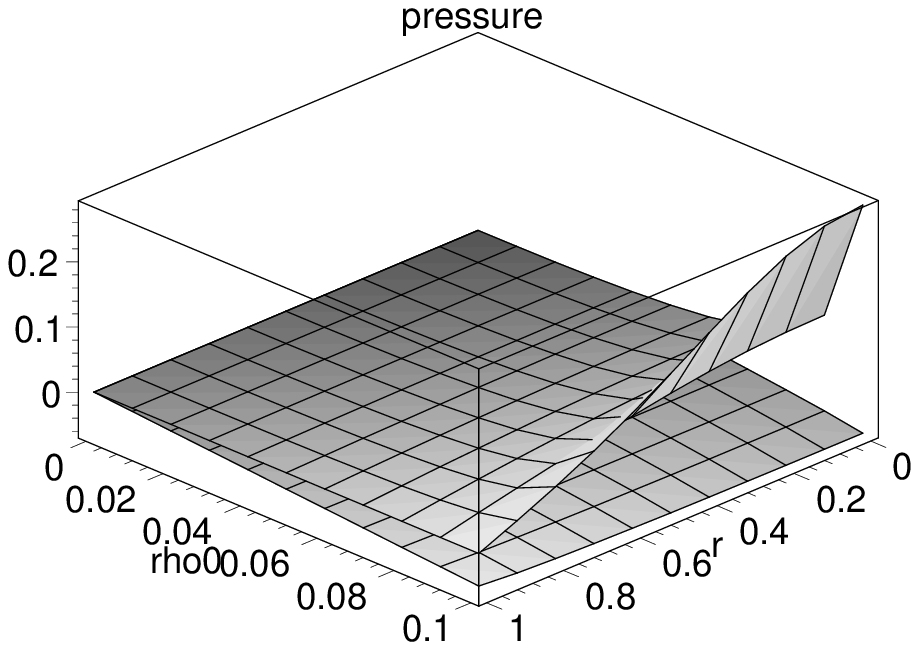}
\end{center}

\emptyline
\noindent
To imagine the space-time geometry it is necessary to consider an
equator section of the star (
$\theta $
=
$\pi $
/2) at fixed time moment in 3-dimensional flat space. The
corresponding procedure is named as "\underline{embedding}". From
the metric tensor, the 2-dimensional line element is 
$\mathit{ds}^{2}$
 = 
$\frac {\mathit{dr}^{2}}{1 - \frac {2\,m}{r}}$
  + 
$r^{2}$
$\mathit{dphi}^{2}$
 and for Euclidean 3-space we have 
$\mathit{ds}^{2}$
 = 
$\mathit{dz}^{2}$
 + 
$\mathit{dr}^{2}$
 +
$r^{2}$
$\mathit{dphi}^{2}$
. We will investigate the 2-surface \textit{z=z}(\textit{r}). As
\textit{dz=
$\frac {\mathit{dz}}{\mathit{dr}}$
 dr,} one can obtain Euclidian line element: 
$\mathit{ds}^{2}$
 = [1+
$(\frac {\mathit{dz}}{\mathit{dr}})^{2}$
] 
$\mathit{dr}^{2}$
 +
$r^{2}$
$\mathit{dphi}^{2}$

\emptyline
\begin{mapleinput}
\mapleinline{active}{1d}{subs(theta=Pi/2,get_compts(sch));#Schwarzschild metric\\
 \indent dr^2*\%[2,2]  + dphi^2*\%[3,3] =  \\
 (1+diff(z(r),r)^2)*dr^2 + r^2*dphi^2;#equality of\\
  intervals of flat and embedded spaces
  \indent \indent diff(z(r),r) = solve(\%,diff(z(r),r))[1];
   \indent \indent \indent dsolve(\%,z(r));# embedding}{%
}
\end{mapleinput}

\mapleresult
\begin{maplelatex}
\[
 \left[ 
{\begin{array}{cccc}
 - 1 + {\displaystyle \frac {2\,M}{r}}  & 0 & 0 & 0 \\ [2ex]
0 & {\displaystyle \frac {1}{1 - {\displaystyle \frac {2\,M}{r}} 
}}  & 0 & 0 \\ [2ex]
0 & 0 & r^{2} & 0 \\
0 & 0 & 0 & r^{2}\,\mathrm{sin}({\displaystyle \frac {1}{2}} \,
\pi )^{2}
\end{array}}
 \right] 
\]
\end{maplelatex}

\begin{maplelatex}
\[
{\displaystyle \frac {\mathit{dr}^{2}}{1 - {\displaystyle \frac {
2\,M}{r}} }}  + r^{2}\,\mathit{dphi}^{2}=(1 + ({\frac {\partial 
}{\partial r}}\,\mathrm{z}(r))^{2})\,\mathit{dr}^{2} + r^{2}\,
\mathit{dphi}^{2}
\]
\end{maplelatex}

\begin{maplelatex}
\[
{\frac {\partial }{\partial r}}\,\mathrm{z}(r)={\displaystyle 
\frac {\sqrt{2}\,\sqrt{(r - 2\,M)\,M}}{r - 2\,M}} 
\]
\end{maplelatex}

\begin{maplelatex}
\[
\mathrm{z}(r)=2\,\sqrt{2\,r\,M - 4\,M^{2}} + \mathit{\_C1}
\]
\end{maplelatex}

\emptyline
\noindent
Now let's take the penultimate equation and to express \textit{M}
through 
${\rho _{0}}$
 = const:

\emptyline
\begin{mapleinput}
\mapleinline{active}{1d}{rho_sol :=\\
 solve(M = 4/3*Pi*rho0*R^3,rho0);#density from full mass
 \indent fun1 :=\\
  Int(1/sqrt(r/((4/3)*Pi*rho0*r^3)-1),r);# inside space
  \indent \indent fun2 :=\\
   Int(1/sqrt(r/((4/3)*Pi*rho0)-1),r);# outside space}{%
}
\end{mapleinput}

\mapleresult
\begin{maplelatex}
\[
\mathit{rho\_sol} := {\displaystyle \frac {3}{4}} \,
{\displaystyle \frac {M}{\pi \,R^{3}}} 
\]
\end{maplelatex}

\begin{maplelatex}
\[
\mathit{fun1} := {\displaystyle \int } 2\,{\displaystyle \frac {1
}{\sqrt{3\,{\displaystyle \frac {1}{r^{2}\,\pi \,\rho 0}}  - 4}}
} \,dr
\]
\end{maplelatex}

\begin{maplelatex}
\[
\mathit{fun2} := {\displaystyle \int } 2\,{\displaystyle \frac {1
}{\sqrt{3\,{\displaystyle \frac {r}{\pi \,\rho 0}}  - 4}}} \,dr
\]
\end{maplelatex}

\begin{mapleinput}
\mapleinline{active}{1d}{fun3 := value(subs(rho0=rho_sol,fun1));# inside
 fun4 := value(subs(rho0=rho_sol,fun2));# outside}{%
}
\end{mapleinput}

\mapleresult
\begin{maplelatex}
\[
\mathit{fun3} := {\displaystyle \frac { - R^{3} + r^{2}\,M}{
\sqrt{ - {\displaystyle \frac { - R^{3} + r^{2}\,M}{r^{2}\,M}} }
\,r\,M}} 
\]
\end{maplelatex}

\begin{maplelatex}
\[
\mathit{fun4} := {\displaystyle \frac {\sqrt{4\,{\displaystyle 
\frac {r\,R^{3}}{M}}  - 4}\,M}{R^{3}}} 
\]
\end{maplelatex}

\emptyline
\noindent
The resulting embedding for equatorial and vertical sections is
presented below (Newtonian case corresponds to horizontal surface,
that is the asymptote for \textit{r--}\TEXTsymbol{>}
$\infty $
). The outside space lies out of outside ring representing star
border.

\emptyline
\begin{mapleinput}
\mapleinline{active}{1d}{\indent \indent fig1 :=\\
plot3d(\\
subs(\\
M=1,subs(R=2.66*M,subs(r=sqrt(x^2+y^2),fun3))),\\
x=-5..5,y=-5..5,\\
grid=[100,100],style=PATCHCONTOUR):# the inside\\
 of star [r in units of M]

\indent \indent fig2 :=\\
plot3d(\\
subs(M=1,subs(R=2.66*M,subs(r=sqrt(x^2+y^2),fun4))),\\
x=-5..5,y=-5..5,style=PATCHCONTOUR):# the outside\\
 of star

\indent \indent display(fig1,fig2,axes=boxed);}{%
}
\end{mapleinput}

\mapleresult
\begin{center}
\mapleplot{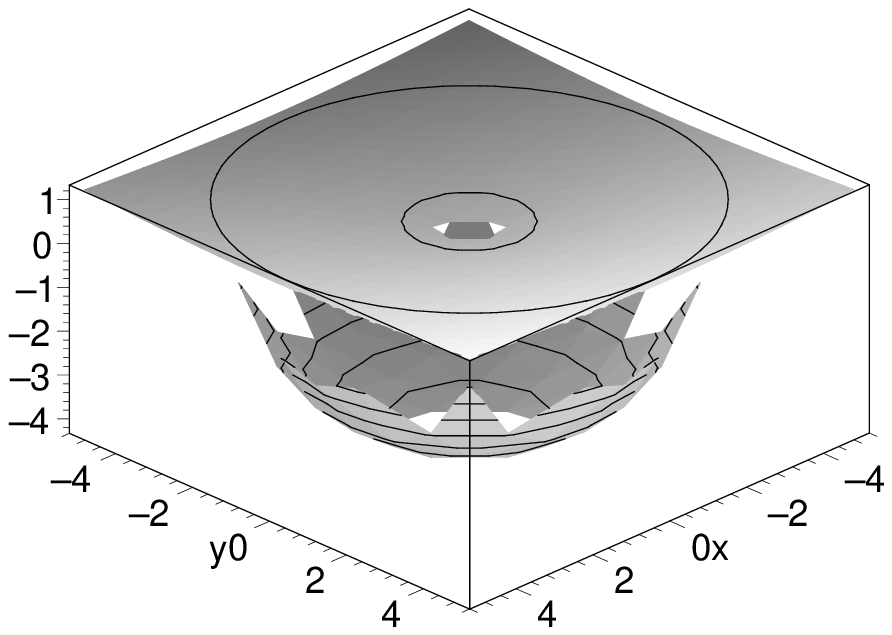}
\end{center}

\emptyline
\noindent
We can see, that the change of radial coordinate \textit{dr} versus
the variation of interval \textit{dl= 
$\frac {\mathit{dr}}{\sqrt{1 - \frac {2\,M}{r}}}$
} (\textit{dl} is the length of segment of curve
on the depicted surface) in vicinity of the star is smaller in compare
with Newtonian case (plane \textit{z}=0, where \textit{dr=dl}). The
star "rolls" the space so that an observer moves away from the star,
but the radial coordinate increases slowly. But what is a hole in the
vicinity of the center of surface, which illustrates the outside
space? What happens if the star radius becomes equal or less than the
radius of this hole \textit{R=}2\textit{M}? Such unique object is the
so-called \underline{black hole} (see below).

\emptyline
\noindent
But before consideration of the main features of black hole let us
consider the motion of probe particle in space with Schwarzschild
metric. We will base our consideration on the relativistic low of
motion: 
$g^{(\alpha , \,\beta )}$
${p_{\alpha }}$
${p_{\beta }}$
 + 
$\mu ^{2}$
 = 0 (\textit{p} is 4-momentum, 
$\mu $
 is the rest mass). Then (if \textit{p}  =[\textit{
$ - E, \,{\frac {\partial }{\partial \lambda }}\,r, \,0, \,L$
}], where \textit{L} corresponds to angle momentum, 
$\lambda $
 is the affine parameter, second term describes the radial velocity):

\emptyline
\begin{mapleinput}
\mapleinline{active}{1d}{-E^2/(1-2*M/r(tau))\\
 + diff(r(tau),tau)^2/(1-2*M/r(tau)) +\\
  L^2/r(tau)^2+1 = 0;# tau=lambda*mu, E=E/mu, L=L/mu
 \indent eq6 := diff(r(tau),tau)^2  =\\
   expand(solve(\%,diff(r(tau),tau)^2));#integral of motion
   \indent \indent V := sqrt(-factor(op(2,rhs(eq6)) + op(3,rhs(eq6)) +\\
    op(4,rhs(eq6)) + op(5,rhs(eq6))));# effective potential}{%
}
\end{mapleinput}

\mapleresult
\begin{maplelatex}
\[
 - {\displaystyle \frac {E^{2}}{1 - {\displaystyle \frac {2\,M}{
\mathrm{r}(\tau )}} }}  + {\displaystyle \frac {({\frac {
\partial }{\partial \tau }}\,\mathrm{r}(\tau ))^{2}}{1 - 
{\displaystyle \frac {2\,M}{\mathrm{r}(\tau )}} }}  + 
{\displaystyle \frac {L^{2}}{\mathrm{r}(\tau )^{2}}}  + 1=0
\]
\end{maplelatex}

\begin{maplelatex}
\[
\mathit{eq6} := ({\frac {\partial }{\partial \tau }}\,\mathrm{r}(
\tau ))^{2}=E^{2} + {\displaystyle \frac {2\,M}{\mathrm{r}(\tau )
}}  - {\displaystyle \frac {L^{2}}{\mathrm{r}(\tau )^{2}}}  + 
{\displaystyle \frac {2\,L^{2}\,M}{\mathrm{r}(\tau )^{3}}}  - 1
\]
\end{maplelatex}

\begin{maplelatex}
\[
V := \sqrt{{\displaystyle \frac {(\mathrm{r}(\tau )^{2} + L^{2})
\,(\mathrm{r}(\tau ) - 2\,M)}{\mathrm{r}(\tau )^{3}}} }
\]
\end{maplelatex}

\emptyline
\noindent
The effective potential determining the orbital motion is (we vary the
angular momentum \textit{L} and \textit{r}):

\emptyline
\begin{mapleinput}
\mapleinline{active}{1d}{subs(\{M=1,r(tau)=x,L=y\},V):
 \indent plot(\{subs(y=3,\%),subs(y=4,\%),subs(y=6,\%),1\},\\
 x=0..30,axes=boxed,color=[red,green,blue,brown],\\
 title=`potentials`,\\
 view=0.6..1.4);#level E=1 corresponds\\
  to a particle, which was initially motionless}{%
}
\end{mapleinput}

\mapleresult
\begin{center}
\mapleplot{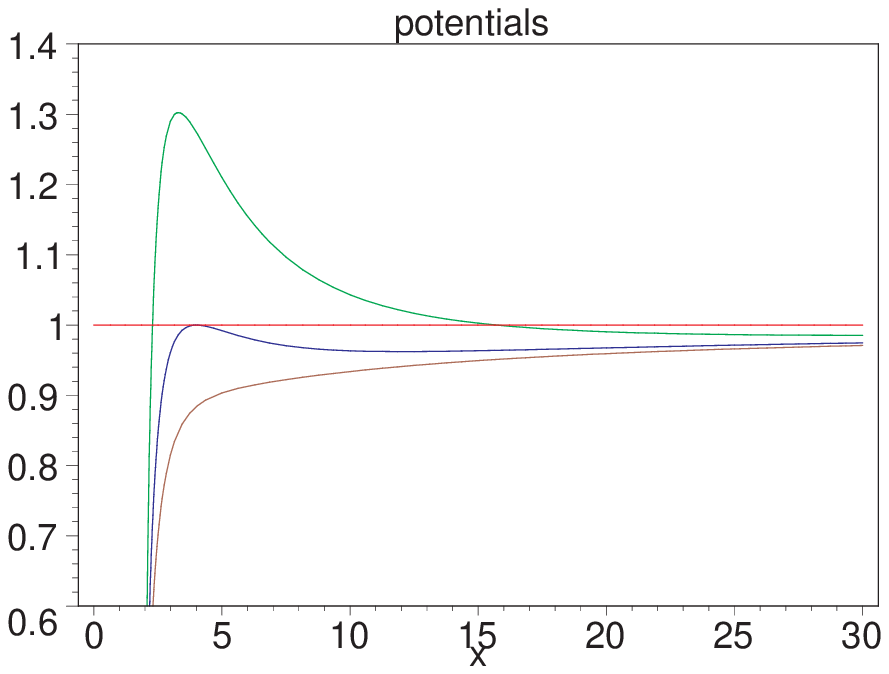}
\end{center}

\emptyline
\noindent
We considered the motion in the relativistic potential in the first
part, but only in the framework of linear approximation (weak field).
As result of such motion the perihelion rotation and light ray
deflection appear. But now, in the more general case, one can see a
very interesting feature of the relativistic orbital motion: unlike
Newtonian case there is the maximum 
${V_{\mathit{max}}}$
 with the following potential decrease as result of radial coordinate
decrease. If 
${V_{\mathit{max}}}$
\TEXTsymbol{>}\textit{E }\TEXTsymbol{>} 1 the motion is infinite (it
is analogue of the hyperbolical motion in Newtonian case). 
${V_{\mathit{max}}}$
\TEXTsymbol{>}\textit{E}=1 corresponds to parabolic motion. When
E lies in the potential hole or \textit{E }= \textit{V}, we
have the finite motion. The motion with energy, which is equal to
extremal values of potential, corresponds to circle orbits (the stable
motion for potential minimum, unstable one for the maximum). The existence
of the extremes is defined by expressions:

\emptyline
\begin{mapleinput}
\mapleinline{active}{1d}{numer( simplify( diff( subs(r(tau)=r,V), r) ) ) =\\
 0:# zeros of potential's derivative
 \indent solve(\%,r);}{%
}
\end{mapleinput}

\mapleresult
\begin{maplelatex}
\[
{\displaystyle \frac {1}{2}} \,{\displaystyle \frac {(L + \sqrt{L
^{2} - 12\,M^{2}})\,L}{M}} , \,{\displaystyle \frac {1}{2}} \,
{\displaystyle \frac {(L - \sqrt{L^{2} - 12\,M^{2}})\,L}{M}} 
\]
\end{maplelatex}

\emptyline
\noindent
As consequence, the circle motion is possible if \textit{L}
$\geq$ 2\textit{M
$\sqrt{3}$
}, when \textit{r} $\geq$ 3(2\textit{M}).  The minimal
distance, which allows the circular unstable motion, is defined by

\emptyline
\begin{mapleinput}
\mapleinline{active}{1d}{limit(1/2*(L-sqrt(L^2-12*M^2))*L/M, L=infinity);}{%
}
\end{mapleinput}

\mapleresult
\begin{maplelatex}
\[
3\,M
\]
\end{maplelatex}

\emptyline
\noindent
Lack of extremes (red curve) or transition to \textit{r}
\TEXTsymbol{<} 3\textit{M} results in the falling motion (the similar
case is \textit{E} \TEXTsymbol{>} 
${V_{\mathit{max}}}$
). This is the so-called gravitational capture and has no analogy in
Newtonian mechanics of point-like masses. The finite orbital motion in
the case of the right local minimum existence (blue curve) is the analog
of one in the Newtonian case, but has no elliptical character (see
first part).

\noindent
As the particular case, let's consider the radial (\textit{L}=0) fall
of particle. The proper time of falling particle is defined by next
expression

\emptyline
\begin{mapleinput}
\mapleinline{active}{1d}{tau = Int(\\ 1/sqrt(subs(\{L=0,r(tau)=r\},rhs(eq6))),r);
 \indent tau = Int(1/(sqrt(2*M/r-2*M/R)),r);#R=2*M/(1-E^2) -\\
  apoastr, i.e. radius of zero velocity
  \indent \indent limit( value( rhs(\%) ),r=2*M ):
   \indent simplify(\%);#So, this is convergent integral\\
    for r-->2*M}{%
}
\end{mapleinput}

\mapleresult
\begin{maplelatex}
\[
\tau ={\displaystyle \int } {\displaystyle \frac {1}{\sqrt{E^{2}
 + {\displaystyle \frac {2\,M}{r}}  - 1}}} \,dr
\]
\end{maplelatex}

\begin{maplelatex}
\[
\tau ={\displaystyle \int } {\displaystyle \frac {1}{\sqrt{2\,
{\displaystyle \frac {M}{r}}  - {\displaystyle \frac {2\,M}{R}} }
}} \,dr
\]
\end{maplelatex}

\begin{maplelatex}
\maplemultiline{
{\displaystyle \frac {1}{4}} \sqrt{{\displaystyle \frac { - 2\,M
 + R}{R}} }\,R ( - 2\,\sqrt{2}\,\sqrt{ - ( - 2\,M + R)\,M} + R\,\mathrm{ln}(2)\\
 - R\,\mathrm{ln}( - R + 4\,M + 2\,\sqrt{2}\,\sqrt{ - ( - 2\,M + 
R)\,M}))\,\sqrt{2} \\
 \left/ {\vrule height0.41em width0em depth0.41em} \right. \! 
 \! \sqrt{ - ( - 2\,M + R)\,M} }
\end{maplelatex}

\emptyline
\noindent
For the remote exterior observer we have 
${\frac {\partial }{\partial \tau }}\,r$
 = 
${\frac {\partial }{\partial t}}\,r$
${\frac {\partial }{\partial \tau }}\,t$
=
${\frac {\partial }{\partial t}}\,r$
$\frac {E}{1 - \frac {2\,M}{r}}$
  =\\
   \textit{E ${\frac {\partial }{\partial t}}\,{r_{o}}$
}
\noindent
(here 
${r_{0}}$
 is the time coordinate relatively infinitely remote observer):

\emptyline
\begin{mapleinput}
\mapleinline{active}{1d}{r[o] := Int(1/(1-2*M/r),r) = int(1/(1-2*M/r),r);
 limit( value( rhs(\%) ),r=2*M );#this is\\
  divergent integral for r-->2*M}{%
}
\end{mapleinput}

\mapleresult
\begin{maplelatex}
\[
{r_{o}} := {\displaystyle \int } {\displaystyle \frac {1}{1 - 
{\displaystyle \frac {2\,M}{r}} }} \,dr=r + 2\,M\,\mathrm{ln}(r
 - 2\,M)
\]
\end{maplelatex}

\begin{maplelatex}
\[
 - \mathrm{signum}(M)\,\infty 
\]
\end{maplelatex}

\emptyline
\noindent
These equations will be utilized below.

\emptyline

\subsection{Degeneracy stars and gravitational collapse}

\emptyline
The previously obtained expressions for the time of radial fall have an
interesting peculiarity if the radius of star is less or equal to
2\textit{M }(
$\frac {2\,G\,M}{c^{2}}$
 in dimensional case, that is the so-called "\underline{gravitational
radius}" 
${R_{g}}$
). The finite proper time 
$\tau $
 of fall corresponds to the infinite "remote" time 
${r_{o}}$
, when \textit{r--}\TEXTsymbol{>}2\textit{M. }This means, that for
the remote observer the fall does not finish never.\textit{ }It
results from the relativistic time's slowing-down. The consequence of
this phenomena is the infinite red shift of "falling" photons because
of the red shift in Schwarzschild metric is defined by the next
frequencies relation: 
$\frac {{\omega _{1}}}{{\omega _{2}}}$
 =
$\frac {\Delta \,{\tau _{2}}}{\Delta \,{\tau _{1}}}$
=
$\sqrt{\frac {{g_{0, \,0}}({r_{2}})}{{g_{0, \,0}}({r_{1}})}}$
= 
$\frac {\sqrt{1 - \frac {2\,M}{{r_{2}}}}}{\sqrt{1 - \frac {2\,M}{
{r_{1}}}}}$
  (here 
$\Delta $
$\tau $
 are the proper time intervals between light flash in different
radial points). Simultaneously, the escape velocity is equal to 
$\sqrt{\frac {2\,G\,M}{R}}$
 that is the velocity of light for sphere with \textit{R}=
${R_{g}}$
. So, we cannot receive any information from interior of black hole.

\emptyline
\noindent
How can appear the object with the radius, which is less than 
${R_{g}}$
? In the beginning we consider the pressure free radial (i.e. angular
part of metric is equal to 0) collapse of dust sphere with mass
\textit{M}.

\emptyline
\begin{mapleinput}
\mapleinline{active}{1d}{r := 'r':
 \indent E := 'E':
  \indent \indent subs( r=r(t),get_compts(sch) ):
   \indent d(s)^2 =\\
    \%[1,1]*d(t)^2 + \%[2,2]*d(r)^2;#Schwarzschild metric
    -d(tau)^2 = collect(\\
    subs( d(r)=diff(r(t),t)*d(t),rhs(\%)),d(t));#tau is\\
     the proper time for the observer on the surface\\
      of sphere
      \indent \%/d(tau)^2;
     \indent \indent subs(\{d(t)=E/(1-2*M/r(t)),d(tau)=1\},\%);#we used \\
     d(t)/d(tau) = E/(1-2*M/r)
    \indent pot_1 :=\\
     factor( solve(\%,(diff(r(t),t))^2) );#"potential"}{%
}
\end{mapleinput}

\mapleresult
\begin{maplelatex}
\[
\mathrm{d}(s)^{2}=( - 1 + {\displaystyle \frac {2\,M}{\mathrm{r}(
t)}} )\,\mathrm{d}(t)^{2} + {\displaystyle \frac {\mathrm{d}(r)^{
2}}{1 - {\displaystyle \frac {2\,M}{\mathrm{r}(t)}} }}
\]
\end{maplelatex}

\begin{maplelatex}
\[
 - \mathrm{d}(\tau )^{2}= \left(  \!  - 1 + {\displaystyle 
\frac {2\,M}{\mathrm{r}(t)}}  + {\displaystyle \frac {({\frac {
\partial }{\partial t}}\,\mathrm{r}(t))^{2}}{1 - {\displaystyle 
\frac {2\,M}{\mathrm{r}(t)}} }}  \!  \right) \,\mathrm{d}(t)^{2}
\]
\end{maplelatex}

\begin{maplelatex}
\[
-1={\displaystyle \frac { \left(  \!  - 1 + {\displaystyle 
\frac {2\,M}{\mathrm{r}(t)}}  + {\displaystyle \frac {({\frac {
\partial }{\partial t}}\,\mathrm{r}(t))^{2}}{1 - {\displaystyle 
\frac {2\,M}{\mathrm{r}(t)}} }}  \!  \right) \,\mathrm{d}(t)^{2}
}{\mathrm{d}(\tau )^{2}}} 
\]
\end{maplelatex}

\begin{maplelatex}
\[
-1={\displaystyle \frac { \left(  \!  - 1 + {\displaystyle 
\frac {2\,M}{\mathrm{r}(t)}}  + {\displaystyle \frac {({\frac {
\partial }{\partial t}}\,\mathrm{r}(t))^{2}}{1 - {\displaystyle 
\frac {2\,M}{\mathrm{r}(t)}} }}  \!  \right) \,E^{2}}{(1 - 
{\displaystyle \frac {2\,M}{\mathrm{r}(t)}} )^{2}}} 
\]
\end{maplelatex}

\begin{maplelatex}
\[
\mathit{pot\_1} := {\displaystyle \frac {(\mathrm{r}(t) - 2\,M)^{
2}\,( - \mathrm{r}(t) + E^{2}\,\mathrm{r}(t) + 2\,M)}{E^{2}\,
\mathrm{r}(t)^{3}}} 
\]
\end{maplelatex}

\begin{mapleinput}
\mapleinline{active}{1d}{plot(\{subs(\{E=0.5,M=1,r(t)=r\},pot_1),0*r\},\\
r=1.7..3,axes=boxed,title=`(dr/dt)^2 vs r`);}{%
}
\end{mapleinput}

\mapleresult
\begin{center}
\mapleplot{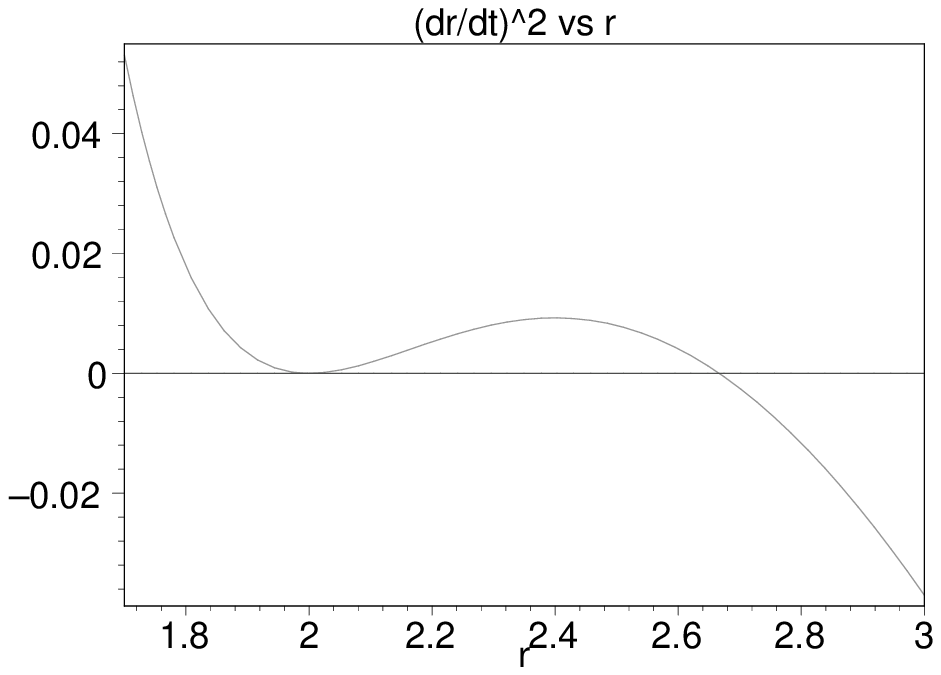}
\end{center}

\emptyline
\noindent
The collapse is the motion from the right point of zero velocity 
${\frac {\partial }{\partial t}}\,r$
 (this is the velocity for remote observer) to the left point of zero
velocity. These points are

\emptyline
\begin{mapleinput}
\mapleinline{active}{1d}{solve(subs(r(t)=r,pot_1)=0,r);}{%
}
\end{mapleinput}

\mapleresult
\begin{maplelatex}
\[
2\,M, \,2\,M, \, - 2\,{\displaystyle \frac {M}{ - 1 + E^{2}}} 
\]
\end{maplelatex}

\emptyline
\noindent
that are the apoastr (see above) and the gravitational radius. Hence
(from the value for apoastr and \textit{pot\_1})\\ 
$({\frac {\partial }{\partial t}}\,r)^{2}$
 = 
$\frac {(1 - \frac {2\,M}{r})^{2}\,(\frac {2\,M}{r} - 1 + E^{2})
}{E^{2}}$
 = 
$\frac {(1 - \frac {2\,M}{r})^{2}\,(\frac {2\,M}{r} - \frac {2\,M
}{R})}{1 - \frac {2\,M}{R}}$
 = 
$(1 - \frac {{R_{g}}}{r})^{2}\, \left(  \! 1 - \frac {1 - \frac {
{R_{g}}}{r}}{1 - \frac {{R_{g}}}{R}} \!  \right)$. 

\noindent
The slowing-down of the collapse for remote observer in vicinity of 
${R_{g}}$
 results from the time slowing-down (see above). But the collapsing
observer has the proper velocity 
${\frac {\partial }{\partial \tau }}\,r$
: 

\emptyline
\begin{mapleinput}
\mapleinline{active}{1d}{pot_2 := simplify(pot_1*(E/(1-2*M/r(t)))^2);#
\indent d/d(t)=(d/d(tau))*(1-2*M/r)/E}{%
}
\end{mapleinput}

\mapleresult
\begin{maplelatex}
\[
\mathit{pot\_2} := {\displaystyle \frac { - \mathrm{r}(t) + E^{2}
\,\mathrm{r}(t) + 2\,M}{\mathrm{r}(t)}} 
\]
\end{maplelatex}

\begin{mapleinput}
\mapleinline{active}{1d}{plot(\{subs(\{E=0.5,M=1,r(t)=r\},pot_2),0*r\},\\
r=1.7..3,axes=boxed,title=`(dr/dtau)^2 vs r`);}{%
}
\end{mapleinput}

\mapleresult
\begin{center}
\mapleplot{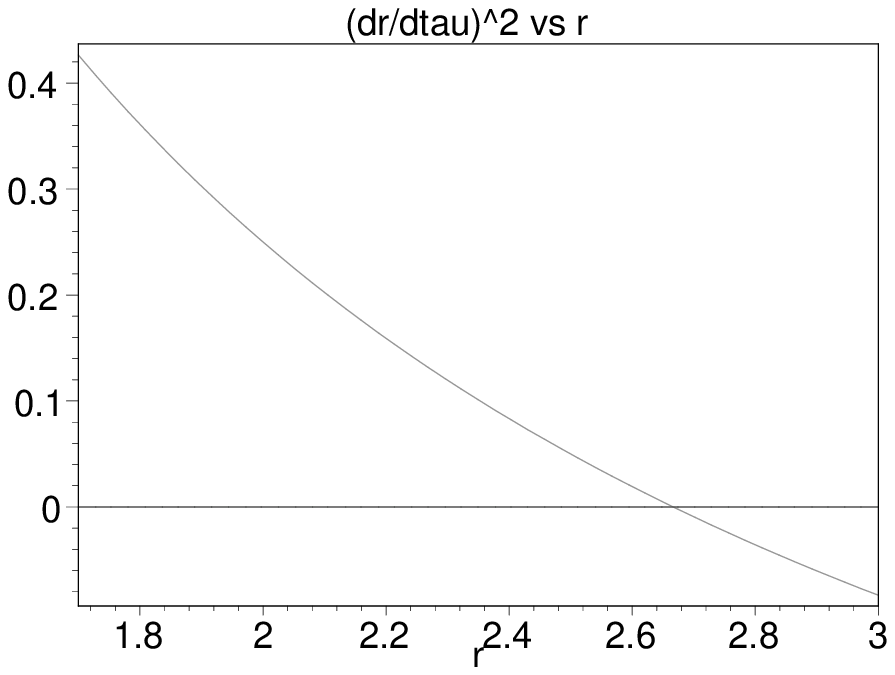}
\end{center}

\emptyline
\noindent
As result, the collapsing surface crosses the gravitational radius at
finite time moment and 
$\lim _{r\rightarrow {R_{g}}}\,{\frac {\partial }{\partial t}}\,r
$
 =1 (that is the velocity of light), when \textit{E }= 1 (\textit{R}=
$\infty $
). Time of collapse for falling observer is

\emptyline
\begin{mapleinput}
\mapleinline{active}{1d}{Int(1/sqrt(R/r-1),r)/sqrt(1-E^2);#from pot_2,\\
 R is apoastr simplify( value(\%), radical, symbolic);}{%
}
\end{mapleinput}

\mapleresult
\begin{maplelatex}
\[
{\displaystyle \frac {{\displaystyle \int } {\displaystyle 
\frac {1}{\sqrt{{\displaystyle \frac {R}{r}}  - 1}}} \,dr}{\sqrt{
1 - E^{2}}}} 
\]
\end{maplelatex}

\begin{maplelatex}
\[
 - {\displaystyle \frac {1}{2}} \,{\displaystyle \frac {2\,\sqrt{
r\,(R - r)} + R\,\mathrm{arctan}({\displaystyle \frac {1}{2}} \,
{\displaystyle \frac {R - 2\,r}{\sqrt{r\,(R - r)}}} )}{\sqrt{1 - 
E^{2}}}} 
\]
\end{maplelatex}

\emptyline
\noindent
It is obvious, that the obtained integral is 
$\frac {1}{\sqrt{1 - E^{2}}}$
$\int _{r}^{R}\frac {1}{\sqrt{\frac {R}{r} - 1}}\,dr$
  = 
$\frac {\pi \,R}{\sqrt{1 - E^{2}}}$
 = 
$\frac {\pi \,M}{(1 - E^{2})^{(\frac {3}{2})}}$
.      

\emptyline
\emptyline
\noindent
It is natural, that the real stars are not clouds of dust and the
equilibrium is supported by nuclear reactions. But when the nuclear
reactions in the star finish, the "fall" of the star's surface can
be prevented only by pressure of electron or baryons. The equilibrium
state corresponds to the minimum of the net-energy composed of
gravitational energy 
$\frac { - G\,M^{2}}{R}$
 and thermal kinetic energy of composition. Let consider the hydrogen
ball. When the temperature tends to zero, the kinetic energy of
electrons is not equal to zero due to quantum degeneracy. In this case
one electron occupies the cell with volume \symbol{126}
$\lambda ^{3}$
, where 
$\lambda $
=
$\frac {h}{2\,\pi \,{p_{e}}}$
 is the Compton wavelength (
${p_{e}}$
 is the electron's momentum). For non-relativistic electron:

\emptyline
\begin{mapleinput}
\mapleinline{active}{1d}{E[e] := p[e]^2/m[e];#kinetic energy of electron
 \indent E[k] := simplify( n[e]*R^3*subs( p[e]=\\
 h*n[e]^(1/3)/(2*pi),E[e] ) );#full kinetic energy,\\
  n[e] is the number density of electrons
  \indent \indent E[k] := subs( n[e]=M/R^3/m[p],\%);# m[p] is the mass\\
   of proton. We supposed m[e]<<m[p] and n[p]=n[e]
   \indent \indent \indent E[g] := -G*M^2/R:# gravitational energy
    \indent \indent \indent \indent E := E[g] + E[k];# full energy}{%
}
\end{mapleinput}

\mapleresult
\begin{maplelatex}
\[
{E_{e}} := {\displaystyle \frac {{p_{e}}^{2}}{{m_{e}}}} 
\]
\end{maplelatex}

\begin{maplelatex}
\[
{E_{k}} := {\displaystyle \frac {1}{4}} \,{\displaystyle \frac {{
n_{e}}^{(5/3)}\,R^{3}\,h^{2}}{\pi ^{2}\,{m_{e}}}} 
\]
\end{maplelatex}

\begin{maplelatex}
\[
{E_{k}} := {\displaystyle \frac {1}{4}} \,{\displaystyle \frac {(
{\displaystyle \frac {M}{R^{3}\,{m_{p}}}} )^{(5/3)}\,R^{3}\,h^{2}
}{\pi ^{2}\,{m_{e}}}} 
\]
\end{maplelatex}

\begin{maplelatex}
\[
E :=  - {\displaystyle \frac {G\,M^{2}}{R}}  + {\displaystyle 
\frac {{\displaystyle \frac {1}{4}} \,({\displaystyle \frac {M}{R
^{3}\,{m_{p}}}} )^{(5/3)}\,R^{3}\,h^{2}}{\pi ^{2}\,{m_{e}}}} 
\]
\end{maplelatex}

\begin{mapleinput}
\mapleinline{active}{1d}{pot := -1/R+1/R^2:# the dependence of energy on R
 \indent plot(pot,R=0.5..10, \\
 title=`energy of degenerated star`);}{%
}
\end{mapleinput}

\mapleresult
\begin{center}
\mapleplot{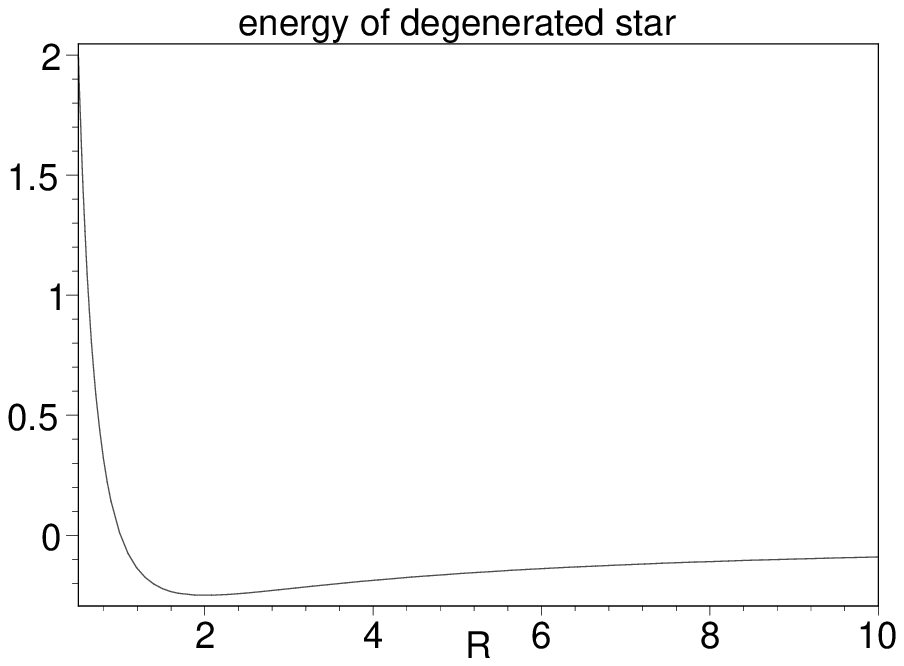}
\end{center}

\emptyline
\noindent
One can see that the dependence of energy on radius has the minimum and,
as consequence, there is the equilibrium state.

\emptyline
\begin{mapleinput}
\mapleinline{active}{1d}{solve( diff(E, R) = 0, R);# minimum of energy\\
 corresponding to equilibrium state}{%
}
\end{mapleinput}

\mapleresult
\begin{maplelatex}
\maplemultiline{
{\displaystyle \frac {1}{2}} \,{\displaystyle \frac {\mathrm{\%1}
\,h^{2}}{M\,{m_{p}}^{2}\,\pi ^{2}\,{m_{e}}\,G}} , \,
{\displaystyle \frac {1}{2}} \,{\displaystyle \frac { \left(  \! 
 - {\displaystyle \frac {1}{2}} \,{\displaystyle \frac {\mathrm{
\%1}}{M\,{m_{p}}}}  + {\displaystyle \frac {{\displaystyle 
\frac {1}{2}} \,I\,\sqrt{3}\,\mathrm{\%1}}{M\,{m_{p}}}}  \! 
 \right) \,h^{2}}{\pi ^{2}\,{m_{p}}\,{m_{e}}\,G}} , \,\\
{\displaystyle \frac {1}{2}} \,{\displaystyle \frac { \left(  \! 
 - {\displaystyle \frac {1}{2}} \,{\displaystyle \frac {\mathrm{
\%1}}{M\,{m_{p}}}}  + {\displaystyle \frac {{\displaystyle 
\frac {-1}{2}} \,I\,\sqrt{3}\,\mathrm{\%1}}{M\,{m_{p}}}}  \! 
 \right) \,h^{2}}{\pi ^{2}\,{m_{p}}\,{m_{e}}\,G}}  \\
\mathrm{\%1} := (M^{2}\,{m_{p}})^{(1/3)} }
\end{maplelatex}

\emptyline
\noindent
So, we have the equilibrium state with 
${R_{\mathit{min}}}$
 \symbol{126} 
$\frac {h^{2}}{G\,M^{(\frac {1}{3})}\,{m_{e}}\,{m_{p}}^{(\frac {5
}{3})}}$
  and the star, which exists in such equilibrium state as result of
the termination of nuclear reactions with the following radius
decrease, is the \underline{white dwarf}. The value of number density
in this state is

\emptyline
\begin{mapleinput}
\mapleinline{active}{1d}{simplify( subs(\\
R[min]=h^2/(G*M^(1/3)*m[e]*m[p]^(5/3)),\\
n[e]=M/R[min]^3/m[p]));#equilibrium number density}{%
}
\end{mapleinput}

\mapleresult
\begin{maplelatex}
\[
{n_{e}}={\displaystyle \frac {M^{2}\,G^{3}\,{m_{e}}^{3}\,{m_{p}}
^{4}}{h^{6}}} 
\]
\end{maplelatex}

\emptyline
\noindent
So, the mass's increase decreases the equilibrium radius (unlike model
of liquid drop, see above) and to increase the number density (the maximal
density of solid or liquid defined by atomic packing is
\symbol{126}20 g/
$\mathit{cm}^{3}$
, for white dwarf this value is \symbol{126}
$10^{7}$
 g/
$\mathit{cm}^{3}$
!). But, in compliance with principle of uncertainty, the last causes
the electron momentum increase (
${p_{e}}\,{n_{e}}^{( - \frac {1}{3})}$
 \symbol{126}\textit{ h}). Our nonrelativistic approximation implies 
${p_{e}}$
\TEXTsymbol{<}\TEXTsymbol{<} 
${m_{e}}\,c$
  or

\emptyline
\begin{mapleinput}
\mapleinline{active}{1d}{simplify( rhs(\%)^(1/3)*h ) - c*m[e];#this must be\\
 large negative value
 \indent solve(\%=0,M);}{%
}
\end{mapleinput}

\mapleresult
\begin{maplelatex}
\[
({\displaystyle \frac {M^{2}\,G^{3}\,{m_{e}}^{3}\,{m_{p}}^{4}}{h
^{6}}} )^{(1/3)}\,h - c\,{m_{e}}
\]
\end{maplelatex}

\begin{maplelatex}
\[
{\displaystyle \frac {\sqrt{G\,c\,h}\,h\,c}{G^{2}\,{m_{p}}^{2}}} 
, \, - {\displaystyle \frac {\sqrt{G\,c\,h}\,h\,c}{G^{2}\,{m_{p}}
^{2}}} 
\]
\end{maplelatex}

\emptyline
\noindent
The last result gives the nonrelativistic criterion \textit{M}
\TEXTsymbol{<}\TEXTsymbol{<} 
$\frac {(h\,c)^{(\frac {3}{2})}}{{m_{p}}^{2}\,G^{(\frac {3}{2})}}
$
. Therefore in the massive white dwarf the electrons have to be the relativistic
particles:

\emptyline
\begin{mapleinput}
\mapleinline{active}{1d}{E[e] := h*c*n[e]^(1/3);#E=p*c for relativistic particle
 \indent E[k] := simplify( n[e]*R^3*E[e] );
  \indent \indent E[k] := subs( n[e]=M/R^3/m[p],\%);
   \indent \indent \indent E[g] := -G*M^2/R:
    \indent \indent E := E[g] + E[k];#this is dependence -a/R+b/R
#equilibrium state:
\indent simplify( diff(E, R), radical ):
 numer(\%);#there is not dependence on R therefore\\
  there is not stable configuration with energy minimum
  \indent solve( \%=0, M );}{%
}
\end{mapleinput}

\mapleresult
\begin{maplelatex}
\[
{E_{e}} := h\,c\,{n_{e}}^{(1/3)}
\]
\end{maplelatex}

\begin{maplelatex}
\[
{E_{k}} := {n_{e}}^{(4/3)}\,R^{3}\,h\,c
\]
\end{maplelatex}

\begin{maplelatex}
\[
{E_{k}} := ({\displaystyle \frac {M}{R^{3}\,{m_{p}}}} )^{(4/3)}\,
R^{3}\,h\,c
\]
\end{maplelatex}

\begin{maplelatex}
\[
E :=  - {\displaystyle \frac {G\,M^{2}}{R}}  + ({\displaystyle 
\frac {M}{R^{3}\,{m_{p}}}} )^{(4/3)}\,R^{3}\,h\,c
\]
\end{maplelatex}

\begin{maplelatex}
\[
 - ( - G\,M\,{m_{p}} + ({\displaystyle \frac {M}{R^{3}\,{m_{p}}}
} )^{(1/3)}\,h\,c\,R)\,M
\]
\end{maplelatex}

\begin{maplelatex}
\[
0, \,{\displaystyle \frac {\sqrt{G\,c\,h}\,h\,c}{G^{2}\,{m_{p}}^{
2}}} , \, - {\displaystyle \frac {\sqrt{G\,c\,h}\,h\,c}{G^{2}\,{m
_{p}}^{2}}} 
\]
\end{maplelatex}

\emptyline
\noindent
We obtained the estimation for so-called critical mass 
${M_{c}}$
\symbol{126}
$\frac {(h\,c)^{(\frac {3}{2})}}{{m_{p}}^{2}\,G^{(\frac {3}{2})}}
$
  =1.4
${M_{\mathit{sun}}}$
 (so-called \underline{Chandrasekhar limit} for white dwarfs). The
smaller masses produce the white dwarf with non-relativistic electrons
but larger masses causes \underline{collapse} of star, which can not
be prevented by pressure of degeneracy electrons. Such collapse can
form a \underline{neutron star} for 1.4
${M_{\mathit{sun}}}$
 \TEXTsymbol{<}\textit{ M \TEXTsymbol{<}} 3
${M_{\mathit{sun}}}$. Collapse for larger masses has to result in the black hole formation.

\emptyline

\subsection{Schwarzschild black hole}

\emptyline
\noindent
Now we return to Schwarzschild metric.

\emptyline
\begin{mapleinput}
\mapleinline{active}{1d}{get_compts(sch);}{%
}
\end{mapleinput}

\mapleresult
\begin{maplelatex}
\[
 \left[ 
{\begin{array}{cccc}
 - 1 + {\displaystyle \frac {2\,M}{r}}  & 0 & 0 & 0 \\ [2ex]
0 & {\displaystyle \frac {1}{1 - {\displaystyle \frac {2\,M}{r}} 
}}  & 0 & 0 \\ [2ex]
0 & 0 & r^{2} & 0 \\
0 & 0 & 0 & r^{2}\,\mathrm{sin}(\theta )^{2}
\end{array}}
 \right] 
\]
\end{maplelatex}

\emptyline
\noindent
One can see two singularities: \textit{r}=2\textit{M} and
\textit{r}=0. What is a sense of first singularity? When we cross the
horizon, 
${\mathit{sh}_{1, \,1}}$
 and 
${\mathit{sh}_{2, \,2}}$
  (i.e. 
${g_{0, \,0}}$
 and 
${g_{1, \,1}}$
) change signs. The space and time exchange the roles! The fall gets
inevitable as the time flowing. As consequence, when particle or
signal cross the gravitational radius, they cannot escape the falling
on \textit{r}=0. This fact can be illustrated by infinite value of
acceleration on \textit{r=}2\textit{M}, which is \textit{-
$\frac {{(\Gamma ^{\alpha })_{0, \,0}}}{{g_{0, \,0}}}$
} (
$\Gamma $
 are the Christoffel symbols):

\emptyline
\begin{mapleinput}
\mapleinline{active}{1d}{D1sch := d1metric( sch, coord ):
 \indent Cf1 := Christoffel1 ( D1sch ):
  \indent \indent displayGR(Christoffel1,\%);}{%
}
\end{mapleinput}

\mapleresult
\begin{maplelatex}
\[
\mathit{The\ Christoffel\ Symbols\ of\ the\ First\ Kind}
\]
\end{maplelatex}

\begin{maplelatex}
\[
\mathit{non-zero\ components\ :}
\]
\end{maplelatex}

\begin{maplelatex}
\[
\mathit{\ [11,2]}={\displaystyle \frac {M}{r^{2}}} 
\]
\end{maplelatex}

\begin{maplelatex}
\[
\mathit{\ [12,1]}= - {\displaystyle \frac {M}{r^{2}}} 
\]
\end{maplelatex}

\begin{maplelatex}
\[
\mathit{\ [22,2]}= - {\displaystyle \frac {M}{(r - 2\,M)^{2}}} 
\]
\end{maplelatex}

\begin{maplelatex}
\[
\mathit{\ [23,3]}=r
\]
\end{maplelatex}

\begin{maplelatex}
\[
\mathit{\ [24,4]}=r\,\mathrm{sin}(\theta )^{2}
\]
\end{maplelatex}

\begin{maplelatex}
\[
\mathit{\ [33,2]}= - r
\]
\end{maplelatex}

\begin{maplelatex}
\[
\mathit{\ [34,4]}=r^{2}\,\mathrm{sin}(\theta )\,\mathrm{cos}(
\theta )
\]
\end{maplelatex}

\begin{maplelatex}
\[
\mathit{\ [44,2]}= - r\,\mathrm{sin}(\theta )^{2}
\]
\end{maplelatex}

\begin{maplelatex}
\[
\mathit{\ [44,3]}= - r^{2}\,\mathrm{sin}(\theta )\,\mathrm{cos}(
\theta )
\]
\end{maplelatex}

\begin{mapleinput}
\mapleinline{active}{1d}{-get_compts(Cf1)[1,1,2]/get_compts(sch)[1,1];#radial\\
 component of acceleration}{%
}
\end{mapleinput}

\mapleresult
\begin{maplelatex}
\[
 - {\displaystyle \frac {M}{r^{2}\,( - 1 + {\displaystyle \frac {
2\,M}{r}} )}} 
\]
\end{maplelatex}

\emptyline
\noindent
Such particles and signals will be expelled from the cause-effect
chain of universe. Therefore an imaginary surface \textit{r}=
${R_{g}}$
 is named "\underline{event horizon}".\\

\noindent
The absence of true physical singularity for \textit{r}=
${R_{g}}$
 can be illustrated in the following way. The invariant of Riemann
tensor 
${R_{\alpha , \,\beta , \,\gamma , \,\delta }}\,R^{(\alpha , \,
\beta , \,\gamma , \,\delta )}$
 is

\emptyline
\begin{mapleinput}
\mapleinline{active}{1d}{schinv := invert( sch, 'detg' ):
 \indent D2sch := d2metric( D1sch, coord ):
  \indent \indent Cf1 := Christoffel1 ( D1sch ):
   \indent \indent \indent RMN := Riemann( schinv, D2sch, Cf1 ):
    \indent \indent raise(schinv,RMN,1):#raise of indexes\\
     in Riemann tensor
     \indent raise(schinv,\%,2):
    raise(schinv,\%,3):
   \indent RMNinv := raise(schinv,\%,4):
  \indent \indent prod(RMN,RMNinv,[1,1],[2,2],[3,3],[4,4]);}{%
}
\end{mapleinput}

\mapleresult
\begin{maplelatex}
\[
\mathrm{table(}[\mathit{compts}=48\,{\displaystyle \frac {M^{2}}{
r^{6}}} , \,\mathit{index\_char}=[]])
\]
\end{maplelatex}

\emptyline
\noindent
and has no singularity on horizon. This is only coordinate singularity
and its sense is the lack of rigid coordinates inside horizon. True
physical singularity is \textit{r}=0 and has the character of the so-called \underline{space-like singularity} (the inavitable singularity for the observer crossing horizon).\\

\noindent 
Now try to embed the instant equator section of curved space into flat
space (see above):

\emptyline
\begin{mapleinput}
\mapleinline{active}{1d}{z(r)[1] = int(sqrt(2*r*M-4*M^2)/(-r+2*M),r);
 \indent z(r)[2] = int(-sqrt(2*r*M-4*M^2)/(-r+2*M),r);}{%
}
\end{mapleinput}

\mapleresult
\begin{maplelatex}
\[
{\mathrm{z}(r)_{1}}= - 2\,\sqrt{2\,r\,M - 4\,M^{2}}
\]
\end{maplelatex}

\begin{maplelatex}
\[
{\mathrm{z}(r)_{2}}=2\,\sqrt{2\,r\,M - 4\,M^{2}}
\]
\end{maplelatex}

\begin{mapleinput}
\mapleinline{active}{1d}{plot3d(\{subs(\{M=1,r=sqrt(x^2+y^2)\},rhs(\%\%)),\\
 subs(\{M=1,r=sqrt(x^2+y^2)\},rhs(\%))\},\\
  x=-10..10,y=-10..10,axes=boxed,style=PATCHCONTOUR,\\
  grid=[100,100],title=`Schwarzschild black hole`);}{%
}
\end{mapleinput}

\mapleresult
\begin{center}
\mapleplot{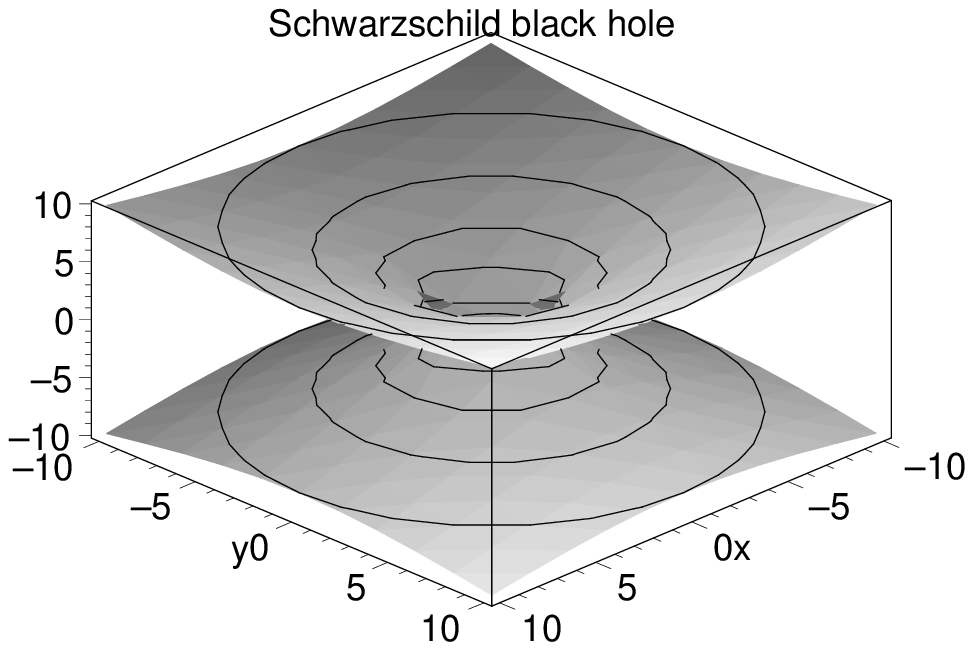}
\end{center}

\emptyline
\noindent
This is the black hole, which looks as a neck of battle
(Einstein-Rosen bridge): the inside way to horizon and the outside way
to horizon are the ways between different but identically
asymptotically flat universes ("wormhole" through 2-dimensional sphere
with minimal radius 2\textit{M}). But since the static geometry is not
valid upon horizon (note, that here \textit{t--\TEXTsymbol{>}t+dt} is
not time translation), this scheme is not stable.

\emptyline
\noindent
Let's try to exclude the above-mentioned coordinate singularity on the
horizon by means of coordinate change. For example, we consider the
null geodesics (
$\mathit{ds}^{2}$
=0) in Schwarzschild space-time corresponding to radial motion of
photons: 
$\mathit{dt}^{2}$
 = 
$\frac {\mathit{dr}^{2}}{(1 - \frac {{R_{g}}}{r})^{2}}$
. Hence (see above considered expression for 
${r_{0}}$
):  

\begin{mapleinput}
\mapleinline{active}{1d}{t = combine( int(-1/(1-R[g]/r),r) );}{%
}
\end{mapleinput}

\mapleresult
\begin{maplelatex}
\[
t= - r - {R_{g}}\,\mathrm{ln}(r - {R_{g}})
\]
\end{maplelatex}

\emptyline
\noindent
Then if \textit{v} is the constant, which defines the radial
coordinate for fixed \textit{t}, we have

\emptyline
\begin{mapleinput}
\mapleinline{active}{1d}{t = -r - R[g]*ln( abs(-r/R[g]+1) ) + v;#module\\
 allows to extend the expression for r<R[g]}{%
}
\end{mapleinput}

\mapleresult
\begin{maplelatex}
\[
t= - r - {R_{g}}\,\mathrm{ln}( \left|  \! \,{\displaystyle 
\frac {r}{{R_{g}}}}  - 1\, \!  \right| ) + v
\]
\end{maplelatex}

\emptyline
\noindent
Differentiation of this equation with subsequent substitution of 
$\mathit{dt}^{2}$
 in expression for interval in Schwarzschild metric produce

\emptyline
\begin{mapleinput}
\mapleinline{active}{1d}{defform(f=0,w1=1,w2=1,w3=1,v=1,R[g]=0,r=0,v=0);
  \indent d(t)^2 = expand(\\
   subs( d(R[g])=0,\\
   d( -r-R[g]*ln( r/R[g]-1 )+v ) )^2);# differentials\\
    in new coordinates
   \indent \indent subs( d(t)^2=rhs(\%),sch_compts[1,1]*d(t)^2 ) +\\
sch_compts[2,2]*d(r)^2 + sch_compts[3,3]*d(theta)^2 +\\
sch_compts[4,4]*d(phi)^2:
    \indent collect( simplify( subs(R[g]=2*M,\%) ),\\
    \{d(r)^2,d(v)^2\});#new metric}{%
}
\end{mapleinput}

\mapleresult
\begin{maplelatex}
\[
\mathrm{d}(t)^{2}=\mathrm{d}(v)^{2} - {\displaystyle \frac {2\,
\mathrm{d}(v)\,r\,\mathrm{d}(r)}{r - {R_{g}}}}  + {\displaystyle 
\frac {r^{2}\,\mathrm{d}(r)^{2}}{(r - {R_{g}})^{2}}} 
\]
\end{maplelatex}

\begin{maplelatex}
\[
 - {\displaystyle \frac {(r - 2\,M)\,\mathrm{d}(v)^{2}}{r}}  + 2
\,\mathrm{d}(r)\,\mathrm{d}(v) - {\displaystyle \frac { - r^{3}\,
\mathrm{d}(\theta )^{2} - r^{3}\,\mathrm{d}(\phi )^{2} + r^{3}\,
\mathrm{d}(\phi )^{2}\,\mathrm{cos}(\theta )^{2}}{r}} 
\]
\end{maplelatex}

\emptyline
\noindent
So, we have a new linear element (in the so-called
Eddington - Finkelstein coordinates) 
$\mathit{ds}^{2}$
= - (
$1 - \frac {{R_{g}}}{r}$
)
$\mathit{dV}^{2}$
 + 2\textit{dvdr + 
$r^{2}$
$\mathrm{d}(\Omega )^{2}$
}  (
$\mathrm{d}(\Omega )$
 is the spherical part). The corresponding metric has the regular
character in all region of \textit{r} (except for \textit{r}=0). It
should be noted, that the regularization was made by transition to
"light coordinate" \textit{v} therefore such coordinates can not be
realized physically, but formally we continued analytically the
coordinates to all \textit{r}\TEXTsymbol{>}0. We can see, that for
future directed (i.e. 
$\mathit{dv}^{2}$
\TEXTsymbol{>}0) null (
$\mathit{ds}^{2}$
=0) or time like (
$\mathit{ds}^{2}$
\TEXTsymbol{<}0) worldlines \textit{dr}\TEXTsymbol{<}0 for
\textit{r}\TEXTsymbol{<}
${R_{g}}$
 that corresponds to above-mentioned conclusion about inevitable
fall on singularity.\\

\noindent 
In the conclusion, we consider the problem of the deviation from
spherical symmetry for static black hole. Such deviation can be
described by the characteristic of quadrupole momentum \textit{q}. Erez
and Rosen found the corresponding static metric with axial symmetry:

\emptyline
\begin{mapleinput}
\mapleinline{active}{1d}{coord := [t, lambda, mu, phi]:
 \indent er_compts :=\\
  array(symmetric,sparse,1..4,1..4):# metric components
  \indent \indent er_compts[1,1] :=\\
   -exp(2*psi):# coefficient of d(t)^2 in interval
   \indent \indent \indent er_compts[2,2] :=\\
M^2*exp(2*gamma-2*psi)*(lambda^2-mu^2)/\\
(lambda^2-1):# coefficient of d(lambda)^2 in interval
     \indent \indent er_compts[3,3] :=\\
M^2*exp(2*gamma-2*psi)*(lambda^2-mu^2)/\\
(1-mu^2):# coefficient of d(mu)^2 in interval  
       \indent er_compts[4,4] :=\\
        M^2*exp(-2*psi)*(lambda^2-1)*(1-mu^2):#coefficient\\
         of d(phi)^2 in interval
      er := create([-1,-1], eval(er_compts));# axially\\
       symmetric metric}{%
}
\end{mapleinput}

\mapleresult
\begin{maplelatex}
\maplemultiline{
\mathit{er} := \mathrm{table(}[\mathit{compts}=\\
 \left[ 
{\begin{array}{cccc}
 - e^{(2\,\psi )} & 0 & 0 & 0 \\
0 & {\displaystyle \frac {M^{2}\,e^{(2\,\gamma  - 2\,\psi )}\,(
\lambda ^{2} - \mu ^{2})}{\lambda ^{2} - 1}}  & 0 & 0 \\ [2ex]
0 & 0 & {\displaystyle \frac {M^{2}\,e^{(2\,\gamma  - 2\,\psi )}
\,(\lambda ^{2} - \mu ^{2})}{1 - \mu ^{2}}}  & 0 \\ [2ex]
0 & 0 & 0 & M^{2}\,e^{( - 2\,\psi )}\,(\lambda ^{2} - 1)\,(1 - 
\mu ^{2})
\end{array}}
 \right]  \\
,
\mathit{index\_char}=[-1, \,-1] \\
]) }
\end{maplelatex}

\emptyline
\noindent
where

\emptyline
\begin{mapleinput}
\mapleinline{active}{1d}{f1 := psi = 1/2*(
(1+q*(3*lambda^2-1)*(3*mu^2-1)/4)*ln((lambda-1)/\\
(lambda+1))+3/2*q*lambda*(3*mu^2-1) );
 \indent f2 := gamma =\\
1/2*(1+q+q^2)*ln((lambda^2-1)/(lambda^2-mu^2))\\
-3/2*q*(1-mu^2)*(lambda*ln((lambda-1)/(lambda+1))+2)\\
+9/4*q^2*(1-mu^2)*((lambda^2+mu^2-1-9*lambda^2*mu^2)*\\
(lambda^2-1)/16*ln((lambda-1)/(lambda+1))^2+\\
(lambda^2+7*mu^2-5/3-9*mu^2*lambda^2)*lambda*\\
ln((lambda-1)/(lambda+1))/4+1/4*lambda^2*(1-9*mu^2)+\\
(mu^2-1/3) );
   \indent \indent f3 := lambda = r/M-1;
    \indent \indent \indent f4 := mu = cos(theta);}{%
}
\end{mapleinput}

\mapleresult
\begin{maplelatex}
\maplemultiline{
\mathit{f1} :=\\
 \psi ={\displaystyle \frac {1}{2}} \,(1 + 
{\displaystyle \frac {1}{4}} \,q\,(3\,\lambda ^{2} - 1)\,(3\,\mu 
^{2} - 1))\,\mathrm{ln}({\displaystyle \frac {\lambda  - 1}{
\lambda  + 1}} ) + {\displaystyle \frac {3}{4}} \,q\,\lambda \,(3
\,\mu ^{2} - 1)
}
\end{maplelatex}

\begin{maplelatex}
\maplemultiline{
\mathit{f2} := \gamma ={\displaystyle \frac {1}{2}} \,(1 + q + q
^{2})\,\mathrm{ln}({\displaystyle \frac {\lambda ^{2} - 1}{
\lambda ^{2} - \mu ^{2}}} ) -\\
 {\displaystyle \frac {3}{2}} \,q\,(
1 - \mu ^{2})\,(\lambda \,\mathrm{ln}({\displaystyle \frac {
\lambda  - 1}{\lambda  + 1}} ) + 2) + {\displaystyle \frac {9}{4}
} \,q^{2}\,(1 - \mu ^{2})( \\
{\displaystyle \frac {1}{16}} \,(\lambda ^{2} + \mu ^{2} - 1 - 9
\,\lambda ^{2}\,\mu ^{2})\,(\lambda ^{2} - 1)\,\mathrm{ln}(
{\displaystyle \frac {\lambda  - 1}{\lambda  + 1}} )^{2} \\
\mbox{} + {\displaystyle \frac {1}{4}} \,(\lambda ^{2} + 7\,\mu 
^{2} - {\displaystyle \frac {5}{3}}  - 9\,\lambda ^{2}\,\mu ^{2})
\,\lambda \,\mathrm{ln}({\displaystyle \frac {\lambda  - 1}{
\lambda  + 1}} ) +\\
 {\displaystyle \frac {1}{4}} \,\lambda ^{2}\,(
1 - 9\,\mu ^{2}) + \mu ^{2} - {\displaystyle \frac {1}{3}} ) }
\end{maplelatex}

\begin{maplelatex}
\[
\mathit{f3} := \lambda ={\displaystyle \frac {r}{M}}  - 1
\]
\end{maplelatex}

\begin{maplelatex}
\[
\mathit{f4} := \mu =\mathrm{cos}(\theta )
\]
\end{maplelatex}

\emptyline
\noindent
In the case \textit{q}=0 we have

\emptyline
\begin{mapleinput}
\mapleinline{active}{1d}{get_compts(er):
 \indent map2(subs,\{psi=rhs(f1),gamma=rhs(f2)\},\%):
  \indent \indent map2(subs,q=0,\%):
   \indent map2(subs,\{lambda=rhs(f3),mu=rhs(f4)\},\%):
    map(simplify,\%);}{%
}
\end{mapleinput}

\mapleresult
\begin{maplelatex}
\[
 \left[ 
{\begin{array}{cccc}
 - {\displaystyle \frac {r - 2\,M}{r}}  & 0 & 0 & 0 \\ [2ex]
0 & {\displaystyle \frac {r\,M^{2}}{r - 2\,M}}  & 0 & 0 \\ [2ex]
0 & 0 &  - {\displaystyle \frac {r^{2}}{ - 1 + \mathrm{cos}(
\theta )^{2}}}  & 0 \\ [2ex]
0 & 0 & 0 &  - r^{2}\,( - 1 + \mathrm{cos}(\theta )^{2})
\end{array}}
 \right] 
\]
\end{maplelatex}

\emptyline
\noindent
That is the Schwarzschild metric with regard to \textit{f3} and
\textit{f4}.\\ 

\noindent
Now let's find the horizon in the general case of nonzero quadrupole
moment. For static field (
${\frac {\partial }{\partial t}}\,{g_{\alpha , \,\beta }}$
=0) the corresponding condition is 
${g_{0, \,0}}$
=0. Then

\emptyline
\begin{mapleinput}
\mapleinline{active}{1d}{map2(subs,psi=rhs(f1),er):
 \indent er2 := map2(subs,gamma=rhs(f2),\%):
  \indent \indent get_compts(er2)[1,1];}{%
}
\end{mapleinput}

\mapleresult
\begin{maplelatex}
\[
 - e^{((1 + 1/4\,q\,(3\,\lambda ^{2} - 1)\,(3\,\mu ^{2} - 1))\,
\mathrm{ln}(\frac {\lambda  - 1}{\lambda  + 1}) + 3/2\,q\,\lambda
 \,(3\,\mu ^{2} - 1))}
\]
\end{maplelatex}

\emptyline
\noindent 
That is \textit{r}=2\textit{M} (
$\lambda $
=1). But the result for invariant of curvature is:

\emptyline
\begin{mapleinput}
\mapleinline{active}{1d}{erinv := invert( er2, 'detg' ):
 \indent D1er := d1metric( er2, coord ):
  \indent \indent D2er := d2metric( D1er, coord ):
   \indent \indent \indent Cf1 := Christoffel1 ( D1er ):
    \indent \indent RMN := Riemann( erinv, D2er, Cf1 ):
     \indent raise(erinv,RMN,1):#raise of indexes\\
      in Riemann tensor
      raise(erinv,\%,2):
       \indent raise(erinv,\%,3):
        \indent \indent RMNinv := raise(erinv,\%,4):
         \indent \indent \indent prod(RMN,RMNinv,[1,1],[2,2],[3,3],[4,4]):
          \indent \indent \indent \indent get_compts(\%):
           \indent \indent \indent series(\%,q=0,2):#expansion on q
            \indent \indent convert(\%,polynom):
             \indent res := simplify(\%):}{%
}
\end{mapleinput}

\emptyline
\noindent 
In the spherical case the result corresponds to above obtained:

\emptyline
\begin{mapleinput}
\mapleinline{active}{1d}{factor( subs(q=0,res) ):#spherical symmetry
 \indent subs(lambda=1,\%);# this is 48*M^2/r^6}{%
}
\end{mapleinput}

\mapleresult
\begin{maplelatex}
\[
{\displaystyle \frac {3}{4}} \,{\displaystyle \frac {1}{M^{4}}} 
\]
\end{maplelatex}

\emptyline
\noindent 
There is no singularity. But for nonzero \textit{q} (let's choose $\mu $ = 0 for sake of simplification):

\emptyline
\begin{mapleinput}
\mapleinline{active}{1d}{subs(mu=0,res):
 #L'Hospital's rule for calculation of limit
 \indent limit(diff(numer(\%),lambda),lambda=1);
  \indent \indent limit(diff(denom(\%),lambda),lambda=1);}{%
}
\end{mapleinput}

\mapleresult
\begin{maplelatex}
\[
 - \mathrm{signum}(q)\,\infty 
\]
\end{maplelatex}

\begin{maplelatex}
\[
0
\]
\end{maplelatex}

\emptyline
\noindent
Hence, there is a true singularity on horizon that can be regarded as
the demonstration of the impossibility of static axially symmetric
black hole with nonzero quadrupole momentum. Such momentum will be "taken
away" by the gravitational waves in the process of the black hole
formation. 

\emptyline

\subsection{Reissner-Nordstr\"om black hole (charged black hole)}

\emptyline
\noindent
The generalization of Schwarzschild metric on the case of spherically
symmetric vacuum solution of bounded Einstein-Maxwell equations
results in

\emptyline
\begin{mapleinput}
\mapleinline{active}{1d}{coord := [t, r, theta, phi]:
 \indent rn_compts := \\
 array(symmetric,sparse,1..4,1..4):# metric components
  \indent \indent rn_compts[1,1] :=\\
   -(1-2*M/r+Q^2/r^2):# coefficient of d(t)^2 in interval
   \indent \indent \indent rn_compts[2,2] :=\\
    1/(1-2*M/r+Q^2/r^2):# coefficient of d(r)^2 in interval
    \indent \indent rn_compts[3,3] := \\
    g_matrix[3,3]:# coefficient of d(theta)^2 in interval  
     \indent rn_compts[4,4] :=\\
      g_matrix[4,4]:# coefficient of d(phi)^2 in interval
      rn :=\\
       create([-1,-1], eval(rn_compts));# Reissner-Nordstrom\\
        (RN) metric}{%
}
\end{mapleinput}

\mapleresult
\begin{maplelatex}
\maplemultiline{
\mathit{rn} :=\\
 \mathrm{table(}[\mathit{compts}= \left[ 
{\begin{array}{cccc}
 - 1 + {\displaystyle \frac {2\,M}{r}}  - {\displaystyle \frac {Q
^{2}}{r^{2}}}  & 0 & 0 & 0 \\ [2ex]
0 & {\displaystyle \frac {1}{1 - {\displaystyle \frac {2\,M}{r}} 
 + {\displaystyle \frac {Q^{2}}{r^{2}}} }}  & 0 & 0 \\ [2ex]
0 & 0 & r^{2} & 0 \\
0 & 0 & 0 & r^{2}\,\mathrm{sin}(\theta )^{2}
\end{array}}
 \right] ,  \\
\mathit{index\_char}=[-1, \,-1] \\
]) }
\end{maplelatex}

\emptyline
\noindent 
Here \textit{Q} is the electric charge. The metric has three
singularities: \textit{r}=0 and

\emptyline
\begin{mapleinput}
\mapleinline{active}{1d}{denom(get_compts(rn)[2,2]) = 0;
 \indent r_p := solve(\%, r)[1];
  \indent \indent r_n := solve(\%\%, r)[2];}{%
}
\end{mapleinput}

\mapleresult
\begin{maplelatex}
\[
r^{2} - 2\,r\,M + Q^{2}=0
\]
\end{maplelatex}

\begin{maplelatex}
\[
\mathit{r\_p} := M + \sqrt{M^{2} - Q^{2}}
\]
\end{maplelatex}

\begin{maplelatex}
\[
\mathit{r\_n} := M - \sqrt{M^{2} - Q^{2}}
\]
\end{maplelatex}

\emptyline
\noindent
Let's calculate the invariant of curvature:

\emptyline
\begin{mapleinput}
\mapleinline{active}{1d}{rninv := invert( rn, 'detg' ):
 \indent D1rn := d1metric( rn, coord ):
  \indent \indent D2rn := d2metric( D1rn, coord ):
   \indent \indent \indent Cf1 := Christoffel1 ( D1rn ):
    \indent \indent RMN := Riemann( rninv, D2rn, Cf1 ):
     \indent raise(rninv,RMN,1):#raise of indexes\\
      in Riemann tensor
      raise(rninv,\%,2):
       \indent raise(rninv,\%,3):
        \indent \indent RMNinv := raise(rninv,\%,4):
         \indent \indent \indent prod(RMN,RMNinv,[1,1],[2,2],[3,3],[4,4]);}{%
}
\end{mapleinput}

\mapleresult
\begin{maplelatex}
\[
\mathrm{table(}[\mathit{compts}=8\,{\displaystyle \frac {6\,r^{2}
\,M^{2} - 12\,r\,M\,Q^{2} + 7\,Q^{4}}{r^{8}}} , \,\mathit{
index\_char}=[]])
\]
\end{maplelatex}

\begin{mapleinput}
\mapleinline{active}{1d}{solve(get_compts(\%),r);#nonphysical roots for\\
 numerator with nonzero Q}{%
}
\end{mapleinput}

\mapleresult
\begin{maplelatex}
\[
{\displaystyle \frac {(1 + {\displaystyle \frac {1}{6}} \,I\,
\sqrt{6})\,Q^{2}}{M}} , \,{\displaystyle \frac {(1 - 
{\displaystyle \frac {1}{6}} \,I\,\sqrt{6})\,Q^{2}}{M}} 
\]
\end{maplelatex}

\emptyline
\noindent
That is the situation like to one in Schwarzschild metric and two last
singularities have a coordinate character. Now let us plot the signs of
two first terms in linear element (we plot inverse value for second
term in order to escape the divergence due to coordinate
singularities). 

\emptyline
\begin{mapleinput}
\mapleinline{active}{1d}{plot(\{subs(\{M=1,Q=1/2\},get_compts(rn)[1,1]),\\
 subs(\{M=1,Q=1/2\},1/get_compts(rn)[2,2])\},\\
  r=0.1..2, title=`signs of first and second terms of\\
   linear element`);}{%
}
\end{mapleinput}

\mapleresult
\begin{center}
\mapleplot{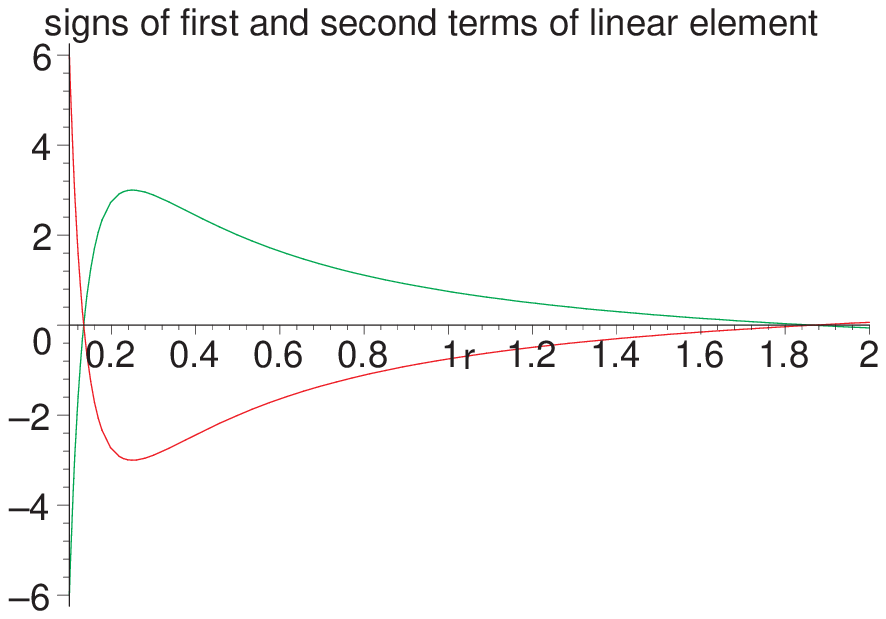}
\end{center}

\emptyline
\noindent
One can see the radical difference from the Schwarzschild black hole.
The space and time terms exchange the roles in region \textit{r\_n}
\TEXTsymbol{<}\textit{ r }\TEXTsymbol{<} \textit{r\_p} (between
so-called inner and outer horizons with 
${g_{0, \,0}}$
=0). But there are the usual signs in the vicinity of physical
singularity, i. e. it has a \underline{time-like} character and the
falling observer can avoid this singularity.\\

\noindent
Next difference is the lack of coordinate singularities for 
$M^{2}$
\TEXTsymbol{<}
$Q^{2}$
. In this case we have the so-called \underline{naked singularity.}
One can demonstrate that there is no such singularity as result of
usual collapse of charged shell with mass \textit{M} and charge
\textit{Q}. The total energy (in Newtonian limit but with correction
in framework of special relativity, \textit{M\_0} is the rest mass) is

\emptyline
\begin{mapleinput}
\mapleinline{active}{1d}{en := M(r) = M_0 + Q^2/r - M(r)^2/r;#we use\\
 the geometric units of charge so that the Coulomb low is\\
  G*Q_1*Q_2/r^2
 \indent sol := solve(en,M(r));}{%
}
\end{mapleinput}

\mapleresult
\begin{maplelatex}
\[
\mathit{en} := \mathrm{M}(r)=\mathit{M\_0} + {\displaystyle 
\frac {Q^{2}}{r}}  - {\displaystyle \frac {\mathrm{M}(r)^{2}}{r}
} 
\]
\end{maplelatex}

\begin{maplelatex}
\[
\mathit{sol} :=  - {\displaystyle \frac {1}{2}} \,r + 
{\displaystyle \frac {1}{2}} \,\sqrt{r^{2} + 4\,\mathit{M\_0}\,r
 + 4\,Q^{2}}, \, - {\displaystyle \frac {1}{2}} \,r - 
{\displaystyle \frac {1}{2}} \,\sqrt{r^{2} + 4\,\mathit{M\_0}\,r
 + 4\,Q^{2}}
\]
\end{maplelatex}

\emptyline
\noindent
The choice of the solution is defined by the correct asymptotic 
$\lim _{r\rightarrow \infty }\,\mathit{sol}$
 = \textit{M\_0}:

\emptyline
\begin{mapleinput}
\mapleinline{active}{1d}{if limit(sol[1],r=infinity)=M_0 then true_sol :=\\
 sol[1] fi:
 if limit(sol[2],r=infinity)=M_0 then true_sol :=\\
  sol[2] fi:
  \indent true_sol;}{%
}
\end{mapleinput}

\mapleresult
\begin{maplelatex}
\[
 - {\displaystyle \frac {1}{2}} \,r + {\displaystyle \frac {1}{2}
} \,\sqrt{r^{2} + 4\,\mathit{M\_0}\,r + 4\,Q^{2}}
\]
\end{maplelatex}

\begin{mapleinput}
\mapleinline{active}{1d}{diff(true_sol,r):
 \indent subs(M_0=solve(en,M_0),\%):
  \indent \indent simplify(\%,radical,symbolic);}{%
}
\end{mapleinput}

\mapleresult
\begin{maplelatex}
\[
{\displaystyle \frac { - Q^{2} + \mathrm{M}(r)^{2}}{(r + 2\,
\mathrm{M}(r))\,r}} 
\]
\end{maplelatex}

\emptyline
\noindent
So, 
${\frac {\partial }{\partial r}}\,\mathrm{M}(r)$
 =
$\frac {\mathrm{M}(r)^{2} - Q^{2}}{r\,(r + 2\,\mathrm{M}(r))}$
. The collapse is possible if \textit{M} decreases with
decreasing\textit{ R} (domination of gravity over the Coulomb
interaction in the process of collapse) that is possible, when 
$M^{2}$
\TEXTsymbol{>}
$Q^{2}$
. It should be noted, that the limit

\emptyline
\begin{mapleinput}
\mapleinline{active}{1d}{limit(true_sol,r=0);}{%
}
\end{mapleinput}

\mapleresult
\begin{maplelatex}
\[
\sqrt{Q^{2}}
\]
\end{maplelatex}

\emptyline
\noindent
resolves the problem of the infinite proper energy of charged
particle.\\

\emptyline
\noindent
Now we will consider the pressure free collapse of charged sphere of dust
by analogy with Schwarzschild metric.

\emptyline
\begin{mapleinput}
\mapleinline{active}{1d}{r := 'r':
 \indent E := 'E':
  \indent \indent subs( r=r(t),get_compts(rn) ):
   \indent d(s)^2 = \%[1,1]*d(t)^2 + \%[2,2]*d(r)^2;#RN metric
    -d(tau)^2 =\\
     collect(subs( d(r)=diff(r(t),t)*d(t),\\
     rhs(\%)),d(t));#tau is the proper time for the observer\\
      on the surface of sphere
     \indent \%/d(tau)^2;
      \indent \indent subs(\{d(t)=E/(1-2*M/r(t)+Q^2/r^2),\\
      d(tau)=1\},\%);#we used d(t)/d(tau)=E/(1-2*M/r+Q^2/r^2)
    \indent pot_1 :=\\
     factor( solve(\%,(diff(r(t),t))^2) ):#"potential"\\
      for remote observer
  pot_2 :=\\
   simplify(pot_1*(E/(1-2*M/r(t)+Q^2/r^2))^2):#"potential"\\
    for collapsing observer\\
     d/d(t)=(d/d(tau))*(1-2*M/r+Q^2/r^2)/E}{%
}
\end{mapleinput}

\mapleresult
\begin{maplelatex}
\[
\mathrm{d}(s)^{2}=( - 1 + {\displaystyle \frac {2\,M}{\mathrm{r}(
t)}}  - {\displaystyle \frac {Q^{2}}{\mathrm{r}(t)^{2}}} )\,
\mathrm{d}(t)^{2} + {\displaystyle \frac {\mathrm{d}(r)^{2}}{1 - 
{\displaystyle \frac {2\,M}{\mathrm{r}(t)}}  + {\displaystyle 
\frac {Q^{2}}{\mathrm{r}(t)^{2}}} }} 
\]
\end{maplelatex}

\begin{maplelatex}
\[
 - \mathrm{d}(\tau )^{2}= \left(  \!  - 1 + {\displaystyle 
\frac {2\,M}{\mathrm{r}(t)}}  - {\displaystyle \frac {Q^{2}}{
\mathrm{r}(t)^{2}}}  + {\displaystyle \frac {({\frac {\partial }{
\partial t}}\,\mathrm{r}(t))^{2}}{1 - {\displaystyle \frac {2\,M
}{\mathrm{r}(t)}}  + {\displaystyle \frac {Q^{2}}{\mathrm{r}(t)^{
2}}} }}  \!  \right) \,\mathrm{d}(t)^{2}
\]
\end{maplelatex}

\begin{maplelatex}
\[
-1={\displaystyle \frac { \left(  \!  - 1 + {\displaystyle 
\frac {2\,M}{\mathrm{r}(t)}}  - {\displaystyle \frac {Q^{2}}{
\mathrm{r}(t)^{2}}}  + {\displaystyle \frac {({\frac {\partial }{
\partial t}}\,\mathrm{r}(t))^{2}}{1 - {\displaystyle \frac {2\,M
}{\mathrm{r}(t)}}  + {\displaystyle \frac {Q^{2}}{\mathrm{r}(t)^{
2}}} }}  \!  \right) \,\mathrm{d}(t)^{2}}{\mathrm{d}(\tau )^{2}}
} 
\]
\end{maplelatex}

\begin{maplelatex}
\[
-1={\displaystyle \frac { \left(  \!  - 1 + {\displaystyle 
\frac {2\,M}{\mathrm{r}(t)}}  - {\displaystyle \frac {Q^{2}}{
\mathrm{r}(t)^{2}}}  + {\displaystyle \frac {({\frac {\partial }{
\partial t}}\,\mathrm{r}(t))^{2}}{1 - {\displaystyle \frac {2\,M
}{\mathrm{r}(t)}}  + {\displaystyle \frac {Q^{2}}{\mathrm{r}(t)^{
2}}} }}  \!  \right) \,E^{2}}{(1 - {\displaystyle \frac {2\,M}{
\mathrm{r}(t)}}  + {\displaystyle \frac {Q^{2}}{r^{2}}} )^{2}}} 
\]
\end{maplelatex}

\begin{mapleinput}
\mapleinline{active}{1d}{plot(\{subs(\{E=0.5,M=1,Q=1/2,r(t)=r\},pot_2),\\
0*r\},r=0.11..3,axes=boxed,title=`(dr/dtau)^2 vs r for\\
 collapsing observer`);}{%
}
\end{mapleinput}

\mapleresult
\begin{center}
\mapleplot{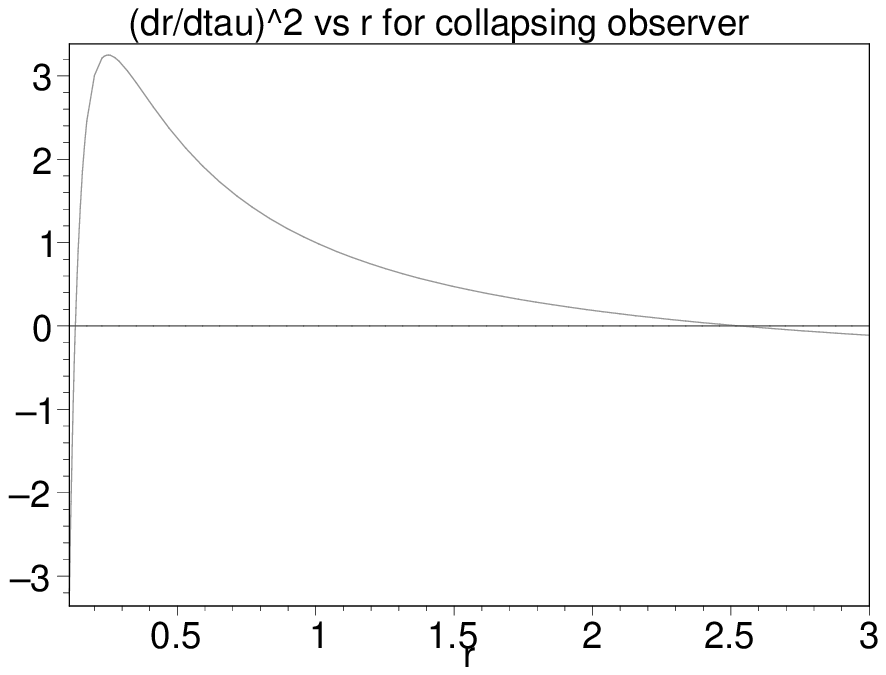}
\end{center}

\emptyline
\noindent
The difference from Schwarzschild collapse is obvious: the observer
crosses the outer and inner horizons but does not reach the
singularity because of the collapsar explodes as \underline{white
hole} due to repulsion with consequent recollapse and so on.\\

\noindent
And at last, we consider the "extreme" case 
$M^{2}$
=
$Q^{2}$
. 

\emptyline
\begin{mapleinput}
\mapleinline{active}{1d}{subs(Q^2=M^2, get_compts(rn));
 \indent factor(\%[1,1]);}{%
}
\end{mapleinput}

\mapleresult
\begin{maplelatex}
\[
 \left[ 
{\begin{array}{cccc}
 - 1 + {\displaystyle \frac {2\,M}{r}}  - {\displaystyle \frac {M
^{2}}{r^{2}}}  & 0 & 0 & 0 \\ [2ex]
0 & {\displaystyle \frac {1}{1 - {\displaystyle \frac {2\,M}{r}} 
 + {\displaystyle \frac {M^{2}}{r^{2}}} }}  & 0 & 0 \\ [2ex]
0 & 0 & r^{2} & 0 \\
0 & 0 & 0 & r^{2}\,\mathrm{sin}(\theta )^{2}
\end{array}}
 \right] 
\]
\end{maplelatex}

\begin{maplelatex}
\[
 - {\displaystyle \frac {(r - M)^{2}}{r^{2}}} 
\]
\end{maplelatex}

\emptyline
\noindent
So, we have one coordinate singularity in \textit{r}=\textit{M}. What
happen with second horizon? Let's find the distance between horizons for
fixed \textit{t} and angular coordinates for RN-metric: 

\emptyline
\begin{mapleinput}
\mapleinline{active}{1d}{get_compts(rn):
 \indent d(s)^2 = \%[2,2]*d(r)^2;#RN metric
# or
   \indent d(s)^2 = d(r)^2/expand( (1-r_p/r)*(1-r_n/r) );#second\\
   representation of expression}{%
}
\end{mapleinput}

\mapleresult
\begin{maplelatex}
\[
\mathrm{d}(s)^{2}={\displaystyle \frac {\mathrm{d}(r)^{2}}{1 - 
{\displaystyle \frac {2\,M}{r}}  + {\displaystyle \frac {Q^{2}}{r
^{2}}} }} 
\]
\end{maplelatex}

\begin{maplelatex}
\[
\mathrm{d}(s)^{2}={\displaystyle \frac {\mathrm{d}(r)^{2}}{1 - 
{\displaystyle \frac {2\,M}{r}}  + {\displaystyle \frac {Q^{2}}{r
^{2}}} }} 
\]
\end{maplelatex}

\emptyline
\noindent
Hence, when \textit{r\_p--\TEXTsymbol{>}r\_n} (
$M^{2} - Q^{2}$
\textit{--\TEXTsymbol{>}}0) 

\emptyline
\begin{mapleinput}
\mapleinline{active}{1d}{r_p := 'r_p':
 \indent r_n := 'r_n':
   \indent \indent s = Int(1/sqrt((1-r_p/r)*(1-r_n/r)),r=r_n..r_p) ;
    \indent \indent \indent simplify( value(rhs(\%)),radical,symbolic );}{%
}
\end{mapleinput}

\mapleresult
\begin{maplelatex}
\[
s={\displaystyle \int _{\mathit{r\_n}}^{\mathit{r\_p}}} 
{\displaystyle \frac {1}{\sqrt{(1 - {\displaystyle \frac {
\mathit{r\_p}}{r}} )\,(1 - {\displaystyle \frac {\mathit{r\_n}}{r
}} )}}} \,dr
\]
\end{maplelatex}

\begin{maplelatex}
\maplemultiline{
{\displaystyle \frac {1}{2}} \,\mathit{r\_n}\,\mathrm{ln}(
\mathit{r\_p} - \mathit{r\_n}) + {\displaystyle \frac {1}{2}} \,
\mathit{r\_p}\,\mathrm{ln}(\mathit{r\_p} - \mathit{r\_n}) - \\
{\displaystyle \frac {1}{2}} \,\mathit{r\_n}\,\mathrm{ln}( - 
\mathit{r\_p} + \mathit{r\_n}) - {\displaystyle \frac {1}{2}} \,
\mathit{r\_p}\,\mathrm{ln}( - \mathit{r\_p} + \mathit{r\_n})
}
\end{maplelatex}

\emptyline
\noindent
we have \textit{s--\TEXTsymbol{>}
$\infty $
}. So, there is the infinitely long Einstein-Rosen bridge (charged
string) between horizons that means a lack of wormhole between
asymptotically flat universes. This fact can be illustrated by means
of embedding of equatorial section of static RN space in flat
Euclidian space (see above).

\emptyline
\begin{mapleinput}
\mapleinline{active}{1d}{d(r)^2/(1-1/r)^2  =\\
  (1+diff(z(r),r)^2)*d(r)^2;#equality of radial elements\\
   of intervals, M=1
  \indent diff(z(r),r) = solve(\%,diff(z(r),r))[1];
   \indent \indent dsolve(\%,z(r));# embedding}{%
}
\end{mapleinput}

\mapleresult
\begin{maplelatex}
\[
{\displaystyle \frac {\mathrm{d}(r)^{2}}{(1 - {\displaystyle 
\frac {1}{r}} )^{2}}} =(1 + ({\frac {\partial }{\partial r}}\,
\mathrm{z}(r))^{2})\,\mathrm{d}(r)^{2}
\]
\end{maplelatex}

\begin{maplelatex}
\[
{\frac {\partial }{\partial r}}\,\mathrm{z}(r)={\displaystyle 
\frac {\sqrt{2\,r - 1}}{r - 1}} 
\]
\end{maplelatex}

\begin{maplelatex}
\[
\mathrm{z}(r)=2\,\sqrt{2\,r - 1} + \mathrm{ln}(\sqrt{2\,r - 1} - 
1) - \mathrm{ln}(\sqrt{2\,r - 1} + 1) + \mathit{\_C1}
\]
\end{maplelatex}

\begin{mapleinput}
\mapleinline{active}{1d}{plot3d(subs(r=sqrt(x^2+y^2),\\
2*sqrt(2*r-1)+ln(sqrt(2*r-1)-1)-ln(sqrt(2*r-1)+1)),\\
x=-10..10,y=-10..10,axes=boxed,style=PATCHCONTOUR,\\
grid=[100,100],title=`extreme RN black hole`);\\
#we expressed arctanh through ln}{%
}
\end{mapleinput}

\mapleresult
\begin{center}
\mapleplot{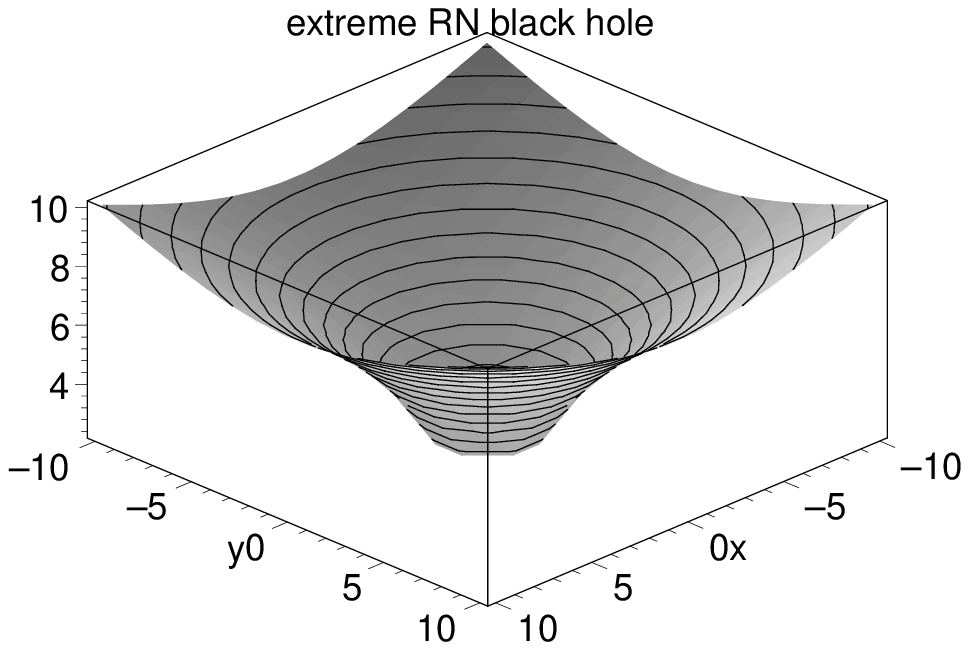}
\end{center}

\emptyline
\noindent
The asymptotic behavior of RN-metric as \textit{r--\TEXTsymbol{>}
$\infty $
} is Minkowski. For investigation of the situation
\textit{r--\TEXTsymbol{>}M} let introduce the new coordinate (see, for
example, 
\cite{P.K. Townsend}): 

\emptyline
\begin{mapleinput}
\mapleinline{active}{1d}{rn_assym := subs(\\
\{r=M*(1+lambda),Q^2=M^2\},get_compts(rn) );}{%
}
\end{mapleinput}

\mapleresult
\begin{maplelatex}
\maplemultiline{
\mathit{rn\_assym} :=  \\
 \left[ 
{\begin{array}{cccc}
 - 1 + {\displaystyle \frac {2}{\lambda  + 1}}  - {\displaystyle 
\frac {1}{(\lambda  + 1)^{2}}}  & 0 & 0 & 0 \\ [2ex]
0 & {\displaystyle \frac {1}{1 - {\displaystyle \frac {2}{\lambda
  + 1}}  + {\displaystyle \frac {1}{(\lambda  + 1)^{2}}} }}  & 0
 & 0 \\ [2ex]
0 & 0 & M^{2}\,(\lambda  + 1)^{2} & 0 \\
0 & 0 & 0 & M^{2}\,(\lambda  + 1)^{2}\,\mathrm{sin}(\theta )^{2}
\end{array}}
 \right]  }
\end{maplelatex}

\begin{mapleinput}
\mapleinline{active}{1d}{m1 :=\\
 series(rn_assym[1,1],lambda=0,3);# we keep only\\
  leading term in lambda
 \indent m2 :=\\
  series(rn_assym[2,2],lambda=0,3);# we keep only\\
   leading term in lambda
  \indent \indent d(s)^2 = convert(m1,polynom)*d(t)^2\\
+ M^2*convert(m2,polynom)*d(lambda)^2 +\\
 M^2*d(Omega)^2;#d(Omega) is spherical part }{%
}
\end{mapleinput}

\mapleresult
\begin{maplelatex}
\[
\mathit{m1} :=  - \lambda ^{2} + \mathrm{O}(\lambda ^{3})
\]
\end{maplelatex}

\begin{maplelatex}
\[
\mathit{m2} := \lambda ^{-2} + \mathrm{O}(\lambda ^{-1})
\]
\end{maplelatex}

\begin{maplelatex}
\[
\mathrm{d}(s)^{2}= - \lambda ^{2}\,\mathrm{d}(t)^{2} + 
{\displaystyle \frac {M^{2}\,\mathrm{d}(\lambda )^{2}}{\lambda ^{
2}}}  + M^{2}\,\mathrm{d}(\Omega )^{2}
\]
\end{maplelatex}

\emptyline
\noindent
This is the \underline{Robinson-Bertotti metric}. The last term
describes two - dimensional sphere with radius \textit{M} (these
dimensions are compactified in the vicinity of horizon) and the first
terms corresponds to anti-de Sitter space-time (see below) with
constant negative curvature. 

\emptyline

\subsection{Kerr black hole (rotating black hole)}

\emptyline
\noindent
In the general form the stationary rotating black hole is described by
so-called \underline{Kerr-Newman} metric, which in the Boyer-Linquist
coordinates can be presented as:

\emptyline
\begin{mapleinput}
\mapleinline{active}{1d}{coord := [t, r, theta, phi]:
\indent kn_compts :=\\ 
array(sparse,1..4,1..4):# metric components,\\
 a=J/M, J is the angular momentum

\indent \indent kn_compts[1,1] :=\\
-(Delta-a^2*sin(theta)^2)/Sigma:#coefficient of d(t)^2

\indent \indent \indent kn_compts[1,4] :=\\
-2*a*sin(theta)^2*(r^2+a^2-Delta)/\\
Sigma:# coefficient of d(t)*dphi

\indent \indent kn_compts[2,2] :=\\ 
Sigma/Delta:# coefficient of d(r)^2

\indent kn_compts[3,3] :=\\ 
Sigma:# coefficient of d(theta)^2

kn_compts[4,4] :=\\ 
(((r^2+a^2)^2-Delta*a^2*sin(theta)^2)/\\
Sigma)*sin(theta)^2:#coefficient of d(phi)^2
      
\indent kn := create([-1,-1], eval(kn_compts));# Kerr-Newman\\
 (KN) metric

\indent \indent #where

\indent sub_1 := Sigma = r^2+a^2*cos(theta)^2;
sub_2 := Delta = r^2-2*M*r+a^2+sqrt(Q^2+P^2);# P is\\
 the magnetic (monopole) charge}{%
}
\end{mapleinput}

\mapleresult
\begin{maplelatex}
\maplemultiline{
\mathit{kn} := \mathrm{table(}[\mathit{compts}=\\
 \left[ 
{\begin{array}{cccc}
 - {\displaystyle \frac {\Delta  - a^{2}\,\mathrm{sin}(\theta )^{
2}}{\Sigma }}  & 0 & 0 &  - 2\,{\displaystyle \frac {a\,\mathrm{
sin}(\theta )^{2}\,(r^{2} + a^{2} - \Delta )}{\Sigma }}  \\ [2ex]
0 & {\displaystyle \frac {\Sigma }{\Delta }}  & 0 & 0 \\ [2ex]
0 & 0 & \Sigma  & 0 \\
0 & 0 & 0 & {\displaystyle \frac {((r^{2} + a^{2})^{2} - \Delta 
\,a^{2}\,\mathrm{sin}(\theta )^{2})\,\mathrm{sin}(\theta )^{2}}{
\Sigma }} 
\end{array}}
 \right] ,  \\
\mathit{index\_char}=[-1, \,-1] \\
]) }
\end{maplelatex}

\begin{maplelatex}
\[
\mathit{sub\_1} := \Sigma =r^{2} + a^{2}\,\mathrm{cos}(\theta )^{
2}
\]
\end{maplelatex}

\begin{maplelatex}
\[
\mathit{sub\_2} := \Delta =r^{2} - 2\,r\,M + a^{2} + \sqrt{Q^{2}
 + P^{2}}
\]
\end{maplelatex}

\emptyline
\noindent
In the absence of charges, this results in Kerr metric. The obvious
singularities are (except for an usual singularity of spherical
coordinates 
$\theta $
=0):

\emptyline
\begin{mapleinput}
\mapleinline{active}{1d}{r_p :=\\
 solve( subs(\{Q=0,P=0\},\\
 rhs(sub_2))=0,r )[1];#outer horizon
 \indent r_n :=\\
  solve( subs(\{Q=0,P=0\},\\
  rhs(sub_2))=0,r )[2];#inner horizon
  \indent \indent solve( subs(\{Q=0,P=0\},rhs(sub_1))=0,theta );}{%
}
\end{mapleinput}

\mapleresult
\begin{maplelatex}
\[
\mathit{r\_p} := M + \sqrt{M^{2} - a^{2}}
\]
\end{maplelatex}

\begin{maplelatex}
\[
\mathit{r\_n} := M - \sqrt{M^{2} - a^{2}}
\]
\end{maplelatex}

\begin{maplelatex}
\[
{\displaystyle \frac {1}{2}} \,\pi  - I\,\mathrm{arcsinh}(
{\displaystyle \frac {r}{a}} ), \,{\displaystyle \frac {1}{2}} \,
\pi  + I\,\mathrm{arcsinh}({\displaystyle \frac {r}{a}} )
\]
\end{maplelatex}

\emptyline
\noindent
The last produces \textit{r}=0, 
$\theta $
=
$\frac {\pi }{2}$
.\\

\noindent
As it was in the case of charged static black hole, there are three
different situations: 
$M^{2}$
 \TEXTsymbol{<} 
$a^{2}$
, 
$M^{2}$
 = 
$a^{2}$
, 
$M^{2}$
 \TEXTsymbol{>} 
$a^{2}$
.\\ 

\noindent
Let us consider the signs of  
${g_{0, \,0}}$
, 
${g_{1, \,1}}$
, 
${g_{4, \,4}}$
  (
$M^{2}$
 \TEXTsymbol{>} 
$a^{2}$
).   

\begin{mapleinput}
\mapleinline{active}{1d}{plot3d(subs(\{M=1,a=1/2,P=0,Q=0\},subs(\\
\{Sigma=rhs(sub_1),Delta=rhs(sub_2)\},\\
get_compts(kn)[1,1])),r=0.1..4,theta=0..Pi,color=red):

\indent plot3d(subs(\{M=1,a=1/2,P=0,Q=0,theta=2*Pi/3\},\\
subs(\{Sigma=rhs(sub_1),Delta=rhs(sub_2)\},\\
1/get_compts(kn)[2,2])),\\
r=0.1..4,theta=0..Pi,color=green):

\indent \indent display(\%,\%\%,\\
title=`signs of first and second diagonal elements\\
 of metric`,axes=boxed);

\indent plot3d(subs(\{M=1,a=1/2,P=0,Q=0\},subs(\\
\{Sigma=rhs(sub_1),Delta=rhs(sub_2)\},\\
get_compts(kn)[4,4])),r=-0.01..0.01,\\
theta=Pi/2-0.01..Pi/2+0.01,color=blue,\\
 title=`four diagonal element of metric in vicinity of\\
 singularity`,axes=boxed);}{%
}
\end{mapleinput}

\mapleresult
\begin{center}
\mapleplot{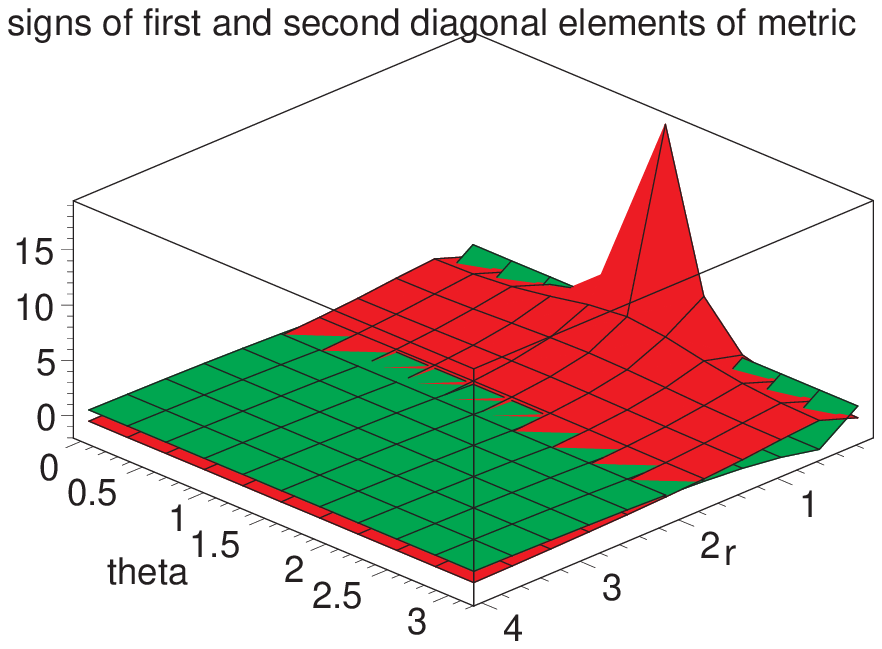}
\end{center}

\begin{center}
\mapleplot{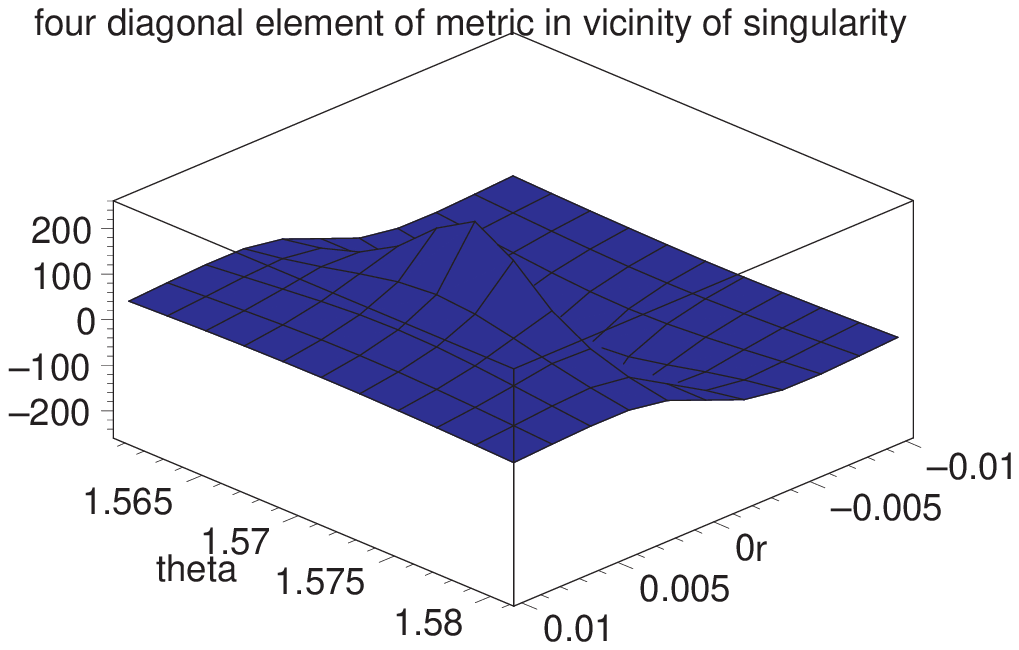}
\end{center}

\emptyline
\noindent
One can see, that the approach to \textit{r}=0 in the line, which
differs from 
$\theta $
=
$\frac {\pi }{2}$, corresponds to usual signs of diagonal elements of metric, i.e. the
observer crosses \textit{r}=0 and comes into region
\textit{r}\TEXTsymbol{<}0 without collision with singularity. But from
the second picture we can see the change of the 
${g_{3, \,3}}$
-sign for \textit{r}\TEXTsymbol{<}0. Now 
$\phi $
 is the time like coordinate. But this coordinate has circle
character and, as consequence, we find oneself in the world with
closed time lines. The approach to \textit{r}=0 in the line of 
$\theta $
=
$\frac {\pi }{2}$
 produces the change of 
${g_{0, \,0}}$
-sign, i.e. we find a true singularity in this direction. These facts
demonstrate that the singularity in Kerr black hole has a more
complicated character than in above considered black holes.

\emptyline
\noindent
The more careful consideration gives the following results:

\emptyline
\noindent
1) 
$M^{2}$
 \TEXTsymbol{<} 
$a^{2}$
. There exist no horizon (\textit{r\_p} and \textit{r\_n} are
complex), but the singularity in \textit{r}=0, 
$\theta $
=
$\frac {\pi }{2}$
 keeps. To remove the coordinate singularity in 
$\theta $
=0 we introduce the \underline{Kerr-Schild coordinates} with linear
element  

\emptyline
\begin{mapleinput}
\mapleinline{active}{1d}{macro(ts=`t*`):
 \indent ks_le := d(s)^2 = -d(ts)^2 + d(x)^2 + d(y)^2\\
  + d(z)^2 + (2*M*r^3/(r^4+a^2*z^2))*\\
  ((r*(x*d(x)+y*d(y))-a*(x*d(y)-y*d(x)))/\\
  (r^2+a^2)+z*d(z)/r+d(ts) )^2;
  \indent \indent x + I*y = (r + I*a)*sin(theta)*\\
  exp(I*(Int(1,phi)+Int(a/Delta,r)));
   \indent z = r*cos(theta);
    ts = Int(1,t) + Int((r^2+a^2/Delta),r) - r;}{%
}
\end{mapleinput}

\mapleresult
\begin{maplelatex}
\maplemultiline{
\mathit{ks\_le} := \mathrm{d}(s)^{2}= - \mathrm{d}(\mathit{t*})^{
2} + \mathrm{d}(x)^{2} + \mathrm{d}(y)^{2} + \mathrm{d}(z)^{2}
 \\
\mbox{} + {\displaystyle \frac {2\,M\,r^{3}\,({\displaystyle 
\frac {r\,(x\,\mathrm{d}(x) + y\,\mathrm{d}(y)) - a\,(x\,\mathrm{
d}(y) - y\,\mathrm{d}(x))}{r^{2} + a^{2}}}  + {\displaystyle 
\frac {z\,\mathrm{d}(z)}{r}}  + \mathrm{d}(\mathit{t*}))^{2}}{r^{
4} + a^{2}\,z^{2}}}  }
\end{maplelatex}

\begin{maplelatex}
\[
x + I\,y=(r + I\,a)\,\mathrm{sin}(\theta )\,e^{ \left(  \! I\,
 \left(  \! \int 1\,d\phi  + \int \frac {a}{\Delta }\,dr \! 
 \right)  \!  \right) }
\]
\end{maplelatex}

\begin{maplelatex}
\[
z=r\,\mathrm{cos}(\theta )
\]
\end{maplelatex}

\begin{maplelatex}
\[
\mathit{t*}={\displaystyle \int } 1\,dt + {\displaystyle \int } r
^{2} + {\displaystyle \frac {a^{2}}{\Delta }} \,dr - r
\]
\end{maplelatex}

\emptyline
\noindent
which is reduced to Minkowski metric by \textit{M--\TEXTsymbol{>}}0.

\emptyline
\begin{mapleinput}
\mapleinline{active}{1d}{int(subs( \{P=0,Q=0\},\\
subs(Delta=rhs(sub_2),a/Delta) ),r):
 \indent x+I*y=(r+I*a)*sin(theta)*exp(I*(phi+\%));}{%
}
\end{mapleinput}

\mapleresult
\begin{maplelatex}
\[
x + I\,y=(r + I\,a)\,\mathrm{sin}(\theta )\,e^{ \left(  \! I\,
 \left(  \! \phi  + \frac {a\,\mathrm{arctan}(1/2\,\frac {2\,r - 
2\,M}{\sqrt{a^{2} - M^{2}}})}{\sqrt{a^{2} - M^{2}}} \!  \right) 
 \!  \right) }
\]
\end{maplelatex}

\noindent
When \textit{r}=0, 
$\theta $
=
$\frac {\pi }{2}$
 the singularity is the ring 
$x^{2} + y^{2}$
 = 
$a^{2}$
, \textit{z }= 0.\\

\noindent
2) 
$M^{2}$
\TEXTsymbol{>}
$a^{2}$
. As before we have the ring singularity, but there are the horizons
\textit{r\_p} and \textit{r\_n}. As an additional feature of Kerr metric
we note here the existence of coordinate singularity:

\begin{mapleinput}
\mapleinline{active}{1d}{get_compts(kn)[1,1]=0;
 \indent subs( Delta=rhs( sub_2 ),numer( lhs(\%) ) ):
  \indent \indent solve( subs(\{Q=0,P=0\},\%) = 0,r );}{%
}
\end{mapleinput}

\mapleresult
\begin{maplelatex}
\[
 - {\displaystyle \frac {\Delta  - a^{2}\,\mathrm{sin}(\theta )^{
2}}{\Sigma }} =0
\]
\end{maplelatex}

\begin{maplelatex}
\[
M + \sqrt{M^{2} - a^{2} + a^{2}\,\mathrm{sin}(\theta )^{2}}, \,M
 - \sqrt{M^{2} - a^{2} + a^{2}\,\mathrm{sin}(\theta )^{2}}
\]
\end{maplelatex}

\emptyline
\noindent
The crossing of ellipsoid \textit{r}\_\textit{1}= 
$M + \sqrt{M^{2} - a^{2}\,\mathrm{cos}(\theta )^{2}}$
 produces the change of 
${g_{0, \,0}}$
 -sign. As it was for Schwarzschild black hole this fact demonstrates
the lack of static coordinates under this surface, which is called
\underline{"ergosphere}" and the region between \textit{r\_p} and
\textit{r\_1} is the ergoregion. The absence of singularity for 
${r_{r, \,r}}$
 suggests that the nonstatic behavior results from entrainment of
observer by black hole rotation, but not from fall on singularity, as
it takes place for observer under horizon in Schwarzschild black hole.

\emptyline

\subsection{Conclusion}

\emptyline
\noindent
So, the elementary analysis by means of basic Maple 6 functions allows
to obtain the main results of black hole physics including conditions
of collapsar formation, the space-time structure of static spherically
symmetric charged and uncharged black holes and stationary axially
symmetric black hole.

\emptyline

\section{Cosmological models}

\emptyline

\subsection{Introduction}

\emptyline
\noindent
I suppose that the most impressive achievements of the GR-theory
belong to the cosmology. The description of the global structure of
the universe changes our opinion about reality and radically widens
the horizon of knowledge. The considerable progress in the
observations and measures allows to say that we are living in a gold
age of cosmology. Here we will consider some basic conceptions of
relativistic cosmology by means of analytical capabilities of Maple 6.

\emptyline

\subsection{Robertson-Walker metric}

\emptyline
\noindent
We will base on the conception of foliation of 4-dimensional manifold
on 3-dimensional space-like hyperplanes with a time-dependent
geometry. Now let's restrict oneself to the isotropic and homogeneous
evolution models, i.e. the models without dependence of curvature on
the observer's location or orientation. We will suppose also, that such foliation with
isotropic and homogeneous geometry and energy-momentum distribution
exists at each time moment. These space-like foliations are the
so-called \underline{hyperplanes of homogeneity}. The time vectors are
defined by the proper time course in each space point.\\

\noindent
Let us consider 3-geometry at fixed time moment. As it is known, in
3-dimensional space the curvature tensor has the following form:

\emptyline
\begin{center}
${R_{\mathit{iklm}}}$
 = \\
${R_{\mathit{il}}}$
${g_{\mathit{km}}}$
\textit{ - 
${R_{\mathit{im}}}$
${g_{\mathit{kl}}}$
} + 
${R_{\mathit{km}}}$
${g_{\mathit{il}}}$
\textit{ - 
${R_{\mathit{kl}}}$
${g_{\mathit{im}}}$
} + 
$\frac {R}{2}$
 (
${g_{\mathit{im}}}$
${g_{\mathit{kl}}}$
 \textit{-} 
${g_{\mathit{il}}}$
${g_{\mathit{km}}}$
).    
\end{center}

\emptyline
\noindent
But if (as it takes a place in our case) there is not the dependence
of curvature on space-direction in arbitrary point, our space has a
constant curvature (\textit{Schur's} theorem) that results in

\emptyline
\begin{center}
   
${R_{\mathit{iklm}}}$
 = 
$\frac {R}{2}$
 (
${g_{\mathit{im}}}$
${g_{\mathit{kl}}}$
 \textit{-} 
${g_{\mathit{il}}}$
${g_{\mathit{km}}}$
),     (1)
\end{center}

\emptyline
\noindent
where \textit{R} is the constant Ricci scalar.

\emptyline
\noindent
Now we will consider the so-called \underline{Robertson-Walker
metric}:

\emptyline
\begin{center}
${g_{\mathit{ij}}}$
=
$\frac {{\delta _{\mathit{ij}}}}{(1 + \frac {R\,{\delta _{
\mathit{ij}}}\,x^{i}\,x^{j}}{8})^{2}}$
    (2)
\end{center}

\emptyline
\noindent
For this metric the spatial curvature tensor is:

\emptyline
\begin{mapleinput}
\mapleinline{active}{1d}{restart:
 \indent with(tensor):
  \indent \indent with(linalg):
    \indent \indent \indent with(difforms):
         
coord := [x, y, z]:# coordinates
 \indent g_compts :=\\
  array(symmetric,sparse,1..3,1..3):# metric components
  \indent \indent g_compts[1,1] := 1/\\
  (1+R*(x^2+y^2+z^2)/8)^2:# component of interval\\
  attached to d(x)^2 
   \indent \indent \indent g_compts[2,2] :=\\
    1/(1+R*(x^2+y^2+z^2)/8)^2:# component of interval\\
    attached to d(y)^2
     \indent \indent g_compts[3,3] :=\\
      1/(1+R*(x^2+y^2+z^2)/8)^2:# component of\\
      interval attached to d(z)^2 

\indent g := create([-1,-1], eval(g_compts));# covariant\\
 metric tensor 
ginv := invert( g, 'detg' ):# contravariant\\
 metric tensor

\indent D1g := d1metric( g, coord ):# calculation of \\
curvature tensor
\indent \indent D2g := d2metric( D1g, coord ):
\indent \indent \indent Cf1 := Christoffel1 ( D1g ):
\indent \indent \indent \indent RMN := Riemann( ginv, D2g, Cf1 ):
\indent \indent \indent \indent \indent displayGR(Riemann, RMN);}{%
}
\end{mapleinput}

\begin{maplelatex}
\maplemultiline{
g := \mathrm{table(}[\mathit{index\_char}=[-1, \,-1], \,\mathit{
compts}= \left[ 
{\begin{array}{ccc}
\mathrm{\%1} & 0 & 0 \\
0 & \mathrm{\%1} & 0 \\
0 & 0 & \mathrm{\%1}
\end{array}}
 \right] ]) \\
\mathrm{\%1} := {\displaystyle \frac {1}{(1 + {\displaystyle 
\frac {1}{8}} \,R\,(x^{2} + y^{2} + z^{2}))^{2}}}  }
\end{maplelatex}

\begin{maplelatex}
\[
\mathit{The\ Riemann\ Tensor}
\]
\end{maplelatex}

\begin{maplelatex}
\[
\mathit{non-zero\ components\ :}
\]
\end{maplelatex}

\begin{maplelatex}
\[
\mathit{\ R1212}=2048\,{\displaystyle \frac {R}{(8 + R\,x^{2} + R
\,y^{2} + R\,z^{2})^{4}}} 
\]
\end{maplelatex}

\begin{maplelatex}
\[
\mathit{\ R1313}=2048\,{\displaystyle \frac {R}{(8 + R\,x^{2} + R
\,y^{2} + R\,z^{2})^{4}}} 
\]
\end{maplelatex}

\begin{maplelatex}
\[
\mathit{\ R2323}=2048\,{\displaystyle \frac {R}{(8 + R\,x^{2} + R
\,y^{2} + R\,z^{2})^{4}}} 
\]
\end{maplelatex}

\begin{maplelatex}
\[
\mathit{character\ :\ [-1,\ -1,\ -1,\ -1]}
\]
\end{maplelatex}

\emptyline
\noindent
Now we try to perform the direct calculations from Eq. (1):

\emptyline
\begin{mapleinput}
\mapleinline{active}{1d}{A := get_compts(g):
\indent B := array(1..3, 1..3, 1..3, 1..3):
\indent \indent for i from 1 to 3 do
 \indent \indent \indent for j from 1 to 3 do
  \indent \indent for k from 1 to 3 do
   \indent for l from 1 to 3 do
    B[i,j,k,l] :=\\
     eval(R*(A[i,k]*A[j,l] - A[i,l]*A[j,k])/2):# Eq. 1.\\
These calculations can be performed also by means of\\
 tensor[prod] and tensor[permute_indices]
\indent if B[i,j,k,l]<>0 then \\
print([i,j,k,l],B[i,j,k,l]);# non-zero components
\indent \indent else
\indent \indent \indent fi:
     \indent \indent \indent \indent od: od: od: od:}{%
}
\end{mapleinput}

\mapleresult
\begin{maplelatex}
\[
[1, \,2, \,1, \,2], \,{\displaystyle \frac {1}{2}} \,
{\displaystyle \frac {R}{(1 + {\displaystyle \frac {1}{8}} \,R\,(
x^{2} + y^{2} + z^{2}))^{4}}} 
\]
\end{maplelatex}

\begin{maplelatex}
\[
[1, \,2, \,2, \,1], \, - {\displaystyle \frac {1}{2}} \,
{\displaystyle \frac {R}{(1 + {\displaystyle \frac {1}{8}} \,R\,(
x^{2} + y^{2} + z^{2}))^{4}}} 
\]
\end{maplelatex}

\begin{maplelatex}
\[
[1, \,3, \,1, \,3], \,{\displaystyle \frac {1}{2}} \,
{\displaystyle \frac {R}{(1 + {\displaystyle \frac {1}{8}} \,R\,(
x^{2} + y^{2} + z^{2}))^{4}}} 
\]
\end{maplelatex}

\begin{maplelatex}
\[
[1, \,3, \,3, \,1], \, - {\displaystyle \frac {1}{2}} \,
{\displaystyle \frac {R}{(1 + {\displaystyle \frac {1}{8}} \,R\,(
x^{2} + y^{2} + z^{2}))^{4}}} 
\]
\end{maplelatex}

\begin{maplelatex}
\[
[2, \,1, \,1, \,2], \, - {\displaystyle \frac {1}{2}} \,
{\displaystyle \frac {R}{(1 + {\displaystyle \frac {1}{8}} \,R\,(
x^{2} + y^{2} + z^{2}))^{4}}} 
\]
\end{maplelatex}

\begin{maplelatex}
\[
[2, \,1, \,2, \,1], \,{\displaystyle \frac {1}{2}} \,
{\displaystyle \frac {R}{(1 + {\displaystyle \frac {1}{8}} \,R\,(
x^{2} + y^{2} + z^{2}))^{4}}} 
\]
\end{maplelatex}

\begin{maplelatex}
\[
[2, \,3, \,2, \,3], \,{\displaystyle \frac {1}{2}} \,
{\displaystyle \frac {R}{(1 + {\displaystyle \frac {1}{8}} \,R\,(
x^{2} + y^{2} + z^{2}))^{4}}} 
\]
\end{maplelatex}

\begin{maplelatex}
\[
[2, \,3, \,3, \,2], \, - {\displaystyle \frac {1}{2}} \,
{\displaystyle \frac {R}{(1 + {\displaystyle \frac {1}{8}} \,R\,(
x^{2} + y^{2} + z^{2}))^{4}}} 
\]
\end{maplelatex}

\begin{maplelatex}
\[
[3, \,1, \,1, \,3], \, - {\displaystyle \frac {1}{2}} \,
{\displaystyle \frac {R}{(1 + {\displaystyle \frac {1}{8}} \,R\,(
x^{2} + y^{2} + z^{2}))^{4}}} 
\]
\end{maplelatex}

\begin{maplelatex}
\[
[3, \,1, \,3, \,1], \,{\displaystyle \frac {1}{2}} \,
{\displaystyle \frac {R}{(1 + {\displaystyle \frac {1}{8}} \,R\,(
x^{2} + y^{2} + z^{2}))^{4}}} 
\]
\end{maplelatex}

\begin{maplelatex}
\[
[3, \,2, \,2, \,3], \, - {\displaystyle \frac {1}{2}} \,
{\displaystyle \frac {R}{(1 + {\displaystyle \frac {1}{8}} \,R\,(
x^{2} + y^{2} + z^{2}))^{4}}} 
\]
\end{maplelatex}

\begin{maplelatex}
\[
[3, \,2, \,3, \,2], \,{\displaystyle \frac {1}{2}} \,
{\displaystyle \frac {R}{(1 + {\displaystyle \frac {1}{8}} \,R\,(
x^{2} + y^{2} + z^{2}))^{4}}} 
\]
\end{maplelatex}

\emptyline
\noindent
The direct calculation on the basis of Eq. (1) and metric (2) results
in the tensor with symmetries
[\textit{i,j,k,l}] = [\textit{k,l,i,j}] = [\textit{j,i,k,l}] = -[
\textit{i,j,l,k}] and
[\textit{i,j,k,l}] + [\textit{i,l,j,k}] + [\textit{i,k,l,j}] = 0. These
symmetries and values of nonzero components correspond to curvature
tensor, which was calculated previously from (2). As consequence,
metric (2) satisfies to condition of constant curvature (Schur's
theorem, Eq. (1)). We will base our further calculations on this
metric.

\emptyline
\noindent
The next step is the choice of the appropriate coordinates. We
consider two types of coordinates: spherical and pseudospherical.\\

\noindent
In the case of the spherical coordinates the transition of Eq. (2) to
a new basis results in:

\emptyline
\begin{mapleinput}
\mapleinline{active}{1d}{g_sp :=\\
simplify(subs(\\
\{x=r*cos(phi)*sin(theta),y=r*sin(phi)*sin(theta),\\
z=r*cos(theta)\},get_compts(g))):# This is transformation\\
 of coordinates in metric tensor
\indent Sp_compts := array(1..3,1..3):# matrix of\\
 conversion from (NB!) spherical coordinates
  \indent \indent Sp_compts[1,1] := cos(phi)*sin(theta): 
   \indent \indent \indent Sp_compts[2,1] := sin(phi)*sin(theta):
     \indent \indent \indent \indent Sp_compts[3,1] := cos(theta):
      \indent \indent \indent Sp_compts[1,2] := r*cos(phi)*cos(theta):
       \indent \indent Sp_compts[2,2] := r*sin(phi)*cos(theta):
        \indent Sp_compts[3,2] := -r*sin(theta):
         Sp_compts[1,3] := -r*sin(phi)*sin(theta):
          \indent Sp_compts[2,3] := r*cos(phi)*sin(theta):
           \indent \indent Sp_compts[3,3] := 0: 

\indent Sp := eval(Sp_compts):

simplify(\\
 multiply(transpose(Sp), g_sp, Sp));#transition\\
 to new coordinates in metric }{%
}
\end{mapleinput}

\mapleresult
\begin{maplelatex}
\maplemultiline{
 \left[ 
{\begin{array}{ccc}
64\,{\displaystyle \frac {1}{\mathrm{\%1}}}  & 0 & 0 \\ [2ex]
0 & 64\,{\displaystyle \frac {r^{2}}{\mathrm{\%1}}}  & 0 \\ [2ex]
0 & 0 &  - 64\,{\displaystyle \frac {r^{2}\,( - 1 + \mathrm{cos}(
\theta )^{2})}{\mathrm{\%1}}} 
\end{array}}
 \right]  \\
\mathrm{\%1} := (8 + R\,r^{2}\,\mathrm{cos}(\phi )^{2}\,\mathrm{
sin}(\theta )^{2} + R\,r^{2}\,\mathrm{sin}(\phi )^{2}\,\mathrm{
sin}(\theta )^{2} + R\,r^{2}\,\mathrm{cos}(\theta )^{2})^{2} }
\end{maplelatex}

\emptyline
\noindent
The denominator is

\emptyline
\begin{mapleinput}
\mapleinline{active}{1d}{denom(\%[1,1]):
 \indent expand(\%):
  \indent \indent simplify(\%,trig):
   \indent \indent \indent factor(\%);}{%
}
\end{mapleinput}

\mapleresult
\begin{maplelatex}
\[
(R\,r^{2} + 8)^{2}
\]
\end{maplelatex}

\emptyline
\noindent
After substitution \textit{r* }= \textit{r 
$\sqrt{\frac {R}{2}}$
} (we will omit asterix and suppose \textit{R} \TEXTsymbol{>} 0), the
metric is converted as:

\emptyline
\begin{mapleinput}
\mapleinline{active}{1d}{coord := [r, theta, phi]:# spherical coordinates
 \indent g_compts_sp :=\\
  array(symmetric,sparse,1..3,1..3):# metric components
  \indent \indent g_compts_sp[1,1] :=\\
   (2/R)/(1+r^2/4)^2:# component of interval\\
    attached to d(r)^2 
   \indent g_compts_sp[2,2] :=\\
    (2/R)*r^2/(1+r^2/4)^2:# component of interval\\
    attached to d(theta)^2
     g_compts_sp[3,3] :=\\
      (2/R)*r^2*sin(theta)^2/(1+r^2/4)^2:# component of\\
       interval attached to d(phi)^2 

\indent g_sp := create([-1,-1], eval(g_compts_sp));# covariant\\
 metric tensor}{%
}
\end{mapleinput}

\mapleresult
\begin{maplelatex}
\maplemultiline{
\mathit{g\_sp} := \mathrm{table(}[\mathit{index\_char}=[-1, \,-1]
,  \\
\mathit{compts}= \left[ 
{\begin{array}{ccc}
2\,{\displaystyle \frac {1}{R\,(1 + {\displaystyle \frac {1}{4}} 
\,r^{2})^{2}}}  & 0 & 0 \\ [2ex]
0 & 2\,{\displaystyle \frac {r^{2}}{R\,(1 + {\displaystyle 
\frac {1}{4}} \,r^{2})^{2}}}  & 0 \\ [2ex]
0 & 0 & 2\,{\displaystyle \frac {r^{2}\,\mathrm{sin}(\theta )^{2}
}{R\,(1 + {\displaystyle \frac {1}{4}} \,r^{2})^{2}}} 
\end{array}}
 \right]  \\
]) }
\end{maplelatex}

\emptyline
\noindent
The denominator's form suggests the transformation of radial
coordinate: \textit{sin(
$\chi $
) = r/}(\textit{1 +
 $r^{2}$
}/4). Then for 
$\mathit{dr}^{2}$
-component we have

\emptyline
\begin{mapleinput}
\mapleinline{active}{1d}{diff(sin(chi),chi)*d(chi) =\\
simplify(diff(r/(1+r^2/4),r))*d(r);
 \indent lhs(\%)^2 = rhs(\%)^2;
  \indent \indent subs(cos(chi)^2=1-r^2/(1+r^2/4)^2,lhs(\%))\\
   = rhs(\%):
   \indent \indent \indent factor(\%);}{%
}
\end{mapleinput}

\mapleresult
\begin{maplelatex}
\[
\mathrm{cos}(\chi )\,\mathrm{d}(\chi )= - 4\,{\displaystyle 
\frac {( - 4 + r^{2})\,\mathrm{d}(r)}{(4 + r^{2})^{2}}} 
\]
\end{maplelatex}

\begin{maplelatex}
\[
\mathrm{cos}(\chi )^{2}\,\mathrm{d}(\chi )^{2}=16\,
{\displaystyle \frac {( - 4 + r^{2})^{2}\,\mathrm{d}(r)^{2}}{(4
 + r^{2})^{4}}} 
\]
\end{maplelatex}

\begin{maplelatex}
\[
{\displaystyle \frac {(r - 2)^{2}\,(r + 2)^{2}\,\mathrm{d}(\chi )
^{2}}{(4 + r^{2})^{2}}} =16\,{\displaystyle \frac {(r - 2)^{2}\,(
r + 2)^{2}\,\mathrm{d}(r)^{2}}{(4 + r^{2})^{4}}} 
\]
\end{maplelatex}

\emptyline
\noindent
that leads to a new metric (\textit{d
$\chi $
} = 
$\frac {\mathit{dr}}{(1 + \frac {r^{2}}{4})^{2}}$%
):

\emptyline
\begin{mapleinput}
\mapleinline{active}{1d}{coord := [r, theta, phi]:# spherical coordinates
 \indent g_compts_sp :=\\
  array(symmetric,sparse,1..3,1..3):# metric components
  \indent \indent g_compts_sp[1,1] := \\
  2/R:# component of interval attached to d(chi)^2

   \indent \indent \indent g_compts_sp[2,2] :=\\
    (2/R)*sin(chi)^2:# component of interval\\
    attached to d(theta)^2
     \indent \indent g_compts_sp[3,3] :=\\
      (2/R)*sin(chi)^2*sin(theta)^2:# component of\\
      interval attached to d(phi)^2 

\indent g_sp :=\\
 create([-1,-1], eval(g_compts_sp));# covariant\\
  metric tensor}{%
}
\end{mapleinput}

\mapleresult
\begin{maplelatex}
\maplemultiline{
\mathit{g\_sp} := \mathrm{table(}[\mathit{index\_char}=[-1, \,-1]
, \,\mathit{compts}=\\
 \left[ 
{\begin{array}{ccc}
2\,{\displaystyle \frac {1}{R}}  & 0 & 0 \\ [2ex]
0 & 2\,{\displaystyle \frac {\mathrm{sin}(\chi )^{2}}{R}}  & 0 \\
 [2ex]
0 & 0 & 2\,{\displaystyle \frac {\mathrm{sin}(\chi )^{2}\,
\mathrm{sin}(\theta )^{2}}{R}} 
\end{array}}
 \right] ]) }
\end{maplelatex}

\emptyline
\noindent
Here \textit{R}/2 is the Gaussian curvature \textit{K
}(\textit{K}\TEXTsymbol{>}0) so that the linear element is: 

\emptyline
\begin{mapleinput}
\mapleinline{active}{1d}{l_sp := d(s)^2 = a^2*(d(chi)^2 +\\
sin(chi)^2*(d(theta)^2+sin(theta)^2*d(phi)^2));}{%
}
\end{mapleinput}

\mapleresult
\begin{maplelatex}
\[
\mathit{l\_sp} := \mathrm{d}(s)^{2}=a^{2}\,(\mathrm{d}(\chi )^{2}
 + \mathrm{sin}(\chi )^{2}\,(\mathrm{d}(\theta )^{2} + \mathrm{
sin}(\theta )^{2}\,\mathrm{d}(\phi )^{2}))
\]
\end{maplelatex}

\emptyline
\noindent
Here \textit{a} = 
$\sqrt{\frac {1}{K}}$
 is the \underline{scaling factor}. Note, that substitution \textit{r
= a 
$\mathrm{sin}(\chi )$
} produces the alternative form of metric:  

\emptyline
\begin{mapleinput}
\mapleinline{active}{1d}{subs(\\
\{sin(chi)=r/a,d(chi)^2=d(r)^2/(a^2*(1-r^2/a^2))\},\\
l_sp):#we use d(r)=a*cos(chi)*d(chi)
 \indent expand(\%);}{%
}
\end{mapleinput}

\mapleresult
\begin{maplelatex}
\[
\mathrm{d}(s)^{2}={\displaystyle \frac {\mathrm{d}(r)^{2}}{1 - 
{\displaystyle \frac {r^{2}}{a^{2}}} }}  + r^{2}\,\mathrm{d}(
\theta )^{2} + r^{2}\,\mathrm{sin}(\theta )^{2}\,\mathrm{d}(\phi 
)^{2}
\]
\end{maplelatex}

\emptyline
\noindent
For negative curvature:

\emptyline
\begin{mapleinput}
\mapleinline{active}{1d}{coord := [r, theta, phi]:# spherical coordinates
 \indent g_compts_hyp :=\\
  array(symmetric,sparse,1..3,1..3):# metric components
  \indent \indent g_compts_hyp[1,1] :=\\
    (-2/R)/(1-r^2/4)^2:# component of interval\\
    attached to d(r)^2 
   \indent \indent \indent g_compts_hyp[2,2] :=\\
     (-2/R)*r^2/(1-r^2/4)^2:# component of interval\\
     attached to d(theta)^2
     \indent \indent \indent \indent g_compts_hyp[3,3] :=\\
      (-2/R)*r^2*sin(theta)^2/(1-r^2/4)^2:#component of\\
     interval attached to d(phi)^2 

\indent \indent \indent \indent \indent g_hyp :=\\
 create([-1,-1], eval(g_compts_hyp));# covariant metric\\
 tensor}{%
}
\end{mapleinput}

\mapleresult
\begin{maplelatex}
\maplemultiline{
\mathit{g\_hyp} := \mathrm{table(}[\mathit{index\_char}=[-1, \,-1
], \mathit{compts}=\\
 \left[ 
{\begin{array}{ccc}
 - 2\,{\displaystyle \frac {1}{R\,(1 - {\displaystyle \frac {1}{4
}} \,r^{2})^{2}}}  & 0 & 0 \\ [2ex]
0 &  - 2\,{\displaystyle \frac {r^{2}}{R\,(1 - {\displaystyle 
\frac {1}{4}} \,r^{2})^{2}}}  & 0 \\ [2ex]
0 & 0 &  - 2\,{\displaystyle \frac {r^{2}\,\mathrm{sin}(\theta )
^{2}}{R\,(1 - {\displaystyle \frac {1}{4}} \,r^{2})^{2}}} 
\end{array}}
 \right]  \\
]) }
\end{maplelatex}

\emptyline
\noindent
where \textit{R} is the module of Ricci scalar and new radial
coordinate is \textit{I r 
$\sqrt{\frac {R}{2}.}$
} Let's try to use the substitution \textit{sh(
$\chi $
) = r/}(\textit{1-
$r^{2}$
}/4). Then for 
$\mathit{dr}^{2}$
-component we have

\emptyline
\begin{mapleinput}
\mapleinline{active}{1d}{diff(sinh(chi),chi)*d(chi) =\\
simplify(diff(r/(1-r^2/4),r))*d(r);
 \indent lhs(\%)^2 = rhs(\%)^2;
  \indent \indent subs(cosh(chi)^2=1+r^2/(1-r^2/4)^2,lhs(\%))\\
   = rhs(\%):
   \indent \indent \indent factor(\%);}{%
}
\end{mapleinput}

\mapleresult
\begin{maplelatex}
\[
\mathrm{cosh}(\chi )\,\mathrm{d}(\chi )=4\,{\displaystyle \frac {
(4 + r^{2})\,\mathrm{d}(r)}{( - 4 + r^{2})^{2}}} 
\]
\end{maplelatex}

\begin{maplelatex}
\[
\mathrm{cosh}(\chi )^{2}\,\mathrm{d}(\chi )^{2}=16\,
{\displaystyle \frac {(4 + r^{2})^{2}\,\mathrm{d}(r)^{2}}{( - 4
 + r^{2})^{4}}} 
\]
\end{maplelatex}

\begin{maplelatex}
\[
{\displaystyle \frac {(4 + r^{2})^{2}\,\mathrm{d}(\chi )^{2}}{(r
 - 2)^{2}\,(r + 2)^{2}}} =16\,{\displaystyle \frac {(4 + r^{2})^{
2}\,\mathrm{d}(r)^{2}}{(r - 2)^{4}\,(r + 2)^{4}}} 
\]
\end{maplelatex}

\emptyline
\noindent
And finally

\emptyline
\begin{mapleinput}
\mapleinline{active}{1d}{coord := [r, theta, phi]:# spherical coordinates
 \indent g_compts_hyp :=\\
  array(symmetric,sparse,1..3,1..3):# metric components
  \indent \indent g_compts_hyp[1,1] :=\\
   -2/R:# component of interval\\
    attached to d(chi)^2 
   \indent \indent \indent g_compts_hyp[2,2] :=\\
    (-2/R)*sinh(chi)^2:# component of interval\\
    attached to d(theta)^2
     \indent \indent \indent \indent g_compts_hyp[3,3] :=\\
     (-2/R)*sinh(chi)^2*sin(theta)^2:# component\\
     of interval attached to d(phi)^2 

\indent \indent \indent \indent \indent g_hyp :=\\
 create([-1,-1], eval(g_compts_hyp));# covariant\\
  metric tensor}{%
}
\end{mapleinput}

\mapleresult
\begin{maplelatex}
\maplemultiline{
\mathit{g\_hyp} :=  \\
\mathrm{table(}[\mathit{index\_char}=[-1, \,-1], \,\mathit{compts
}=\\
 \left[ 
{\begin{array}{ccc}
 - 2\,{\displaystyle \frac {1}{R}}  & 0 & 0 \\ [2ex]
0 &  - 2\,{\displaystyle \frac {\mathrm{sinh}(\chi )^{2}}{R}}  & 
0 \\ [2ex]
0 & 0 &  - 2\,{\displaystyle \frac {\mathrm{sinh}(\chi )^{2}\,
\mathrm{sin}(\theta )^{2}}{R}} 
\end{array}}
 \right] ]) }
\end{maplelatex}

\emptyline
\noindent
Here \textit{R}/2 is the Gaussian curvature \textit{K }
(\textit{K}\TEXTsymbol{<}0). So, the linear element is

\emptyline
\begin{mapleinput}
\mapleinline{active}{1d}{l_hyp := d(s)^2 = a^2*(d(chi)^2 +\\
sinh(chi)^2*(d(theta)^2+sin(theta)^2*d(phi)^2));}{%
}
\end{mapleinput}

\mapleresult
\begin{maplelatex}
\[
\mathit{l\_hyp} := \mathrm{d}(s)^{2}=a^{2}\,(\mathrm{d}(\chi )^{2
} + \mathrm{sinh}(\chi )^{2}\,(\mathrm{d}(\theta )^{2} + \mathrm{
sin}(\theta )^{2}\,\mathrm{d}(\phi )^{2}))
\]
\end{maplelatex}

\emptyline
\noindent
where \textit{a} = 
$\sqrt{\frac {1}{K}}$
 is the imaginary scale factor. The substitution \textit{r }=
\textit{a }sinh(
$\chi $
) results in:\textit{ }

\emptyline
\begin{mapleinput}
\mapleinline{active}{1d}{subs(\\
\{sinh(chi)=r/a,d(chi)^2=d(r)^2/(a^2*(1+r^2/a^2))\},\\
l_hyp):#we use d(r)=a*cosh(chi)*d(chi)
 \indent expand(\%);}{%
}
\end{mapleinput}

\mapleresult
\begin{maplelatex}
\[
\mathrm{d}(s)^{2}={\displaystyle \frac {\mathrm{d}(r)^{2}}{1 + 
{\displaystyle \frac {r^{2}}{a^{2}}} }}  + r^{2}\,\mathrm{d}(
\theta )^{2} + r^{2}\,\mathrm{sin}(\theta )^{2}\,\mathrm{d}(\phi 
)^{2}
\]
\end{maplelatex}

\emptyline
\noindent
At last, for flat space:

\emptyline
\begin{mapleinput}
\mapleinline{active}{1d}{l_flat := d(s)^2 = a^2*(d(chi)^2 +\\
chi^2*(d(theta)^2+sin(theta)^2*d(phi)^2));}{%
}
\end{mapleinput}

\mapleresult
\begin{maplelatex}
\[
\mathit{l\_flat} := \mathrm{d}(s)^{2}=a^{2}\,(\mathrm{d}(\chi )^{
2} + \chi ^{2}\,(\mathrm{d}(\theta )^{2} + \mathrm{sin}(\theta )
^{2}\,\mathrm{d}(\phi )^{2}))
\]
\end{maplelatex}

\emptyline
\noindent
Here \textit{a} is the arbitrary scaling factor.

\emptyline
\emptyline
\noindent
To imagine the considered geometries we can use the usual technique of
embedding. In the general case the imbedding of 3-space is not
possible if the dimension of enveloping flat space is only 4 (we have
6 functions 
${g_{i, \,j}}$
, but the number of free parameters in 4-dimensional space is only
4). However, the constant nonzero curvature allows to embed our curved
space in 4-dimensional flat space as result of next transformation of
coordinates (for spherical geometry):

\emptyline
\begin{mapleinput}
\mapleinline{active}{1d}{defform(f=0,w1=1,w2=1,w3=1,v=1,chi=0,theta=0,phi=0);
\indent e_1 := Dw^2 = a^2*d(cos(chi))^2;
\indent \indent e_2 := Dx^2 =\\
 (a*d(sin(chi)*sin(theta)*cos(phi)))^2;
\indent \indent \indent e_3 := Dy^2 =\\
 (a*d(sin(chi)*sin(theta)*sin(phi)))^2;
\indent \indent \indent \indent e_4 := Dz^2 =\\
 (a*d(sin(chi)*cos(theta)))^2; }{%
}
\end{mapleinput}

\mapleresult
\begin{maplelatex}
\[
\mathit{e\_1} := \mathit{Dw}^{2}=a^{2}\,\mathrm{sin}(\chi )^{2}\,
\mathrm{d}(\chi )^{2}
\]
\end{maplelatex}

\begin{maplelatex}
\maplemultiline{
\mathit{e\_2} := \mathit{Dx}^{2}=a^{2} \\
(\mathrm{sin}(\theta )\,\mathrm{cos}(\phi )\,\mathrm{cos}(\chi )
\,\mathrm{d}(\chi ) + \mathrm{sin}(\chi )\,\mathrm{cos}(\phi )\,
\mathrm{cos}(\theta )\,\mathrm{d}(\theta ) -\\
 \mathrm{sin}(\chi )
\,\mathrm{sin}(\theta )\,\mathrm{sin}(\phi )\,\mathrm{d}(\phi ))
^{2} }
\end{maplelatex}

\begin{maplelatex}
\maplemultiline{
\mathit{e\_3} := \mathit{Dy}^{2}=a^{2} \\
(\mathrm{sin}(\theta )\,\mathrm{sin}(\phi )\,\mathrm{cos}(\chi )
\,\mathrm{d}(\chi ) + \mathrm{sin}(\chi )\,\mathrm{sin}(\phi )\,
\mathrm{cos}(\theta )\,\mathrm{d}(\theta ) +\\
 \mathrm{sin}(\chi )
\,\mathrm{sin}(\theta )\,\mathrm{cos}(\phi )\,\mathrm{d}(\phi ))
^{2} }
\end{maplelatex}

\begin{maplelatex}
\[
\mathit{e\_4} := \mathit{Dz}^{2}=a^{2}\,(\mathrm{cos}(\theta )\,
\mathrm{cos}(\chi )\,\mathrm{d}(\chi ) - \mathrm{sin}(\chi )\,
\mathrm{sin}(\theta )\,\mathrm{d}(\theta ))^{2}
\]
\end{maplelatex}

\begin{mapleinput}
\mapleinline{active}{1d}{#so we have for Euclidian 4-metric:
\indent simplify(\\
 rhs(e_1) + rhs(e_2) + rhs(e_3) + rhs(e_4), trig);}{%
}
\end{mapleinput}

\mapleresult
\begin{maplelatex}
\maplemultiline{
a^{2}(\mathrm{d}(\phi )^{2}\,\mathrm{cos}(\chi )^{2}\,\mathrm{cos
}(\theta )^{2} + \mathrm{d}(\theta )^{2} + \mathrm{d}(\phi )^{2}
 - \mathrm{d}(\theta )^{2}\,\mathrm{cos}(\chi )^{2} -\\
  \mathrm{d}(
\phi )^{2}\,\mathrm{cos}(\theta )^{2} - \mathrm{d}(\phi )^{2}\,
\mathrm{cos}(\chi )^{2} \mbox{} + \mathrm{d}(\chi )^{2}) }
\end{maplelatex}

\emptyline
\begin{mapleinput}
\mapleinline{active}{1d}{collect(\%,d(phi)^2);}{%
}
\end{mapleinput}

\mapleresult
\begin{maplelatex}
\maplemultiline{
a^{2}\,(1 - \mathrm{cos}(\theta )^{2} - \mathrm{cos}(\chi )^{2}
 + \mathrm{cos}(\chi )^{2}\,\mathrm{cos}(\theta )^{2})\,\mathrm{d
}(\phi )^{2} +\\
 a^{2}\,(\mathrm{d}(\theta )^{2} + \mathrm{d}(\chi 
)^{2} - \mathrm{d}(\theta )^{2}\,\mathrm{cos}(\chi )^{2})
}
\end{maplelatex}

\emptyline
\noindent
This expression can be easily converted in \textit{l\_sp}, i. e. the
imbedding is correct.\\

\noindent
Since

\emptyline
\begin{mapleinput}
\mapleinline{active}{1d}{w^2 + x^2 + y^2 + z^2 =\\
simplify(\\
(a*cos(chi))^2 +(a*sin(chi)*sin(theta)*cos(phi))^2 +\\
(a*sin(chi)*sin(theta)*sin(phi))^2 +\\
 (a*sin(chi)*cos(theta))^2); }{%
}
\end{mapleinput}

\mapleresult
\begin{maplelatex}
\[
w^{2} + x^{2} + y^{2} + z^{2}=a^{2}
\]
\end{maplelatex}

\emptyline
\noindent
the considered hypersurface is the 3-sphere in flat 4-space. The scale
factor is the radius of the closed spherical universe.

\emptyline
\noindent
In the case \textit{l\_hyp} there is not the imbedding in flat
Euclidian 4-space, but there is the imbedding in flat hyperbolical
flat space with interval

\begin{center}
\textit{-d
$w^{2}$
+ d
$x^{2}$
+ d
$y^{2}$
+ d
$z^{2}$
}
\end{center}

\noindent
As result we have

\emptyline
\begin{mapleinput}
\mapleinline{active}{1d}{w^2 - x^2 - y^2 - z^2 =\\
simplify(\\
(a*cosh(chi))^2 - (a*sinh(chi)*sin(theta)*cos(phi))^2 -\\
(a*sinh(chi)*sin(theta)*sin(phi))^2 -\\
 (a*sinh(chi)*cos(theta))^2);}{%
}
\end{mapleinput}

\mapleresult
\begin{maplelatex}
\[
w^{2} - x^{2} - y^{2} - z^{2}=a^{2}
\]
\end{maplelatex}

\emptyline
\noindent 
This is expression for 3-hyperboloid in flat 4-space.

\emptyline

\subsection{Standard models}

\emptyline
\noindent
Now we turn to some specific isotropic and homogeneous
models. The assumption of time-dependent character of the geometry
results in the time-dependence of scaling factor \textit{a(t)}. Then
4-metric (spherical geometry) is

\emptyline
\begin{mapleinput}
\mapleinline{active}{1d}{ coord := [t, chi, theta, phi]:
      \indent g_compts := array(symmetric,sparse,1..4,1..4):
     \indent \indent g_compts[1,1] := -1:#dt^2 
    \indent \indent \indent g_compts[2,2] := a(t)^2:#chi^2
   \indent \indent g_compts[3,3]:=a(t)^2*sin(chi)^2:#dtheta^2 
  \indent g_compts[4,4]:=a(t)^2*sin(chi)^2*sin(theta)^2:#dphi^2
 g := create([-1,-1], eval(g_compts));
\indent ginv := invert( g, 'detg' ):}{%
}
\end{mapleinput}

\mapleresult
\begin{maplelatex}
\maplemultiline{
g := \mathrm{table(}[\mathit{index\_char}=[-1, \,-1], \,\mathit{
compts}=\\
 \left[ 
{\begin{array}{rccc}
-1 & 0 & 0 & 0 \\
0 & \mathrm{a}(t)^{2} & 0 & 0 \\
0 & 0 & \mathrm{a}(t)^{2}\,\mathrm{sin}(\chi )^{2} & 0 \\
0 & 0 & 0 & \mathrm{a}(t)^{2}\,\mathrm{sin}(\chi )^{2}\,\mathrm{
sin}(\theta )^{2}
\end{array}}
 \right] ])
}
\end{maplelatex}

\emptyline
\noindent
The Einstein tensor is

\emptyline
\begin{mapleinput}
\mapleinline{active}{1d}{D1g := d1metric( g, coord ):
 \indent D2g := d2metric( D1g, coord ):
  \indent \indent Cf1 := Christoffel1 ( D1g ):
   \indent \indent \indent RMN := Riemann( ginv, D2g, Cf1 ):
    \indent \indent RICCI := Ricci( ginv, RMN ):
     \indent RS := Ricciscalar( ginv, RICCI ):
      G := Einstein( g, RICCI, RS );}{%
}
\end{mapleinput}

\mapleresult
\begin{maplelatex}
\maplemultiline{
G := \mathrm{table(}[\mathit{index\_char}=[-1, \,-1],  \\
\mathit{compts}= \\
 \left[ 
{\begin{array}{c}
{\displaystyle \frac { - 2\,\mathrm{sin}(\chi )^{2} - 3\,(
{\frac {\partial }{\partial t}}\,\mathrm{a}(t))^{2}\,\mathrm{sin}
(\chi )^{2} - 1 + \mathrm{cos}(\chi )^{2}}{\mathrm{a}(t)^{2}\,
\mathrm{sin}(\chi )^{2}}} \,, \,0\,, \,0\,, \,0 \\ [2ex]
0\,, \, - {\displaystyle \frac { - 2\,\mathrm{a}(t)\,\mathrm{sin}
(\chi )^{2}\,({\frac {\partial ^{2}}{\partial t^{2}}}\,\mathrm{a}
(t)) - ({\frac {\partial }{\partial t}}\,\mathrm{a}(t))^{2}\,
\mathrm{sin}(\chi )^{2} - 1 + \mathrm{cos}(\chi )^{2}}{\mathrm{
sin}(\chi )^{2}}} \,, \,0\,, \,0 \\ [2ex]
0\,, \,0\,, \,2\,\mathrm{a}(t)\,\mathrm{sin}(\chi )^{2}\,(
{\frac {\partial ^{2}}{\partial t^{2}}}\,\mathrm{a}(t)) + 
\mathrm{sin}(\chi )^{2} + ({\frac {\partial }{\partial t}}\,
\mathrm{a}(t))^{2}\,\mathrm{sin}(\chi )^{2}\,, \,0 \\
0\,, \,0\,, \,0\,, \,2\,\mathrm{a}(t)\,\mathrm{sin}(\chi )^{2}\,
\mathrm{sin}(\theta )^{2}\,({\frac {\partial ^{2}}{\partial t^{2}
}}\,\mathrm{a}(t)) + \mathrm{sin}(\theta )^{2}\,\mathrm{sin}(\chi
 )^{2} + ({\frac {\partial }{\partial t}}\,\mathrm{a}(t))^{2}\,
\mathrm{sin}(\chi )^{2}\,\mathrm{sin}(\theta )^{2}
\end{array}}
 \right]  \\
]) }
\end{maplelatex}

\emptyline
\noindent
Let's the matter is defined by the energy-momentum tensor\textit{} 
${T_{\mu , \,\nu }}$
 = (\textit{p+
$\rho $
})
${u_{\mu }}$
${u_{\nu }}$
\textit{ +p
${g_{\mu , \,\nu }}$
}:    

\emptyline
\begin{mapleinput}
\mapleinline{active}{1d}{T_compts :=\\
 array(symmetric,sparse,1..4,1..4):# energy-momentum\\
 tensor for drop of liquid
 \indent T_compts[1,1] := rho(t): 
  \indent \indent T_compts[2,2] := a(t)^2*p(t):
   \indent \indent \indent T_compts[3,3] := a(t)^2*sin(chi)^2*p(t): 
    \indent \indent T_compts[4,4] :=\\
     a(t)^2*sin(chi)^2*sin(theta)^2*p(t):
     \indent T := create([-1,-1], eval(T_compts));}{%
}
\end{mapleinput}

\mapleresult
\begin{maplelatex}
\maplemultiline{
T := \mathrm{table(}[\mathit{index\_char}=[-1, \,-1], \mathit{compts}=\\
 \left[ 
{\begin{array}{cccc}
\rho (t) & 0 & 0 & 0 \\
0 & \mathrm{a}(t)^{2}\,\mathrm{p}(t) & 0 & 0 \\
0 & 0 & \mathrm{a}(t)^{2}\,\mathrm{sin}(\chi )^{2}\,\mathrm{p}(t)
 & 0 \\
0 & 0 & 0 & \mathrm{a}(t)^{2}\,\mathrm{sin}(\chi )^{2}\,\mathrm{
sin}(\theta )^{2}\,\mathrm{p}(t)
\end{array}}
 \right]  \\
]) }
\end{maplelatex}

\emptyline
\noindent
Then the first Einstein equation 
${G_{0, \,0}}$
 - 
$\Lambda $
${g_{0, \,0}}$
= - 8
$\pi $
${T_{0, \,0}}$
  (we add the so-called \underline{cosmological constant} 
$\Lambda $
, which can be considered as the energy density of vacuum):

\emptyline
\begin{mapleinput}
\mapleinline{active}{1d}{get_compts(G):
 \indent \%[1,1]:
  \indent \indent e1 := simplify(\%):
\indent \indent \indent get_compts(g):
 \indent \indent \indent \indent \%[1,1]:
  \indent \indent \indent e2 := simplify(\%):
\indent \indent get_compts(T):
 \indent \%[1,1]:
  e3 := simplify(\%):
\indent E[1] := expand(e1/(-3)) = -8*Pi*expand(e3/(-3)) +\\
expand(e2*Lambda/(-3));#first Einstein equation}{%
}
\end{mapleinput}

\mapleresult
\begin{maplelatex}
\[
{E_{1}} := {\displaystyle \frac {1}{\mathrm{a}(t)^{2}}}  + 
{\displaystyle \frac {({\frac {\partial }{\partial t}}\,\mathrm{a
}(t))^{2}}{\mathrm{a}(t)^{2}}} ={\displaystyle \frac {8}{3}} \,
\pi \,\rho (t) + {\displaystyle \frac {1}{3}} \,\Lambda 
\]
\end{maplelatex}

\emptyline
\noindent
Second equation is:

\emptyline
\begin{mapleinput}
\mapleinline{active}{1d}{get_compts(G):
 \indent \%[2,2]:
  \indent \indent e1 := simplify(\%):
\indent \indent \indent get_compts(g):
 \indent \indent \indent \indent \%[2,2]:
  \indent \indent \indent e2 := simplify(\%):
\indent \indent get_compts(T):
 \indent \%[2,2]:
  e3 := simplify(\%):
\indent E[2] :=\\
 expand(e1/a(t)^2) = -8*Pi*expand(e3/a(t)^2) +\\
 expand(e2*Lambda/a(t)^2);#second Einstein equation}{%
}
\end{mapleinput}

\mapleresult
\begin{maplelatex}
\[
{E_{2}} := 2\,{\displaystyle \frac {{\frac {\partial ^{2}}{
\partial t^{2}}}\,\mathrm{a}(t)}{\mathrm{a}(t)}}  + 
{\displaystyle \frac {({\frac {\partial }{\partial t}}\,\mathrm{a
}(t))^{2}}{\mathrm{a}(t)^{2}}}  + {\displaystyle \frac {1}{
\mathrm{a}(t)^{2}}} = - 8\,\pi \,\mathrm{p}(t) + \Lambda 
\]
\end{maplelatex}

\emptyline
\noindent
Two other equations are tautological.\\

\noindent
After elementary transformation we have for the second equation:

\emptyline
\begin{mapleinput}
\mapleinline{active}{1d}{expand(simplify(E[2]-E[1])/2);}{%
}
\end{mapleinput}

\mapleresult
\begin{maplelatex}
\[
{\displaystyle \frac {{\frac {\partial ^{2}}{\partial t^{2}}}\,
\mathrm{a}(t)}{\mathrm{a}(t)}} = - 4\,\pi \,\mathrm{p}(t) + 
{\displaystyle \frac {1}{3}} \,\Lambda  - {\displaystyle \frac {4
}{3}} \,\pi \,\rho (t)
\]
\end{maplelatex}

\emptyline
\noindent
Similarly, for the hyperbolical geometry we have:

\emptyline
\begin{mapleinput}
\mapleinline{active}{1d}{coord := [t, chi, theta, phi]:
 \indent g_compts := array(symmetric,sparse,1..4,1..4):
  \indent \indent g_compts[1,1] := -1:#dt^2 
   \indent \indent \indent g_compts[2,2] := a(t)^2:#chi^2
    \indent \indent g_compts[3,3]:=a(t)^2*sinh(chi)^2:#dtheta^2 
   \indent g_compts[4,4]:=\\
   a(t)^2*sinh(chi)^2*sin(theta)^2:#dphi^2
  g := create([-1,-1], eval(g_compts));
 \indent ginv := invert( g, 'detg' ):}{%
}
\end{mapleinput}

\mapleresult
\begin{maplelatex}
\maplemultiline{
g := \mathrm{table(}[\mathit{index\_char}=[-1, \,-1], \mathit{compts}=\\
 \left[ 
{\begin{array}{rccc}
-1 & 0 & 0 & 0 \\
0 & \mathrm{a}(t)^{2} & 0 & 0 \\
0 & 0 & \mathrm{a}(t)^{2}\,\mathrm{sinh}(\chi )^{2} & 0 \\
0 & 0 & 0 & \mathrm{a}(t)^{2}\,\mathrm{sinh}(\chi )^{2}\,\mathrm{
sin}(\theta )^{2}
\end{array}}
 \right]  \\
]) }
\end{maplelatex}

\begin{mapleinput}
\mapleinline{active}{1d}{D1g := d1metric( g, coord ):
 \indent D2g := d2metric( D1g, coord ):
  \indent \indent Cf1 := Christoffel1 ( D1g ):
   \indent \indent \indent RMN := Riemann( ginv, D2g, Cf1 ):
    \indent \indent RICCI := Ricci( ginv, RMN ):
     \indent RS := Ricciscalar( ginv, RICCI ):
      G := Einstein( g, RICCI, RS );}{%
}
\end{mapleinput}

\mapleresult
\begin{maplelatex}
\maplemultiline{
G := \mathrm{table(}[\mathit{index\_char}=[-1, \,-1], \mathit{compts}=\\
 \left[ 
{\begin{array}{c}
 - {\displaystyle \frac { - 2\,\mathrm{sinh}(\chi )^{2} + 3\,(
{\frac {\partial }{\partial t}}\,\mathrm{a}(t))^{2}\,\mathrm{sinh
}(\chi )^{2} + 1 - \mathrm{cosh}(\chi )^{2}}{\mathrm{a}(t)^{2}\,
\mathrm{sinh}(\chi )^{2}}} \,, \,0\,, \,0\,, \,0 \\ [2ex]
0\,, \,{\displaystyle \frac {2\,\mathrm{a}(t)\,\mathrm{sinh}(\chi
 )^{2}\,({\frac {\partial ^{2}}{\partial t^{2}}}\,\mathrm{a}(t))
 + ({\frac {\partial }{\partial t}}\,\mathrm{a}(t))^{2}\,\mathrm{
sinh}(\chi )^{2} + 1 - \mathrm{cosh}(\chi )^{2}}{\mathrm{sinh}(
\chi )^{2}}} \,, \,0\,, \,0 \\ [2ex]
0\,, \,0\,, \,2\,\mathrm{a}(t)\,\mathrm{sinh}(\chi )^{2}\,(
{\frac {\partial ^{2}}{\partial t^{2}}}\,\mathrm{a}(t)) - 
\mathrm{sinh}(\chi )^{2} + ({\frac {\partial }{\partial t}}\,
\mathrm{a}(t))^{2}\,\mathrm{sinh}(\chi )^{2}\,, \,0 \\
0\,, \,0\,, \,0\,, \,2\,\mathrm{a}(t)\,\mathrm{sinh}(\chi )^{2}\,
\mathrm{sin}(\theta )^{2}\,({\frac {\partial ^{2}}{\partial t^{2}
}}\,\mathrm{a}(t)) - \mathrm{sin}(\theta )^{2}\,\mathrm{sinh}(
\chi )^{2} + ({\frac {\partial }{\partial t}}\,\mathrm{a}(t))^{2}
\,\mathrm{sinh}(\chi )^{2}\,\mathrm{sin}(\theta )^{2}
\end{array}}
 \right]  \\
]) }
\end{maplelatex}

\begin{mapleinput}
\mapleinline{active}{1d}{T_compts :=\\
 array(symmetric,sparse,1..4,1..4):# energy-momentum\\
 tensor for drop of liquid
 \indent T_compts[1,1] := rho(t): 
  \indent \indent T_compts[2,2] := a(t)^2*p(t):
   \indent T_compts[3,3] := a(t)^2*sinh(chi)^2*p(t): 
    T_compts[4,4] := a(t)^2*sinh(chi)^2*sin(theta)^2*p(t):
     \indent T := create([-1,-1], eval(T_compts));}{%
}
\end{mapleinput}

\mapleresult
\begin{maplelatex}
\maplemultiline{
T := \mathrm{table(}[\mathit{index\_char}=[-1, \,-1],  \mathit{compts}=\\
 \left[ 
{\begin{array}{cccc}
\rho (t) & 0 & 0 & 0 \\
0 & \mathrm{a}(t)^{2}\,\mathrm{p}(t) & 0 & 0 \\
0 & 0 & \mathrm{a}(t)^{2}\,\mathrm{sinh}(\chi )^{2}\,\mathrm{p}(t
) & 0 \\
0 & 0 & 0 & \mathrm{a}(t)^{2}\,\mathrm{sinh}(\chi )^{2}\,\mathrm{
sin}(\theta )^{2}\,\mathrm{p}(t)
\end{array}}
 \right]  \\
]) }
\end{maplelatex}

\begin{mapleinput}
\mapleinline{active}{1d}{get_compts(G):
 \indent \%[1,1]:
  \indent \indent e1 := simplify(\%):
\indent \indent \indent get_compts(g):
 \indent \indent \indent \indent \%[1,1]:
  \indent \indent \indent e2 := simplify(\%):
\indent \indent get_compts(T):
 \indent \%[1,1]:
  e3 := simplify(\%):
\indent E[1] :=\\
 expand(e1/(-3)) = -8*Pi*expand(e3/(-3)) +\\
 expand(e2*Lambda/(-3));#first Einstein equation}{%
}
\end{mapleinput}

\mapleresult
\begin{maplelatex}
\[
{E_{1}} :=  - {\displaystyle \frac {1}{\mathrm{a}(t)^{2}}}  + 
{\displaystyle \frac {({\frac {\partial }{\partial t}}\,\mathrm{a
}(t))^{2}}{\mathrm{a}(t)^{2}}} ={\displaystyle \frac {8}{3}} \,
\pi \,\rho (t) + {\displaystyle \frac {1}{3}} \,\Lambda 
\]
\end{maplelatex}

\begin{mapleinput}
\mapleinline{active}{1d}{get_compts(G):
 \indent \%[2,2]:
  \indent \indent e1 := simplify(\%):
\indent \indent \indent get_compts(g):
 \indent \indent \indent \indent \%[2,2]:
  \indent \indent \indent e2 := simplify(\%):
\indent \indent get_compts(T):
 \indent \%[2,2]:
  e3 := simplify(\%):
\indent E[2] :=\\
 expand(e1/a(t)^2) = -8*Pi*expand(e3/a(t)^2) +\\
 expand(e2*Lambda/a(t)^2):#second Einstein equation
\indent \indent expand(simplify(E[2]-E[1])/2);}{%
}
\end{mapleinput}

\mapleresult
\begin{maplelatex}
\[
{\displaystyle \frac {{\frac {\partial ^{2}}{\partial t^{2}}}\,
\mathrm{a}(t)}{\mathrm{a}(t)}} = - 4\,\pi \,\mathrm{p}(t) + 
{\displaystyle \frac {1}{3}} \,\Lambda  - {\displaystyle \frac {4
}{3}} \,\pi \,\rho (t)
\]
\end{maplelatex}

\emptyline
\noindent
The second equation is identical with one in spherical geometry.\\

\noindent
In general, the first and second differential equations for scaling
factor in our ideal universe:

\emptyline
\begin{mapleinput}
\mapleinline{active}{1d}{basic_1 :=\\
 (diff(a(t),t)/a(t))^2 = -K/a(t)^2 + Lambda/3 +\\
 8*Pi*rho(t)/3;#first basic equation for\\
  homogeneous universe
\indent basic_2 :=\\
 diff(a(t),`\$`(t,2))/a(t) =\\
 -4*Pi*(p(t)+rho(t)/3)+1/3*Lambda;# second\\
 basic equation for homogeneous universe}{%
}
\end{mapleinput}

\mapleresult
\begin{maplelatex}
\[
\mathit{basic\_1} := {\displaystyle \frac {({\frac {\partial }{
\partial t}}\,\mathrm{a}(t))^{2}}{\mathrm{a}(t)^{2}}} = - 
{\displaystyle \frac {K}{\mathrm{a}(t)^{2}}}  + {\displaystyle 
\frac {1}{3}} \,\Lambda  + {\displaystyle \frac {8}{3}} \,\pi \,
\rho (t)
\]
\end{maplelatex}

\begin{maplelatex}
\[
\mathit{basic\_2} := {\displaystyle \frac {{\frac {\partial ^{2}
}{\partial t^{2}}}\,\mathrm{a}(t)}{\mathrm{a}(t)}} = - 4\,\pi \,(
\mathrm{p}(t) + {\displaystyle \frac {1}{3}} \,\rho (t)) + 
{\displaystyle \frac {1}{3}} \,\Lambda 
\]
\end{maplelatex}

\emptyline
\noindent
Here \textit{K}=+1 for spherical, -1 for hyperbolical and 0 for flat
geometries. The left-hand side of the first equation is the square of
\underline{Hubble constant} \textit{H}. If we denote recent value of
this parameter as 
${H_{0}}$
, then 
${H_{0}}$
 = 65 
$\frac {\mathit{km}}{s\,\mathit{Mpc}}$
 (\textit{ps} is the parallax second corresponding to distance about
of 3.26 \textit{light-years} or 3*
$10^{13}$
 \textit{km}) \cite{J. Garcia-Bellido}.\\

\noindent
When \textit{K=
$\Lambda $
=}0, the universe is defined by the so-called critical density 
${\rho _{c}}$
= 
$\frac {3\,{H_{0}}^{2}}{8\,\pi }$
 =7.9*
$10^{( - 30)}$
$\frac {g}{\mathit{cm}^{3}}$
. Let's denote the recent relative values 
${\Omega _{i}}$
= 
$\frac {{\rho _{i}}}{{\rho _{c}}}$
 of the matter, radiation, cosmological constant and curvature as:

\begin{center}
${\Omega _{M}}$
 = 
$\frac {8\,\pi \,{\rho _{M}}}{3\,{H_{0}}^{2}}$
, 
${\Omega _{R}}$
 = 
$\frac {8\,\pi \,{\rho _{R}}}{3\,{H_{0}}^{2}}$
, 
${\Omega _{\Lambda }}$
 = 
$\frac {\Lambda }{3\,{H_{0}}^{2}}$
, 
${\Omega _{K}}$
=\textit{ - 
$\frac {K}{{a_{0}}^{2}\,{H_{0}}^{2}}$
}.
\end{center}

\noindent
Here 
${a_{0}}$
 is the recent scaling factor. At this moment the contribution of the
radiation to density is negligible 
${\Omega _{R}}$
\TEXTsymbol{<}\TEXTsymbol{<}
${\Omega _{M}}$
. To rewrite the evolutionary equation in terms of 
${\Omega _{i}}$
 we have to take into account the time dependence of these
parameters: 
$\Lambda $
 does not depend on \textit{a}(\textit{t}) (in framework of standard
model), the dependence of curvature \symbol{126}
$\frac {1}{a^{2}}$
, density of matter \symbol{126}
$\frac {1}{a^{3}}$
, but for radiation \symbol{126}
$\frac {1}{a^{4}}$
. The last results from the change of density of quanta \symbol{126}
$\frac {1}{a^{3}}$
 and their energy \symbol{126}
$\frac {1}{a}$
. The resulting equation can be written as:

\emptyline
\begin{mapleinput}
\mapleinline{active}{1d}{basic_3 := H(a)^2 =\\
 H[0]^2*(Omega[R]*a[0]^4/a^4 + Omega[M]*a[0]^3/a^3\\
 + Omega[Lambda] + Omega[K]*a[0]^2/a^2);}{%
}
\end{mapleinput}

\mapleresult
\begin{maplelatex}
\[
\mathit{basic\_3} := \mathrm{H}(a)^{2}={H_{0}}^{2}\,(
{\displaystyle \frac {{\Omega _{R}}\,{a_{0}}^{4}}{a^{4}}}  + 
{\displaystyle \frac {{\Omega _{M}}\,{a_{0}}^{3}}{a^{3}}}  + {
\Omega _{\Lambda }} + {\displaystyle \frac {{\Omega _{K}}\,{a_{0}
}^{2}}{a^{2}}} )
\]
\end{maplelatex}

\emptyline
\noindent
As result of these definitions we have the \underline{cosmic sum
rule}: 

\begin{center}
1 = 
${\Omega _{M}} + {\Omega _{R}}$
 + 
${\Omega _{\Lambda }}$
 + 
${\Omega _{K}}$
.
\end{center}

\noindent
Let us transit to the new variables: \textit{y}=
$\frac {a}{{a_{0}}}$
, 
$\tau $
=
${H_{0}}$
(
$t - {t_{0}}$
), where 
${t_{0}}$
 is the present time moment. Then

\emptyline
\begin{mapleinput}
\mapleinline{active}{1d}{basic_4 :=\\
 diff(y(tau),tau) = sqrt(1 + (1/y(tau)-1)*Omega[M] +\\
 (y(tau)^2-1)*Omega[Lambda]);#we neglect the radiation\\
  contribution and use the cosmic sum rule.\\
   d(tau)=H[0]*d(t) --> a[0]*H[0]*(d(y)/d(tau))/a =\\
    (d(a)/d(t))/a = H}{%
}
\end{mapleinput}

\mapleresult
\begin{maplelatex}
\[
\mathit{basic\_4} := {\frac {\partial }{\partial \tau }}\,
\mathrm{y}(\tau )=\sqrt{1 + ({\displaystyle \frac {1}{\mathrm{y}(
\tau )}}  - 1)\,{\Omega _{M}} + (\mathrm{y}(\tau )^{2} - 1)\,{
\Omega _{\Lambda }}}
\]
\end{maplelatex}

\emptyline
\noindent
Now we will consider some typical standard models. For review see, for
example, \cite{P.J.E. Peebles}, or on-line survey 
\cite{P. B. Pal}.

\emptyline

\subsubsection{Einstein static}

\emptyline
\begin{mapleinput}
\mapleinline{active}{1d}{rhs(basic_1) = 0:
 \indent rhs(basic_2) = 0:
  \indent \indent allvalues(\\
   solve(\{\%, \%\%\},\{a(t), Lambda\}) );# static universe}{%
}
\end{mapleinput}

\mapleresult
\begin{maplelatex}
\maplemultiline{
\{\Lambda =12\,\pi \,\mathrm{p}(t) + 4\,\pi \,\rho (t), \,
\mathrm{a}(t)=\sqrt{{\displaystyle \frac {K}{4\,\pi \,\mathrm{p}(
t) + 4\,\pi \,\rho (t)}} }\},  \\
\{\Lambda =12\,\pi \,\mathrm{p}(t) + 4\,\pi \,\rho (t), \,
\mathrm{a}(t)= - \sqrt{{\displaystyle \frac {K}{4\,\pi \,\mathrm{
p}(t) + 4\,\pi \,\rho (t)}} }\} }
\end{maplelatex}

\emptyline
\noindent
At this moment for the matter \textit{p}=0 and the contribution of the
radiation is small, therefore:

\emptyline
\begin{center}
$a^{2}$
 =
$\frac {1}{4\,\pi \,\rho }, \,\Lambda =4\,\pi \,\rho $
\end{center}

\emptyline
\noindent
This solution has a very simple form and requires \textit{K} = +1
(the closed spherical universe) and positive 
$\Lambda $
 (repulsive action of vacuum), but does not satisfy the observations
of red shift in the spectra of extragalactic sources. The last is the
well-known fact demonstrating the expansion of universe (the increase of scaling factor).

\emptyline

\subsubsection{Einstein-de Sitter}

\emptyline
\noindent
${\Omega _{M}}$
 = 1 and, as consequence 
${\Omega _{\Lambda }}$
 = 
${\Omega _{K}}$
 = 0. This is the flat universe described by equation

\emptyline
\begin{mapleinput}
\mapleinline{active}{1d}{y(tau)^(1/2)*diff(y(tau),tau) = 1;
 \indent dsolve(\{\%, y(0)=1\}, y(tau)):
  \indent \indent allvalues(\%);}{%
}
\end{mapleinput}

\mapleresult
\begin{maplelatex}
\[
\sqrt{\mathrm{y}(\tau )}\,({\frac {\partial }{\partial \tau }}\,
\mathrm{y}(\tau ))=1
\]
\end{maplelatex}

\begin{maplelatex}
\maplemultiline{
\mathrm{y}(\tau )=\\
{\displaystyle \frac {1}{4}} \,(12\,\tau  + 8)
^{(2/3)}, \,\mathrm{y}(\tau )=( - {\displaystyle \frac {1}{4}} \,
(12\,\tau  + 8)^{(1/3)} + {\displaystyle \frac {1}{4}} \,I\,
\sqrt{3}\,(12\,\tau  + 8)^{(1/3)})^{2},  \\
\mathrm{y}(\tau )=\\
( - {\displaystyle \frac {1}{4}} \,(12\,\tau 
 + 8)^{(1/3)} - {\displaystyle \frac {1}{4}} \,I\,\sqrt{3}\,(12\,
\tau  + 8)^{(1/3)})^{2} }
\end{maplelatex}

\emptyline
\begin{mapleinput}
\mapleinline{active}{1d}{plot(1/4*(12*tau+8)^(2/3), \\
tau= -2/3..2, axes=BOXED,\\
 title=`Einstein-de Sitter universe`);}{%
}
\end{mapleinput}

\mapleresult
\begin{center}
\mapleplot{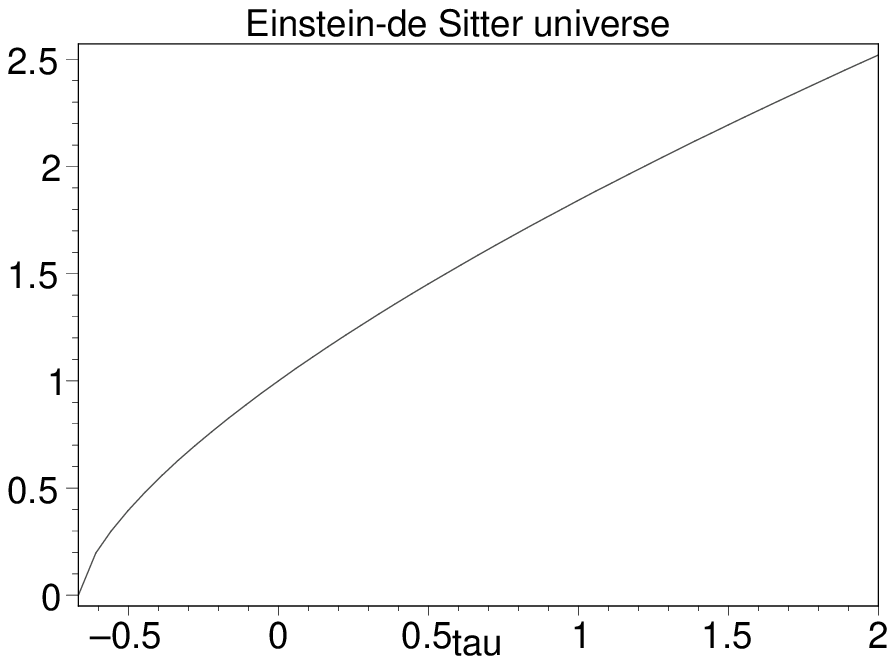}
\end{center}

\emptyline
\noindent
We have the infinite expansion from the initial singularity (Big
Bang).\\

\noindent
From the value of scaling factor 
$\mathrm{y}(\tau )=\frac {(12\,\tau  + 8)^{(\frac {2}{3})}}{4}$
 it is possible to find the universe's age: 

\emptyline
\begin{mapleinput}
\mapleinline{active}{1d}{0 =\\
 (12*H[0]*(0-t[0]) + 8)^(2/3)/4:# y(0)=0
 \indent solve(\%,t[0]);}{%
}
\end{mapleinput}

\mapleresult
\begin{maplelatex}
\[
{\displaystyle \frac {2}{3}} \,{\displaystyle \frac {1}{{H_{0}}}
} 
\]
\end{maplelatex}

\begin{mapleinput}
\mapleinline{active}{1d}{evalf((2/3)*3*10^19/65/60/60/24/365);#[yr]}{%
}
\end{mapleinput}

\mapleresult
\begin{maplelatex}
\[
.9756859072\,10^{10}
\]
\end{maplelatex}

\emptyline
\noindent
The obtained universe's age does not satisfy the observations of the star's age in
older globular clusters.

\emptyline

\subsubsection{de Sitter and anti-de Sitter}

\emptyline
\noindent
The empty universes (
${\Omega _{M}}$
=0) with \textit{R} \TEXTsymbol{>} 0 (de Sitter) and \textit{R
}\TEXTsymbol{<} 0 (anti-de Sitter). Turning to \textit{basic\_1} we
have for de Sitter universe:

\emptyline
\begin{mapleinput}
\mapleinline{active}{1d}{assume(lambda,positive):
 \indent subs(\{rho(t)=0,K=1,Lambda=lambda*3\},\\
 basic_1):#lambda=Lambda/3
  \indent \indent expand(\%*a(t)^2);
   \indent \indent \indent dsolve(\%,a(t));}{%
}
\end{mapleinput}

\mapleresult
\begin{maplelatex}
\[
({\frac {\partial }{\partial t}}\,\mathrm{a}(t))^{2}= - 1 + 
\mathrm{a}(t)^{2}\,\lambda \symbol{126}
\]
\end{maplelatex}

\begin{maplelatex}
\maplemultiline{
\mathrm{a}(t)={\displaystyle \frac {1}{\sqrt{\lambda \symbol{126}
}}} , \,\mathrm{a}(t)= - {\displaystyle \frac {1}{\sqrt{\lambda 
\symbol{126}}}} ,\\
 \,\mathrm{a}(t)={\displaystyle \frac {1}{2}} \,
{\displaystyle \frac {(1 + {\displaystyle \frac {\mathrm{\%2}^{2}
}{\mathrm{\%1}^{2}}} )\,\mathrm{\%1}}{\mathrm{\%2}\,\sqrt{\lambda
 \symbol{126}}}} , \,\mathrm{a}(t)={\displaystyle \frac {1}{2}} 
\,{\displaystyle \frac {(1 + {\displaystyle \frac {\mathrm{\%1}^{
2}}{\mathrm{\%2}^{2}}} )\,\mathrm{\%2}}{\mathrm{\%1}\,\sqrt{
\lambda \symbol{126}}}}  \\
\mathrm{\%1} := e^{(\mathit{\_C1}\,\sqrt{\lambda \symbol{126}})}
 \\
\mathrm{\%2} := e^{(t\,\sqrt{\lambda \symbol{126}})} }
\end{maplelatex}

\emptyline
\noindent
It is obvious, that the first two solutions don't satisfy the
\textit{basic\_2}. Another solutions give:

\begin{center}
   \textit{a}(\textit{t}) = 
$\frac {\mathrm{cosh}(\sqrt{\lambda }\,(t - C))}{2\,\sqrt{\lambda
 }}$
\end{center}

\noindent
where \textit{C} is the positive or negative constant defining the
time moment corresponding to the minimal scaling factor.

\emptyline
\begin{mapleinput}
\mapleinline{active}{1d}{plot(subs(lambda=0.5,\\
cosh(sqrt(lambda)*tau)/2/sqrt(lambda)),\\
tau=-3..3,title=`de Sitter universe`,axes=boxed);}{%
}
\end{mapleinput}

\mapleresult
\begin{center}
\mapleplot{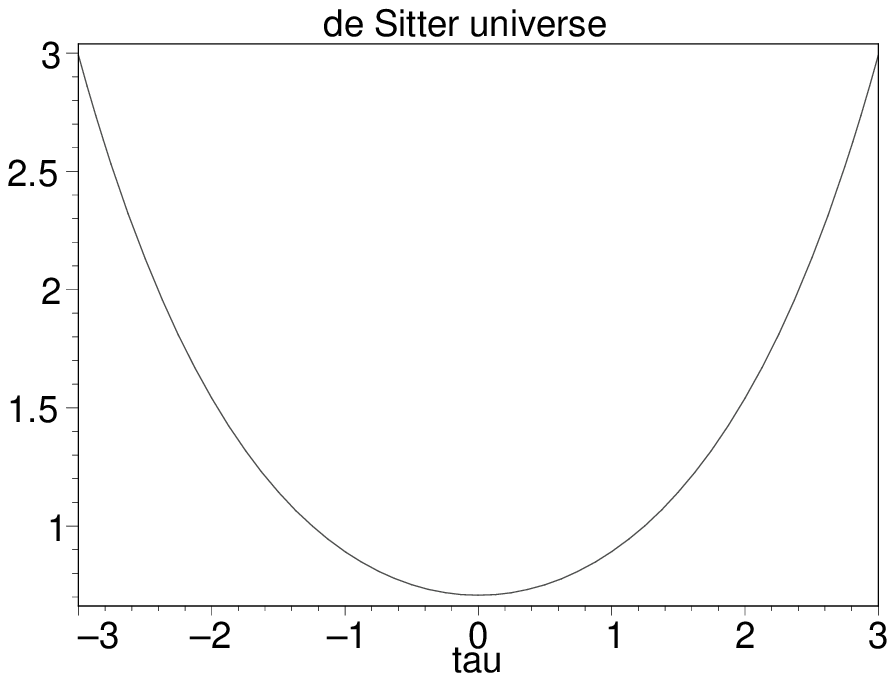}
\end{center}

\emptyline
\noindent
For anti-de Sitter model we have:

\emptyline
\begin{mapleinput}
\mapleinline{active}{1d}{assume(lambda,negative):
\indent subs(\{rho(t)=0,K=-1,Lambda=lambda*3\},\\
basic_1):#lambda=Lambda/3
 \indent \indent expand(\%*a(t)^2);
  \indent \indent \indent dsolve(\%,a(t));}{%
}
\end{mapleinput}

\mapleresult
\begin{maplelatex}
\[
({\frac {\partial }{\partial t}}\,\mathrm{a}(t))^{2}=1 + \mathrm{
a}(t)^{2}\,\lambda \symbol{126}
\]
\end{maplelatex}

\begin{maplelatex}
\maplemultiline{
\mathrm{a}(t)={\displaystyle \frac {\sqrt{ - \lambda \symbol{126}
}}{\lambda \symbol{126}}} , \,\mathrm{a}(t)= - {\displaystyle 
\frac {\sqrt{ - \lambda \symbol{126}}}{\lambda \symbol{126}}} ,\\ 
\,\mathrm{a}(t)= - {\displaystyle \frac {\mathrm{sin}(t\,\sqrt{
 - \lambda \symbol{126}} - \mathit{\_C1}\,\sqrt{ - \lambda 
\symbol{126}})}{\sqrt{ - \lambda \symbol{126}}}} ,  \\
\mathrm{a}(t)={\displaystyle \frac {\mathrm{sin}(t\,\sqrt{ - 
\lambda \symbol{126}} - \mathit{\_C1}\,\sqrt{ - \lambda 
\symbol{126}})}{\sqrt{ - \lambda \symbol{126}}}}  }
\end{maplelatex}

\emptyline
\noindent
De Sitter and anti-de Sitter universes play a crucial role in many
modern issues in GR and cosmology, in particular, in inflationary
scenarios of the beginning of universe evolution. 

\emptyline

\subsubsection{Closed Friedmann-Lemaitre}

\emptyline
${\Omega _{M}}$
 \TEXTsymbol{>} 1 
$\Lambda $
 = 0 and, as consequence 
${\Omega _{K}}$
 \TEXTsymbol{<} 0 (\textit{K}=+1). This is the closed spherical
universe.

\emptyline
\begin{mapleinput}
\mapleinline{active}{1d}{y(tau)*(diff(y(tau),tau)^2+b) =\\
 Omega[M];# b=Omega[M]-1 > 0}{%
}
\end{mapleinput}

\mapleresult
\begin{maplelatex}
\[
\mathrm{y}(\tau )\,(({\frac {\partial }{\partial \tau }}\,
\mathrm{y}(\tau ))^{2} + b)={\Omega _{M}}
\]
\end{maplelatex}

\begin{mapleinput}
\mapleinline{active}{1d}{solve( \%,diff(y(tau),tau) );}{%
}
\end{mapleinput}

\mapleresult
\begin{maplelatex}
\[
{\displaystyle \frac {\sqrt{\mathrm{y}(\tau )\,( - \mathrm{y}(
\tau )\,b + {\Omega _{M}})}}{\mathrm{y}(\tau )}} , \, - 
{\displaystyle \frac {\sqrt{\mathrm{y}(\tau )\,( - \mathrm{y}(
\tau )\,b + {\Omega _{M}})}}{\mathrm{y}(\tau )}} 
\]
\end{maplelatex}

\begin{mapleinput}
\mapleinline{active}{1d}{d(t)=sqrt(y/b/(Omega[M]/b - y))*d(y);}{%
}
\end{mapleinput}

\mapleresult
\begin{maplelatex}
\[
\mathrm{d}(t)=\sqrt{{\displaystyle \frac {y}{b\,({\displaystyle 
\frac {{\Omega _{M}}}{b}}  - y)}} }\,\mathrm{d}(y)
\]
\end{maplelatex}

\emptyline
\noindent
Introducing a new variable 
$\phi $ we have
:

\emptyline
\begin{mapleinput}
\mapleinline{active}{1d}{sqrt(y/(Omega[M]/b - y)) = tan(phi);
 \indent y = solve(\%, y);
  \indent \indent y = convert(rhs(\%),sincos);}{%
}
\end{mapleinput}

\mapleresult
\begin{maplelatex}
\[
\sqrt{{\displaystyle \frac {y}{{\displaystyle \frac {{\Omega _{M}
}}{b}}  - y}} }=\mathrm{tan}(\phi )
\]
\end{maplelatex}

\begin{maplelatex}
\[
y={\displaystyle \frac {\mathrm{tan}(\phi )^{2}\,{\Omega _{M}}}{b
\,(1 + \mathrm{tan}(\phi )^{2})}} 
\]
\end{maplelatex}

\begin{maplelatex}
\[
y={\displaystyle \frac {\mathrm{sin}(\phi )^{2}\,{\Omega _{M}}}{
\mathrm{cos}(\phi )^{2}\,b\,(1 + {\displaystyle \frac {\mathrm{
sin}(\phi )^{2}}{\mathrm{cos}(\phi )^{2}}} )}} 
\]
\end{maplelatex}

\emptyline
\noindent
But this is \textit{y = 
$\frac {{\Omega _{M}}\,\mathrm{sin}(\phi )^{2}}{b}$
} = 
$\frac {{\Omega _{M}}\,(1 - \mathrm{cos}(2\,\phi ))}{2\,b}$
. And

\emptyline
\begin{mapleinput}
\mapleinline{active}{1d}{defform(f=0,w1=0,w2=0,v=1,phi=0,y=0);
 \indent d(y) = d( (Omega[M]/b)*sin(phi)^2 );}{%
}
\end{mapleinput}

\mapleresult
\begin{maplelatex}
\[
\mathrm{d}(y)=2\,\mathrm{sin}(\phi )\,\mathrm{cos}(\phi )\,(
\mathrm{d}(\phi )\,\mathrm{\&\symbol{94}}\,{\displaystyle \frac {
{\Omega _{M}}}{b}} ) + \mathrm{sin}(\phi )^{2}\,\mathrm{d}(
{\displaystyle \frac {{\Omega _{M}}}{b}} )
\]
\end{maplelatex}

\emptyline
\noindent
Then our equation results in

\emptyline
\begin{mapleinput}
\mapleinline{active}{1d}{subs(\\
y=Omega[M]*sin(phi)^2/b,sqrt(y/(b*(Omega[M]/b-y))) ):
 \indent simplify(\%);}{%
}
\end{mapleinput}

\mapleresult
\begin{maplelatex}
\[
\sqrt{ - {\displaystyle \frac { - 1 + \mathrm{cos}(\phi )^{2}}{b
\,\mathrm{cos}(\phi )^{2}}} }
\]
\end{maplelatex}

\begin{mapleinput}
\mapleinline{active}{1d}{d(t) = simplify(\\
(Omega[M]/b)*tan(phi)*2*sin(phi)*cos(phi))\\
*d(phi)/sqrt(b);}{%
}
\end{mapleinput}

\mapleresult
\begin{maplelatex}
\[
\mathrm{d}(t)= - 2\,{\displaystyle \frac {{\Omega _{M}}\,( - 1 + 
\mathrm{cos}(\phi )^{2})\,\mathrm{d}(\phi )}{b^{(3/2)}}} 
\]
\end{maplelatex}

\emptyline
\noindent
Hence, the age of universe is:

\emptyline
\begin{mapleinput}
\mapleinline{active}{1d}{#initial conditions: y(0)=0 --> phi=0
  \indent int( Omega[M]*(1-cos(2*x))/(Omega[M]-1)^(3/2),\\
   x=0..phi):
   \indent \indent sol_1 := t = simplify(\%);#age\\
    of universe vs phi
}{%
}
\end{mapleinput}

\mapleresult
\begin{maplelatex}
\[
\mathit{sol\_1} := t= - {\displaystyle \frac {1}{2}} \,
{\displaystyle \frac {{\Omega _{M}}\,( - 2\,\phi  + \mathrm{sin}(
2\,\phi ))}{({\Omega _{M}} - 1)^{(3/2)}}} 
\]
\end{maplelatex}

\emptyline
\noindent
This equation in the combination with \textit{y}(
$\phi $
) is the parametric representation of cycloid:

\emptyline
\begin{mapleinput}
\mapleinline{active}{1d}{y = Omega[M]*(1-cos(2*phi))/(2*(Omega[M]-1)):
\indent plot([subs(Omega[M]=5,rhs(sol_1)),\\
subs(Omega[M]=5,rhs(\%)),phi=0..Pi], axes=boxed,\\
 title=`closed Friedmann-Lemaitre universe`);}{%
}
\end{mapleinput}

\mapleresult
\begin{center}
\mapleplot{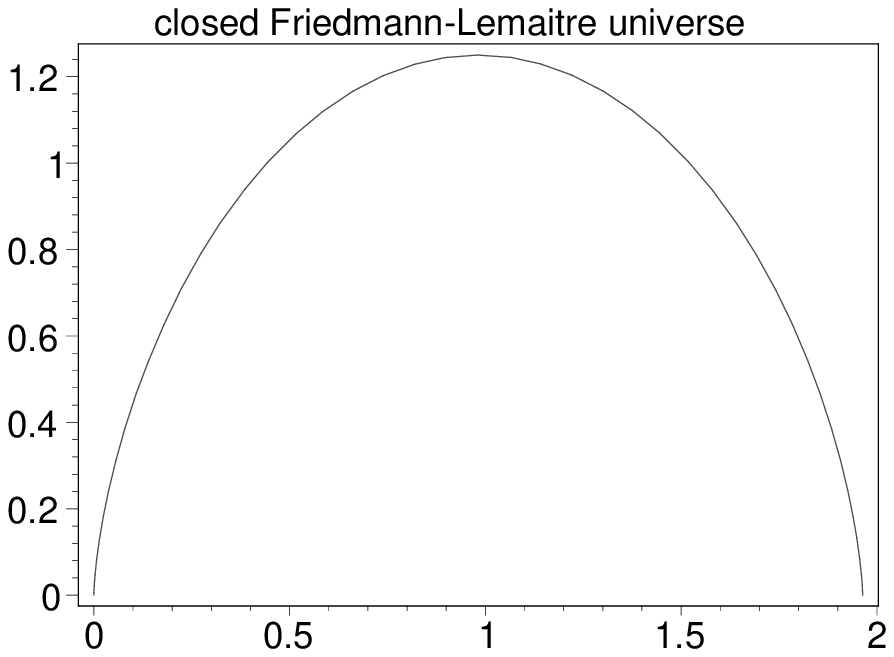}
\end{center}

\emptyline
\noindent
In this model the expansion from the singularity changes into
contraction and terminates in singularity (Big Crunch).\\ 

\noindent
The age of universe is:

\emptyline
\begin{mapleinput}
\mapleinline{active}{1d}{#initial conditions: y(t0)=1 --> phi=?
 \indent 1 = sin(phi)^2*Omega[M]/(Omega[M]-1):
  \indent \indent sol_2 := solve(\%,phi);
   \indent subs(phi=sol_2[1],rhs(sol_1));#age \\
   of universe vs Omega[M]
    plot(\%, Omega[M]=1.01..10, axes=BOXED, \\
    title=`age of universe [1/H[0] ]`);}{%
}
\end{mapleinput}

\mapleresult
\begin{maplelatex}
\[
\mathit{sol\_2} := \mathrm{arcsin}({\displaystyle \frac {\sqrt{{
\Omega _{M}}\,({\Omega _{M}} - 1)}}{{\Omega _{M}}}} ), \, - 
\mathrm{arcsin}({\displaystyle \frac {\sqrt{{\Omega _{M}}\,({
\Omega _{M}} - 1)}}{{\Omega _{M}}}} )
\]
\end{maplelatex}

\begin{maplelatex}
\[
 - {\displaystyle \frac {1}{2}} \,{\displaystyle \frac {{\Omega 
_{M}}\,( - 2\,\mathrm{arcsin}({\displaystyle \frac {\sqrt{{\Omega
 _{M}}\,({\Omega _{M}} - 1)}}{{\Omega _{M}}}} ) + \mathrm{sin}(2
\,\mathrm{arcsin}({\displaystyle \frac {\sqrt{{\Omega _{M}}\,({
\Omega _{M}} - 1)}}{{\Omega _{M}}}} )))}{({\Omega _{M}} - 1)^{(3/
2)}}} 
\]
\end{maplelatex}

\begin{center}
\mapleplot{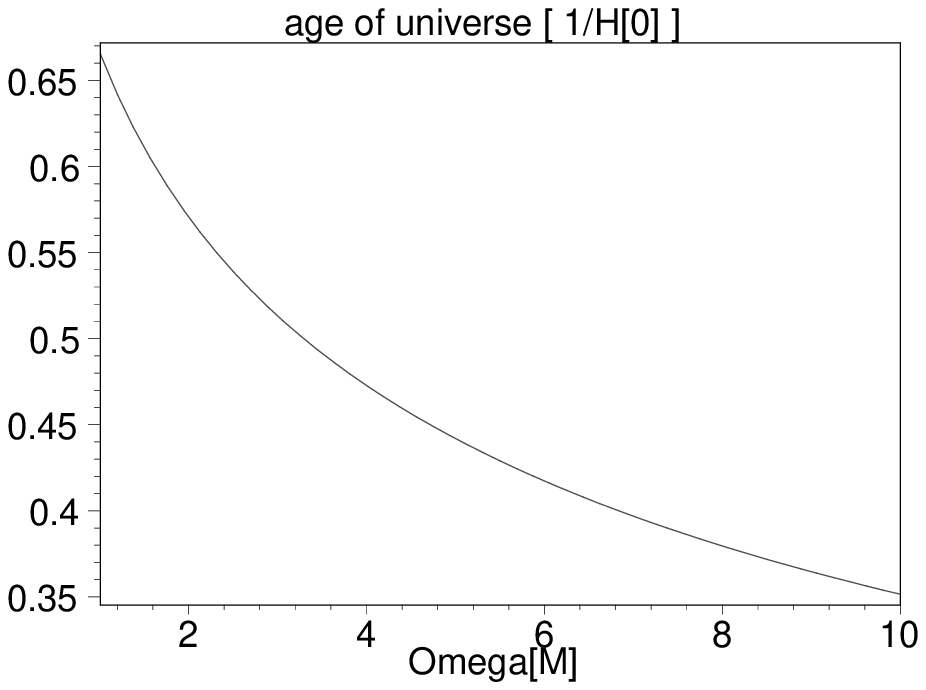}
\end{center}

\emptyline
\noindent
So, the increase of 
${\Omega _{M}}$
 decreases the age of universe, which is less than in Einstein-de
Sitter model. The last and the modern observations of microwave
background suggesting 
${\Omega _{K}}$
\symbol{126}0 do not agree with this model. 

\emptyline

\subsubsection{Open Friedmann-Lemaitre}

\emptyline
\noindent
${\Omega _{M}}$
 \TEXTsymbol{<} 1 
$\Lambda $
 = 0 and, as consequence 
${\Omega _{K}}$
 \TEXTsymbol{>} 0 (\textit{K}= -1). This is the open hyperbolical
spherical universe. From the above introduced equation we have

\emptyline
\begin{mapleinput}
\mapleinline{active}{1d}{y(tau)*(diff(y(tau),tau)^2 - a) =\\
 Omega[M];# a = -b = 1-Omega[M]>0}{%
}
\end{mapleinput}

\mapleresult
\begin{maplelatex}
\[
\mathrm{y}(\tau )\,(({\frac {\partial }{\partial \tau }}\,
\mathrm{y}(\tau ))^{2} - a)={\Omega _{M}}
\]
\end{maplelatex}

\begin{mapleinput}
\mapleinline{active}{1d}{solve( \%,diff(y(tau),tau) );}{%
}
\end{mapleinput}

\mapleresult
\begin{maplelatex}
\[
{\displaystyle \frac {\sqrt{\mathrm{y}(\tau )\,(\mathrm{y}(\tau )
\,a + {\Omega _{M}})}}{\mathrm{y}(\tau )}} , \, - {\displaystyle 
\frac {\sqrt{\mathrm{y}(\tau )\,(\mathrm{y}(\tau )\,a + {\Omega 
_{M}})}}{\mathrm{y}(\tau )}} 
\]
\end{maplelatex}

\begin{mapleinput}
\mapleinline{active}{1d}{diff(y(tau), tau)=sqrt((Omega[M] +\\
a*y(tau))/y(tau));}{%
}
\end{mapleinput}

\mapleresult
\begin{maplelatex}
\[
{\frac {\partial }{\partial \tau }}\,\mathrm{y}(\tau )=\sqrt{
{\displaystyle \frac {\mathrm{y}(\tau )\,a + {\Omega _{M}}}{
\mathrm{y}(\tau )}} }
\]
\end{maplelatex}

\begin{mapleinput}
\mapleinline{active}{1d}{dsolve(\%, y(tau)):
 \indent sol_1 := simplify( subs(a=1-Omega[M],\%),\\
 radical,symbolic);#implicit solution for y(tau)}{%
}
\end{mapleinput}

\mapleresult
\begin{maplelatex}
\maplemultiline{
\mathit{sol\_1} :=  - {\displaystyle \frac {1}{2}} ( - 2\,\tau \,
\mathrm{\%1}\,{\Omega _{M}} + {\Omega _{M}}\,\mathrm{ln}( - {\displaystyle \frac {1}{
2}}\\
 \,{\displaystyle \frac { - {\Omega _{M}} - 2\,\mathrm{y}(\tau
 ) + 2\,\mathrm{y}(\tau )\,{\Omega _{M}} - 2\,\sqrt{ - \mathrm{y}
(\tau )\,( - \mathrm{y}(\tau ) + \mathrm{y}(\tau )\,{\Omega _{M}}
 - {\Omega _{M}})}\,\mathrm{\%1}}{\sqrt{1 - {\Omega _{M}}}}} )
 \\
\mbox{} - 2\,\mathit{\_C1}\,\mathrm{\%1}\,{\Omega _{M}} + 2\,\tau
 \,\mathrm{\%1} - 2\,\sqrt{ - \mathrm{y}(\tau )\,( - \mathrm{y}(
\tau ) + \mathrm{y}(\tau )\,{\Omega _{M}} - {\Omega _{M}})}\,
\mathrm{\%1} +\\
 2\,\mathit{\_C1}\,\mathrm{\%1}) \left/ {\vrule height0.51em width0em depth0.51em} \right. \! 
 \! (\sqrt{1 - {\Omega _{M}}}\,({\Omega _{M}} - 1))=0 \\
\mathrm{\%1} := \sqrt{1 - {\Omega _{M}}} }
\end{maplelatex}

\emptyline
\begin{mapleinput}
\mapleinline{active}{1d}{#definition of _C1
\indent subs(tau=1,sol_1):
 \indent \indent subs(y(1)=1,\%):
  \indent \indent \indent solve(\%,_C1):
   \indent \indent simplify( subs(_C1=\%,sol_1) ):
    \indent subs(y(tau)=y,\%):
     sol_2 := solve(\%,tau);#implicit solution\\
      with defined _C1}{%
}
\end{mapleinput}

\mapleresult
\begin{maplelatex}
\maplemultiline{
\mathit{sol\_2} :=  - {\displaystyle \frac {1}{4}}  \left( 
{\vrule height1.12em width0em depth1.12em} \right. \!  \!  - 4\,
\mathrm{\%1}\,{\Omega _{M}} + {\Omega _{M}}\,\mathrm{ln}( - 
{\displaystyle \frac {({\Omega _{M}} - 2 - 2\,\mathrm{\%1})^{2}}{
{\Omega _{M}} - 1}} ) \\
\mbox{} - \mathrm{ln} \left(  \!  - {\displaystyle \frac {({
\Omega _{M}} + 2\,y - 2\,y\,{\Omega _{M}} + 2\,\sqrt{ - y\,( - y
 + y\,{\Omega _{M}} - {\Omega _{M}})}\,\mathrm{\%1})^{2}}{{\Omega
 _{M}} - 1}}  \!  \right) \,{\Omega _{M}} \\
\mbox{} + 4\,\sqrt{ - y\,( - y + y\,{\Omega _{M}} - {\Omega _{M}}
)}\,\mathrm{\%1} \! \! \left. {\vrule 
height1.12em width0em depth1.12em} \right)  \left/ {\vrule 
height0.51em width0em depth0.51em} \right. \!  \! (\sqrt{1 - {
\Omega _{M}}}\,({\Omega _{M}} - 1)) \\
\mathrm{\%1} := \sqrt{1 - {\Omega _{M}}} }
\end{maplelatex}

\emptyline
\noindent
This solution can be plotted in the following form:

\emptyline
\begin{mapleinput}
\mapleinline{active}{1d}{subs(Omega[M]=0.3,sol_2):
 \indent plot(\%, y=0.01..2, axes=BOXED, colour=RED,\\
  title=`hyperbolical universe\\
   (time vs scaling factor)` );}{%
}
\end{mapleinput}

\mapleresult
\begin{center}
\mapleplot{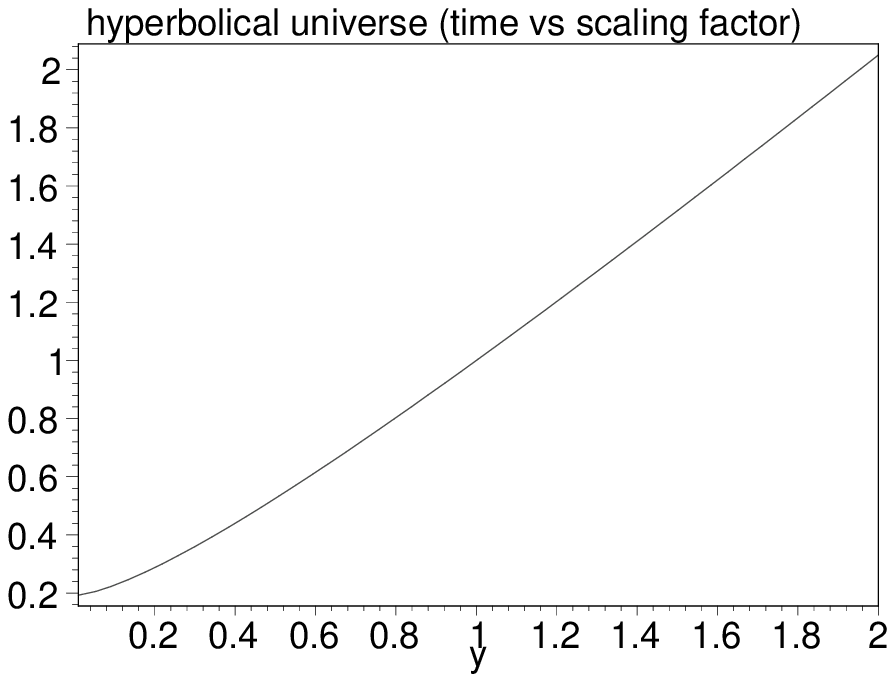}
\end{center}

\emptyline
\noindent
The extreme case 
${\Omega _{M}}$
\textit{--}\TEXTsymbol{>}0 (almost empty world)

\emptyline
\begin{mapleinput}
\mapleinline{active}{1d}{limit(sol_2,Omega[M]=0);}{%
}
\end{mapleinput}

\mapleresult
\begin{maplelatex}
\[
\sqrt{y^{2}}
\]
\end{maplelatex}

\emptyline
\noindent
results in the linear low of universe expansion and gives the
estimation for maximal age of universe in this model:

\emptyline
\begin{mapleinput}
\mapleinline{active}{1d}{evalf(3*10^19/65/60/60/24/365);#[yr]}{%
}
\end{mapleinput}

\mapleresult
\begin{maplelatex}
\[
.1463528861\,10^{11}
\]
\end{maplelatex}

\emptyline
\noindent
The dependence on 
${\Omega _{M}}$
 looks as:

\emptyline
\begin{mapleinput}
\mapleinline{active}{1d}{-(subs(y=0,sol_2)-subs(y=1,sol_2)):#age\\
 of universe [ 1/H[0] ]
  \indent plot(\%, Omega[M]=0..1,axes=boxed,\\
   title=`age of universe [ 1/H[0]]`);}{%
}
\end{mapleinput}

\mapleresult
\begin{center}
\mapleplot{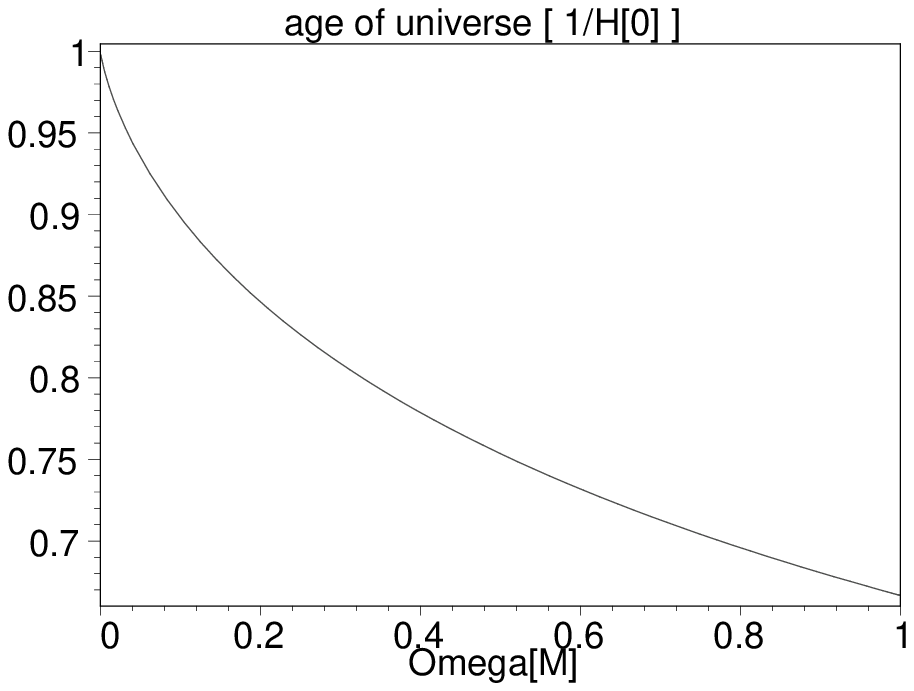}
\end{center}

\emptyline
\noindent
It is natural, the growth of matter density decreases the universe
age. If estimations of 
${\Omega _{M}}$
 give 0.3 (that results in the age $\approx$ 12 gigayears for
considered model), then this model does not satisfies the
observations. Moreover, it is possible, that the nonzero curvature is
not appropriate (see below).

\emptyline

\subsubsection{Expanding spherical and recollapsing hyperbolical
universes}

\emptyline
\noindent
This model demonstrates the possibility of the infinite expansion of
universe with spherical geometry in the presence of nonzero cosmological
constant. 

\emptyline
\begin{mapleinput}
\mapleinline{active}{1d}{with(DEtools):
 \indent DEplot(subs(\{Omega[M]=1,Omega[Lambda]=2.59\},\\
 basic_4),[y(tau)],tau=-2.5..1,[[y(0)=1]],stepsize=0.01,\\
 axes=boxed,linecolor=red,arrows=none,\\
 title=`loitering expansion`);
}{%
}
\end{mapleinput}

\begin{center}
\mapleplot{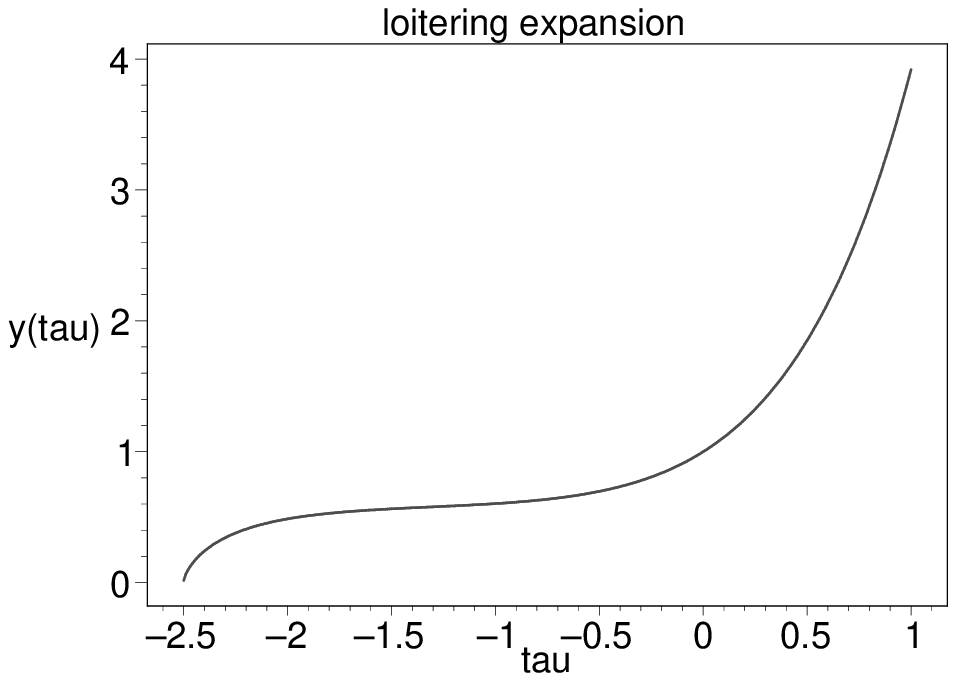}
\end{center}

\emptyline
\noindent
We selected the parameters, which are close to ones for static
universe (${\Omega _{M}}$
=1) but the cosmological term slightly dominates 
${\Omega _{\Lambda }}$
 \TEXTsymbol{>} 
$2\,{\Omega _{M}}$
. There is the delay of expansion, which looks like transition to
collapse, but the domination of 
${\Omega _{\Lambda }}$
 causes the further expansion. This model can explain the large age
of universe for large 
${H_{0}}$
 and to explain the misalignment in the "red shift-distance"
distribution for distant quasars. 

\emptyline
\noindent
The next example demonstrate the collapsing behavior of an open (hyperbolical)
universe:

\begin{mapleinput}
\mapleinline{active}{1d}{p1 :=\\
DEplot(subs(\{Omega[M]=0.5,Omega[Lambda]=-1\},\\
basic_4),[y(tau)],tau=-0.65..0.7,[[y(0)=1]],\\
stepsize=0.01,axes=boxed,linecolor=red,\\
arrows=none):#expansion}{%
}
\end{mapleinput}

\begin{mapleinput}
\mapleinline{active}{1d}{\indent p2 :=\\
DEplot(subs(\{Omega[M]=0.5,Omega[Lambda]=-1\},\\
lhs(basic_4)=-rhs(basic_4)),[y(tau)],\\
tau=0.71..1.98,[[y(0.71)=1.36]],stepsize=0.01,\\
axes=boxed,linecolor=red,arrows=none):#we have to\\
change the sign in right-hand side of basic_4 and\\
 to change the initial condition}{%
}
\end{mapleinput}

\begin{mapleinput}
\mapleinline{active}{1d}{\indent \indent with(plots):
 \indent display(p1,p2,title=`recollapsing open universe`);}{%
}
\end{mapleinput}

\begin{center}
\mapleplot{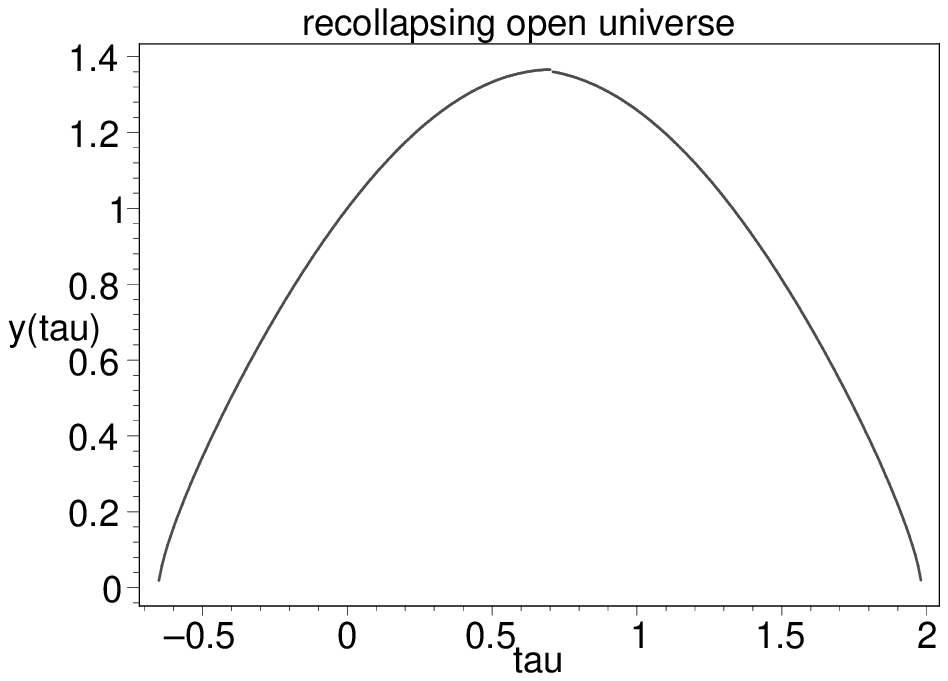}
\end{center}

\subsubsection{Bouncing model}

\emptyline
\noindent
Main feature of the above considered nonstatic models (except for de Sitter model) is the presence
of singularity at the beginning of expansion and, it is possible, at
the end of evolution. On the one hand, the singularity is not
desirable, but, on the other hand, the standard model of Big Bang
demands a high-density and hot beginning of expansion. The last
corresponds to the conditions in the vicinity of the singularity. The choice of
the parameters allows the behavior without singularity:

\emptyline
\begin{mapleinput}
\mapleinline{active}{1d}{p1 :=\\
DEplot(subs(\{Omega[M]=0.1,Omega[Lambda]=1.5\},basic_4),\\
[y(tau)],tau=-1.13..1,[[y(0)=1]],stepsize=0.01,\\
axes=boxed,linecolor=red,arrows=none):#expansion
 \indent p2 :=\\
DEplot(subs(\{Omega[M]=0.1,Omega[Lambda]=1.5\},\\
lhs(basic_4)=-rhs(basic_4)),[y(tau)],tau=-3..-1.14,\\
[[y(-3)=2.3]],stepsize=0.01,axes=boxed,linecolor=red,\\
arrows=none):#contraction
  \indent \indent display(p1,p2,title=`bouncing universe`);}{%
}
\end{mapleinput}

\mapleresult
\begin{center}
\mapleplot{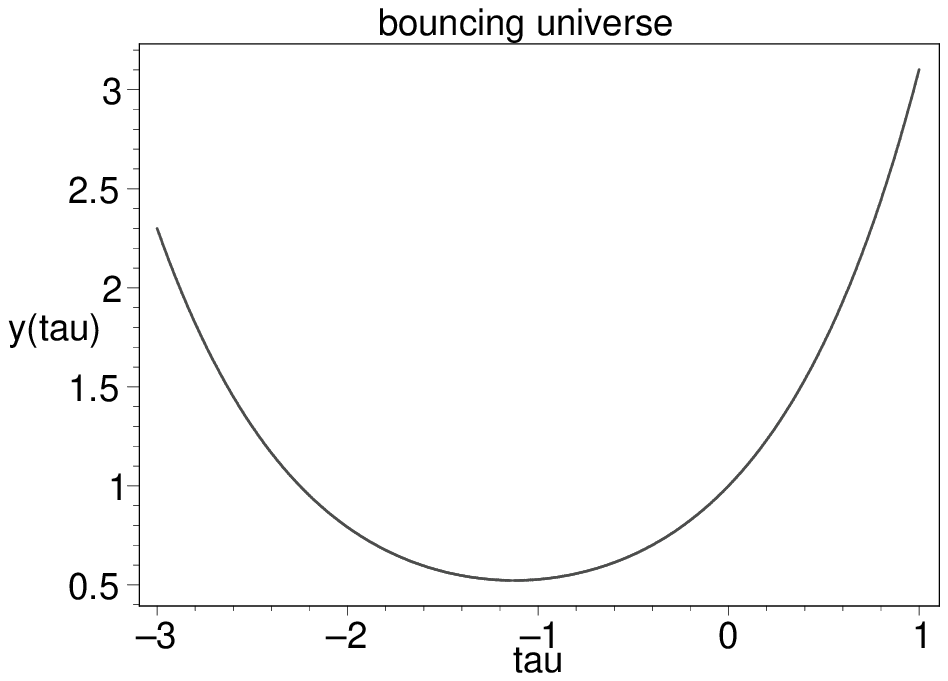}
\end{center}

\subsubsection{Our universe (?)}

\emptyline
\noindent
The modern observations of microwave background suggest 
\cite{J. Garcia-Bellido}: 
${\Omega _{K}}$
 $\approx$ 0 (flat), 
${\Omega _{\Lambda }}$
$\approx$ 0.7, 
${\Omega _{M}}$
 $\approx$ 0.3. The corresponding behavior of scaling factor looks
as:  

\emptyline
\begin{mapleinput}
\mapleinline{active}{1d}{DEplot(subs(\{Omega[M]=0.3,Omega[Lambda]=0.7\},\\
basic_4),[y(tau)],tau=-.96...2,[[y(0)=1]],\\
stepsize=0.01,axes=boxed,linecolor=red,arrows=none,\\
title=` Our universe? `);}{%
}
\end{mapleinput}

\mapleresult
\begin{center}
\mapleplot{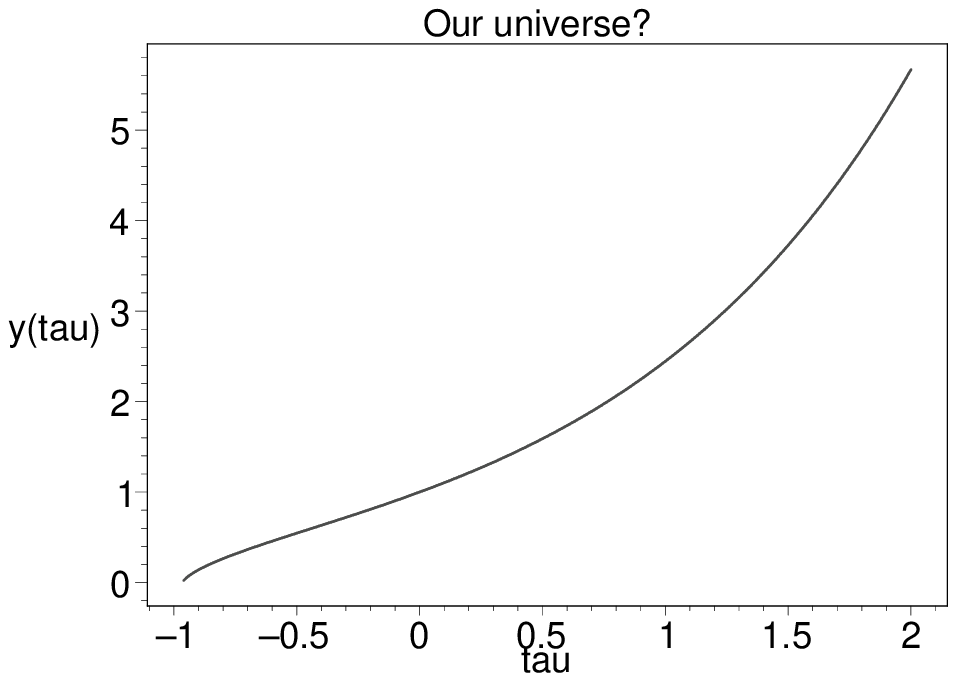}
\end{center}

\emptyline
\noindent
We are to note, that our comments about appropriate (or not
appropriate) character of the considered models are not rigid. The
picture of topologically nontrivial space-time can combine the
universes with the very different local geometries and dynamics in
the framework of a hypothetical Big Universe ("Tree of Universes").

\emptyline

\subsection{Beginning}

\emptyline

\subsubsection{Bianchi models and Mixmaster universe}

\emptyline
\noindent
A good agreement of the isotropic homogeneous models with observations
does not provide for their validity nearby singularity. In particular,
in framework of analytical approach, we can refuse the isotropy of
universe in the beginning of expansion (for review see \cite{Ya.B.
Zel'dovich}). Let us consider the anisotropic homogeneous flat
universe without rotation:

\emptyline
\begin{mapleinput}
\mapleinline{active}{1d}{coord := [t, x, y, z]:
 \indent g_compts :=\\
  array(symmetric,sparse,1..4,1..4):# metric components
  \indent \indent g_compts[1,1] := -1: 
   \indent \indent \indent g_compts[2,2] := a(t)^2:
     \indent \indent g_compts[3,3] := b(t)^2: 
      \indent g_compts[4,4] := c(t)^2:

g := create([-1,-1], eval(g_compts)); 

\indent ginv := invert( g, 'detg' ):}{%
}
\end{mapleinput}

\mapleresult
\begin{maplelatex}
\maplemultiline{
g := \mathrm{table(}[\mathit{index\_char}=[-1, \,-1], \,\mathit{
compts}=\\
 \left[ 
{\begin{array}{rccc}
-1 & 0 & 0 & 0 \\
0 & \mathrm{a}(t)^{2} & 0 & 0 \\
0 & 0 & \mathrm{b}(t)^{2} & 0 \\
0 & 0 & 0 & \mathrm{c}(t)^{2}
\end{array}}
 \right] ])
}
\end{maplelatex}

\emptyline
\noindent
The Einstein tensor is:

\emptyline
\begin{mapleinput}
\mapleinline{active}{1d}{# intermediate values
\indent D1g := d1metric( g, coord ):
 \indent \indent D2g := d2metric( D1g, coord ):
  \indent \indent \indent Cf1 := Christoffel1 ( D1g ):
   \indent \indent \indent \indent RMN := Riemann( ginv, D2g, Cf1 ):
    \indent \indent \indent RICCI := Ricci( ginv, RMN ):
     \indent \indent RS := Ricciscalar( ginv, RICCI ):
\indent Estn := Einstein( g, RICCI, RS ):# Einstein tensor
 displayGR(Einstein,Estn);# Its nonzero components}{%
}
\end{mapleinput}

\mapleresult
\begin{maplelatex}
\[
\mathit{The\ Einstein\ Tensor}
\]
\end{maplelatex}

\begin{maplelatex}
\[
\mathit{non-zero\ components\ :}
\]
\end{maplelatex}

\begin{maplelatex}
\maplemultiline{
\mathit{\ G11}=\\
 - {\displaystyle \frac {({\frac {\partial }{
\partial t}}\,\mathrm{b}(t))\,({\frac {\partial }{\partial t}}\,
\mathrm{a}(t))\,\mathrm{c}(t) + ({\frac {\partial }{\partial t}}
\,\mathrm{c}(t))\,({\frac {\partial }{\partial t}}\,\mathrm{a}(t)
)\,\mathrm{b}(t) + ({\frac {\partial }{\partial t}}\,\mathrm{c}(t
))\,({\frac {\partial }{\partial t}}\,\mathrm{b}(t))\,\mathrm{a}(
t)}{\mathrm{a}(t)\,\mathrm{b}(t)\,\mathrm{c}(t)}} 
}
\end{maplelatex}

\begin{maplelatex}
\maplemultiline{
\mathit{\ G22}=\\
{\displaystyle \frac {\mathrm{a}(t)^{2}\,((
{\frac {\partial ^{2}}{\partial t^{2}}}\,\mathrm{b}(t))\,\mathrm{
c}(t) + ({\frac {\partial ^{2}}{\partial t^{2}}}\,\mathrm{c}(t))
\,\mathrm{b}(t) + ({\frac {\partial }{\partial t}}\,\mathrm{c}(t)
)\,({\frac {\partial }{\partial t}}\,\mathrm{b}(t)))}{\mathrm{b}(
t)\,\mathrm{c}(t)}} 
}
\end{maplelatex}

\begin{maplelatex}
\maplemultiline{
\mathit{\ G33}=\\
{\displaystyle \frac {\mathrm{b}(t)^{2}\,((
{\frac {\partial ^{2}}{\partial t^{2}}}\,\mathrm{a}(t))\,\mathrm{
c}(t) + ({\frac {\partial ^{2}}{\partial t^{2}}}\,\mathrm{c}(t))
\,\mathrm{a}(t) + ({\frac {\partial }{\partial t}}\,\mathrm{c}(t)
)\,({\frac {\partial }{\partial t}}\,\mathrm{a}(t)))}{\mathrm{a}(
t)\,\mathrm{c}(t)}} 
}
\end{maplelatex}

\begin{maplelatex}
\maplemultiline{
\mathit{\ G44}=\\
{\displaystyle \frac {\mathrm{c}(t)^{2}\,((
{\frac {\partial ^{2}}{\partial t^{2}}}\,\mathrm{a}(t))\,\mathrm{
b}(t) + ({\frac {\partial ^{2}}{\partial t^{2}}}\,\mathrm{b}(t))
\,\mathrm{a}(t) + ({\frac {\partial }{\partial t}}\,\mathrm{b}(t)
)\,({\frac {\partial }{\partial t}}\,\mathrm{a}(t)))}{\mathrm{a}(
t)\,\mathrm{b}(t)}} 
}
\end{maplelatex}

\begin{maplelatex}
\[
\mathit{character\ :\ [-1,\ -1]}
\]
\end{maplelatex}

\emptyline
\noindent
In the vacuum state the right-hand sides of Einstein equations are 0.
Then

\emptyline
\begin{mapleinput}
\mapleinline{active}{1d}{E_eqn := get_compts(Estn):
 \indent e1 := numer( E_eqn[1,1] ) = 0;
  \indent \indent e2 :=\\
   expand( numer( E_eqn[2,2] )/a(t)^2 ) = 0;
   \indent \indent \indent e3 :=\\
    expand( numer( E_eqn[3,3] )/b(t)^2 ) = 0;
    \indent \indent \indent \indent e4 :=\\
     expand( numer( E_eqn[4,4] )/c(t)^2 ) = 0;}{%
}
\end{mapleinput}

\mapleresult
\begin{maplelatex}
\maplemultiline{
\mathit{e1} :=  - ({\frac {\partial }{\partial t}}\,\mathrm{b}(t)
)\,({\frac {\partial }{\partial t}}\,\mathrm{a}(t))\,\mathrm{c}(t
) - ({\frac {\partial }{\partial t}}\,\mathrm{c}(t))\,({\frac {
\partial }{\partial t}}\,\mathrm{a}(t))\,\mathrm{b}(t) -\\
 (
{\frac {\partial }{\partial t}}\,\mathrm{c}(t))\,({\frac {
\partial }{\partial t}}\,\mathrm{b}(t))\,\mathrm{a}(t)=0
}
\end{maplelatex}

\begin{maplelatex}
\[
\mathit{e2} := ({\frac {\partial ^{2}}{\partial t^{2}}}\,\mathrm{
b}(t))\,\mathrm{c}(t) + ({\frac {\partial ^{2}}{\partial t^{2}}}
\,\mathrm{c}(t))\,\mathrm{b}(t) + ({\frac {\partial }{\partial t
}}\,\mathrm{c}(t))\,({\frac {\partial }{\partial t}}\,\mathrm{b}(
t))=0
\]
\end{maplelatex}

\begin{maplelatex}
\[
\mathit{e3} := ({\frac {\partial ^{2}}{\partial t^{2}}}\,\mathrm{
a}(t))\,\mathrm{c}(t) + ({\frac {\partial ^{2}}{\partial t^{2}}}
\,\mathrm{c}(t))\,\mathrm{a}(t) + ({\frac {\partial }{\partial t
}}\,\mathrm{c}(t))\,({\frac {\partial }{\partial t}}\,\mathrm{a}(
t))=0
\]
\end{maplelatex}

\begin{maplelatex}
\[
\mathit{e4} := ({\frac {\partial ^{2}}{\partial t^{2}}}\,\mathrm{
a}(t))\,\mathrm{b}(t) + ({\frac {\partial ^{2}}{\partial t^{2}}}
\,\mathrm{b}(t))\,\mathrm{a}(t) + ({\frac {\partial }{\partial t
}}\,\mathrm{b}(t))\,({\frac {\partial }{\partial t}}\,\mathrm{a}(
t))=0
\]
\end{maplelatex}

\emptyline
\noindent
We will search the power-behaved solutions of this system
(\underline{Kasner metric}):

\emptyline
\begin{mapleinput}
\mapleinline{active}{1d}{e5 := simplify(\\
subs(\{a(t)=a_0*t^p_1,b(t)=b_0*t^p_2,c(t)=c_0*t^p_3\},\\
e1) );
 \indent e6 := simplify(\\
 subs(\{a(t)=a_0*t^p_1,b(t)=b_0*t^p_2,c(t)=c_0*t^p_3\},\\
 e2) );
  \indent \indent e7 := simplify(\\
  subs(\{a(t)=a_0*t^p_1,b(t)=b_0*t^p_2,c(t)=c_0*t^p_3\},\\
  e3) );
   \indent \indent \indent e8 := simplify(\\
   subs(\{a(t)=a_0*t^p_1,b(t)=b_0*t^p_2,c(t)=c_0*t^p_3\},\\
   e4) );}{%
}
\end{mapleinput}

\mapleresult
\begin{maplelatex}
\maplemultiline{
\mathit{e5} :=  - \mathit{b\_0}\,t^{(\mathit{p\_2} - 2 + \mathit{
p\_1} + \mathit{p\_3})}\,\mathit{p\_2}\,\mathit{a\_0}\,\mathit{
p\_1}\,\mathit{c\_0} \\
\mbox{} - \mathit{c\_0}\,t^{(\mathit{p\_2} - 2 + \mathit{p\_1} + 
\mathit{p\_3})}\,\mathit{p\_3}\,\mathit{a\_0}\,\mathit{p\_1}\,
\mathit{b\_0} \\
\mbox{} - \mathit{c\_0}\,t^{(\mathit{p\_2} - 2 + \mathit{p\_1} + 
\mathit{p\_3})}\,\mathit{p\_3}\,\mathit{b\_0}\,\mathit{p\_2}\,
\mathit{a\_0}=0 }
\end{maplelatex}

\begin{maplelatex}
\maplemultiline{
\mathit{e6} := \mathit{b\_0}\,t^{(\mathit{p\_3} - 2 + \mathit{
p\_2})}\,\mathit{p\_2}^{2}\,\mathit{c\_0} - \mathit{b\_0}\,t^{(
\mathit{p\_3} - 2 + \mathit{p\_2})}\,\mathit{p\_2}\,\mathit{c\_0}
 \\
\mbox{} + \mathit{c\_0}\,t^{(\mathit{p\_3} - 2 + \mathit{p\_2})}
\,\mathit{p\_3}^{2}\,\mathit{b\_0} - \mathit{c\_0}\,t^{(\mathit{
p\_3} - 2 + \mathit{p\_2})}\,\mathit{p\_3}\,\mathit{b\_0} \\
\mbox{} + \mathit{c\_0}\,t^{(\mathit{p\_3} - 2 + \mathit{p\_2})}
\,\mathit{p\_3}\,\mathit{b\_0}\,\mathit{p\_2}=0 }
\end{maplelatex}

\begin{maplelatex}
\maplemultiline{
\mathit{e7} := \mathit{a\_0}\,t^{(\mathit{p\_3} - 2 + \mathit{
p\_1})}\,\mathit{p\_1}^{2}\,\mathit{c\_0} - \mathit{a\_0}\,t^{(
\mathit{p\_3} - 2 + \mathit{p\_1})}\,\mathit{p\_1}\,\mathit{c\_0}
 \\
\mbox{} + \mathit{c\_0}\,t^{(\mathit{p\_3} - 2 + \mathit{p\_1})}
\,\mathit{p\_3}^{2}\,\mathit{a\_0} - \mathit{c\_0}\,t^{(\mathit{
p\_3} - 2 + \mathit{p\_1})}\,\mathit{p\_3}\,\mathit{a\_0} \\
\mbox{} + \mathit{c\_0}\,t^{(\mathit{p\_3} - 2 + \mathit{p\_1})}
\,\mathit{p\_3}\,\mathit{a\_0}\,\mathit{p\_1}=0 }
\end{maplelatex}

\begin{maplelatex}
\maplemultiline{
\mathit{e8} := \mathit{a\_0}\,t^{(\mathit{p\_1} - 2 + \mathit{
p\_2})}\,\mathit{p\_1}^{2}\,\mathit{b\_0} - \mathit{a\_0}\,t^{(
\mathit{p\_1} - 2 + \mathit{p\_2})}\,\mathit{p\_1}\,\mathit{b\_0}
 \\
\mbox{} + \mathit{b\_0}\,t^{(\mathit{p\_1} - 2 + \mathit{p\_2})}
\,\mathit{p\_2}^{2}\,\mathit{a\_0} - \mathit{b\_0}\,t^{(\mathit{
p\_1} - 2 + \mathit{p\_2})}\,\mathit{p\_2}\,\mathit{a\_0} \\
\mbox{} + \mathit{b\_0}\,t^{(\mathit{p\_1} - 2 + \mathit{p\_2})}
\,\mathit{p\_2}\,\mathit{a\_0}\,\mathit{p\_1}=0 }
\end{maplelatex}

\emptyline
\noindent
If \textit{t} is not equal to zero, we have:

\emptyline
\begin{mapleinput}
\mapleinline{active}{1d}{e9 := factor( e5/t^(p_2-2+p_1+p_3) );
 \indent e10 := factor( e6/t^(p_3-2+p_2) );
  \indent \indent e11 := factor( e7/t^(p_3-2+p_1) );
   \indent \indent \indent e12 := factor( e8/t^(p_2-2+p_1) );
}{%
}
\end{mapleinput}

\mapleresult
\begin{maplelatex}
\[
\mathit{e9} :=  - \mathit{b\_0}\,\mathit{a\_0}\,\mathit{c\_0}\,(
\mathit{p\_2}\,\mathit{p\_1} + \mathit{p\_3}\,\mathit{p\_1} + 
\mathit{p\_3}\,\mathit{p\_2})=0
\]
\end{maplelatex}

\begin{maplelatex}
\[
\mathit{e10} := \mathit{b\_0}\,\mathit{c\_0}\,(\mathit{p\_2}^{2}
 - \mathit{p\_2} + \mathit{p\_3}^{2} - \mathit{p\_3} + \mathit{
p\_3}\,\mathit{p\_2})=0
\]
\end{maplelatex}

\begin{maplelatex}
\[
\mathit{e11} := \mathit{a\_0}\,\mathit{c\_0}\,(\mathit{p\_1}^{2}
 - \mathit{p\_1} + \mathit{p\_3}^{2} - \mathit{p\_3} + \mathit{
p\_3}\,\mathit{p\_1})=0
\]
\end{maplelatex}

\begin{maplelatex}
\[
\mathit{e12} := \mathit{a\_0}\,\mathit{b\_0}\,(\mathit{p\_1}^{2}
 - \mathit{p\_1} + \mathit{p\_2}^{2} - \mathit{p\_2} + \mathit{
p\_2}\,\mathit{p\_1})=0
\]
\end{maplelatex}

\begin{mapleinput}
\mapleinline{active}{1d}{e13 := e9/(-b_0*a_0*c_0);
 \indent e14 := e10/(b_0*c_0);
  \indent \indent e15 := e11/(a_0*c_0);
   \indent \indent \indent e16 := e12/(a_0*b_0);}{%
}
\end{mapleinput}

\mapleresult
\begin{maplelatex}
\[
\mathit{e13} := \mathit{p\_2}\,\mathit{p\_1} + \mathit{p\_3}\,
\mathit{p\_1} + \mathit{p\_3}\,\mathit{p\_2}=0
\]
\end{maplelatex}

\begin{maplelatex}
\[
\mathit{e14} := \mathit{p\_2}^{2} - \mathit{p\_2} + \mathit{p\_3}
^{2} - \mathit{p\_3} + \mathit{p\_3}\,\mathit{p\_2}=0
\]
\end{maplelatex}

\begin{maplelatex}
\[
\mathit{e15} := \mathit{p\_1}^{2} - \mathit{p\_1} + \mathit{p\_3}
^{2} - \mathit{p\_3} + \mathit{p\_3}\,\mathit{p\_1}=0
\]
\end{maplelatex}

\begin{maplelatex}
\[
\mathit{e16} := \mathit{p\_1}^{2} - \mathit{p\_1} + \mathit{p\_2}
^{2} - \mathit{p\_2} + \mathit{p\_2}\,\mathit{p\_1}=0
\]
\end{maplelatex}

\emptyline
\noindent
The following manipulation gives a connection between parameters:

\emptyline
\begin{mapleinput}
\mapleinline{active}{1d}{((e14 + e15 + e16) - e13)/2;}{%
}
\end{mapleinput}

\mapleresult
\begin{maplelatex}
\[
\mathit{p\_2}^{2} - \mathit{p\_2} + \mathit{p\_3}^{2} - \mathit{
p\_3} + \mathit{p\_1}^{2} - \mathit{p\_1}=0
\]
\end{maplelatex}

\emptyline
\noindent
That is 
$\mathit{p\_1}^{2}$
 + 
$\mathit{p\_2}^{2}$
 + 
$\mathit{p\_3}^{2}$
 = \textit{p\_1} + \textit{p\_2} + \textit{p\_3}      (3)\textit{.} 

\emptyline
\begin{mapleinput}
\mapleinline{active}{1d}{collect(e15-e16,\{p_2,p_3\});
 \indent collect(e14-e15,\{p_1,p_2\});}{%
}
\end{mapleinput}

\mapleresult
\begin{maplelatex}
\[
\mathit{p\_3}^{2} + (\mathit{p\_1} - 1)\,\mathit{p\_3} - \mathit{
p\_2}^{2} + (1 - \mathit{p\_1})\,\mathit{p\_2}=0
\]
\end{maplelatex}

\begin{maplelatex}
\[
\mathit{p\_2}^{2} + (\mathit{p\_3} - 1)\,\mathit{p\_2} - \mathit{
p\_1}^{2} + (1 - \mathit{p\_3})\,\mathit{p\_1}=0
\]
\end{maplelatex}

\emptyline
\noindent
Hence

\begin{center}
   \textit{p\_3 }(\textit{p\_3 }+ \textit{p\_1} \textit{- }1) =
\textit{p\_2} (\textit{p\_2} + \textit{p\_1 -} 1),
\end{center}

\begin{center}
\textit{p\_2 }(\textit{p\_3 }+ \textit{p\_2} \textit{- }1) =
\textit{p\_1} (\textit{p\_3} + \textit{p\_1 -} 1).
\end{center}

\noindent
The last expressions suggest the simple solution:

\begin{center}
(\textit{p\_3 }+ \textit{p\_1} \textit{- }1) = \textit{p\_2},
(\textit{p\_2} + \textit{p\_1 -} 1) = \textit{p\_3} and
\end{center}

\begin{center}
(\textit{p\_3 }+ \textit{p\_2} \textit{- }1) = \textit{p\_1},
(\textit{p\_3} + \textit{p\_1 -} 1) = \textit{p\_2}, 
\end{center}

\begin{center}
or
\end{center}

\begin{center}
(\textit{p\_3 }+ \textit{p\_1} \textit{- }1) = \textit{-}\textit{
p\_2}, (\textit{p\_2} + \textit{p\_1 -} 1) = \textit{- p\_3} and
\end{center}

\begin{center}
(\textit{p\_3 }+ \textit{p\_2} \textit{- }1) = \textit{- p\_1},
(\textit{p\_3} + \textit{p\_1 -} 1) = \textit{- p\_2}
\end{center}

\noindent
The first results in

\emptyline
\begin{mapleinput}
\mapleinline{active}{1d}{p_3 + p_1 - 1 + p_2 + p_1 - 1 + p_3 + p_2\\
 - 1 - (p_1+p_2+p_3);}{%
}
\end{mapleinput}

\mapleresult
\begin{maplelatex}
\[
\mathit{p\_3} + \mathit{p\_1} - 3 + \mathit{p\_2}
\]
\end{maplelatex}

\emptyline
\noindent
The second produces

\emptyline
\begin{mapleinput}
\mapleinline{active}{1d}{(p_3 + p_1 - 1 + p_2 + p_1 - 1 +\\
 p_3 + p_2 - 1 + (p_1+p_2+p_3))/3;}{%
}
\end{mapleinput}

\mapleresult
\begin{maplelatex}
\[
\mathit{p\_3} + \mathit{p\_1} - 1 + \mathit{p\_2}
\]
\end{maplelatex}

\emptyline
\noindent
These expressions in the combination with Eq.(3) suggest that there is
only one independent parameter \textit{p}:

\emptyline
\begin{mapleinput}
\mapleinline{active}{1d}{solve(\{p_1 + p_2 + p =\\
 1, p_1^2 + p_2^2 + p^2 = 1\},\{p_1,p_2\}):
 \indent sol_1 := allvalues(\%);#first choice
solve(\{p_1 + p_2 + p = 3, p_1^2 + p_2^2 + p^2 = 3\},\\
\{p_1,p_2\}):
 \indent sol_2 := allvalues(\%);#second choice}{%
}
\end{mapleinput}

\mapleresult
\begin{maplelatex}
\maplemultiline{
\mathit{sol\_1} := \{\mathit{p\_1}= - {\displaystyle \frac {1}{2}
} \,p + {\displaystyle \frac {1}{2}}  - {\displaystyle \frac {1}{
2}} \,\mathrm{\%1}, \,\mathit{p\_2}= - {\displaystyle \frac {1}{2
}} \,p + {\displaystyle \frac {1}{2}}  + {\displaystyle \frac {1
}{2}} \,\mathrm{\%1}\},  \\
\{\mathit{p\_1}= - {\displaystyle \frac {1}{2}} \,p + 
{\displaystyle \frac {1}{2}}  + {\displaystyle \frac {1}{2}} \,
\mathrm{\%1}, \,\mathit{p\_2}= - {\displaystyle \frac {1}{2}} \,p
 + {\displaystyle \frac {1}{2}}  - {\displaystyle \frac {1}{2}} 
\,\mathrm{\%1}\} \\
\mathrm{\%1} := \sqrt{ - 3\,p^{2} + 2\,p + 1} }
\end{maplelatex}

\begin{maplelatex}
\maplemultiline{
\mathit{sol\_2} :=\\
 \{\mathit{p\_2}= - {\displaystyle \frac {1}{2}
} \,p + {\displaystyle \frac {3}{2}}  + {\displaystyle \frac {1}{
2}} \,I\,(p - 1)\,\sqrt{3}, \,\mathit{p\_1}= - {\displaystyle 
\frac {1}{2}} \,p + {\displaystyle \frac {3}{2}}  - 
{\displaystyle \frac {1}{2}} \,I\,(p - 1)\,\sqrt{3}\},  \\
\{\mathit{p\_2}= - {\displaystyle \frac {1}{2}} \,p + 
{\displaystyle \frac {3}{2}}  - {\displaystyle \frac {1}{2}} \,I
\,(p - 1)\,\sqrt{3}, \,\mathit{p\_1}= - {\displaystyle \frac {1}{
2}} \,p + {\displaystyle \frac {3}{2}}  + {\displaystyle \frac {1
}{2}} \,I\,(p - 1)\,\sqrt{3}\} }
\end{maplelatex}

\emptyline
\noindent
So, if the parameters \textit{p\_1}, \textit{p\_2}, \textit{p\_3 }are
real, we have 
$\mathit{p\_1}^{2}$
 + 
$\mathit{p\_2}^{2}$
 + 
$\mathit{p\_3}^{2}$
 = \textit{p\_1} + \textit{p\_2} + \textit{p\_3} =1 and

\emptyline
\begin{mapleinput}
\mapleinline{active}{1d}{sol_p_1 := factor(subs(sol_1,p_1));
 \indent sol_p_2 := factor(subs(sol_1,p_2));
  \indent \indent sol_p_3 := 1 - \%\% - \%;}{%
}
\end{mapleinput}

\mapleresult
\begin{maplelatex}
\[
\mathit{sol\_p\_1} :=  - {\displaystyle \frac {1}{2}} \,p + 
{\displaystyle \frac {1}{2}}  - {\displaystyle \frac {1}{2}} \,
\sqrt{ - (3\,p + 1)\,(p - 1)}
\]
\end{maplelatex}

\begin{maplelatex}
\[
\mathit{sol\_p\_2} :=  - {\displaystyle \frac {1}{2}} \,p + 
{\displaystyle \frac {1}{2}}  + {\displaystyle \frac {1}{2}} \,
\sqrt{ - (3\,p + 1)\,(p - 1)}
\]
\end{maplelatex}

\begin{maplelatex}
\[
\mathit{sol\_p\_3} := p
\]
\end{maplelatex}

\emptyline
\noindent
The values of the analyzed parameters are shown in the next figure:

\emptyline
\begin{mapleinput}
\mapleinline{active}{1d}{plot(\{sol_p_1,sol_p_2,sol_p_3\},p=-1/3..1,\\
 axes=boxed, title=`powers of t`);}{%
}
\end{mapleinput}

\mapleresult
\begin{center}
\mapleplot{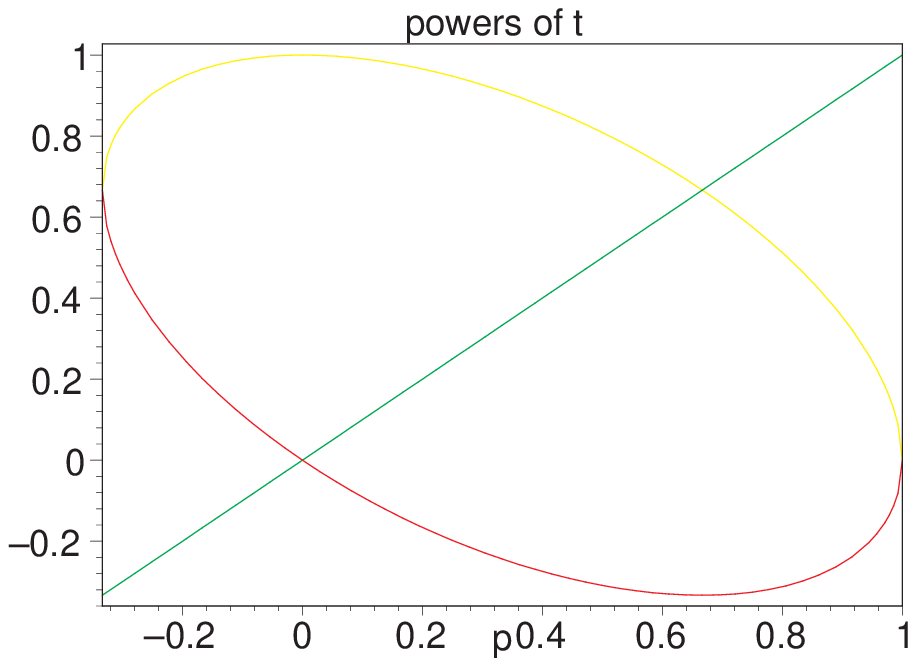}
\end{center}

\emptyline
\noindent
Hence, there are two directions of expansion and one direction of
contraction for \textit{t--}\TEXTsymbol{>}
$\infty $
 (so-called \underline{"pancake" singularity}) or one direction of
expansion and two direction of contraction for
\textit{t--}\TEXTsymbol{>}0 (\underline{"sigar" singularity}).

\emptyline
\noindent
Now let's consider more complicated situation with homogeneous
anisotropic metric.\\

\noindent
We will use the tetrad notation of Einstein equations that allows to
avoid the explicit manipulations with coordinates. In the synchronous
frame we have for the homogeneous geometry (here Greek indexes change
from 1 to 3) \cite{L.D. Landau}:

\begin{center}
${g_{0, \,0}}$
 = \textit{-}1, 
${g_{\alpha , \,\beta }}$
 = 
${\eta _{a, \,b}}$
${(e^{a})_{\alpha }}$
${(e^{b})_{\beta }}$
, 
\end{center}

\noindent
where 
${\eta _{a, \,b}}$
 is the time dependent matrix, 
${(e^{a})_{\alpha }}$
 is the frame vector (\textit{a} changes from 1 to 3, that is the
number of space triad). We will use the following representation for 
$\eta $
:    

\emptyline
\begin{mapleinput}
\mapleinline{active}{1d}{eta :=\\
 array(\\
 [[a(t)^2,0,0],[0,b(t)^2,0],[0,0,c(t)^2]]);#a, b, c\\
  are the time-dependent coefficients of anisotropic\\
   deformation
 \indent eta_inv := inverse(eta):
  \indent \indent kappa1 :=\\
   map(diff,eta,t):#kappa1[a,b]=diff(eta[a,b],t)
   \indent kappa2 := multiply(\\
    map(diff,eta,t),eta_inv);#kappa2[a]^b=\\
    diff(eta[a,b],t)*eta_inv, i.e. we raise the index\\
     by eta_inv}{%
}
\end{mapleinput}

\mapleresult
\begin{maplelatex}
\[
\eta  :=  \left[ 
{\begin{array}{ccc}
\mathrm{a}(t)^{2} & 0 & 0 \\
0 & \mathrm{b}(t)^{2} & 0 \\
0 & 0 & \mathrm{c}(t)^{2}
\end{array}}
 \right] 
\]
\end{maplelatex}

\begin{maplelatex}
\[
\kappa 2 :=  \left[ 
{\begin{array}{ccc}
2\,{\displaystyle \frac {{\frac {\partial }{\partial t}}\,
\mathrm{a}(t)}{\mathrm{a}(t)}}  & 0 & 0 \\ [2ex]
0 & 2\,{\displaystyle \frac {{\frac {\partial }{\partial t}}\,
\mathrm{b}(t)}{\mathrm{b}(t)}}  & 0 \\ [2ex]
0 & 0 & 2\,{\displaystyle \frac {{\frac {\partial }{\partial t}}
\,\mathrm{c}(t)}{\mathrm{c}(t)}} 
\end{array}}
 \right] 
\]
\end{maplelatex}

\emptyline
\noindent
The first vacuum Einstein equation has a form

\begin{center}
${R_{0}}^{0}$
 = \textit{-
$\frac {1}{2}$
${\frac {\partial }{\partial t}}\,{\kappa _{a}}^{a}$
} \textit{-} 
$\frac {1}{4}$
${\kappa _{a}}^{b}$
${\kappa _{b}}^{a}$
 = 0     
\end{center}

\emptyline
\begin{mapleinput}
\mapleinline{active}{1d}{eq_1 :=\\
 evalm( -trace(map(diff,kappa2,t))/2 -\\
  trace(multiply(kappa2,transpose(kappa2))/4) )\\
   = 0;#first Einstein equation}{%
}
\end{mapleinput}

\mapleresult
\begin{maplelatex}
\[
\mathit{eq\_1} :=  - {\displaystyle \frac {{\frac {\partial ^{2}
}{\partial t^{2}}}\,\mathrm{a}(t)}{\mathrm{a}(t)}}  - 
{\displaystyle \frac {{\frac {\partial ^{2}}{\partial t^{2}}}\,
\mathrm{b}(t)}{\mathrm{b}(t)}}  - {\displaystyle \frac {{\frac {
\partial ^{2}}{\partial t^{2}}}\,\mathrm{c}(t)}{\mathrm{c}(t)}} =
0
\]
\end{maplelatex}

\emptyline
\noindent
The next Einstein equations demand the definition of Bianchi
structured constants:
$C^{\mathit{ab}}$
 = 
$n^{\mathit{ab}} + e^{\mathit{abc}}\,{a_{c}}$
, 
${(C^{c})_{\mathit{ab}}}$
 =
${e_{\mathit{abd}}}$
$C^{\mathit{dc}}$
. Here 
${e_{\mathit{abc}}}$
 =
$e^{\mathit{abc}}$
 is the unit antisymmetric symbol, 
$n^{\mathit{ab}}$
 is the symmetric "tensor", which can be expressed through its
principal values 
${n_{i}}$
, 
${a_{c}}$
 is the vector. The Bianchi types are:  

\emptyline

\begin{table*}[h]
	\begin{center}
		\begin{tabular}{|c|c|c|c|c|}
		\hline
			Type & \textit{a} & $n_1$ & \textit{$n_2$} & $n_3$ \\ 
			\hline
			I & 0 & 0 & 0 & 0 \\
			\hline
			II & 0 & 1 & 0 & 0 \\
			\hline
			VII & 0 & 1 & 1 & 0 \\
			\hline
			VI & 0 & 1 & -1 & 0 \\
			\hline
			IX & 0 & 1 & 1 & 1 \\
			\hline
			VIII & 0 & 1 & 1 & -1 \\
			\hline
			V & 1 & 0 & 0 & 0 \\
			\hline
			IV & 1 & 0 & 0 & 1 \\
			\hline
			VII & $\zeta$ & 0 & 1 & 1 \\
			\hline
			III ($\zeta$=1), VI ($\zeta$ $\neq$ 0) & $\zeta$ & 0 & 1 & -1 \\
			\hline
		\end{tabular}
	\end{center}

	\caption{Bianchi constants}
	\label{tab:BianchiConstants}
\end{table*}

\emptyline
\noindent
We will analyze type IX model, where 
$C^{11}$
 = 
$C^{22}$
 = 
$C^{33}$
 = 
${C_{23}}^1$
 = 
${C_{31}}^2$
 = 
${C_{12}}^3$
 = 1. The curvature 
${P_{a}}^{b}$
 = 
$\frac {1}{2\,\eta }$
 \{2
$C^{\mathit{bd}}$
${C_{\mathit{ad}}}$
  + 
$C^{\mathit{db}}$
${C_{\mathit{ad}}}$
 + 
$C^{\mathit{bd}}$
${C_{\mathit{da}}}$
\textit{ - 
${C_{d}}^{d}$
}(
${(C^{b})_{a}} + {C_{a}}^{b}$
) + 
${\delta _{a}}^{b}$
[
${(C^{d})_{d}}^{2}$
\textit{-}2
$C^{\mathit{df}}$
${C_{\mathit{df}}}$
]\} is \textit{   }

\emptyline
\begin{mapleinput}
\mapleinline{active}{1d}{Cab :=\\
 array([[1,0,0],[0,1,0],[0,0,1]]);#C^(ab)
 \indent C_ab := multiply(eta,eta,Cab);#C[ab]
  \indent \indent Ca_b := multiply(eta,Cab);#C[a]^b
   \indent \indent \indent C_a_b := multiply(Cab,eta);#C^b[a];
    \indent \indent delta := array([[1,0,0],[0,1,0],[0,0,1]]):
     \indent evalm( (4*multiply(Cab,C_ab) -\\
      trace(C_a_b)*evalm(C_a_b+Ca_b) +\\
      delta*(trace(C_a_b)^2 -\\
       2*trace(multiply(Cab,C_ab))))/(2*det(eta))):#here\\
        we use the diagonal form of eta and symmetry\\
         of structured constants
       \indent \indent P := map(simplify,\%);}{%
}
\end{mapleinput}

\mapleresult
\begin{maplelatex}
\[
\mathit{Cab} :=  \left[ 
{\begin{array}{rrr}
1 & 0 & 0 \\
0 & 1 & 0 \\
0 & 0 & 1
\end{array}}
 \right] 
\]
\end{maplelatex}

\begin{maplelatex}
\[
\mathit{C\_ab} :=  \left[ 
{\begin{array}{ccc}
\mathrm{a}(t)^{4} & 0 & 0 \\
0 & \mathrm{b}(t)^{4} & 0 \\
0 & 0 & \mathrm{c}(t)^{4}
\end{array}}
 \right] 
\]
\end{maplelatex}

\begin{maplelatex}
\[
\mathit{Ca\_b} :=  \left[ 
{\begin{array}{ccc}
\mathrm{a}(t)^{2} & 0 & 0 \\
0 & \mathrm{b}(t)^{2} & 0 \\
0 & 0 & \mathrm{c}(t)^{2}
\end{array}}
 \right] 
\]
\end{maplelatex}

\begin{maplelatex}
\[
\mathit{C\_a\_b} :=  \left[ 
{\begin{array}{ccc}
\mathrm{a}(t)^{2} & 0 & 0 \\
0 & \mathrm{b}(t)^{2} & 0 \\
0 & 0 & \mathrm{c}(t)^{2}
\end{array}}
 \right] 
\]
\end{maplelatex}

\begin{maplelatex}
\[
P :=  \left[ 
{\begin{array}{c}
 - {\displaystyle \frac {1}{2}} \,{\displaystyle \frac { - 
\mathrm{a}(t)^{4} + \mathrm{b}(t)^{4} - 2\,\mathrm{b}(t)^{2}\,
\mathrm{c}(t)^{2} + \mathrm{c}(t)^{4}}{\mathrm{a}(t)^{2}\,
\mathrm{b}(t)^{2}\,\mathrm{c}(t)^{2}}} \,, \,0\,, \,0 \\ [2ex]
0\,, \, - {\displaystyle \frac {1}{2}} \,{\displaystyle \frac {
 - \mathrm{b}(t)^{4} + \mathrm{a}(t)^{4} - 2\,\mathrm{a}(t)^{2}\,
\mathrm{c}(t)^{2} + \mathrm{c}(t)^{4}}{\mathrm{a}(t)^{2}\,
\mathrm{b}(t)^{2}\,\mathrm{c}(t)^{2}}} \,, \,0 \\ [2ex]
0\,, \,0\,, \,{\displaystyle \frac {1}{2}} \,{\displaystyle 
\frac {\mathrm{c}(t)^{4} - \mathrm{a}(t)^{4} + 2\,\mathrm{a}(t)^{
2}\,\mathrm{b}(t)^{2} - \mathrm{b}(t)^{4}}{\mathrm{a}(t)^{2}\,
\mathrm{b}(t)^{2}\,\mathrm{c}(t)^{2}}} 
\end{array}}
 \right] 
\]
\end{maplelatex}

\emptyline
\noindent
So, in the degenerate case \textit{a} = \textit{b} =
\textit{c},\textit{ t = const} this model describes the closed
spherical universe with a positive scalar curvature.\\

\indent
Hence we can obtained the next group of vacuum Einstein equations:

\begin{center}
 
${R_{a}}^{b}$
 = \textit{-
$\frac {1}{2\,\sqrt{\eta }}$
} [(
${\frac {\partial }{\partial t}}\,\sqrt{\eta }\,{\kappa _{a}}^{b}
$
)\textit{ -} 
${P_{a}}^{b}$
] = 0   
\end{center}

\emptyline
\begin{mapleinput}
\mapleinline{active}{1d}{evalm(\\
 -map(diff,evalm(sqrt(det(eta))*kappa2),t)/\\
 (2*sqrt(det(eta)))):
 \indent evalm( map(simplify,\%) - P );}{%
}
\end{mapleinput}

\mapleresult
\begin{maplelatex}
\maplemultiline{
 \left[ {\vrule height1.38em width0em depth1.38em} \right. \! 
 \!  - {\displaystyle \frac {({\frac {\partial ^{2}}{\partial t^{
2}}}\,\mathrm{a}(t))\,\mathrm{b}(t)\,\mathrm{c}(t) + ({\frac {
\partial }{\partial t}}\,\mathrm{b}(t))\,({\frac {\partial }{
\partial t}}\,\mathrm{a}(t))\,\mathrm{c}(t) + ({\frac {\partial 
}{\partial t}}\,\mathrm{c}(t))\,({\frac {\partial }{\partial t}}
\,\mathrm{a}(t))\,\mathrm{b}(t)}{\mathrm{a}(t)\,\mathrm{b}(t)\,
\mathrm{c}(t)}}  \\
\mbox{} + {\displaystyle \frac {{\displaystyle \frac {1}{2}} \,(
 - \mathrm{a}(t)^{4} + \mathrm{b}(t)^{4} - 2\,\mathrm{b}(t)^{2}\,
\mathrm{c}(t)^{2} + \mathrm{c}(t)^{4})}{\mathrm{a}(t)^{2}\,
\mathrm{b}(t)^{2}\,\mathrm{c}(t)^{2}}} , \,0\,, \,0 \! \! \left. 
{\vrule height1.38em width0em depth1.38em} \right]  \\
 \left[ {\vrule height1.38em width0em depth1.38em} \right. \! 
 \! 0\,,  - {\displaystyle \frac {({\frac {\partial ^{2}}{
\partial t^{2}}}\,\mathrm{b}(t))\,\mathrm{a}(t)\,\mathrm{c}(t) + 
({\frac {\partial }{\partial t}}\,\mathrm{b}(t))\,({\frac {
\partial }{\partial t}}\,\mathrm{a}(t))\,\mathrm{c}(t) + (
{\frac {\partial }{\partial t}}\,\mathrm{c}(t))\,({\frac {
\partial }{\partial t}}\,\mathrm{b}(t))\,\mathrm{a}(t)}{\mathrm{a
}(t)\,\mathrm{b}(t)\,\mathrm{c}(t)}}  \\
\mbox{} + {\displaystyle \frac {{\displaystyle \frac {1}{2}} \,(
 - \mathrm{b}(t)^{4} + \mathrm{a}(t)^{4} - 2\,\mathrm{a}(t)^{2}\,
\mathrm{c}(t)^{2} + \mathrm{c}(t)^{4})}{\mathrm{a}(t)^{2}\,
\mathrm{b}(t)^{2}\,\mathrm{c}(t)^{2}}} , \,0 \! \! \left. 
{\vrule height1.38em width0em depth1.38em} \right]  \\
 \left[ {\vrule height1.38em width0em depth1.38em} \right. \! 
 \! 0\,, \,0\,,\\
\mbox{} - {\displaystyle \frac {({\frac {\partial ^{2}}{
\partial t^{2}}}\,\mathrm{c}(t))\,\mathrm{a}(t)\,\mathrm{b}(t) + 
({\frac {\partial }{\partial t}}\,\mathrm{c}(t))\,({\frac {
\partial }{\partial t}}\,\mathrm{a}(t))\,\mathrm{b}(t) + (
{\frac {\partial }{\partial t}}\,\mathrm{c}(t))\,({\frac {
\partial }{\partial t}}\,\mathrm{b}(t))\,\mathrm{a}(t)}{\mathrm{a
}(t)\,\mathrm{b}(t)\,\mathrm{c}(t)}}  \\
\mbox{} - {\displaystyle \frac {1}{2}} \,{\displaystyle \frac {
\mathrm{c}(t)^{4} - \mathrm{a}(t)^{4} + 2\,\mathrm{a}(t)^{2}\,
\mathrm{b}(t)^{2} - \mathrm{b}(t)^{4}}{\mathrm{a}(t)^{2}\,
\mathrm{b}(t)^{2}\,\mathrm{c}(t)^{2}}}  \! \! \left. {\vrule 
height1.38em width0em depth1.38em} \right]  }
\end{maplelatex}

\emptyline
\noindent
The obtained expression gives three Einstein equations:

\begin{mapleinput}
\mapleinline{active}{1d}{eq_2 :=\\
 Diff( (diff(a(t),t)*b(t)*c(t)),t )/(a(t)*b(t)*c(t))\\
  = ((b(t)^2-c(t)^2)^2 - a(t)^4 )/\\
  (2*a(t)^2*b(t)^2*c(t)^2);
 \indent eq_3 :=\\
  Diff( (diff(b(t),t)*a(t)*c(t)),t )/(a(t)*b(t)*c(t))\\
   = ((a(t)^2-c(t)^2)^2 - b(t)^4 )/\\
   (2*a(t)^2*b(t)^2*c(t)^2);
  \indent \indent eq_4 :=\\
   Diff( (diff(c(t),t)*b(t)*a(t)),t )/(a(t)*b(t)*c(t))\\
    = ((a(t)^2-b(t)^2)^2 - c(t)^4 )/\\
    (2*a(t)^2*b(t)^2*c(t)^2);}{%
}
\end{mapleinput}

\mapleresult
\begin{maplelatex}
\[
\mathit{eq\_2} := {\displaystyle \frac {{\frac {\partial }{
\partial t}}\,({\frac {\partial }{\partial t}}\,\mathrm{a}(t))\,
\mathrm{b}(t)\,\mathrm{c}(t)}{\mathrm{a}(t)\,\mathrm{b}(t)\,
\mathrm{c}(t)}} ={\displaystyle \frac {1}{2}} \,{\displaystyle 
\frac {(\mathrm{b}(t)^{2} - \mathrm{c}(t)^{2})^{2} - \mathrm{a}(t
)^{4}}{\mathrm{a}(t)^{2}\,\mathrm{b}(t)^{2}\,\mathrm{c}(t)^{2}}} 
\]
\end{maplelatex}

\begin{maplelatex}
\[
\mathit{eq\_3} := {\displaystyle \frac {{\frac {\partial }{
\partial t}}\,({\frac {\partial }{\partial t}}\,\mathrm{b}(t))\,
\mathrm{a}(t)\,\mathrm{c}(t)}{\mathrm{a}(t)\,\mathrm{b}(t)\,
\mathrm{c}(t)}} ={\displaystyle \frac {1}{2}} \,{\displaystyle 
\frac {(\mathrm{a}(t)^{2} - \mathrm{c}(t)^{2})^{2} - \mathrm{b}(t
)^{4}}{\mathrm{a}(t)^{2}\,\mathrm{b}(t)^{2}\,\mathrm{c}(t)^{2}}} 
\]
\end{maplelatex}

\begin{maplelatex}
\[
\mathit{eq\_4} := {\displaystyle \frac {{\frac {\partial }{
\partial t}}\,({\frac {\partial }{\partial t}}\,\mathrm{c}(t))\,
\mathrm{b}(t)\,\mathrm{a}(t)}{\mathrm{a}(t)\,\mathrm{b}(t)\,
\mathrm{c}(t)}} ={\displaystyle \frac {1}{2}} \,{\displaystyle 
\frac {(\mathrm{a}(t)^{2} - \mathrm{b}(t)^{2})^{2} - \mathrm{c}(t
)^{4}}{\mathrm{a}(t)^{2}\,\mathrm{b}(t)^{2}\,\mathrm{c}(t)^{2}}} 
\]
\end{maplelatex}

\emptyline
\noindent
The last group of Einstein equations results from

\emptyline
\begin{center}
${R_{a}}^{0}$
 = \textit{-
$\frac {1}{2}$
${\kappa _{b}}^{c}$
}(
${(C^{b})_{\mathit{ca}}}$
 \textit{-} 
${\delta _{a}}^{b}$
${(C^{d})_{\mathit{dc}}}$
) = 0     
\end{center}

\noindent
It is obvious, that in our case this equation is identically zero (see
expression for 
${(C^{b})_{\mathit{ca}}}$
). The substitutions of \textit{a}=
$e^{\alpha }$
, \textit{b}=
$e^{\beta }$
, \textit{c}=
$e^{\gamma }$
 (do not miss these exponents with frame vectors!),
\textit{dt}=\textit{a b c }(\textit{d
$\tau $
}) and the expressions 
${\frac {\partial }{\partial t}}\,a$
 = 
${\frac {\partial }{\partial \tau }}\,a$
${\frac {\partial }{\partial t}}\,\tau $
 = 
$\frac {{\frac {\partial }{\partial \tau }}\,a}{a\,b\,c}$
 = 
$\frac {{\frac {\partial }{\partial \tau }}\,\alpha }{b\,c}$
 and  
${\frac {\partial ^{2}}{\partial t^{2}}}\,a$
 = 
$\frac {1}{a\,b\,c}$
${\frac {\partial }{\partial \tau }}\,\frac {{\frac {\partial }{
\partial \tau }}\,\alpha }{b\,c}$
 = 
$\frac {1}{a\,b\,c}$
[
$\frac {{\frac {\partial ^{2}}{\partial \tau ^{2}}}\,\alpha }{b\,
c}$
 \textit{-} (
$\frac {({\frac {\partial }{\partial \tau }}\,\gamma ) + (
{\frac {\partial }{\partial \tau }}\,\beta )}{b\,c}$
)
${\frac {\partial }{\partial \tau }}\,\alpha $
] result in the simplification of the equations form:  

\emptyline
\begin{mapleinput}
\mapleinline{active}{1d}{eq_1 :=\\
 diff((alpha(tau)+beta(tau)+gamma(tau)), tau\$2)/2 =\\
 diff(alpha(tau), tau)*diff(beta(tau),tau) +\\
 diff(alpha(tau),tau)*diff(gamma(tau),tau) +\\
 diff(beta(tau),tau)*diff(gamma(tau),tau); 
 \indent eq_2 :=\\
  2*diff(alpha(tau),tau\$2) = (b(t)^2-c(t)^2)^2-a(t)^4;
  \indent \indent eq_3 :=\\
   2*diff(beta(tau),tau\$2) = (a(t)^2-c(t)^2)^2-b(t)^4;
   \indent \indent \indent eq_4 :=\\
    2*diff(gamma(tau),tau\$2) = (a(t)^2-b(t)^2)^2-c(t)^4;}{%
}
\end{mapleinput}

\mapleresult
\begin{maplelatex}
\maplemultiline{
\mathit{eq\_1} := {\displaystyle \frac {1}{2}} \,({\frac {
\partial ^{2}}{\partial \tau ^{2}}}\,\alpha (\tau )) + 
{\displaystyle \frac {1}{2}} \,({\frac {\partial ^{2}}{\partial 
\tau ^{2}}}\,\beta (\tau )) + {\displaystyle \frac {1}{2}} \,(
{\frac {\partial ^{2}}{\partial \tau ^{2}}}\,\gamma (\tau ))= \\
({\frac {\partial }{\partial \tau }}\,\alpha (\tau ))\,({\frac {
\partial }{\partial \tau }}\,\beta (\tau )) + ({\frac {\partial 
}{\partial \tau }}\,\alpha (\tau ))\,({\frac {\partial }{
\partial \tau }}\,\gamma (\tau )) + ({\frac {\partial }{\partial 
\tau }}\,\beta (\tau ))\,({\frac {\partial }{\partial \tau }}\,
\gamma (\tau )) }
\end{maplelatex}

\begin{maplelatex}
\[
\mathit{eq\_2} := 2\,({\frac {\partial ^{2}}{\partial \tau ^{2}}}
\,\alpha (\tau ))=(\mathrm{b}(t)^{2} - \mathrm{c}(t)^{2})^{2} - 
\mathrm{a}(t)^{4}
\]
\end{maplelatex}

\begin{maplelatex}
\[
\mathit{eq\_3} := 2\,({\frac {\partial ^{2}}{\partial \tau ^{2}}}
\,\beta (\tau ))=(\mathrm{a}(t)^{2} - \mathrm{c}(t)^{2})^{2} - 
\mathrm{b}(t)^{4}
\]
\end{maplelatex}

\begin{maplelatex}
\[
\mathit{eq\_4} := 2\,({\frac {\partial ^{2}}{\partial \tau ^{2}}}
\,\gamma (\tau ))=(\mathrm{a}(t)^{2} - \mathrm{b}(t)^{2})^{2} - 
\mathrm{c}(t)^{4}
\]
\end{maplelatex}

\emptyline
\noindent
Comparison of obtained equations with above considered Kasner's type
equations suggests that these systems are identical if the right-hand
sides of \textit{eq\_2}, \textit{eq\_3} and \textit{eq\_4} are equal
to zero. Therefore we will consider the system's evolution as dynamics
of Kasner's solution perturbation. In this case 
$\tau $
 =
$\mathrm{ln}(t)$
 + \textit{const} (\textit{dt}=\textit{a b c d
$\tau $
 }= 
$t^{({p_{1}} + {p_{2}} + {p_{3}})}$
\textit{d
$\tau $
}, as result \textit{d
$\tau $
} = 
$\frac {\mathit{dt}}{t}$
) and 
${\frac {\partial }{\partial \tau }}\,\alpha $
 = 
${p_{1}}$
, 
${\frac {\partial }{\partial \tau }}\,\beta $
 = 
${p_{2}}$
, 
${\frac {\partial }{\partial \tau }}\,\gamma $
 = 
${p_{3}}$
. We assume that the right-hand sides are close to zero, but
contribution of 
$e^{(4\,\alpha )}$
 - terms dominates. Hence:

\emptyline
\begin{mapleinput}
\mapleinline{active}{1d}{eq_5 := diff(alpha(tau),`\$`(tau,2)) =\\
 -exp(4*alpha(tau))/2;
 \indent eq_6 := diff(beta(tau),`\$`(tau,2)) =\\
  exp(4*alpha(tau))/2;
  eq_7 := diff(gamma(tau),`\$`(tau,2)) =\\
   exp(4*alpha(tau))/2;}{%
}
\end{mapleinput}

\mapleresult
\begin{maplelatex}
\[
\mathit{eq\_5} := {\frac {\partial ^{2}}{\partial \tau ^{2}}}\,
\alpha (\tau )= - {\displaystyle \frac {1}{2}} \,e^{(4\,\alpha (
\tau ))}
\]
\end{maplelatex}

\begin{maplelatex}
\[
\mathit{eq\_6} := {\frac {\partial ^{2}}{\partial \tau ^{2}}}\,
\beta (\tau )={\displaystyle \frac {1}{2}} \,e^{(4\,\alpha (\tau 
))}
\]
\end{maplelatex}

\begin{maplelatex}
\[
\mathit{eq\_7} := {\frac {\partial ^{2}}{\partial \tau ^{2}}}\,
\gamma (\tau )={\displaystyle \frac {1}{2}} \,e^{(4\,\alpha (\tau
 ))}
\]
\end{maplelatex}

\emptyline
\noindent
The first equation is analog of equation describing the
one-dimensional motion (
$\alpha $
 is the coordinate) in the presence of exponential barrier:

\emptyline
\begin{mapleinput}
\mapleinline{active}{1d}{p:=\\
 dsolve(\{eq_5,alpha(5)=-2,D(alpha)(5)=-0.5\},alpha(tau),\\
type=numeric):
}{%
}
\end{mapleinput}

\begin{mapleinput}
\mapleinline{active}{1d}{with(plots):
\indent odeplot(p,[tau,diff(alpha(tau),tau)],-5..5,\\
color=green):
 plot(exp(4*(-0.5)*tau)/2,tau=-5..5,color=red):
  \indent display(\%,\%\%,view=-1..1,axes=boxed,\\
   title=`reflection on barrier`);}{%
}
\end{mapleinput}

\mapleresult
\begin{center}
\mapleplot{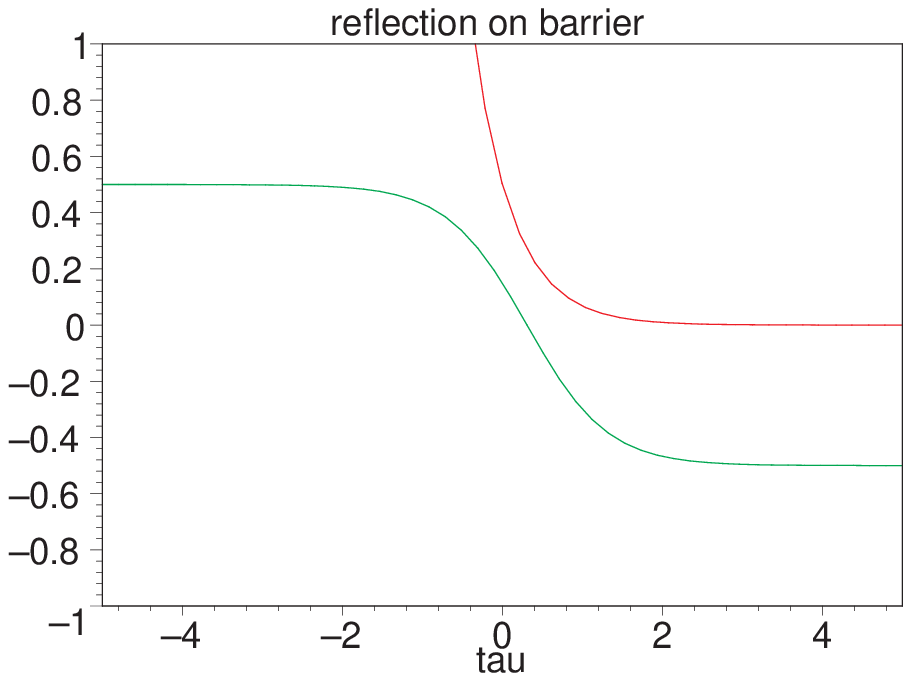}
\end{center}

\emptyline
\noindent
We can see, that the"particle" with initial velocity 
${p_{1}}$
= -0.5 (green curve) is reflected on barrier (red curve) with change
of velocity sign. But \textit{eq\_5}, \textit{eq\_6}, \textit{eq\_7}
results in 
${p_{1}} + {p_{2}}$
 = \textit{const}, 
${p_{1}} + {p_{3}}$
 = \textit{const} therefore

\begin{center}
${(p^{\mathit{new}})_{2}}$
 = 
${p_{2}}$
 + 
${p_{1}}$
\textit{ -} 
${(p^{\mathit{new}})_{1}}$
 = 
${p_{2}}$
 + 2 
${p_{1}}$
,  
${(p^{\mathit{new}})_{3}}$
 = 
${p_{3}}$
 + 
${p_{1}}$
\textit{ -} 
${(p^{\mathit{new}})_{1}}$
 = 
${p_{3}}$
 + 2 
${p_{1}}$
,
\end{center}

\begin{center}
    \textit{t} = \textit{a b c = 
$e^{((1 + 2\,{p_{1}})\,\tau )}$
} ,
\end{center}

\noindent
and the result is:

\begin{center}
\textit{a }= 
$t^{( - \frac {{p_{1}}}{1 + 2\,{p_{1}}})}$
, \textit{b} = 
$t^{(\frac {{p_{2}} + 2\,{p_{1}}}{1 + 2\,{p_{1}}})}$
, \textit{c} = 
$t^{(\frac {{p_{3}} + 2\,{p_{1}}}{1 + 2\,{p_{1}}})}$
\end{center}

\noindent
The following procedure gives a numerical algorithm for iteration of
powers of \textit{t}.

\emptyline
\begin{mapleinput}
\mapleinline{active}{1d}{iterations := proc(iter,p) 
  \indent p_1_old :=\\
   evalhf( -1/2*p+1/2-1/2*sqrt(-(3*p+1)*(p-1)) );
   \indent \indent p_2_old :=\\
    evalhf( -1/2*p+1/2+1/2*sqrt(-(3*p+1)*(p-1)) );
    \indent \indent \indent p_3_old := evalhf( p ):
     
\indent \indent \indent \indent for m from 1 to iter do
 \indent \indent \indent \indent \indent if(p_1_old<0) then
  \indent \indent \indent \indent pp := evalhf(1+2*p_1_old):
   \indent \indent \indent p_1_n := evalhf(-p_1_old/pp):
    \indent \indent p_2_n := evalhf((p_2_old+2*p_1_old)/pp):
     \indent p_3_n := evalhf((p_3_old+2*p_1_old)/pp):
  else fi:
        
 \indent if(p_2_old<0) then
  \indent \indent pp := evalhf(1+2*p_2_old):
   \indent \indent \indent p_1_n := evalhf((p_1_old+2*p_2_old)/pp): 
    \indent \indent p_2_n := evalhf(-p_2_old/pp):
     \indent p_3_n := evalhf((p_3_old+2*p_2_old)/pp):
 else fi:

 \indent if(p_3_old<0) then
  \indent \indent pp := evalhf(1+2*p_3_old):
   \indent \indent \indent p_1_n := evalhf((p_1_old+2*p_3_old)/pp): 
    \indent \indent p_2_n := evalhf((p_2_old+2*p_3_old)/pp):
     \indent p_3_n := evalhf(-p_3_old/pp):
 else fi:

 \indent p_1_old := p_1_n:
  \indent \indent p_2_old := p_2_n:
   \indent p_3_old := p_3_n:
    if m = iter then pts :=\\
     [p_1_old, p_2_old,p_3_old] fi;
     \indent od:
      \indent \indent pts
       \indent \indent \indent end:}{%
}
\end{mapleinput}

\begin{mapleinput}
\mapleinline{active}{1d}{with(plots):
 \indent pointplot3d(\\
 \{seq(iterations(i,0.6), i=1 ..8)\},symbol=diamond,\\
 axes=BOXED,labels=[p_1,p_2,p_3]);
  \indent \indent pointplot3d(\\
  \{seq(iterations(i,0.6), i=7 ..12)\},symbol=diamond,\\
  axes=BOXED,labels=[p_1,p_2,p_3]);}{%
}
\end{mapleinput}

\mapleresult
\begin{center}
\mapleplot{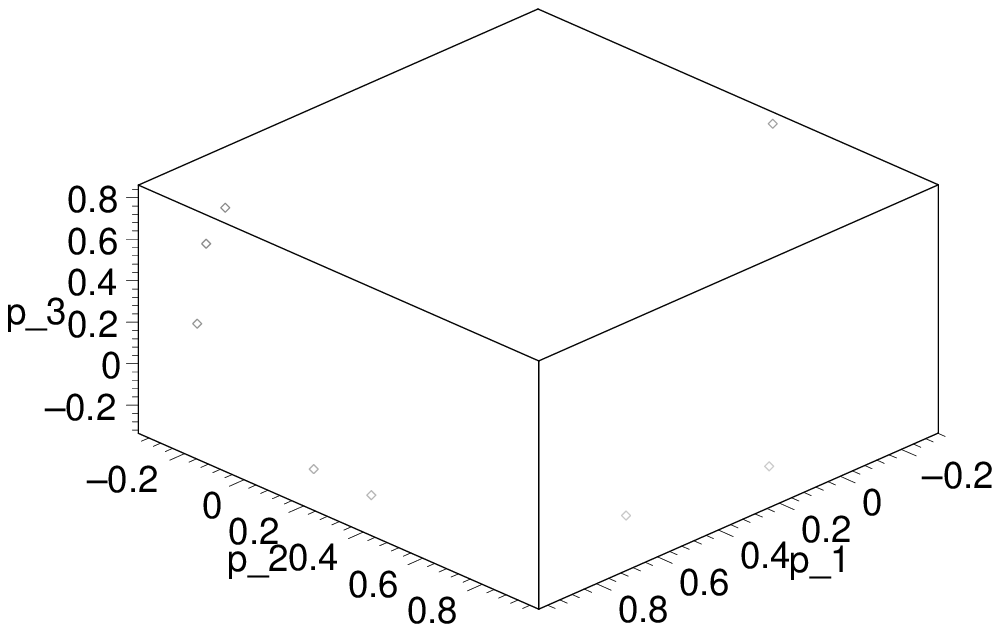}
\end{center}

\begin{center}
\mapleplot{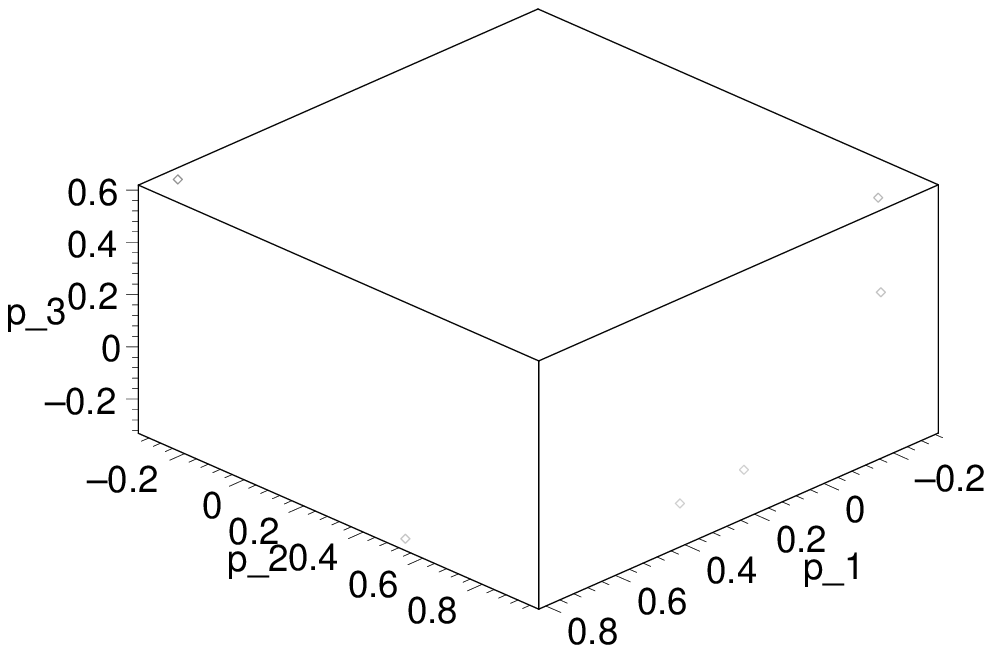}
\end{center}

\emptyline
\noindent
One can see, that the evolution has character of switching between
different Kasner's epochs (all points lie on the section of sphere 
$\mathit{p\_1}^{2}$
 + 
$\mathit{p\_2}^{2}$
 + 
$\mathit{p\_3}^{2}$
 =1 by plane \textit{p\_1} + \textit{p\_2} + \textit{p\_3} =1). The
negative power of \textit{t} increases the corresponding scaling
factor if \textit{t--}\TEXTsymbol{>}0. In the upper figure the
periodical change of signs takes place between \textit{y}- and
\textit{z}- directions. The universe oscillates in these directions
and monotone contracts in \textit{x-}direction
(\textit{t--}\TEXTsymbol{>}0). Next the evolution changes (lower
figure): \textit{x-}z oscillations are accompanied by the contraction
in \textit{y}-direction. The whole picture is the switching between
different oscillations as the result of logarithmical approach to
singularity. It is important, that there is the strong dependence on
initial conditions that can cause the chaotical scenario of
oscillations:    

\emptyline
\begin{mapleinput}
\mapleinline{active}{1d}{pointplot3d(\\
\{seq(iterations(i,0.601),i=1..100)\},symbol=diamond,\\
axes=BOXED,labels=[p_1,p_2,p_3],\\
title=`Kasner's epochs (I)`);
 \indent pointplot3d(\\
\{seq(iterations(i,0.605),i=1..100)\},symbol=diamond,\\
axes=BOXED,labels=[p_1,p_2,p_3],\\
title=`Kasner's epochs (II)`);}{%
}
\end{mapleinput}

\mapleresult
\begin{center}
\mapleplot{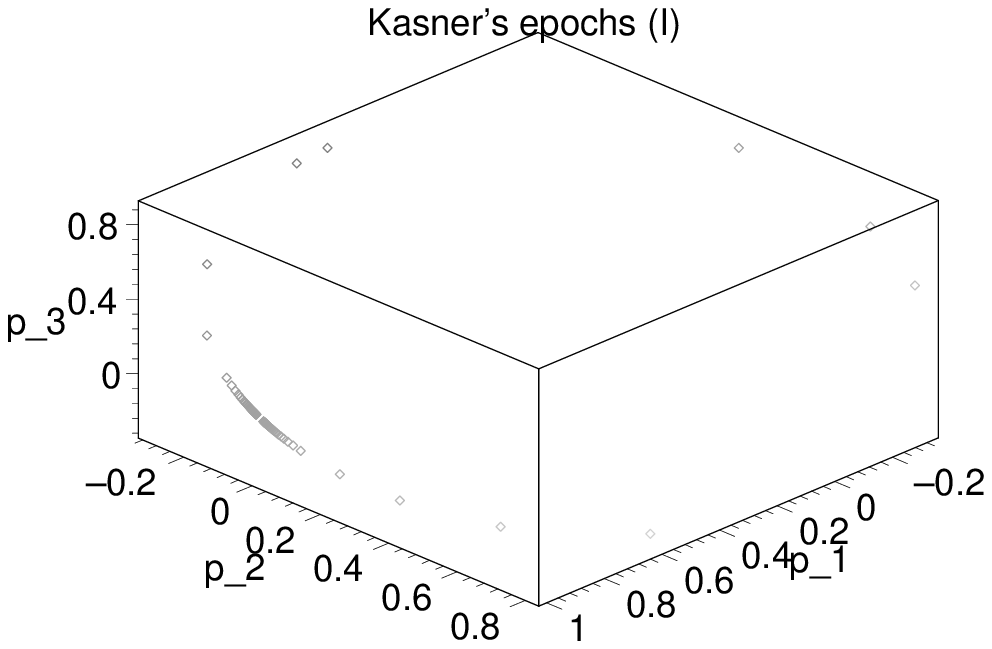}
\end{center}

\begin{center}
\mapleplot{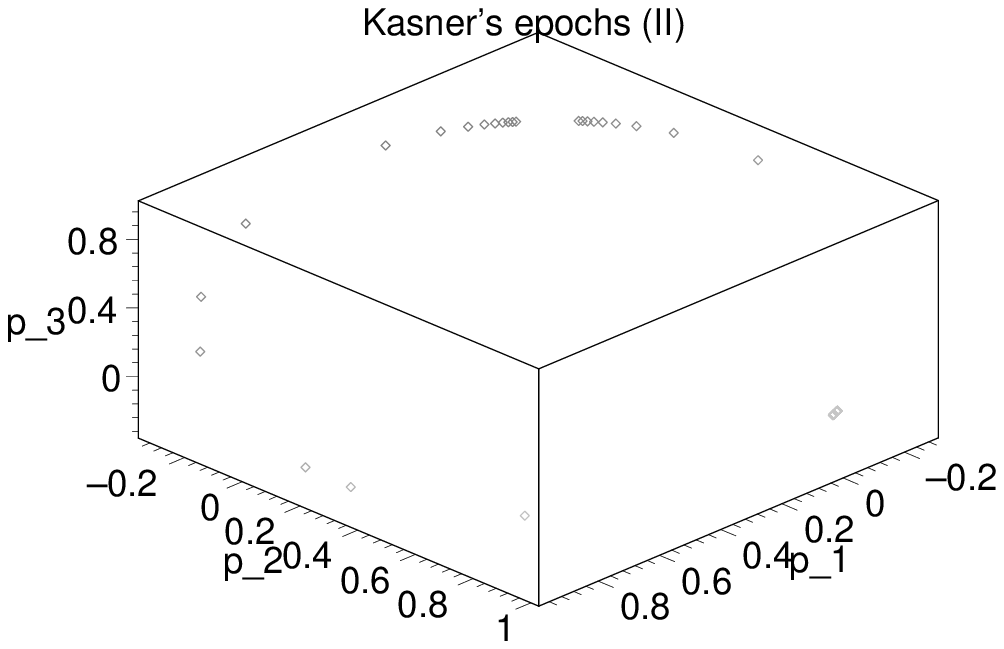}
\end{center}

\emptyline
\noindent
So, we can conclude that the deflection from isotropy changes the
scenario of the universe's evolution in the vicinity of singularity,
so that the dynamics of the geometry looks like chaotical behavior of
the deterministic nonlinear systems (so-called \underline{Mixmaster
universe}, see \cite{J. Wainwright}). 

\emptyline

\subsubsection{Inflation}

\emptyline
The above-considered standard homogeneous models face the challenge of
so-called \underline{horizon}. The homogeneity of the universe results from
the causal connection between its different regions. But the distance
between these regions is defined by the time of expansion. Let's find
the maximal distance of the light signal propagation in the expanding
region. As 
$\mathit{ds}^{2}$
 = 0 the accompanying radial coordinate of horizon is  

\emptyline
\begin{mapleinput}
\mapleinline{active}{1d}{r = Int(1/a(t),t=0..t0);# t0 is the age\\
 of the universe
 #or
  \indent r = Int(1/y(tau),tau=-t0*H0..0)/(a0*H0);}{%
}
\end{mapleinput}

\mapleresult
\begin{maplelatex}
\[
r={\displaystyle \int _{0}^{\mathit{t0}}} {\displaystyle \frac {1
}{\mathrm{a}(t)}} \,dt
\]
\end{maplelatex}

\begin{maplelatex}
\[
r={\displaystyle \frac {{\displaystyle \int _{ - \mathit{t0}\,
\mathit{H0}}^{0}} {\displaystyle \frac {1}{\mathrm{y}(\tau )}} \,
d\tau }{\mathit{a0}\,\mathit{H0}}} 
\]
\end{maplelatex}

\emptyline
\noindent
For instance, in the Einstein-de Sitter model we have

\emptyline
\begin{mapleinput}
\mapleinline{active}{1d}{int(\\
 subs(y(tau)=1/4*(12*tau+8)^(2/3),1/y(tau)),\\
 tau=-2/3..0)/(a0*H0):# we use the above obtained\\
  expressions for this model
 \indent simplify( subs(a0=1/4*(12*0+8)^(2/3),\%),radical );}{%
}
\end{mapleinput}

\mapleresult
\begin{maplelatex}
\[
2\,{\displaystyle \frac {1}{\mathit{H0}}} 
\]
\end{maplelatex}

\emptyline
\noindent
and in the Friedmann-Lemaitre:

\emptyline
\begin{mapleinput}
\mapleinline{active}{1d}{int(\\
subs(y(phi)=Omega[M]*(1-cos(2*phi))/\\
(2*b),1/2*Omega[M]*(2-2*cos(2*phi))/\\
(b^(3/2))/y(phi)),phi )/a0/H0;# we use \\
the above obtained expressions for this model
 \indent simplify( subs(K=-1,subs(b=-K/a0^2/H0^2,\%)),\\
 radical,symbolic);#b=Omega[M]-1=Omega[K] and K=-1\\
  (spherical universe)}{%
}
\end{mapleinput}

\mapleresult
\begin{maplelatex}
\[
2\,{\displaystyle \frac {\phi }{\sqrt{b}\,\mathit{a0}\,\mathit{H0
}}} 
\]
\end{maplelatex}

\begin{maplelatex}
\[
2\,\phi 
\]
\end{maplelatex}

\emptyline
\noindent
The last example is of interest. The increase of 
$\phi $
 from 0 to 
$\pi $
/2 corresponds to expansion of universe up to its maximal radius (see
above). At the moment of maximal expansion the observer can see the
photons from antipole of universe that corresponds to formation of
full causal connection:

\emptyline
\begin{mapleinput}
\mapleinline{active}{1d}{plot([2*phi-sin(2*phi),2*phi],\\
phi=0..Pi, title=`age of universe and time of\\
 photon travel (a.u.)`);}{%
}
\end{mapleinput}

\mapleresult
\begin{center}
\mapleplot{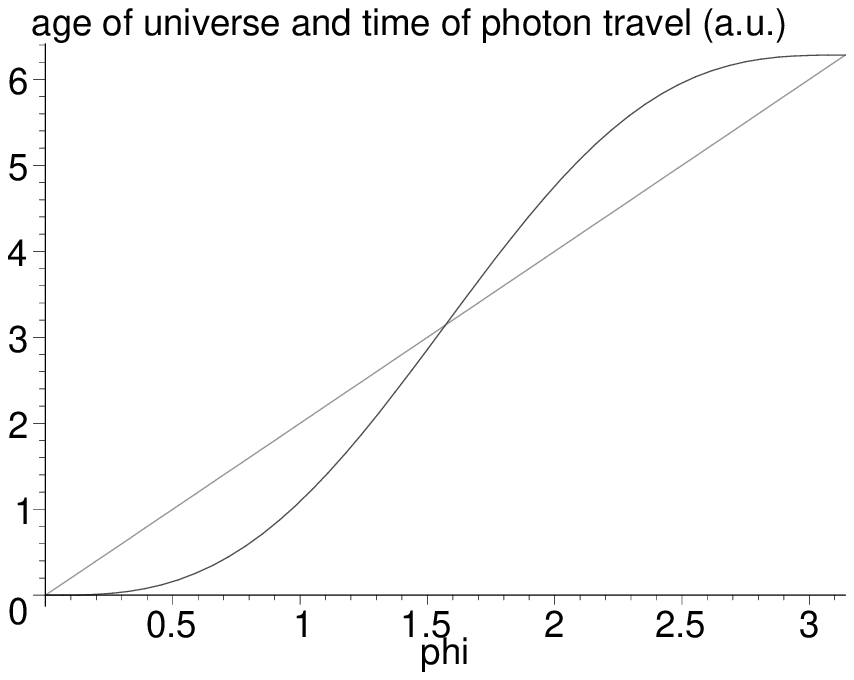}
\end{center}

\emptyline
\noindent
This figure shows the age of Friedmann-Lemaitre universe in comparison
with the time of photon propagation from outermost point of universe.
For 
$\phi $
\TEXTsymbol{<}
$\pi $
/2 there are the regions, which don't connect with us. After this
points there is the causal connection. At the moment of recollapse the
photons complete revolutions around universe.\\

\noindent
The different approaches were developed in order to overcome the
problem of horizon (see above considered loitering and Mixmaster
models). But the most popular models are based on the so-called
\underline{inflation scenario}, which can explain also the global
flatness of universe (see, for example, \cite{A.D. Linde}, on-line reviews 
\cite{A. Linde,R.H. Brandenberger}).\\

\noindent
For escape the horizon problem we have to suppose the accelerated
expansion with 
${\frac {\partial ^{2}}{\partial t^{2}}}\,\mathrm{a}(t)$
  \TEXTsymbol{>} 0. As result of this condition, the originally
causally connected regions will expand faster, than the horizon
expands. So, we observe the light from the regions, which was
connected at the early stage of universe's evolution. In the
subsection "Standard models" we derived from Einstein equations the
energy conservation low \textit{basic\_2}:

\begin{center}
$\frac {{\frac {\partial ^{2}}{\partial t^{2}}}\,\mathrm{a}(t)}{
\mathrm{a}(t)}= - 4\,\pi \,(\mathrm{p}(t) + \frac {\rho (t)}{3})
 + \frac {\Lambda }{3}$ 
\end{center}

\noindent
This equation demonstrates that in the usual conditions (\textit{p}
\TEXTsymbol{>} 0) the pressure is a source of gravitation, but if
\textit{p} \TEXTsymbol{<} \textit{-
$\frac {\rho }{3}$
} (
$\Lambda $
 = 0) or 
$\Lambda $
 \TEXTsymbol{>} 4
$\pi $
$\rho $
 (\textit{p}=0) the repulsion dominates and the expansion is
accelerated. As the source of the repulsion, the scalar "inflaton" field 
$\phi $
 is considered.\\

\noindent
Let us introduce the simple Lagrangian for this homogeneous field:
\textit{}

\begin{center}
  \textit{L}=-
$\frac {1}{2}$
${\frac {\partial }{\partial {x_{i}}}}\,\phi $
${\frac {\partial }{\partial {x_{j}}}}\,\phi $
$g^{\mathit{ij}}$
 \textit{-} \textit{V}(
$\phi $
),
\end{center}

\noindent
where \textit{V} is the potential energy. The procedure,
which was considered in first section, allows to derive from this
Lagrangian the field equation:

\begin{center}
$\frac {1}{\sqrt{ - g}}$
${\frac {\partial }{\partial {x_{i}}}}\,\sqrt{ - g}\,g^{\mathit{
ij}}\,({\frac {\partial }{\partial {x_{j}}}}\,\phi )$
 + 
${\frac {\partial }{\partial \phi }}\,V$
 = 0.
 \end{center}   

\noindent
In the case of flat RW metric this equation results in:

\emptyline
\begin{mapleinput}
\mapleinline{active}{1d}{coord := [t, r, theta, phi]:
 \indent g_compts :=\\
  array(symmetric,sparse,1..4,1..4):
  \indent \indent g_compts[1,1] := -1: 
   \indent \indent \indent g_compts[2,2] := a(t)^2:
    \indent \indent \indent \indent g_compts[3,3]:= a(t)^2*r^2: 
   \indent \indent \indent g_compts[4,4]:= a(t)^2*r^2*sin(theta)^2:
  \indent \indent g := create([-1,-1], eval(g_compts)):#definition\\
   of flat RW metric
\indent d1g:= d1metric(g, coord):
 g_inverse:= invert( g, 'detg'):
  \indent Cf1:= Christoffel1 ( d1g ):
   \indent \indent Cf2:= Christoffel2 ( g_inverse, Cf1 ):
    \indent \indent \indent det_scal := simplify(\\
     sqrt( -det( get_compts(g) )),radical,symbolic ):
    \indent \indent g_det_sq :=\\
     create([], det_scal):#sqrt(-g)
     \indent field :=\\
      create([], phi(t)):#field 
      #calculation of first term in the field equation
\indent cov_diff( field, coord, Cf2 ):
 \indent \indent T := prod(g_inverse,\%):
  \indent \indent \indent contract(T,[2,3]):
   \indent \indent \indent \indent Tt := prod(g_det_sq,\%):
    \indent \indent \indent Ttt := cov_diff( Tt, coord, Cf2 ):
     \indent \indent get_compts(\%)[1,1]/det_scal:
       \indent expand(-\%):
     field_eq := \% + diff(V(phi),phi) = 0;
     }{%
}
\end{mapleinput}

\mapleresult
\begin{maplelatex}
\[
\mathit{field\_eq} := 3\,{\displaystyle \frac {({\frac {\partial 
}{\partial t}}\,\phi (t))\,({\frac {\partial }{\partial t}}\,
\mathrm{a}(t))}{\mathrm{a}(t)}}  + ({\frac {\partial ^{2}}{
\partial t^{2}}}\,\phi (t)) + ({\frac {\partial }{\partial \phi 
}}\,\mathrm{V}(\phi ))=0
\]
\end{maplelatex}

\emptyline
\noindent
Now let's define the energy-momentum tensor for first Einstein
equation:

\begin{center}
${T_{\mathit{ij}}}$
 = 
${g_{\mathit{ij}}}$
$L$
 \textit{- 
${g_{\mathit{jl}}}$
${f_{\mathit{x\_i}}}$
${\frac {\partial }{\partial {f_{\mathit{x\_l}}}}}\,L$
,}
\end{center}

\noindent
where 
${f_{\mathit{x\_i}}}$
 =
${\frac {\partial }{\partial {x_{i}}}}\,\phi $
. As consequence, 
$\rho $
 = [-
$\frac {1}{2}$
$({\frac {\partial }{\partial t}}\,\phi )^{2}$
 + \textit{V}(
$\phi $
)] + 
$({\frac {\partial }{\partial t}}\,\phi )^{2}$
 = 
$\frac {1}{2}$
$({\frac {\partial }{\partial t}}\,\phi )^{2}$
+ \textit{V}(
$\phi $
)\textit{, }and\textit{ }(for  \textit{i},  \textit{j }=
1,2,3) 
${T_{\mathit{ij}}}$
 = 
${g_{\mathit{ij}}}$
(
$\frac {1}{2}$
$({\frac {\partial }{\partial t}}\,\phi )^{2}$
 - \textit{V}(
$\phi $
)) $\Rightarrow$ \textit{p} = 
$\frac {1}{2}$
$({\frac {\partial }{\partial t}}\,\phi )^{2}$
 - \textit{V}(
$\phi $
). Last expression emplies that the necessary condition
\textit{p} \TEXTsymbol{<} \textit{-
$\frac {\rho }{3}$
} may result from large \textit{V}.

\noindent
The dynamics of our system is guided by coupled system of equations:

\emptyline
\begin{mapleinput}
\mapleinline{active}{1d}{basic_inf_1 :=\\
 K/(a(t)^2)+diff(a(t),t)^2/(a(t)^2) =\\
8/3*Pi*(1/2*diff(phi(t),t)^2+V(phi));#see above E[1]
 \indent basic_inf_2 := field_eq;}{%
}
\end{mapleinput}

\mapleresult
\begin{maplelatex}
\[
\mathit{basic\_inf\_1} := {\displaystyle \frac {K}{\mathrm{a}(t)
^{2}}}  + {\displaystyle \frac {({\frac {\partial }{\partial t}}
\,\mathrm{a}(t))^{2}}{\mathrm{a}(t)^{2}}} ={\displaystyle \frac {
8}{3}} \,\pi \,({\displaystyle \frac {1}{2}} \,({\frac {\partial 
}{\partial t}}\,\phi (t))^{2} + \mathrm{V}(\phi ))
\]
\end{maplelatex}

\begin{maplelatex}
\[
\mathit{basic\_inf\_2} := 3\,{\displaystyle \frac {({\frac {
\partial }{\partial t}}\,\phi (t))\,({\frac {\partial }{\partial 
t}}\,\mathrm{a}(t))}{\mathrm{a}(t)}}  + ({\frac {\partial ^{2}}{
\partial t^{2}}}\,\phi (t)) + ({\frac {\partial }{\partial \phi 
}}\,\mathrm{V}(\phi ))=0
\]
\end{maplelatex}

\emptyline
\noindent
The meaning of \textit{K} is considered earlier, here we will suppose
\textit{K}=0 (local flatness). In the slow-rolling approximation 
${\frac {\partial ^{2}}{\partial t^{2}}}\,\phi (t)$
, 
$({\frac {\partial }{\partial t}}\,\phi (t))^{2}$
 \textit{--}\TEXTsymbol{>}0 and for potential \textit{V}(
$\phi $
) =
$\frac {m\,\phi ^{2}}{2}$
 we have:

\emptyline
\begin{mapleinput}
\mapleinline{active}{1d}{basic_inf_1_n :=\\
 subs(K=0,K/(a(t)^2)+diff(a(t),t)^2/(a(t)^2)) =\\
8/3*Pi*m^2*phi(t)^2/2;#see above E[1]
 \indent basic_inf_2_n :=  op(1,lhs(field_eq)) +\\
subs(phi=phi(t),expand(subs(V(phi)=m^2*phi^2/2,\\
op(3,lhs(field_eq)))))=0;}{%
}
\end{mapleinput}

\mapleresult
\begin{maplelatex}
\[
\mathit{basic\_inf\_1\_n} := {\displaystyle \frac {({\frac {
\partial }{\partial t}}\,\mathrm{a}(t))^{2}}{\mathrm{a}(t)^{2}}} 
={\displaystyle \frac {4}{3}} \,\pi \,m^{2}\,\phi (t)^{2}
\]
\end{maplelatex}

\begin{maplelatex}
\[
\mathit{basic\_inf\_2\_n} := 3\,{\displaystyle \frac {({\frac {
\partial }{\partial t}}\,\phi (t))\,({\frac {\partial }{\partial 
t}}\,\mathrm{a}(t))}{\mathrm{a}(t)}}  + m^{2}\,\phi (t)=0
\]
\end{maplelatex}

\begin{mapleinput}
\mapleinline{active}{1d}{dsolve(\{basic_inf_1_n,basic_inf_2_n,\\
phi(0)=phi0,a(0)=a0\},\{phi(t),a(t)\}):
 \indent expand(\%);}{%
}
\end{mapleinput}

\mapleresult
\begin{maplelatex}
\[
 \left\{  \! \mathrm{a}(t)={\displaystyle \frac {e^{( - 1/6\,m^{2
}\,t^{2})}\,\mathit{a0}}{e^{-(2/3\,\phi 0\,\sqrt{\pi }\,m\,\sqrt{3
}\,t)}}} , \,\phi (t)={\displaystyle -\frac {1}{6}} \,
{\displaystyle \frac {\sqrt{3}\,m\,t}{\sqrt{\pi }}}  + \phi 0 \! 
 \right\} 
\]
\end{maplelatex}

\emptyline
\noindent
One can see, that for 
${\phi _{0}}$
\TEXTsymbol{>}\TEXTsymbol{>}1 there is an quasi-exponential expansion
(inflation) of universe:

\begin{center}
   \textit{a}(\textit{t})=
${a_{0}}$
$e^{(H\,t)}$
, where \textit{H} = 2
$\sqrt{\frac {\pi }{3}}$
${\phi _{0}}$
 \textit{m}
\end{center}

\noindent
When \textit{t} \symbol{126} 
$\frac {2\,\sqrt{3\,\pi }}{m}$
${\phi _{0}}$
 the regime of slow rolling ends and the potential energy of inflaton
field converts into kinetic form that causes the avalanche-like
creation of other particles from vacuum (so-called
"\underline{reheating}"). Only at this moment the scenario become
similar to Big Bang picture. What is the size of universe after
inflation?  

\emptyline
\begin{mapleinput}
\mapleinline{active}{1d}{a(t)=subs(\\
\{H=2*sqrt(Pi/3)*phi[0]*m,t=2*sqrt(3*Pi)/m*phi[0]\},\\
a[0]*exp(H*t));}{%
}
\end{mapleinput}

\mapleresult
\begin{maplelatex}
\[
\mathrm{a}(t)={a_{0}}\,e^{(4\,\pi \,{\phi _{0}}^{2})}
\]
\end{maplelatex}

\emptyline
\noindent
If at the beginning of inflation the energy and size of universe are
defined by Plank density and length, then 
$m^{2}$
${(\phi ^{2})_{0}}$
 \symbol{126}1, 
${a_{0}}$
$\mathit{\symbol{126}10}^{( - 33)}$
 \textit{cm}. The necessary value of \textit{m 
$\mathit{\symbol{126}10}^{( - 6)}$
} is constrained from the value of fluctuation of background
radiation (
$\mathit{\symbol{126}10}^{( - 5)}$
). As result \textit{a \symbol{126} 
$10^{( - 33)}$
$e^{(\frac {4\,\pi }{m^{2}})}$
 =
$10^{( - 33)}$
$e^{(4\,\pi \,10^{12})}$
 cm }, that is much larger of the observable universe. This is the
reason why observable part of universe is flat, homogeneous and
isotropic.\\

\noindent
Let's return to \textit{basic\_inf\_2}. The reheating results from
oscillation of inflaton field around potential minimum that creates
new boson particles. The last damps the oscillations that can be
taking into consideration by introducing of dumping term in
\textit{basic\_inf\_2}:

\emptyline
\begin{mapleinput}
\mapleinline{active}{1d}{reh_inf :=\\
3*diff(phi(t),t)*H+Gamma*diff(phi(t),t) +\\
 diff(phi(t),`\$`(t,2)) = -m^2*phi(t);}{%
}
\end{mapleinput}

\mapleresult
\begin{maplelatex}
\[
\mathit{reh\_inf} := 3\,({\frac {\partial }{\partial t}}\,\phi (t
))\,H + \Gamma \,({\frac {\partial }{\partial t}}\,\phi (t)) + (
{\frac {\partial ^{2}}{\partial t^{2}}}\,\phi (t))= - m^{2}\,\phi
 (t)
\]
\end{maplelatex}

\emptyline
\noindent
Here \textit{H} is the Hubble constant, 
$\Gamma $
 is the decay rate, which is defined by coupling with material boson
field. When \textit{H }\TEXTsymbol{>} 
$\Gamma $
, there are the coherent oscillations:

\emptyline
\begin{mapleinput}
\mapleinline{active}{1d}{dsolve(subs(Gamma=0,reh_inf),phi(t));
 \indent diff(\%,t);}{%
}
\end{mapleinput}

\emptyline
\mapleresult
\begin{maplelatex}
\maplemultiline{
\phi (t)=\mathit{\_C1}\,e^{( - 1/2\,(3\,H + \sqrt{(3\,H - 2\,m)\,
(3\,H + 2\,m)})\,t)} +\\
 \mathit{\_C2}\,e^{( - 1/2\,(3\,H - \sqrt{(
3\,H - 2\,m)\,(3\,H + 2\,m)})\,t)}
}
\end{maplelatex}

\begin{maplelatex}
\maplemultiline{
{\frac {\partial }{\partial t}}\,\phi (t)=\mathit{\_C1}\,( - 
{\displaystyle \frac {3}{2}} \,H - {\displaystyle \frac {1}{2}} 
\,\mathrm{\%1})\,e^{( - 1/2\,(3\,H + \mathrm{\%1})\,t)} +\\ 
\mathit{\_C2}\,( - {\displaystyle \frac {3}{2}} \,H + 
{\displaystyle \frac {1}{2}} \,\mathrm{\%1})\,e^{( - 1/2\,(3\,H
 - \mathrm{\%1})\,t)} \\
\mathrm{\%1} := \sqrt{(3\,H - 2\,m)\,(3\,H + 2\,m)} }
\end{maplelatex}

\emptyline
\noindent
The last expression demonstrates the decrease of energy density via
time increase (
$\rho $
\symbol{126}
$({\frac {\partial }{\partial t}}\,\phi )^{2}$
) and, as result, the decrease of \textit{H} (see
\textit{basic\_inf\_1}). This leads to \textit{H} \TEXTsymbol{<} 
$\Gamma $
, that denotes the reheating beginning.

\emptyline
\noindent
In the conclusion of this short and elementary description it should
be noted, that the modern inflation scenarios modify the standard
model essentially. The Universe here looks like fractal tree of
bubbles-universes, inflating and recollapsing foams, topological
defects etc. The main features of this picture wait for thorough
investigations. 

\emptyline

\section{Conclusion}

\emptyline
\noindent
Our consideration was brief and I omitted some important topics like
perturbation theory for black holes and cosmological models, Penrose
diagrams and conformal mappings for investigation of causal structure,
gravitational waves etc. But I suppose to advance this worksheet
hereafter.

\end{document}